\documentclass{aa}
\RequirePackage{amsmath}
\RequirePackage{amssymb}
\RequirePackage{epsfig}		
\usepackage{graphicx,epsfig,fancyhdr,psfig,rotating,amsmath,epsf,txfonts,natbib,color}
\usepackage [english]{babel}
\usepackage [autostyle, english = american]{csquotes}

\def\etal{{et al.}}

\def\3o{O~{\sc iii}}
\def\4o{O~{\sc iv}}

\def\arcsec{$^{\prime\prime}$}
\def\dem{{\rm DEM}}
\def\tdecay{\tau_{\rm obs}}
\newcommand{\comment}[1]{}

\begin{document}
\title{Energetics of Hi-C EUV Brightenings}
\author{Srividya Subramanian$^{1}$\thanks{Previously at Inter-University Centre for Astronomy and Astrophysics, Post Bag-4, Ganeshkhind, Pune 411007, India}, Vinay L.\ Kashyap$^{2}$, Durgesh Tripathi$^{3}$, Maria S.\ Madjarska$^{4}$ and J.\ G.\ Doyle$^{1}$}

\offprints{durgesh@iucaa.in}

\institute{$^{1}$ Armagh Observatory and Planetarium, College Hill, Armagh BT61 9DG, United Kingdom \\
$^{2}$ Harvard-Smithsonian Center for Astrophysics, 60 Garden St, Cambridge, MA 02138, United States\\
$^{3}$ Inter-University Centre for Astronomy and Astrophysics, Post Bag-4, Ganeshkhind, Pune 411007, India \\
$^{4}$ Max Planck Institute for Solar System Research, Justus-von-Liebig-Weg 3, 37077, G\"ottingen, Germany\\
}
\date{Received date, accepted date}
\abstract{ We study the thermal structure and energetics of the point-like EUV brightenings within a system of fan loops observed in the active region \textsl{AR~11520}. These brightenings were simultaneously observed on 2012 July 11 by the HIgh-resolution Coronal (Hi-C) imager and the Atmospheric Imaging Assembly (AIA) on board the Solar Dynamics Observatory (SDO). We identified 27 brightenings by automatically determining intensity enhancements in both Hi-C and AIA ~193~{\AA}  light curves. The energetics of these brightenings were studied by using the Differential Emission Measure (DEM) diagnostics. The DEM weighted temperatures of these transients are in the range $\log{T} (K) = 6.2 - 6.6$ with radiative energies ${\approx}10^{24-25}$~ergs and densities ${\approx}$ a few times $10^{9}$~cm$^{-3}$. To the best of our knowledge, these are the smallest brightenings in EUV ever detected. We used these results to determine the mechanism of energy loss in these brightenings. Our analysis reveals that the dominant mechanism of energy loss for all the identified brightenings is conduction rather than radiation.}

\authorrunning{Subramanian \etal}
\titlerunning{Hi-C EUV Heating}
\maketitle

\section{Introduction} \label{s:intro}

One of the most important unsolved problems in solar and heliospheric physics is that of determining the mechanism by which the corona is heated. It is clear now that coronal heating and magnetic fields are correlated, but the actual mechanism of how magnetic energy is transferred to coronal thermal energy is not well understood \citep[see ][ for a review]{jim06}. \cite{parker88} suggested a heating mechanism wherein the energy that is built-up due to highly turbulent convective motions in the photosphere, in the form of twisting and tangling of magnetic field lines, is transferred to the upper layers and released through magnetic reconnection processes. This process is inherently impulsive in nature and is often referred to as the \textsl{nanoflare model} of coronal heating. It was envisioned by \cite{parker88} that these impulsive events may have an energy content of ${\approx}10^{24}$~ergs, $\approx$9 orders of magnitude lower than standard solar flares. However, there have been no direct observation of individual nanoflares, possibly because they occur at scales unresolvable by the currently available instruments. Thus, their existence has always been questioned.

Numerous small scale energetic events with length-scales ranging from a few arcsec to tens of arcsecs and lifetimes ranging from a few minutes to hours, like explosive events \citep[e.g.,][and the references therein]{dere89,GupT_2015,zhengua17}, EUV blinkers \citep[e.g.,][and the references therein]{harrison97, ss12}, spicules \citep[e.g.,][and the references therein]{roberts45, georgiaspicule12}, macrospicules \citep[e.g.,][and the references therein]{bohlin75, moore77, KaySMT:13} and X-ray \& EUV jets \citep[e.g.,][and the references therein]{shibata92,ChiYIT_2008,ss10,ChaGMT:15,MulTDM_2016}, have been observed in the solar atmosphere. These events are omnipresent in the solar atmosphere. However, their contribution to coronal heating is still inconclusive.

Spatial resolution plays a crucial role in the interpretation of observed coronal plasma. So far, the space-borne extreme ultra-violet (EUV) and X-ray observations have not achieved the resolutions, which would allow us to observe the individual strands that are presumed to make up coronal loops. The High-resolution Coronal \citep[Hi-C; ][]{cirtainhic13} rocket flight has recorded the best resolution images of the solar corona.
These observations have provided us with a spectacular trove of data of a group of active regions (ARs). These observations have unraveled interesting transient features in the moss \citep{testahic13} and inter-moss regions \citep{winebargerhic13}, as well as at the footpoints of a fan loop system associated with \textsl{AR~11520} \citep[EUV bright dots;][]{regnierhic14}. These brightenings were classified as nanoflare-like brightenings by the respective authors. 

The study presented in this paper is focussed on the transients EUV brightenings seen in fan loop systems associated with \textsl{AR~11520}. These brightenings appear as tiny dot-like intensity enhancements. A sample of 8 such events was studied by \citet{regnierhic14}. These events were characterised in four different categories based on their light curve characteristics, as single intensity peak events, double intensity peak events, long duration events, and bursty events with multiple intensity peaks. They have length scales of ${\sim}1$\arcsec\ and lifetime of ${\sim}25$~s. These scales are either comparable or much shorter than the resolvable limits of the best available EUV full-disk imager, the Atmospheric Imaging Assembly \citep[AIA; ][]{lemen12} on board the Solar Dynamic Observatory (SDO). These tiny dots are the smallest EUV brightenings ever observed or reported in the literature and have energies a few orders of magnitude more than the nanoflares energy budget.

Using a potential field extrapolation of line-of-sight magnetograms obtained with Helioseismic and Magnetic Imager \citep[HMI;][]{hmi} on board SDO, \citet{regnierhic14} found that these events are predominantly located at the footpoints of large scale trans-equatorial coronal loops with energy content of $10^{26}$~ergs per brightening. \citet{regnierhic14} concluded that these bright dots are multi-thermal in nature with temperatures ranging between $\log{T} (K) = 5.3 - 6.5$ using the EM loci analysis method \citep[see e.g.,][]{jordan87,TriMDY_2010}.

Since direct observations of the heating mechanisms responsible for these brightenings are still infeasible, a comprehensive understanding of such small scale brightenings is crucial. A quantitative analysis of the energetics involved in these transients is of paramount importance to establish their role in heating the fan loops and the solar corona in general. Here, we study the brightenings as a statistical ensemble, i.e., by analysing all the individual brightenings that occurred near the footpoint of the trans-equatorial fan loop system during the course of the Hi-C observations. Our objective is to identify such brightenings in Hi-C observations and study their energetics by employing Differential Emission Measure (DEM) diagnostics. For such diagnostics, it is mandatory to have multi-filter observations. Therefore, we searched for the same brightening in the AIA~193~{\AA} observations. The DEMs are then derived using AIA filters. Even though the AIA~193~{\AA} channel's effective area and thermal response are similar to that of Hi-C, these events are readily observed in the latter but not in the former. In order to make sure that the events observed in AIA are the same as those seen in Hi-C, we perform a detailed analysis to understand the inter-calibration between Hi-C and AIA including the effects of the spatial and temporal binning, and the point spread functions. 

The paper is structured as follows. In \S\ref{s:obs}, we present the observations taken using Hi-C and AIA that are analysed in this work. In \S\ref{s:analysis}, we describe how the brightenings are detected in both Hi-C and AIA~{193}~{\AA}, and present the inter-calibration of the instruments including spatial pixelisation and PSF effects, and temporal variability. In \S\ref{s:disc}, we discuss the results with reference to the energetics of the events, and finally we summarise and conclude in \S\ref{s:conc}.

\section{Observations and data reduction} \label{s:obs}

\begin{figure*}[htp!]
\centering
\hspace{-1.2cm}
\includegraphics[trim=0cm 4cm 0cm 4cm, clip=true,scale=0.5]{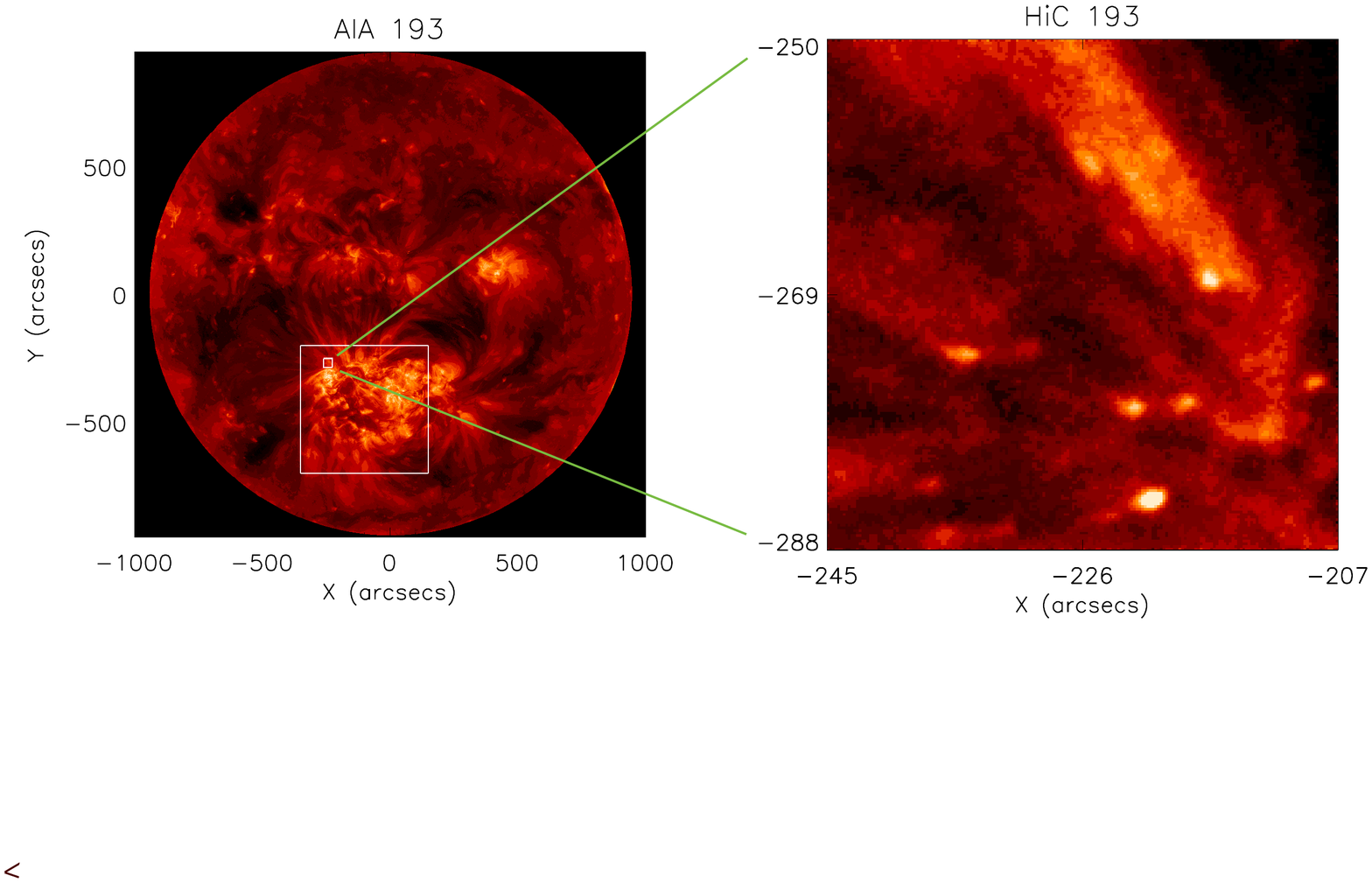}
\caption{Establishing the spatial context of the analysis. Left: Full disk AIA~193~{\AA} image, with a large white-bordered box showing the full Hi-C field-of-view, and a smaller inner box showing the field of interest studied in this work. Right: Co-temporal Hi-C image of the field-of-interest.} \label{f:AIA_HiC_FOV_FOI}
\end{figure*}

Near simultaneous images of \textsl{AR~11520} taken on 2012~July~11, with both Hi-C and AIA instruments, have been used in this study. Being a sounding rocket mission, Hi-C produced data for only about 5 minutes. The telescope field-of-view (FOV) suffered from a considerable jittering. Due to this, the FOVs of each Hi-C images were different. A common FOV covered by all Hi-C images is obtained by co-aligning individual Hi-C images with respect to each other. It is represented by the white outer box in the left panel of Figure~\ref{f:AIA_HiC_FOV_FOI}, which shows the AIA full disk image of the Sun recorded using the 193~{\AA} channel. The left panel also shows the {\sl field of interest} (FOI; inner white box) that encompasses the dot like EUV brightenings and the associated coronal footpoints of the fan loop system studied in this work. This fan loop system is rooted in the plage along the north-west of the active region. Here we have used only the data obtained between 18:52:48 and 18:55:30~UT. The right panel of Figure~\ref{f:AIA_HiC_FOV_FOI} displays the FOI as observed by Hi-C.

The AIA provides continuous full disk images of the solar atmosphere in 10 different UV/EUV passbands with a pixel size of 0.6\arcsec. In this study, the data obtained over a region corresponding to the Hi-C FOI (inner box of Figure~\ref{f:AIA_HiC_FOV_FOI}, left panel), taken between 18:45 and 19:00~UT, {in six EUV channels (131~\AA, 171~\AA, 193~{\AA}, 211~{\AA}, 335~{\AA} and 94~{\AA})} are used. All data have been processed using the standard reduction procedure available in Solar SoftWare (SSW). The Hi-C level 1.0 data are corrected with the proper time stamp and are co-aligned to match with the AIA data.

\section{Data analysis}\label{s:analysis}
\subsection{Brightening identification}\label{ss:brid}

For automatically detecting the EUV brightenings in the fan loop system studied in \citet{regnierhic14} (shown in Figure~\ref{f:HiC_pixsiz}), we use an identification procedure on the light curves obtained from individual pixels \citep[][]{ss10,ss12}. The identifications are carried out independently for both the AIA and Hi-C data. For this purpose, pixelwise light curves are obtained for AIA~193~{\AA} as well as Hi-C. All the light curves, for both AIA and Hi-C, are boxcar smoothed over a window of three frames in time in order to minimise false identifications of background fluctuations as brightenings. Thus, brightenings are identified only when their lifetime exceeds 16.5~s in Hi-C and 36~s in AIA (note that this difference has no effect on the detections of the events in AIA vis-a-vis Hi-C). The background is estimated locally by excluding the identified intensity peaks from the respective light curves and averaging over the rest of the light curves. In the case of Hi-C, an event is considered to be valid only when two or more adjacent pixels are detected by the identification procedure, and these pixels are merged into a single event.

\begin{figure*}[htp!]
\begin{center}
\includegraphics[trim=0cm 3cm 0cm 2cm, clip=true,scale=1]{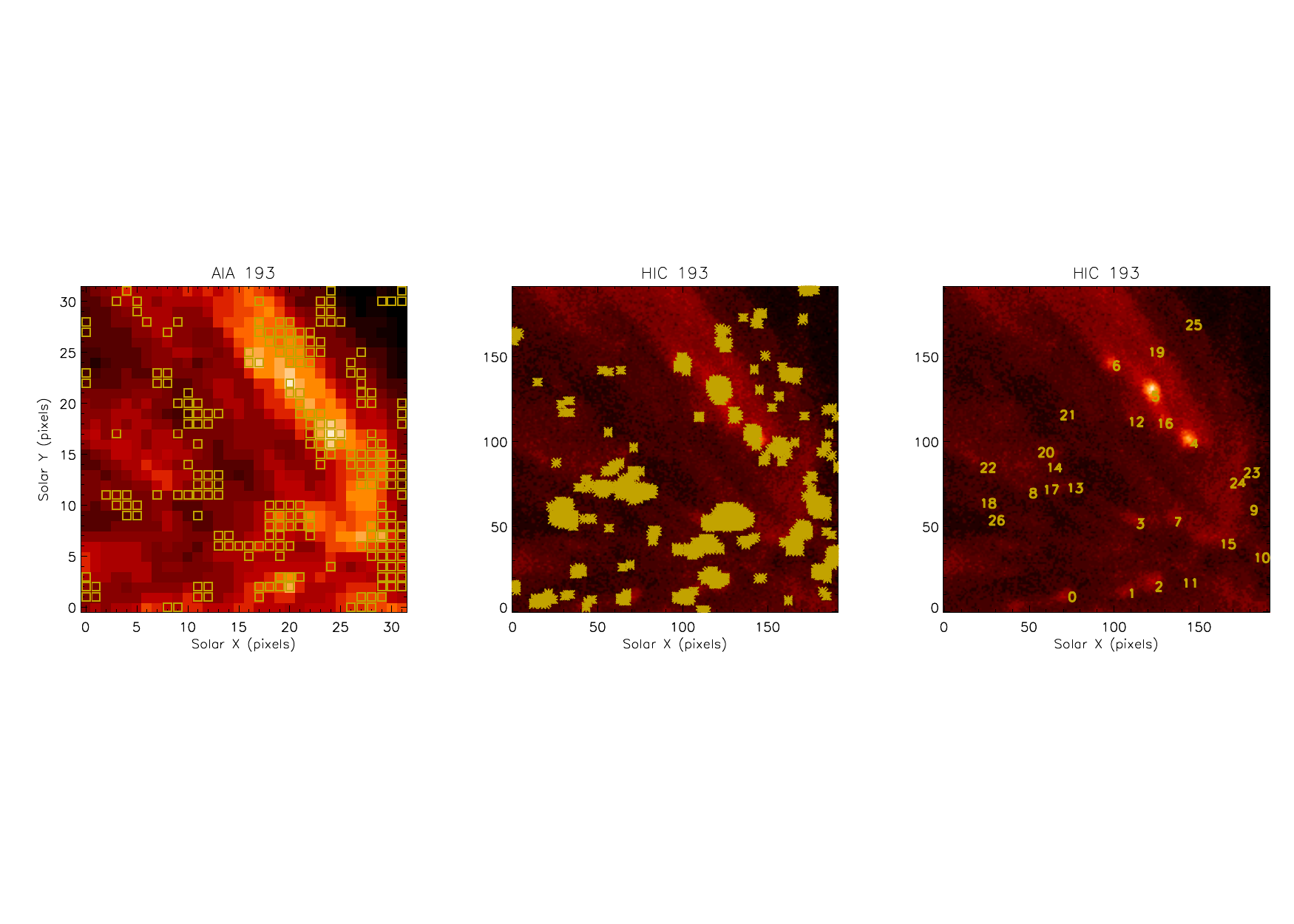}
\caption{Locations of brightenings in the field of interest (FOI); see Figure~\ref{f:AIA_HiC_FOV_FOI}. The FOI with all the identified brightening pixels marked with yellow boxes in AIA~193~{\AA} (left) and Hi-C~193~{\AA} (middle), and with the 27 brightenings considered for further analysis marked by number on the Hi-C image (right).} \label{f:brloc}
\end{center}
\end{figure*}

The algorithm identifies pixel-wise intensity enhancements that are above an user defined intensity threshold in the input light curves. It also computes the start and end times of the events, and their respective positions in pixel coordinates. An intensity threshold of 1.17$\times$ and 1.07$\times$ the local background in Hi-C and AIA, respectively, commensurate with the observed fluctuations in the background. This means that only enhancements larger than $1\sigma$ percentage fluctuations are considered real. Figure~\ref{f:brloc} shows the FOI in AIA~193~{\AA} (left) and Hi-C~193~{\AA} (middle) with all identified pixels over-plotted with yellow boxes. The identified pixels are then spatially grouped together to form brightenings.

Of all the identified brightenings in Hi-C and AIA 193~{\AA} observations, 27 events that fall near the footpoint region of the fan loop system are considered for further analysis. The right panel in Figure~\ref{f:brloc} shows the Hi-C~193~{\AA} image of the FOI, same as in the middle panel, but marked with the locations of the identified 27 brightenings. As already emphasised earlier, the prime aim of this work is to probe the thermal structure and energetics of these events. We accomplish this by computing DEMs using AIA data alone. The AIA~193~{\AA} channel's effective area and thermal response are similar to that of Hi-C~193~{\AA}, and thus including the Hi-C data in DEM analysis is not required. Using the Hi-C filter in the DEM would double the weight on the 193~{\AA} channel in the analysis. However, Hi-C information is crucial in identifying these brightenings in the first place. Since these brightenings are not readily discernible visually as individual events in AIA~193~{\AA} observations, most of these events would be discarded as not real without the input from the Hi-C observations. 

Light curves of these 27 brightenings are then derived by summing $2{\times}2$ pixels in AIA and $6{\times}6$ pixels in Hi-C. In Figure~\ref{f:brlc_7_26}, we show illustrative examples of two brightening events, Br7 and Br26, that compare the Hi-C and AIA~193~{\AA} light curves. Despite the similarity in Hi-C and AIA~193~{\AA} filter responses, the respective light curves show significant disparities and we have to make sure that the events observed in AIA are the same as those seen in Hi-C. Therefore, we next consider different causes that account for these differences and develop a correction that allows us to analyse the events in tandem.

\subsection{Comparison of Hi-C and AIA 193~{\AA}~brightenings}\label{ss:lc-comp}

Even though both Hi-C and AIA~193~{\AA} detectors show increased emission over roughly the same time period, there are notable differences between the two light curves. First, the ratio of the peak intensity to the background is always larger in the Hi-C light curves compared to that in AIA light curves. For the two sample events shown here, the ratios are respectively $\approx$2.8 and 1.5 in Hi-C, versus 1.3 and 1.1 in AIA. It is also evident that the AIA light curves have a background almost twice as high as that in Hi-C. Another important difference is that the brightenings show broader peaks in AIA than in Hi-C. These characteristics are not unique for these two brightenings, but are observed in all the analysed events {(Appendix: Figure~\ref{f:app_lc_DEM_1} \& \ref{f:app_lc_DEM_2})}. 

The question that naturally arises is whether these differences are introduced by the inherent features of the instrumentations, like the difference in the achieved spatial/temporal-resolution and the PSF (point spread function). Therefore, in order to remove any systematic biases in the direct comparison of Hi-C and AIA intensities for these brightenings, it is critical to account for the observed differences between Hi-C and AIA light curves, in the background level and the width, along with the peak to background ratio.

We note that the light curves are compared after normalising by their respective exposure times. Furthermore, only events that last at least 3 individual exposure frames are detectable. Thus, neither temporal resolution nor exposure time differences are likely to explain the observed differences between the light curves. It is also unlikely to be due to differences in pixel scales, since the 6$\times$6 pixel binned Hi-C light curves of these brightenings are similar to the light curves obtained from the original Hi-C images. Below we consider the effect of the different PSFs on the light curves.

\subsection{Effect of AIA PSF}\label{lc-comp-psf}

An astronomical image is the convolution of the true source with the PSF of the telescope, with added statistical noise. Therefore, the comparison of images obtained using different PSFs is non-trivial. Thus, in order to understand the effect of PSFs on the observed images, we convolved the Hi-C images with the PSF of AIA~193~{\AA}. Here we have assumed that the Hi-C instrument has a $\delta$-function PSF, i.e., the Hi-C images are high-fidelity representations of the real Sun. This is a reasonable assumption since the Hi-C PSF is significantly sharper than that of AIA. For this analysis, we used a model PSF of AIA instrument obtained using the {\tt aia\_calc\_psf.pro} routine \citep{boerner12}. 

\begin{figure}[htp!]
\begin{center}
\includegraphics[trim=0.5cm 0cm 0.5cm 0cm, clip=true,scale=0.5]{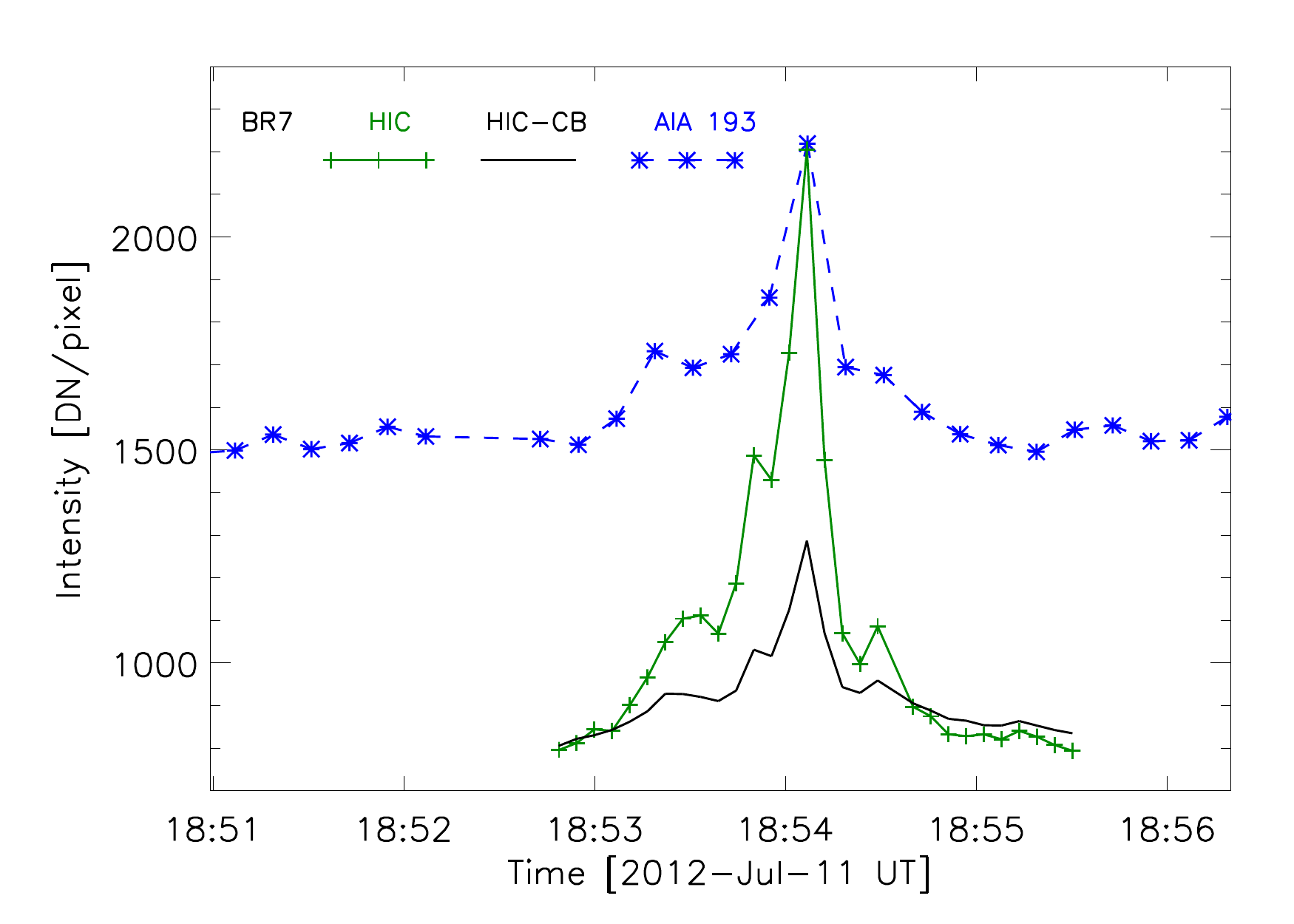}
\includegraphics[trim=0.5cm 0cm 0.5cm 0cm, clip=true,scale=0.5]{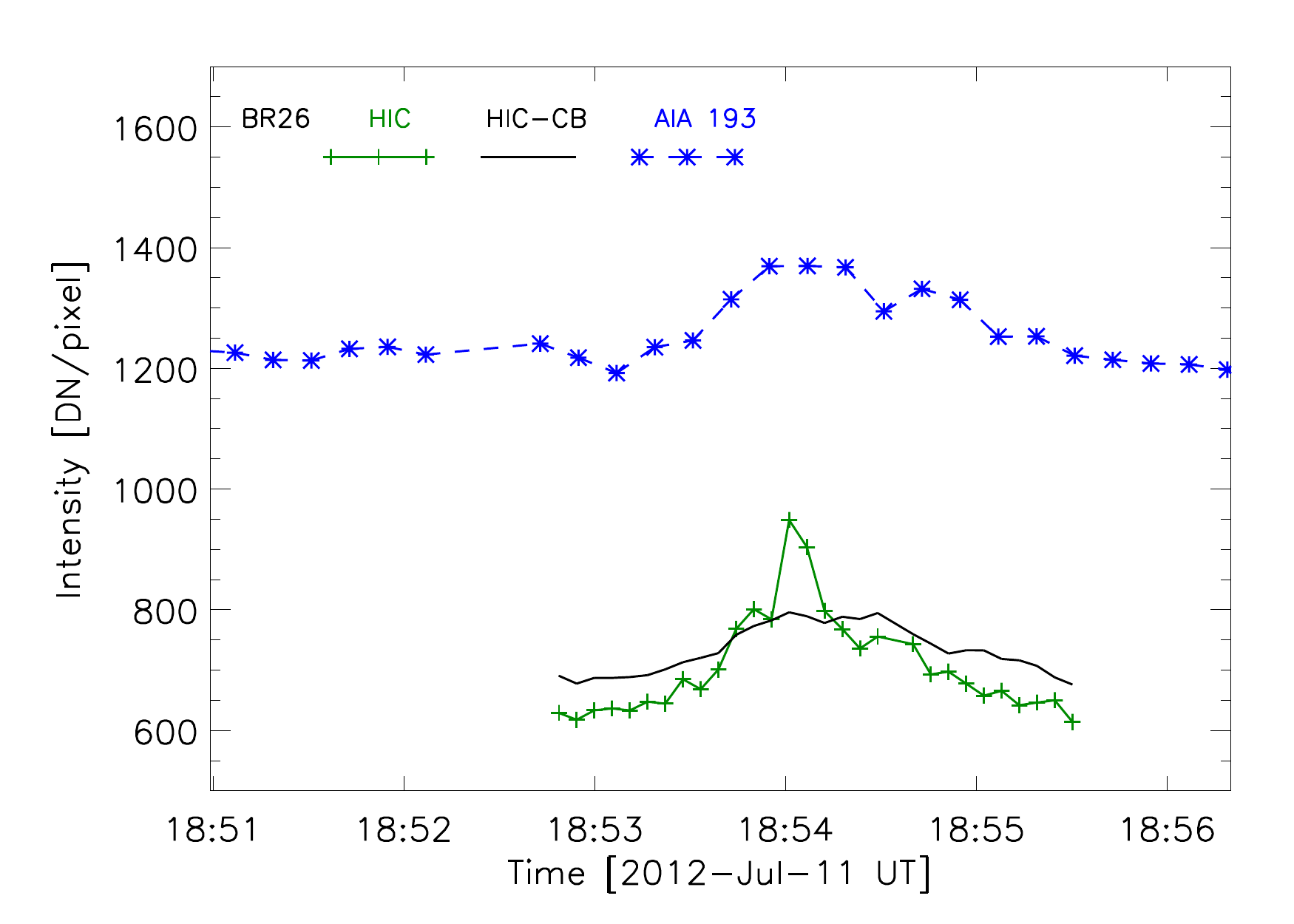}
\caption{Illustrative light curves of sample brightenings Br7 (top) and Br26 (bottom). The original Hi-C light curve (solid green line), the Hi-C convolved-and-binned light curve(Hi-C CB; solid black line), and the AIA light curve (dashed blue curve) are shown in each case.} \label{f:brlc_7_26}
\end{center}
\end{figure}

We then tested different binning schemes and found that a 6$\times$6 pixel binning of the Hi-C convolved images matches the spatial scale of the AIA images (see Appendix~\ref{ap0}). We, henceforward, refer to these AIA-pixel-compatible Hi-C images as Hi-C {\sl convolved-and-binned} (CB) images. A comparison of the light curves of the brightenings derived from the original Hi-C images and the Hi-C CB images clearly shows that convolution (with AIA PSF) and binning improves drastically the resemblance of Hi-C light curves to that of AIA (compare solid green line [HiC], black curve [HiC CB], and dashed blue curve [AIA] in Figure~\ref{f:brlc_7_26}; see also Appendix~\ref{ap2}). It is evident that the broader PSF of the AIA affects the observed evolution of the brightenings, by broadening the width of the peak and reducing the peak-to-background contrast.

Note however that even though the Hi-C convolved-and-binned light curves (solid black line) look similar in shape to the corresponding AIA  (dashed blue line) light curves, the curves remain offset. Thus, the differences between the Hi-C and AIA PSFs alone is insufficient to account for all the observed differences between the respective light curves.

\subsection{Hi-C and AIA 193~{\AA} intensity co-relation}\label{ss:lc-comp-correction}

\begin{figure}[htp!]
\begin{center}
\includegraphics[scale=0.5]{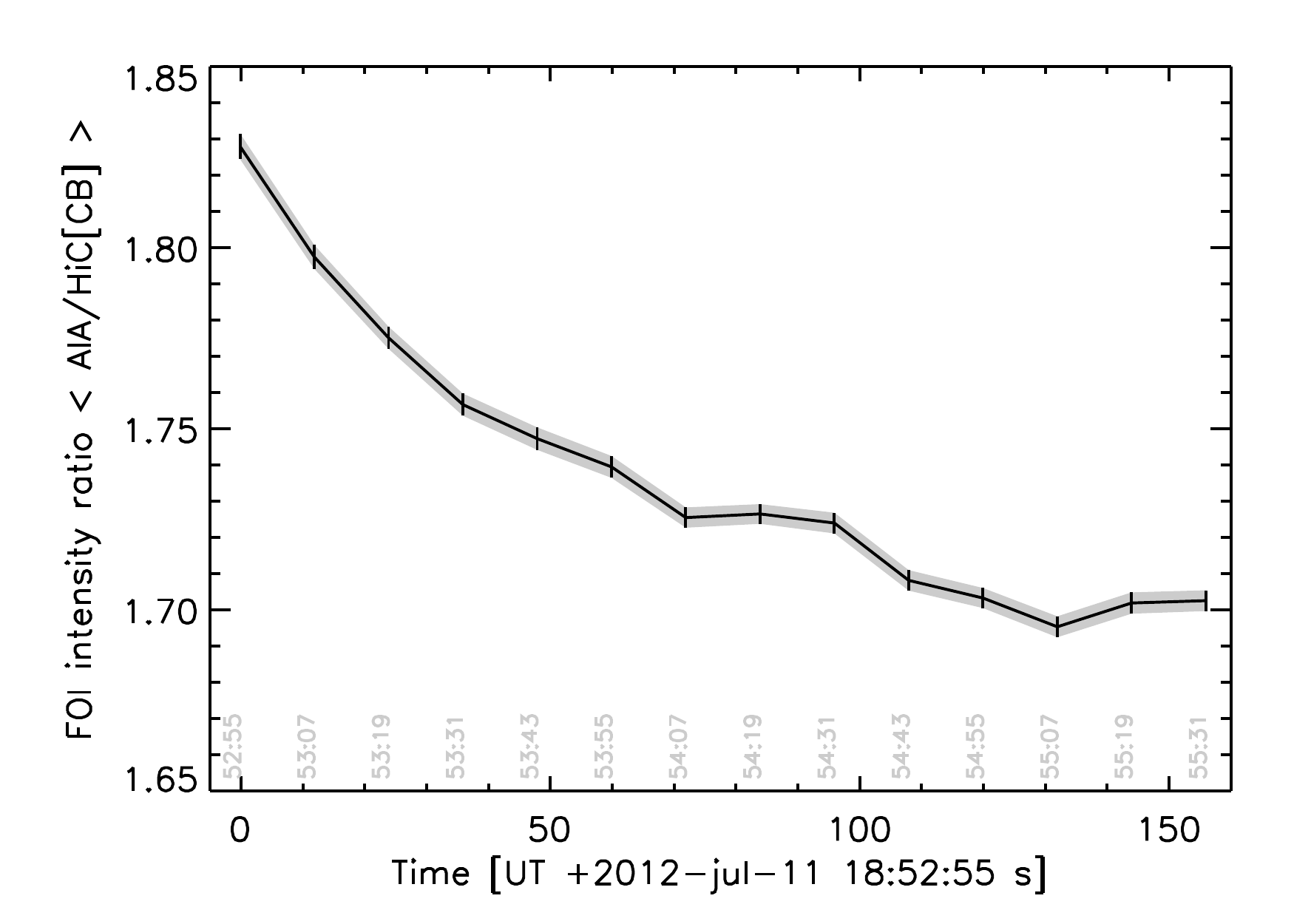}
\caption{
Characterizing the temporal changes between Hi-C and AIA.  Ratio of intensities of AIA and comparable Hi-C convolved and binned (Hi-C [CB]) images, spatially averaged over the field of interest (FOI).  $1\sigma$ standard errors are marked as vertical bars, and the dark gray shaded region represents the error envelope.  The ratios are calculated at the time epochs of the AIA observations (timestamps marked in light gray along the abscissa), using the Hi-C observation nearest in time to it.  A systematic decrease of $\approx8$\% over the duration of the Hi-C observation is clearly visible in the data.
}\label{f:AIAvHiC_FOI}
\end{center}
\end{figure}

Though AIA 193~{\AA} and Hi-C have similar effective areas, the registered intensities show significant differences. Therefore, a quantitative relation between the Hi-C and AIA 193~{\AA} baseline trends must be established. In order to check the relative calibration, we computed the ratio of the intensities in each matched pixel between AIA~193~{\AA} and corresponding Hi-C CB nearest in time. The average of the ratios, from the sample over the pixels in the FOI, is plotted as a function of time in Figure~\ref{f:AIAvHiC_FOI}. A systematic decrease of ${\approx}8\%$ is clearly visible in the data. The standard error\footnote{Standard error ($\sigma_{se}$) of the mean is the standard deviation of the sampling distribution of the mean and expressed as 

\begin{equation}
\sigma_{se}=\frac{\sigma}{\sqrt N}
\end{equation}
where $\sigma$ is the standard deviation of the original distribution and N is the sample size. It determines how precisely the mean of the sample estimates the population mean.} associated with these measurements is estimated to be $0.0028$, too small to be displayed in the Figure~\ref{f:AIAvHiC_FOI}. The error is even negligible when compared to the changes observed over time. In general, the AIA intensities are $69-83\%$ higher than that of Hi-C, with an average of $73\%$, which we attribute to probable differences in the mirrors and slightly different thermal responses of the two filters.

{
As mentioned above, the averaged ratio shows a systematic decrease of ${\approx}$8\% in the relative intensities over the course of the Hi-C observations.} However, no prominent solar activity occurred in the FOI which could account for the observed trend. Intriguingly, this trend has never been reported earlier in other published works on the Hi-C observations. In order to ensure that this is not an artefact of region selection, or due to inopportune plasma flow over the FOI, we investigated this phenomenon in further detail. 

\begin{figure}[htp!]
\begin{center}
\hspace{-0.4cm}
\includegraphics[scale=0.5]{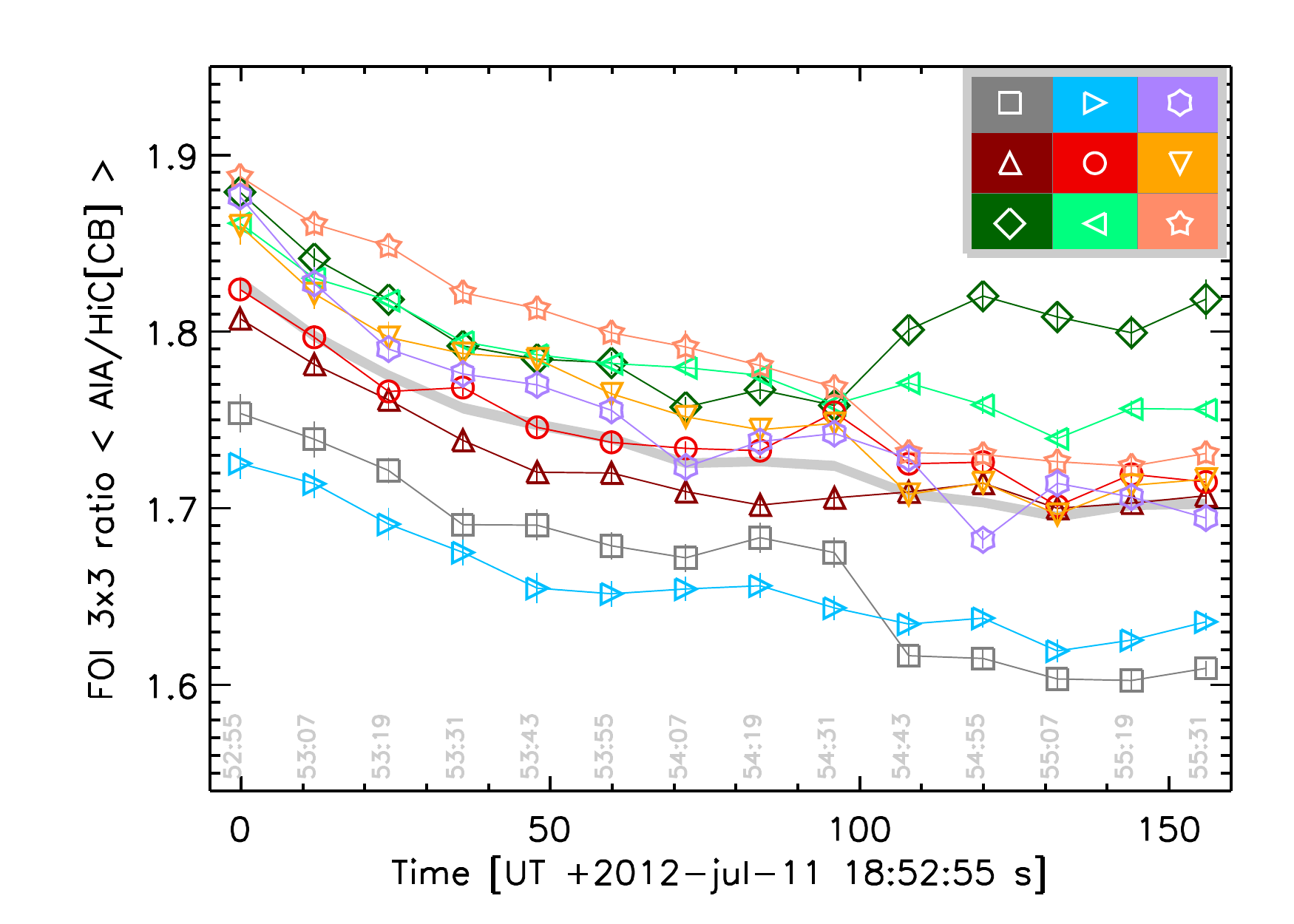}
\hspace{-0.3cm}
\includegraphics[scale=0.5]{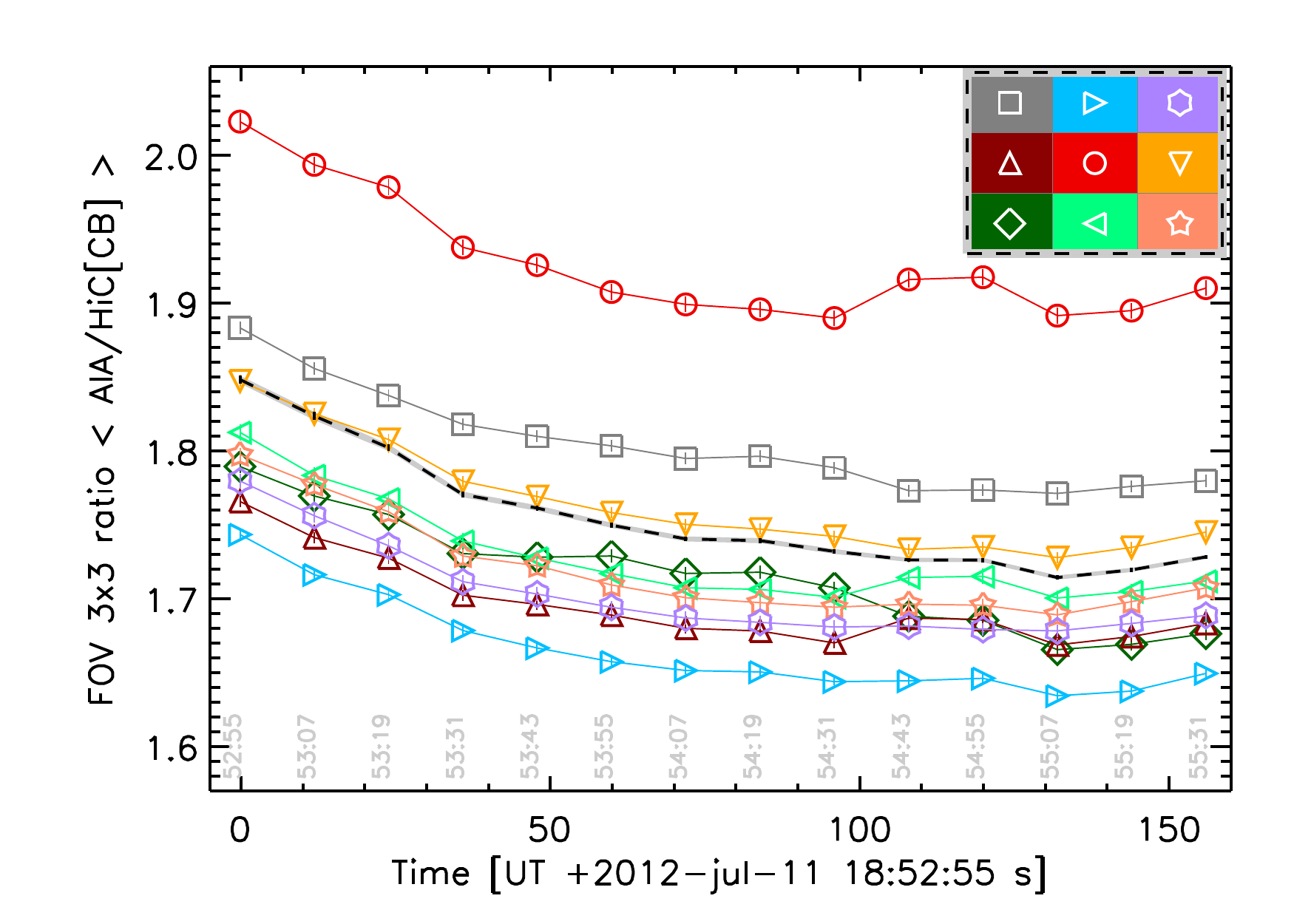}
\caption{
Similar to Figure~\ref{f:AIAvHiC_FOI}, for ratios of AIA and Hi-C[CB] intensities averaged over several sub-regions of the Hi-C field of view.  The field of interest (FOI; top) and the field of view (FOV; bottom) are split into $3{\times}3$ equal-area square grids and the pixel-wise ratios of the light curve intensities are averaged over each cell of the grid are shown.  Each curve is marked with a colour and symbol corresponding to the cell in the grid as shown in the grid key at top right.  In all cases the $1\sigma$ standard errors are marked with thin vertical lines.  The black dashed line in the bottom plot shows the average of the ratios over the full Hi-C FOV.  For comparison, the ratio averaged over the entire FOI from Figure~\ref{f:AIAvHiC_FOI} is shown in the top plot as the gray shaded region.
}\label{f:AIAvHiC_subregs}
\end{center}
\end{figure}

We first subdivide the Hi-C FOI into 3$\times$3 equal-area cells and compute the average of the pixel-wise intensity ratios separately in each cell as above (see Figure~\ref{f:AIAvHiC_subregs}; top). Strikingly, the ratios of the spatially averaged intensities in all the cells show a decreasing trend. We also carry out the same analysis for the entire Hi-C FOV (outer box of Figure~\ref{f:AIA_HiC_FOV_FOI}), by sub-dividing it into 3$\times$3 equal-area cells and again computing the temporal trend of the averaged pixel-wise ratios of the AIA and Hi-C~[CB] intensities. Figure~\ref{f:AIAvHiC_subregs} (bottom) shows the corresponding ratios, and it is evident that the decreasing trend in the ratios persists even in this case. The thick black line in the bottom panel is the average over the entire Hi-C FOV, {\%bf
and it shows the same trend as the ratio of the light curves averaged over all pixels in our FOI (Figure~\ref{f:AIAvHiC_FOI}).}

\begin{figure}[htp!]
\begin{center} 
\includegraphics[scale=0.5]{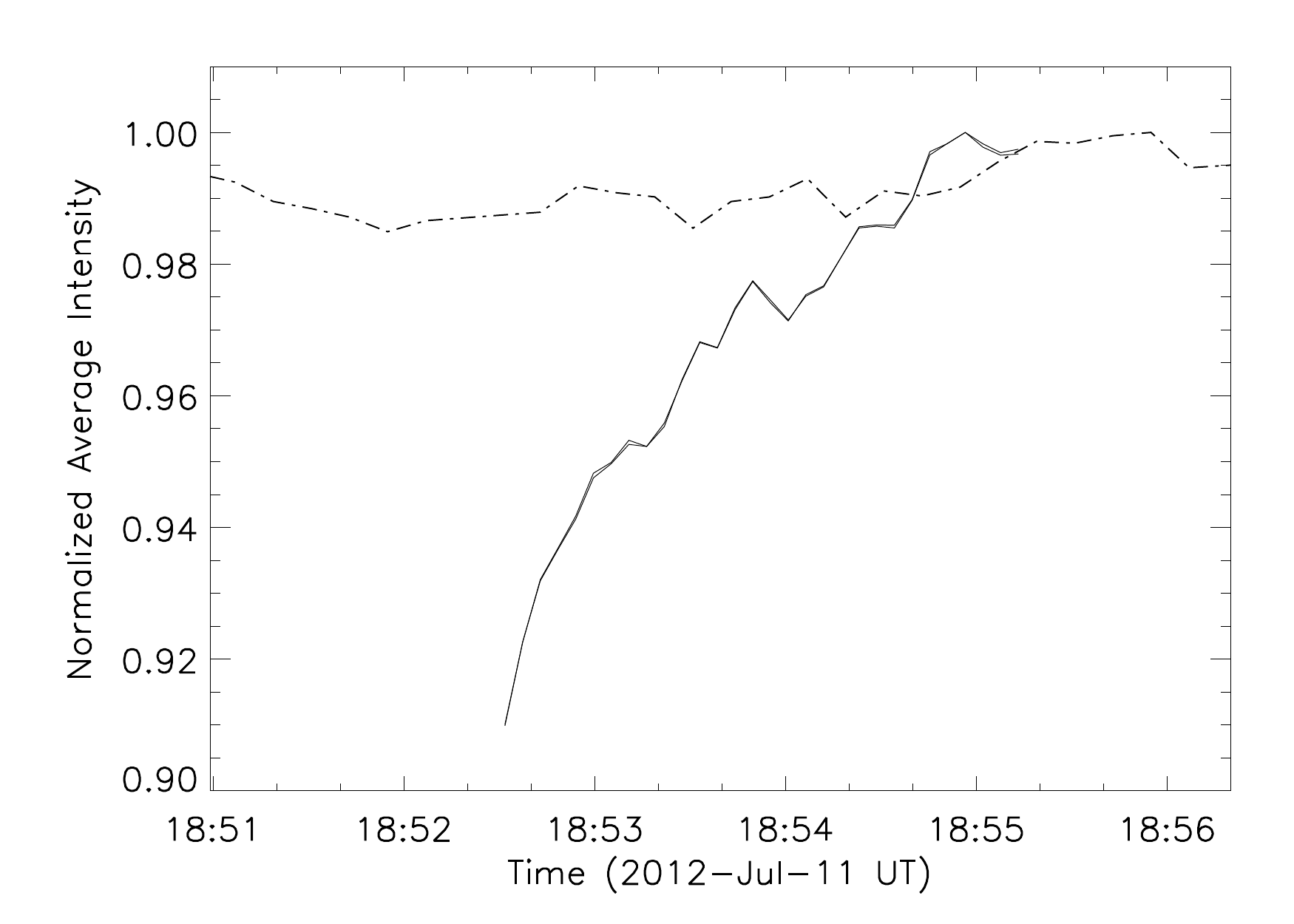}
\caption{Relative intensity variations over the field-of-interest for both Hi-C (solid line) and AIA (dot-dashed line). Both curves have been independently normalised to a maximum of $1.0$ within the duration of interest.} \label{f:AIAvHiC_trend}
\end{center}
\end{figure}

The observed behaviour of {
the spatially averaged ratios of AIA and Hi-C intensities} is fully explained by the intensity variations within the Hi-C alone. Figure~\ref{f:AIAvHiC_trend} shows the averaged AIA and Hi-C intensity curves from the FOI, between 18:50:00{--}18:58:00~UT in the case of AIA (dot-dashed line) and between 18:52:48{--}18:55:30~UT in the case of Hi-C (solid line), normalised to the respective maximum values within this time period. Hi-C shows a nearly monotonic rise by $\approx$8\% over the operational time-scale, while the AIA registers none and only shows a fluctuation of ${\sim}$1\%. In fact, the AIA intensities remain stable beyond the observational duration of the Hi-C flight. This suggests either a large scale, hitherto undetected, flow in the solar corona that preferentially affects the Hi-C filter, or a variation in the calibration characteristics of Hi-C during its flight. We prefer the latter as being a simpler explanation.

Therefore, we attribute this rise in the Hi-C averaged intensity to environmental or instrumental drifts during the Hi-C flight. Regardless of origin, this difference in trends must be corrected for in order to compare the intensity fluctuations between Hi-C and AIA.

\begin{figure}[htp!]
\begin{center}
\includegraphics[trim=0.5cm 0cm 0.5cm 0cm, clip=true,scale=0.5]{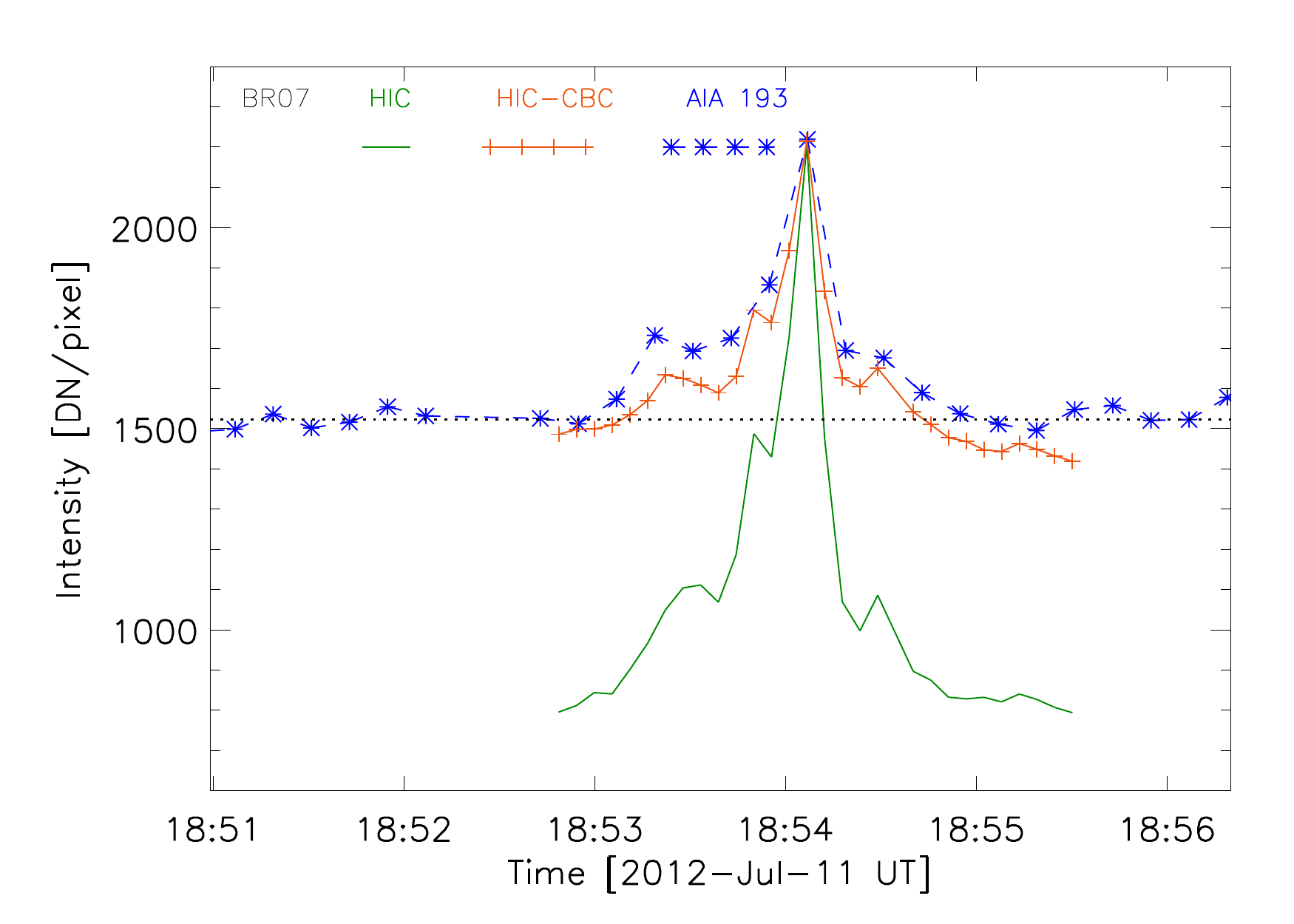}
\includegraphics[trim=0.5cm 0cm 0.5cm 0cm, clip=true,scale=0.5]{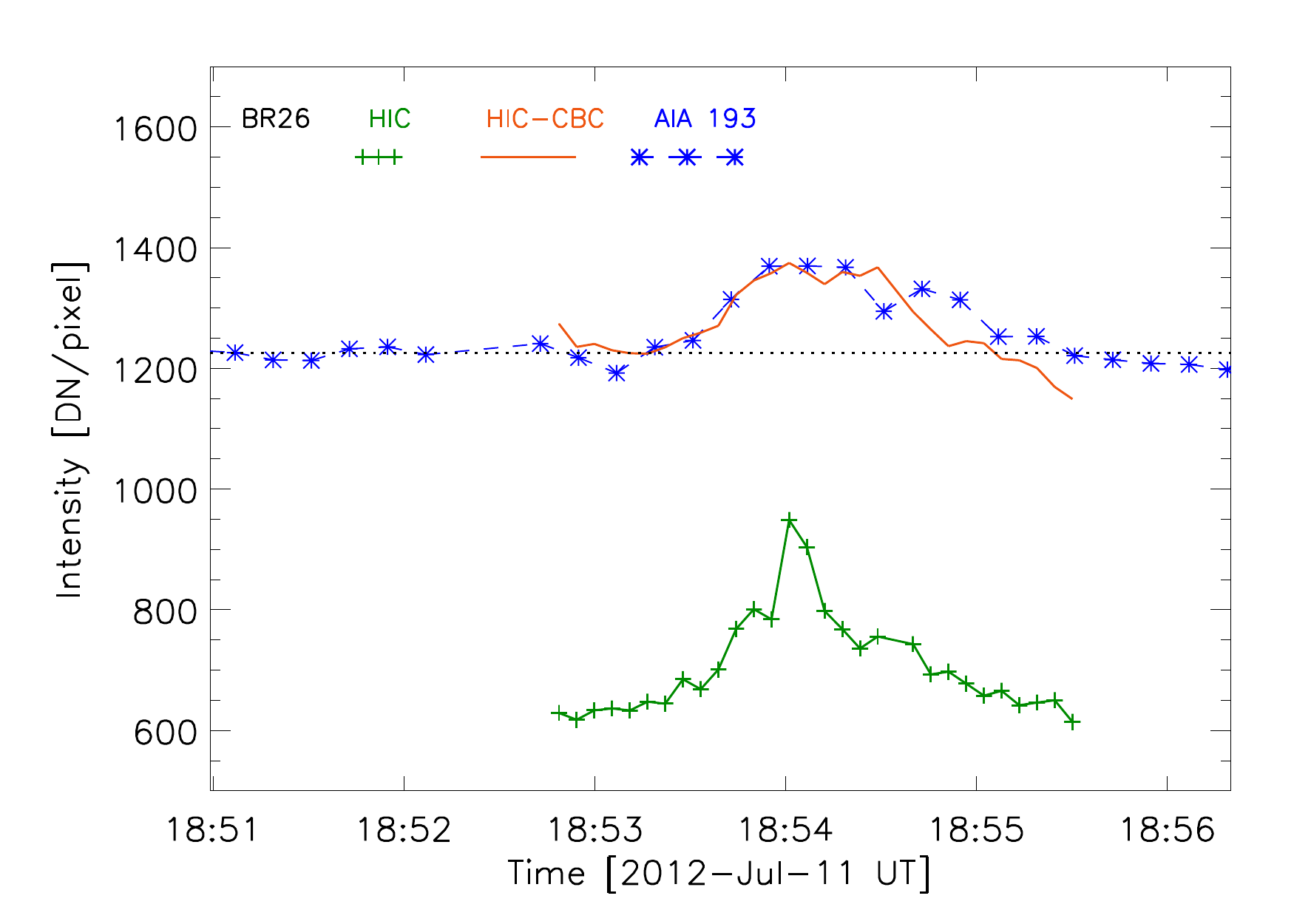}
\caption{As in Fig.~\ref{f:brlc_7_26}, but the original Hi-C convolved-and-binned (CB) light curves are replaced with Hi-C convolved-binned-corrected (CBC) light curves. The horizontal black dotted lines represent the estimated background in AIA.}\label{f:AIAvHiC_7_26_CBC}
\end{center}
\end{figure}

We correct for this apparent trend in the Hi-C sensitivity with a time-dependent correction factor obtained from a spline interpolation of the normalised average ratio curve (Figure~\ref{f:AIAvHiC_FOI}) at the Hi-C observation times. A new set of Hi-C light curves are obtained by multiplying the Hi-C {\sl CB} light curves by this interpolated ratio array. These {\sl convolved-binned-corrected} (CBC) curves are shown in Figure~\ref{f:AIAvHiC_7_26_CBC} (red solid lines), along with the original Hi-C light curve (green line marked with `+' symbols), and the AIA 193~{\AA} light curve (dotted blue line with asterisks), as in Figure~\ref{f:brlc_7_26}. The horizontal black dotted line represents the estimated background in AIA. For this, we interpolated the light curves in seconds. The start and end times are defined as the points in the light curve, respectively on either side of the peak and closest in time with respect to the peak, where the difference in intensity between the light curve and the background is positive and minimum. In two cases, there are no possible intersections between the light curve and the estimated background during the rise phase of the events {
(Br19 and Br22)}, and for these we define the start time visually.

We calculate a $\chi^2$-like measure of similarity between the Hi-C {\sl CBC} curves and the AIA~193~{\AA} light curves in order to understand how well they match. We compute

$$
\chi^2_{\rm average} = \frac{1}{N_{\rm AIA}} \sum_t \frac{(DN_{\rm AIA} - DN_{\rm HiC-CBC})^2}{DN_{\rm AIA} + DN_{\rm HiC-CBC}} \,,
$$

\noindent where $DN_{(.)}$ are the intensities in units of [DN], with assumed $\sqrt{DN}$ errors, and $N_{\rm AIA}=15$ is the number of AIA observations that span the duration of interest. These are reported in Table~\ref{table1} for each case.  

When interpreted as a goodness of similarity, $\chi^2_{\rm average}$ improves from factors of $>10^3$ to factors of a few after the CBC corrections. The visual comparisons also confirm that the Hi-C {\sl CBC} light curves provide a good match for the AIA 193~{\AA} light curves. This combination of pixel-binning, AIA~193~{\AA} PSF convolution, and correction of the intensity for offset and drift, explains the differences between the light curves for most of the brightenings.

As pointed out above, these brightenings are less discernible -- though detectable -- in AIA~193~{\AA} due to the combined effect of the lower AIA resolution and the larger width of the PSF. Hi-C detects these transient brightenings more efficiently by dint of having a smaller spatial spread of the signal and a corresponding lower background.

\begin{table}[ht!]
\small
\centering
\caption{Properties of brightenings derived from Hi-C and AIA~193~{\AA} light curves.}
\label{table1}
\begin{tabular} {ccccc} \\ 
\hline
Br.\ & prominent & Start time~$^a$ & AIA lifetime&	$\chi^2_{\rm average}~^b$  \\
no & hic peaks   &  in AIA      & [Sec]	                   &  \\
\hline \\
    0  &      1  &  18:52:06  &    106  &      4.6\\
    1  &      3  &  18:51:54  &    243  &      4.3\\
    2  &      4  &  18:51:30  &    335  &      3.6\\
    3  &      2  &  18:51:54  &    324  &      1.0\\
    4  &      2  &  18:51:42  &    209  &      5.0\\
    5  &      2  &  18:50:30  &    539  &      1.8\\
    6  &      2  &  18:53:06  &    139  &    10.6\\
    7  &      1  &  18:53:06  &    122  &      6.0\\
    8  &      2  &  18:53:30  &    179  &      2.9\\
    9  &      1  &  18:53:42  &    329  &      8.8\\
   10  &      2  &  18:52:42  &    299  &      3.6\\
   11  &      2  &  18:53:06  &    107  &      3.5\\
   12  &      2  &  18:48:42  &    487  &      2.5\\
   13  &      4  &  18:53:42  &     30  &      1.6\\
   14  &      2  &  18:52:42  &    179  &     1.0\\
   15  &      2  &  18:53:06  &    203  &     10.9\\
   16  &      2  &  18:52:06  &    126  &      3.4\\
   17  &      2  &  18:53:42  &    123  &      1.7\\
   18  &      4  &  18:53:30  &    124  &      1.3\\
   19  &      1  &  18:53:06  &    284  &     7.0\\
   20  &      3  &  18:52:06  &    251  &      1.1\\
   21  &      3  &  18:51:18  &    384  &      0.7\\
   22  &      1  &  18:55:06  &     50  &      3.9\\
   23  &      1  &  18:54:30  &    176  &    19.4\\
   24  &      1  &  18:54:18  &    312  &      8.8\\
   25  &      1  &  18:54:30  &    332  &      0.8\\
   26  &      1  &  18:53:18  &    131  &      2.0\\
\hline
\multicolumn{5}{l}{$^a$ Start time in AIA observed on 11-Jul-2012} \\
\multicolumn{5}{l}{$^b$ Computed by comparing AIA light curves with} \\
\multicolumn{5}{l}{~~Hi-C convolved-binned-corrected (CBC) light curves.} \\
\end{tabular}
\end{table}

\subsection{AIA event intensities}

For each of the brightening events, their intensities are computed in each AIA passband over the lifetime of the event as determined in the AIA~193~{\AA} filter, and the light curves are interpolated to a uniform grid. For the AIA~193~{\AA} light curve, a background level is determined from intensities outside the detectable events, and subtracted from the corresponding event light curve. For other filters, where the contrast between the events and the background vary widely, the minima in the light curves over the event duration are used to set the background. The subtracted light curves are integrated over the event lifetimes, denoting solely the intensities measured in each AIA passband for that event.

The average intensities over the lifetime of the events are given in Table~\ref{t:AIAfluxes}. These intensities are used to compute the emission measure distribution (see Section~\ref{s:dem} below) and infer average temperatures and plasma densities that characterize the brightenings. We compute a nominal statistical error for each filter by estimating the signal over the event lifetime as $\approx\sqrt{{\rm intensity}\times{\rm lifetime}}$. We adopt the larger of this statistical error or $10$\% of the average intensity as the defining uncertainty on the measured fluxes. The adopted values are given Table~\ref{t:AIAfluxes}.

\begin{table*}[ht!]
\small
\centering
\caption{Average intensities over the lifetime of the brightenings, as observed in AIA filters.  The error bars are the larger of estimated statistical or 10\% of the flux.}
\label{t:AIAfluxes}
\begin{tabular} {rrrrrrr} \\
\hline
Br.                             & \multicolumn{6}{c}{Intensities  [DN~s$^{-1}$]} \\
no                              & 193{\AA}      & 94{\AA}       & 131{\AA}      & 171{\AA}      & 211{\AA}      & 335{\AA} \\
\hline \\

0 &       83.8 $\pm$   8.38 &       0.50 $\pm$  0.07 &        7.3 $\pm$   0.73 &      104.3 $\pm$   10.43 &       12.0 $\pm$   1.20 &    2.8 $\pm$   0.28 \\
1 &       39.2 $\pm$   3.92 &       0.93 $\pm$  0.09 &        5.2 $\pm$   0.52 &       61.2 $\pm$   6.12 &        8.7 $\pm$   0.87 &        1.2 $\pm$   0.12 \\
2 &       60.9 $\pm$   6.09 &        1.4 $\pm$   0.14 &        7.3 $\pm$   0.74 &       72.5 $\pm$   7.25 &       12.6 $\pm$   1.26 &        1.0 $\pm$   0.10 \\
3 &       56.5 $\pm$   5.66 &       0.72 $\pm$  0.07 &        4.8 $\pm$   0.48 &      126.0 $\pm$   12.60 &       18.0 $\pm$   1.80 &        1.8 $\pm$   0.17 \\
4 &       79.8 $\pm$   7.98 &        2.2 $\pm$   0.22 &       12.9 $\pm$   1.29 &      103.3 $\pm$   10.33 &       12.1 $\pm$   1.21 &        1.9 $\pm$   0.19 \\
5 &       62.6 $\pm$   6.26 &        1.3 $\pm$   0.13 &        8.5 $\pm$   0.85 &      142.1 $\pm$   14.21 &       24.0 $\pm$   2.40 &        1.7 $\pm$   0.17 \\
6 &       53.6 $\pm$   5.36 &       0.88 $\pm$  0.09 &        9.2 $\pm$   0.92 &       62.7 $\pm$   6.27 &       13.1 $\pm$   1.31 &        0.65 $\pm$  0.07 \\
7 &      101.4 $\pm$   10.14 &       0.89 $\pm$  0.09 &        9.1 $\pm$   0.91 &       88.6 $\pm$   8.86 &        8.7 $\pm$   0.87 &        1.1 $\pm$   0.11 \\
8 &       26.6 $\pm$   2.66 &       0.46 $\pm$  0.05 &        2.6 $\pm$   0.26 &       77.3 $\pm$   7.73 &        9.1 $\pm$   0.91 &        1.6 $\pm$   0.16 \\
9 &       81.6 $\pm$   8.16 &        1.3 $\pm$   0.13 &        3.5 $\pm$   0.35 &       93.2 $\pm$   9.32 &       22.1 $\pm$   2.21 &        1.8 $\pm$   0.17 \\
10 &       44.6 $\pm$   4.46 &       0.98 $\pm$  0.10 &        8.6 $\pm$   0.86 &       73.7 $\pm$   7.37 &       10.0 $\pm$   1.00 &       0.97 $\pm$  0.10 \\
11 &       31.3 $\pm$   3.13 &       0.52 $\pm$  0.07 &        6.2 $\pm$   0.62 &       54.6 $\pm$   5.46 &       11.2 $\pm$   1.12 &         0.52 $\pm$  0.10 \\
12 &       52.0 $\pm$   5.20 &       0.90 $\pm$  0.09 &        5.9 $\pm$   0.59 &       77.9 $\pm$   7.79 &       14.9 $\pm$   1.49 &       2.0 $\pm$   0.20 \\
13 &       23.8 $\pm$   2.38 &        0.5 $\pm$   0.13 &       0.27 $\pm$  0.10 &       38.7 $\pm$   3.87 &        2.3 $\pm$   0.27 &        0.4 $\pm$   0.12 \\
14 &       46.2 $\pm$   4.62 &       0.88 $\pm$  0.09 &        4.1 $\pm$   0.41 &       34.2 $\pm$   3.42 &        9.1 $\pm$   0.91 &        1.7 $\pm$   0.17 \\
15 &       54.4 $\pm$   5.44 &        1.1 $\pm$   0.11 &        4.2 $\pm$   0.42 &       86.7 $\pm$   8.67 &       13.8 $\pm$   1.38 &        1.5 $\pm$   0.15 \\
16 &       75.4 $\pm$   7.54 &       0.79 $\pm$  0.08 &        7.1 $\pm$   0.71 &       78.6 $\pm$   7.86 &       27.7 $\pm$   2.77 &       1.7 $\pm$   0.17 \\
17 &       34.8 $\pm$   3.48 &       0.38 $\pm$  0.06 &        3.2 $\pm$   0.32 &       60.3 $\pm$   6.03 &        8.1 $\pm$   0.81 &        1.2 $\pm$   0.12 \\
18 &       39.2 $\pm$   3.92 &        1.1 $\pm$   0.11 &        3.6 $\pm$   0.36 &       59.6 $\pm$   5.96 &       11.2 $\pm$   1.12 &         0.80 $\pm$  0.10 \\
19 &      130.3 $\pm$  13.03 &        1.4 $\pm$   0.14 &        8.0 $\pm$   0.80 &      184.0 $\pm$   18.40 &       36.6 $\pm$   3.66 &       3.4 $\pm$   0.34 \\
20 &       48.3 $\pm$   4.83 &       0.90 $\pm$  0.09 &        3.3 $\pm$   0.33 &       24.3 $\pm$   2.43 &        9.4 $\pm$   0.94 &       1.0 $\pm$   0.10 \\
21 &       45.0 $\pm$   4.50 &        1.0 $\pm$   0.10 &        4.0 $\pm$   0.40 &       78.5 $\pm$   7.85 &       16.2 $\pm$   1.62 &       0.90 $\pm$  0.10 \\
22 &       26.0 $\pm$   2.60 &       0.39 $\pm$  0.08 &        0.7 $\pm$   0.11 &       45.0 $\pm$   4.50 &        4.8 $\pm$   0.48 &        0.8 $\pm$   0.12 \\
23 &       78.9 $\pm$   7.89 &       0.61 $\pm$  0.06 &        5.4 $\pm$   0.54 &      140.7 $\pm$   14.07 &       22.8 $\pm$   2.28 &        0.77 $\pm$  0.08 \\
24 &       40.8 $\pm$   4.08 &        1.4 $\pm$   0.14 &        3.9 $\pm$   0.39 &       65.7 $\pm$   6.57 &       20.4 $\pm$   2.04 &       1.3 $\pm$   0.13 \\
25 &       37.9 $\pm$   3.79 &       0.81 $\pm$  0.08 &        4.3 $\pm$   0.43 &       51.0 $\pm$   5.10 &        8.5 $\pm$   0.85 &       0.68 $\pm$  0.07 \\
26 &       39.7 $\pm$   3.97 &       0.40 $\pm$  0.06 &        4.6 $\pm$   0.46 &       60.1 $\pm$   6.01 &        5.2 $\pm$   0.52 &        0.64 $\pm$  0.07 \\

\hline
\end{tabular}
\end{table*}
\section{Results and discussion}\label{s:disc}

Having identified 27 brightenings based on Hi-C and AIA, we now turn to study the thermal structure and energetics of these events. An interesting characteristic of our sample of 27 events is that the light curves of all the events are found to have complex structure. They either have multiple peaks, or appear partially as incomplete observations at the beginning or end of the Hi-C observation run. We do not find any event with only one peak, that is, with a monotonic rise followed by a monotonic fall in the Hi-C intensity over their lifetime. 

Table~\ref{table1} lists the properties of the events derived from AIA~193~{\AA} light curves, along with the number of prominent peaks observed in Hi-C light curves. Of the events listed as having one peak, some (Br7, Br9, and Br26) have one prominent peak which could be easily analysed, and have multiple less energetic burstiness during either the rise or decay phases. In some other cases (Br19, Br22, Br23, Br24, and Br25), the events are only partially registered in Hi-C, and their characteristics outside of this epoch cannot be inferred.

\begin{table*}[ht!]
\small
\centering
\caption{Properties of brightenings derived from the DEM analysis.}
\label{t:properties}
\begin{tabular} {lcccrrrrr} \\ 
\hline
Br.\ 			& $\log{T}~^a$   	& EM ~$^b$                      		& Flux, $F_{\rm x}$~$^c$                     		& Energy\\
no           & [$\log$K]	          			&  [10$^{27}$~cm$^{-5}$] 	& [10$^{6}$~ergs s$^{-1}$ cm$^{-2}$]  &[10$^{24}$~ergs] \\
\hline \\
    0  &   6.22$\pm$0.10  &     5.4$\pm$ 1.1  &   1.6$\pm$ 0.1  &   5.0$\pm$ 0.3\\
    1  &   6.51$\pm$0.03  &     7.6$\pm$ 0.9  &   1.1$\pm$ 0.1  &   7.9$\pm$ 0.7\\
    2  &   6.55$\pm$0.05  &     6.9$\pm$ 0.9  &   1.3$\pm$ 0.1  &  12.4$\pm$ 1.4\\
    3  &   6.40$\pm$0.02  &     8.3$\pm$ 0.9  &   1.6$\pm$ 0.1  &  15.4$\pm$ 0.9\\
    4  &   6.55$\pm$0.03  &    11.2$\pm$ 1.6  &   1.7$\pm$ 0.3  &  10.4$\pm$ 1.5\\
    5  &   6.45$\pm$0.04  &    10.2$\pm$ 1.2  &   1.9$\pm$ 0.1  &  30.4$\pm$ 2.1\\
    6  &   6.42$\pm$0.13  &     3.1$\pm$ 0.6  &   0.8$\pm$ 0.1  &   3.4$\pm$ 0.5\\
    7  &   6.34$\pm$0.11  &     3.9$\pm$ 0.9  &   1.1$\pm$ 0.2  &   3.9$\pm$ 0.6\\
    8  &   6.44$\pm$0.01  &     6.1$\pm$ 0.7  &   1.0$\pm$ 0.1  &   5.2$\pm$ 0.3\\
    9  &   6.47$\pm$0.02  &    10.9$\pm$ 1.0  &   1.9$\pm$ 0.1  &  18.3$\pm$ 0.9\\
   10  &   6.50$\pm$0.06  &     5.3$\pm$ 0.8  &   1.0$\pm$ 0.2  &   9.2$\pm$ 1.4\\
   11  &   6.42$\pm$0.09  &     2.8$\pm$ 0.4  &   0.7$\pm$ 0.1  &   2.2$\pm$ 0.2\\
   12  &   6.46$\pm$0.02  &    10.4$\pm$ 1.2  &   1.6$\pm$ 0.1  &  22.6$\pm$ 1.7\\
   13  &   6.25$\pm$0.11  &     0.9$\pm$ 0.2  &   0.3$\pm$ 0.0  &   0.3$\pm$ 0.0\\
   14  &   6.51$\pm$0.02  &     8.7$\pm$ 1.0  &   1.1$\pm$ 0.1  &   5.9$\pm$ 0.5\\
   15  &   6.49$\pm$0.02  &     9.7$\pm$ 1.0  &   1.5$\pm$ 0.1  &   9.0$\pm$ 0.5\\
   16  &   6.37$\pm$0.03  &     8.5$\pm$ 0.9  &   1.8$\pm$ 0.1  &   6.8$\pm$ 0.5\\
   17  &   6.42$\pm$0.03  &     5.0$\pm$ 0.6  &   0.9$\pm$ 0.1  &   3.3$\pm$ 0.2\\
   18  &   6.55$\pm$0.03  &     6.3$\pm$ 0.6  &   1.0$\pm$ 0.1  &   3.8$\pm$ 0.2\\
   19  &   6.41$\pm$0.02  &    16.2$\pm$ 1.8  &   3.1$\pm$ 0.2  &  26.0$\pm$ 1.5\\
   20  &   6.53$\pm$0.02  &     6.7$\pm$ 0.8  &   0.9$\pm$ 0.1  &   7.0$\pm$ 0.6\\
   21  &   6.50$\pm$0.03  &     6.4$\pm$ 0.7  &   1.2$\pm$ 0.1  &  13.7$\pm$ 0.8\\
   22  &   6.37$\pm$0.09  &     1.7$\pm$ 0.4  &   0.5$\pm$ 0.0  &   0.7$\pm$ 0.0\\
   23  &   6.23$\pm$0.05  &     4.4$\pm$ 0.3  &   1.5$\pm$ 0.1  &   7.7$\pm$ 0.4\\
   24  &   6.53$\pm$0.02  &     9.5$\pm$ 0.9  &   1.4$\pm$ 0.1  &  13.2$\pm$ 0.8\\
   25  &   6.52$\pm$0.04  &     4.5$\pm$ 0.6  &   0.8$\pm$ 0.1  &   8.1$\pm$ 0.7\\
   26  &   6.32$\pm$0.10  &     2.5$\pm$ 0.4  &   0.7$\pm$ 0.1  &   2.8$\pm$ 0.3\\
\hline
\multicolumn{8}{l}{$^a$ EM weighted average of $\log{T}$ over the temperature range $\log_{10}{T~{\mathrm[K]}} = 5.5 - 7$.} \\
\multicolumn{8}{l}{$^b$ DEM reconstructions obtained using the AIA filter data.} \\
\multicolumn{8}{l}{$^c$ Based on folding a {\tt Chianti} spectrum with the DEM.} \\
\end{tabular}
\end{table*}
\subsection{Thermal structure}\label{s:dem}

Each filter has a different response $R(T)$ to plasma at a temperature $T$, determined by the emissivities $\varepsilon(T,\lambda)$ of the lines over the wavelength range $\lambda$ that the filter transmits, the transmission sensitivity of the filter, and the sensitivities of the telescope and detector over the corresponding wavelength range, the so-called effective area, $A_{\rm eff}(\lambda)$. Thus, $R(T)~=~\int~\varepsilon(T,\lambda)~A_{\rm eff}(\lambda)~d\lambda$, with $\varLambda(T)~=~\int~\varepsilon(T,\lambda)~d\lambda$, integrated over the domain of $A_{\rm eff}(\lambda)$, being the intrinsic emitted power in the filter passband. We used atomic emissivities from {\tt Chianti} (v7), combined with solar coronal abundances from \cite{SchRB_2012}, and filter calibration available in {\tt SSW} to compute $R(T)$ for all the AIA filters. The observed signal in a pixel for a given filter,

\begin{equation}
{\rm DN} = {\int \dem(T)\,R(T)\,d\log(T)} \,,
\end{equation}
where $\dem(T)$ is the so-called Differential Emission Measure (DEM), defined as
\begin{equation}
\dem(T) = n_e^2 ~ \frac {dh}{d\log{T}} \,, ~~~~~~ \rm{[{cm^{-5}}{\log{K}^{-1}}]}
\label{eq:DEM}
\end{equation}
where $n_e$ is the electron density and $h$ is the column depth of the plasma in the corona.

In general, the signal from a column of plasma is multi-thermal since it views different layers of the corona. However, coronal plasma from isolated structures can be approximated well as isothermal, i.e., $\dem(T)=EM\cdot\delta(T-T_0)$, where $\delta(\cdot)$ represents the Dirac $\delta$-function. When such an approximation is valid, the emission measure (EM) loci predicted for the various filters at different isothermal temperatures $T_0$ will intersect in $EM-T$ space \citep[see e.g.,][]{JorW_1971,TriMDY_2010}. Conversely, the lack of an obvious intersection point for the EM loci indicates that the plasma being observed is multi-thermal, and a detailed DEM analysis is necessary to explain the observed intensities.

\begin{figure*}[htb!]
\begin{center}
\hspace{-0.57cm}
\includegraphics[trim=0cm 0cm 0cm 0cm, clip=true,width=0.32\textwidth]{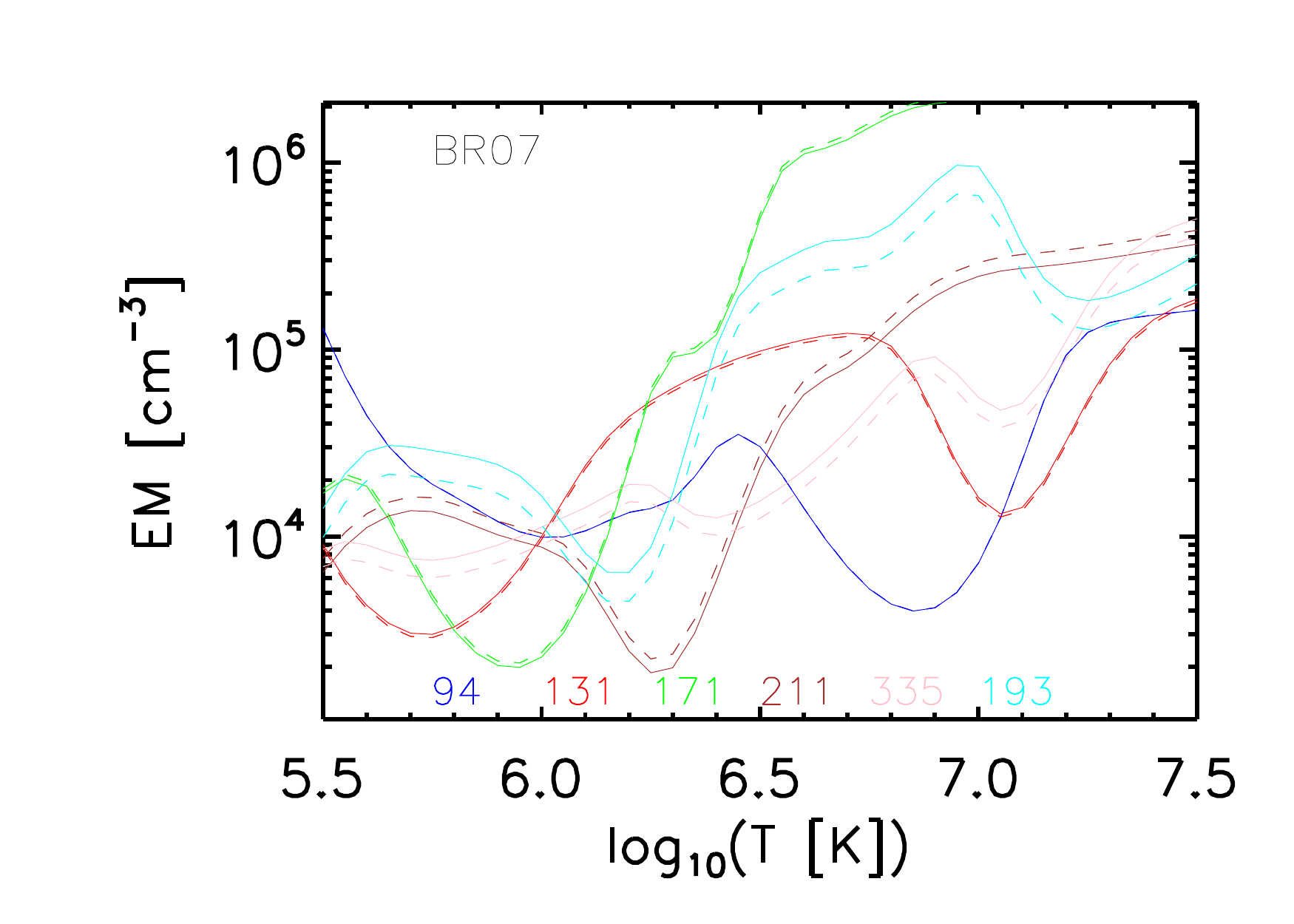}
\includegraphics[trim=0cm 0cm 0cm 0cm, clip=true,width=0.32\textwidth]{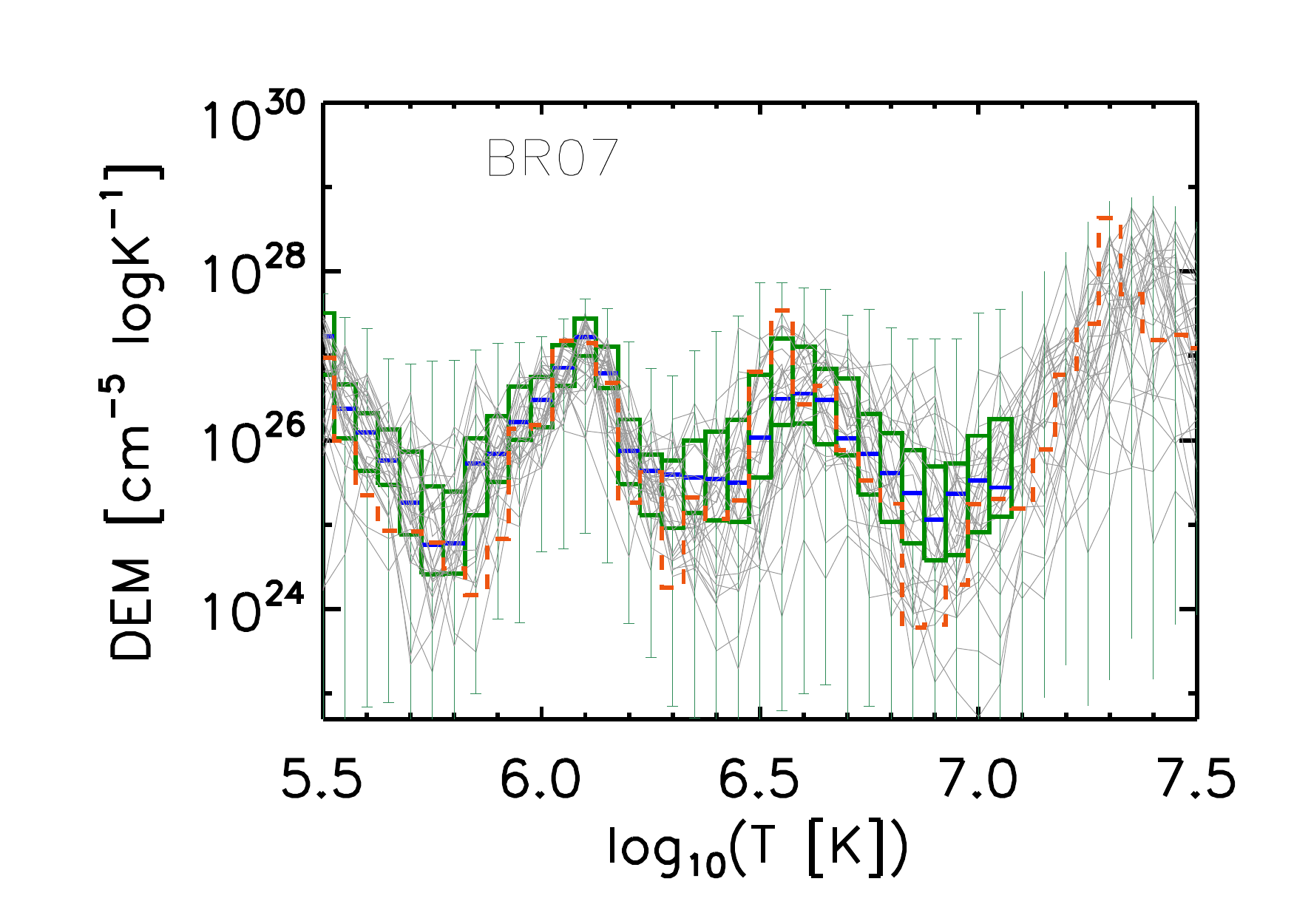}
\includegraphics[trim=0cm 0cm 0cm 0cm, clip=true,width=0.32\textwidth]{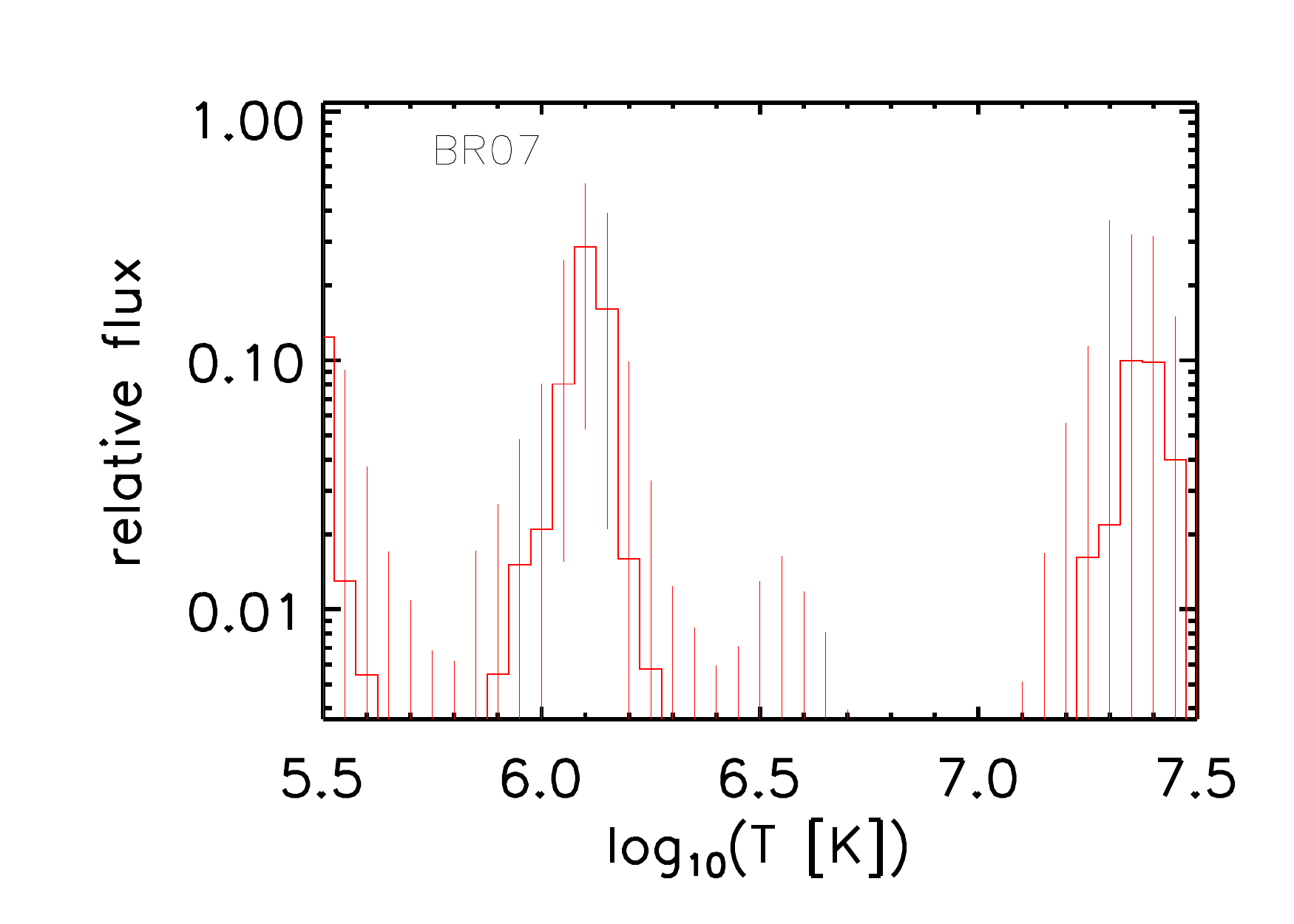}\\
\hspace{-0.57cm}
\includegraphics[trim=0cm 0cm 0cm 0cm, clip=true,width=0.32\textwidth]{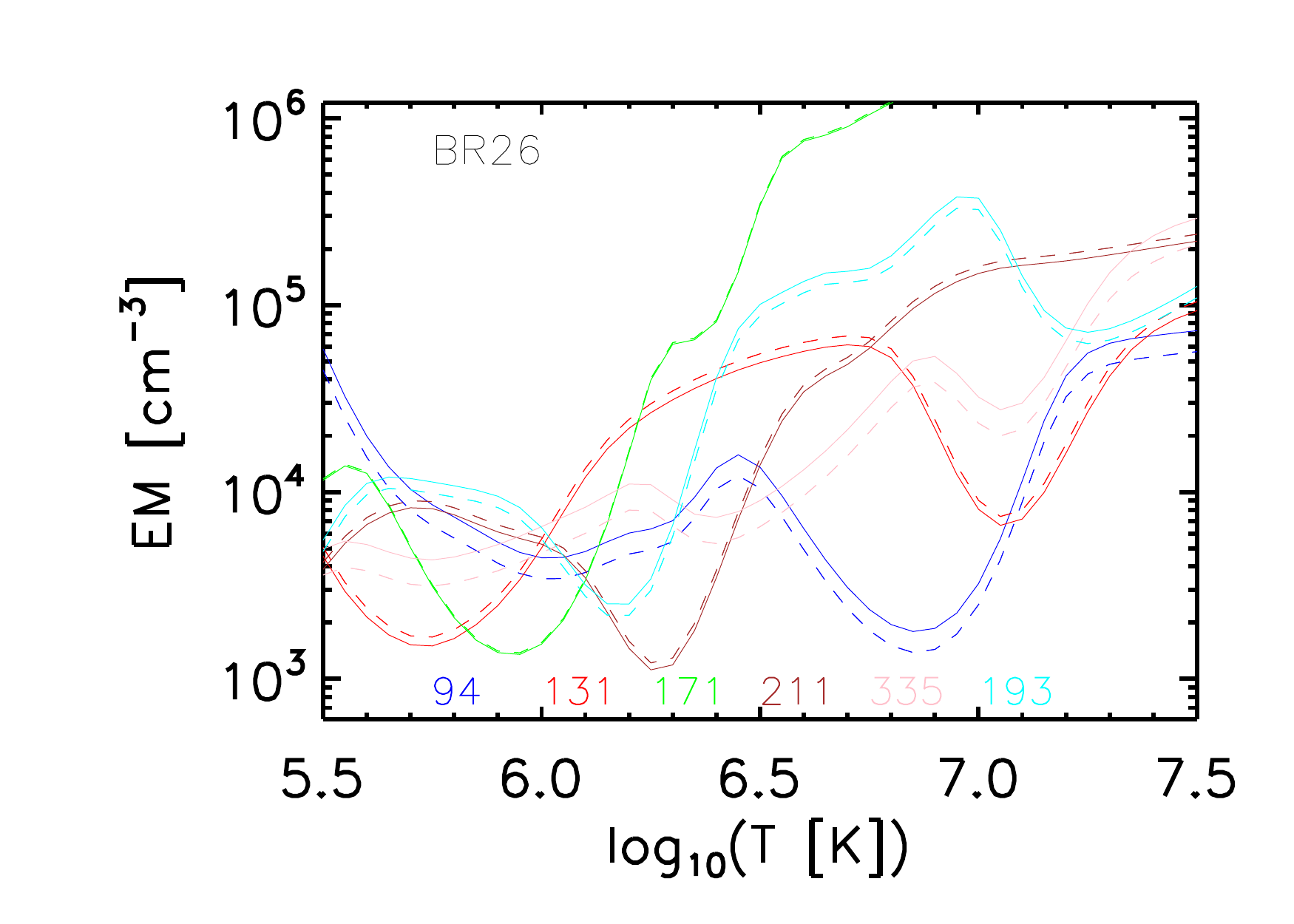}
\includegraphics[trim=0cm 0cm 0cm 0cm, clip=true,width=0.32\textwidth]{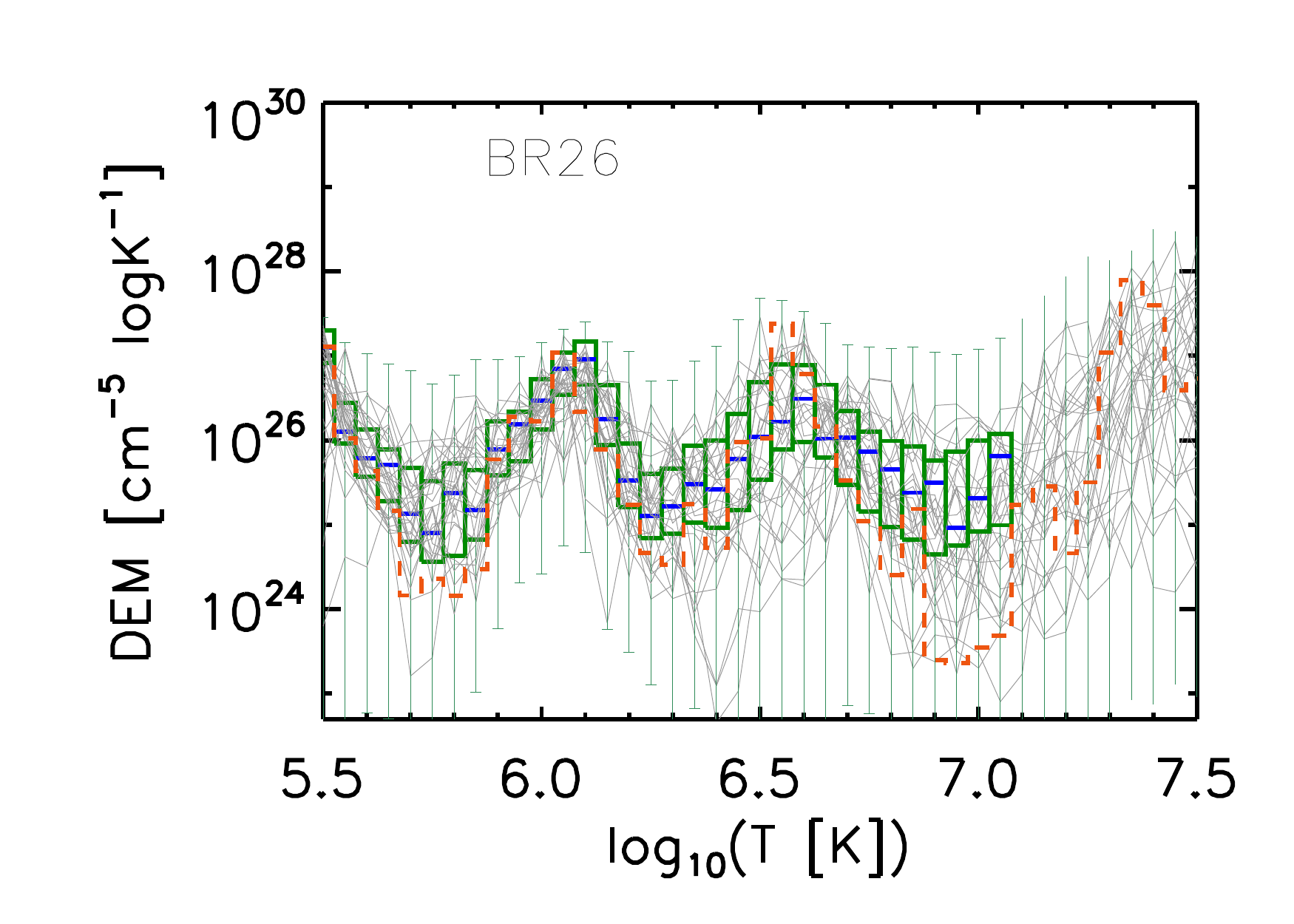}
\includegraphics[trim=0cm 0cm 0cm 0cm, clip=true,width=0.32\textwidth]{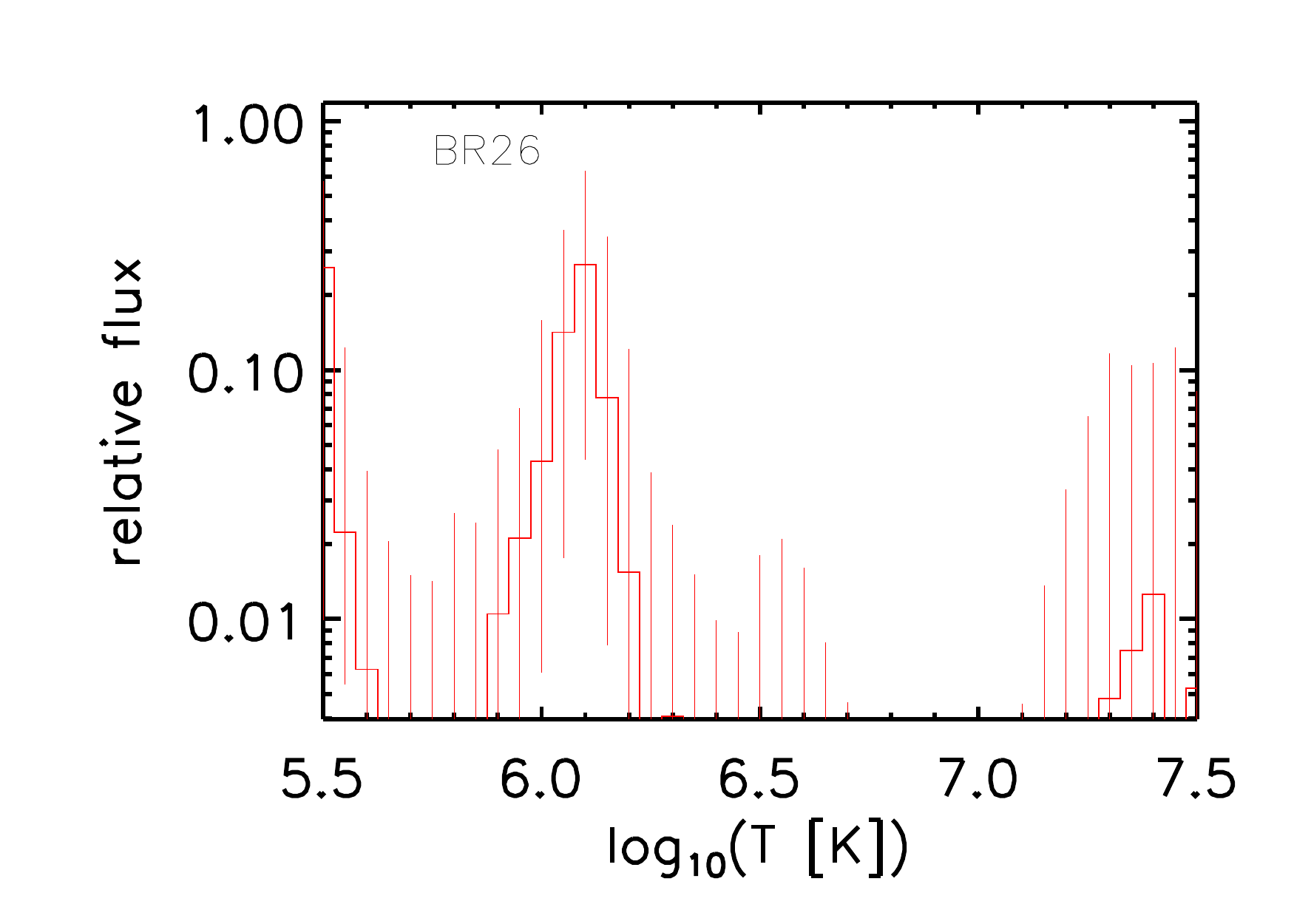}\\
\caption{Analyzing the thermal structure of brightenings.  EM loci curves (left column), DEM distributions (middle column) and an estimate of flux contribution of the DEM at different temperatures (right) are shown for the brightenings Br7 (top row) and Br26 (bottom row).  In the DEM plots (middle column), the blue horizontal segments mark the modes of the distribution of DEMs at each temperature bin, the green boxes indicate 50\% coverage, and the dashed red line is the nominal best-fit (see text).}\label{f:AIADEM_7_26}
\end{center}
\end{figure*}

We show EM loci curves of the sample brightenings, Br7 and Br26, in Figure~\ref{f:AIADEM_7_26} (left panels; the EM loci curves of the rest of the brightenings are given in \ref{ap2} in Figures~\ref{f:app_lc_DEM_1}-\ref{f:app_lc_DEM_4}). The predicted EM loci curves do not show indications of iso-thermality, and we therefore employed a more sophisticated analysis to model the thermal structure of the plasma in each brightening.

 We constructed DEMs for each of these brightenings by applying the Markov Chain Monte Carlo (MCMC) based DEM reconstruction algorithm of \cite{KasD_1998} on the background-subtracted average intensities in each AIA filter. We note that the Hi-C data are excluded from this analysis as they are already mirrored in the AIA~193~{\AA} filter data.

The MCMC iterations produce a set of DEM solutions, one per iteration, each drawn from the posterior probability distribution of the $\dem(T)$. The solutions are locally smoothed to ensure that there are sufficient constraints on the number of parameters that the degrees of freedom is $>$2, sufficient to prevent overfitting, and to usefully characterise a gross shape. The solutions are generated as a combination of isothermal plasma with different EM at over the temperature range $\log{T} (K) = 5.5 - 7$. Even though the thermal response of the AIA filters is defined over a broad temperature range ($\log{T} (K) = 4 - 8$), the magnitudes of the response curves are large and relevant over a much shorter range. The algorithm also provides a measure of the sensitivity of the solutions to the temperature range, as individual temperatures are drawn in each iteration from a sampling distribution that encodes both the overall responses of the AIA filters as well as rapid changes in the response; the former ensures coverage where the thermal response is high, and the latter focuses on regions which provide the most leverage to determining the thermal structure. The number of draws at each temperature then indicates whether that temperature is well covered or not.

In all our analyses, we find that limiting the temperature range to $\log{T} (K) = 5.5 - 7$ ensures coverage at the 95\% level. In all calculations that require integration over temperature, we use this temperature range. The AIA filters are not sensitive to temperatures outside this range and the DEM solutions are effectively unconstrained (the so-called \enquote{toothpaste-tube effect}, a term popularised by L. Golub and J. Schmelz), which causes highly non-linear aliasing effects due to the finite range over which the thermal responses are useful. We ran the code to produce 5000 iterations after burn-in and thinning by a factor of 10, and checked the traces of the fit statistic to ensure that the chain had converged in each case.

Illustrative solutions are shown in Figure~\ref{f:AIADEM_7_26} (second panels from left) for Br7 and Br26 (see Appendix~\ref{ap2} for solutions for all the other brightenings; see third panels from left in Figure~\ref{f:app_lc_DEM_1}-\ref{f:app_lc_DEM_4}). The figures show the range of possible solutions and indicate both point-wise uncertainties (the whisker plots, where the blue horizontal bars show the median solution at the given temperature, the green boxes show the range of 50\% of the solutions at that temperature, and the green vertical line shows the full range of the solutions), as well as the extent of the cross-temperature correlations in the solutions (the grey lines show all the DEM curves that are the modal solution at some temperature). The dashed red line is the nominal best-fit solution, i.e., the solution with the lowest $\chi^2$. Note that the best-fit solution is not required to, and often does not, go through the part of the distributions with the highest probabilities. This is because there is considerable cross-talk in the solutions between the values at different temperatures, leading to different solutions crossing many times.

Once the DEM solutions have been calculated, we can compute which temperatures contribute to the flux observed in the AIA~193~{\AA} filter, and by proxy, to the Hi-C detected brightenings. We fold the DEM solutions with the filter response and plot the relative intensity of the flux at each temperature in Figure~\ref{f:AIADEM_7_26} for Br7 and Br26 (right panels; the right panels in Figure~\ref{f:app_lc_DEM_1}-\ref{f:app_lc_DEM_4} in Appendix~\ref{ap2} show the flux-temperature plot for the other brightenings). Notice that temperatures that contribute most to the flux are generally at $T\approx1-2$~MK, with occasional components appearing at $T\approx3$~MK.

From the DEM solutions, we also estimate the total emission measure, $EM$, and the radiated flux $F_{\rm x}$, computed using the full CHIANTI database over the reliable temperature range (${\log}T(K)=5.5-7$; see above) 

\begin{eqnarray}
EM_{\rm total} &=&	{\int \dem(T)\,d\log{T}} \qquad  \rm{[cm^{-5}]} \\
F_{\rm x} &=&	{\int \dem(T) \varLambda_{\rm total}(T) d\log{T}} \qquad   \rm{[erg~s^{-1}~cm^{-2}]}
\end{eqnarray}

\noindent where $\varLambda_{\rm total}(T)$, the radiative loss function over the EUV passband, is calculated using the procedure {\tt rad\_loss.pro} in the {\tt Chianti} package. The estimated values of these quantities are listed in Table~\ref{t:properties}. { The DEM-weighted average $\log{T}$ are in the range $\log{T} (K) = 6.2 - 6.6$, meaning most of the emission detected in AIA filters is emitted around this temperature range}. The total energy emitted is then the product of the average flux, the lifetime of the event in AIA~193~{\AA}, and the area covered by the event which is 2$\times$2 pixels,
$$ Energy={F_{\rm x}{\times}Lifetime{\times}Area}\qquad\rm{[ergs]}\,.$$

\subsection{Decay time scales and density estimates} \label{s:density}

As with any dynamic solar activity, the EUV brightenings show a rise phase, associated with increase in intensity, followed by a decay phase until they disappear into the background. The cooling phase of the event can provide crucial information about the associated electron density.

We compute the decay timescales $\tdecay$ for each Hi-C event by fitting exponentially decaying models of the form

\begin{equation}
m(t) = \sum_{i=1}^{K} N_i \cdot exp(-(t-t^{(i)}_0)/\tau_i) + B(t) \,, t>t^{(i)}_0
\end{equation}
where the summation is carried out if the brightening has K$>$1 peaks which begin at times $t^{(i)}_0$, $N_i$ are the normalizations of each of the exponential decay components, and $B(t)$ represents a fixed background whose shape is determined from the AIA-to-HiC conversion described in Section~\ref{ss:lc-comp}, scaled to match the Hi-C light curve outside the domain of the brightenings. The fit is carried out over time ranges that show the event to be decaying. The resulting decay timescales are shown in the third column of Table~\ref{t:density}.

In the corona, both the thermal conduction and radiation are effective ways of losing energy gained through any heating process. The relative efficiencies of the two mechanisms is essentially determined by the timescales of the processes, which are dependent on both the temperature and the density. Conduction is expected to be the dominant cooling process in instances of high temperatures and high densities, and during the early phase of flares, while radiative cooling is the primary loss mechanism at low densities and during post-flare cooling phases \citep[see e.g.,][]{CarMA_1995,RafGMK_2009}.

In general, the observed decay timescale can be written as
\begin{equation}
\frac{1}{\tdecay} = \frac{1}{\tau_R} + \frac{1}{\tau_C} \,,
\label{eq:tdecay}
\end{equation}
where $\tau_R$ is the decay timescale for energy loss from optically thin thermal radiation,
\begin{equation}
\tau_{R}=\frac{3\,k_B\,T}{n_{e}\varLambda(T)} \,,
\label{eq:tauR}
\end{equation}
\noindent with $n_{e}$ the electron density, $\varLambda(T)$ the radiative loss rate at the temperature (T) calculated from {\tt Chianti}, and
$\tau_C$ is the decay timescale for energy loss due to conduction (see, e.g., \citep{winebarger03,vinay08}),
\begin{equation}
\tau_{C}=\frac{4\times10^{-10}n_{e}L^2}{T^{5/2}}
\label{eq:tauC0}
\end{equation}
\noindent with $L$ the length scale of the emitting plasma.

We can further simplify Equation~\ref{eq:tauC0} by noting that for small, low-lying, apparently homogeneous events such as those considered here, the length scale can be approximated as $L\sim\frac{EM}{n_e^2}$ (viz.\ Equation~\ref{eq:DEM});  the EM is the integrated DEM as reported in Table~1. Therefore, Equation~\ref{eq:tauC0} reduces to
\begin{equation}
\tau_{C}=\frac{4\times10^{-10}\,EM^2}{T^{5/2}\,n_{e}^3} \,.
\label{eq:tauC1}
\end{equation}

Substituting for $\tau_R$ and $\tau_C$ in Equation~\ref{eq:tdecay}, we obtain a cubic in $n_{e}$,
\begin{equation}
n_{e}^3~\frac{T^{5/2}}{4{\times}10^{-10}\,EM^2} ~+~ n_{e}~\frac{\varLambda(T)}{3\,k_B\,T} ~-~ \frac{1}{\tdecay} ~=~ 0 \,,
\label{eq:necubic}
\end{equation}
which can be solved to determine $n_{e}$ (see Appendix~\ref{ap1}). 

We combine our measurements of the temperatures (Table~\ref{t:properties}) obtained from a DEM reconstruction of AIA filter data, with empirical measurements of decay timescales for each burst identified in Hi-C (see Table~\ref{t:density}), to estimate the plasma density. Densities thus calculated are given in the fourth column of Table~\ref{t:density}.   Uncertainties are estimated through Monte Carlo bootstrap, with the variance in the DEM-weighted temperature, the estimated EM, and the measured decay timescale propagated through to the density. These densities are $\sim 10^{9-10}$~cm$^{-3}$, which is typical of solar active region densities \citep[see e.g.,][]{TriMYD_2008, TriMD_2009, YouWHM_2009}.

 \begin{table}[ht!]
 \small
 \centering
 \caption{Decay time, plasma density estimates and Conductive Loss Importance Factor (CLIF) for each Hi-C burst. }
 \label{t:density}
 \begin{tabular} {ccccc} \\ 
 \hline
 Br.\ & Peak  & Decay Timescale  & Plasma Density~$^a$       & CLIF~$^a$ \\
 No   & \hfil   & $\tdecay$ [sec]                                      & [10$^{9}$~cm$^{-3}$]       & \hfil \\
 \hline
 0 &  0 &  21.1 $\pm$   1.3 &   5.27 $\pm$    0.7 &   1.21 $\pm$    0.1 \\
 1 &  0 &  23.9 $\pm$   2.7 &   3.71 $\pm$    0.3 &   2.21 $\pm$    0.0 \\
 1 &  1 &   9.3 $\pm$   3.1 &   5.08 $\pm$    1.0 &   2.48 $\pm$    0.1 \\
 2 &  0 &  51.5 $\pm$  19.4 &   2.49 $\pm$    0.7 &   2.14 $\pm$    0.2 \\
 2 &  1 &  18.2 $\pm$   7.3 &   3.53 $\pm$    1.2 &   2.44 $\pm$    0.2 \\
 2 &  2 &  45.1 $\pm$  26.8 &   2.61 $\pm$    1.3 &   2.18 $\pm$    0.3 \\
 2 &  3 &  10.2 $\pm$   4.6 &   4.28 $\pm$    1.7 &   2.61 $\pm$    0.2 \\
 3 &  0 &  11.2 $\pm$   0.9 &   6.24 $\pm$    0.5 &   1.98 $\pm$    0.0 \\
 3 &  1 &  34.5 $\pm$   6.0 &   4.28 $\pm$    0.4 &   1.65 $\pm$    0.1 \\
 4 &  0 &  14.6 $\pm$   2.1 &   5.25 $\pm$    0.6 &   2.36 $\pm$    0.1 \\
 4 &  1 &  14.0 $\pm$   3.5 &   5.32 $\pm$    0.7 &   2.37 $\pm$    0.1 \\
 5 &  0 &  16.7 $\pm$   1.6 &   5.70 $\pm$    0.5 &   2.02 $\pm$    0.0 \\
 5 &  1 &  44.3 $\pm$  12.9 &   4.10 $\pm$    0.7 &   1.73 $\pm$    0.1 \\
 6 &  0 &  39.9 $\pm$   4.2 &   2.04 $\pm$    0.3 &   1.98 $\pm$    0.1 \\
 6 &  1 &   9.3 $\pm$   2.0 &   3.32 $\pm$    0.5 &   2.40 $\pm$    0.1 \\
 7 &  0 &   8.0 $\pm$   0.4 &   4.74 $\pm$    0.7 &   2.02 $\pm$    0.1 \\
 7 &  1 &  15.0 $\pm$   3.5 &   3.84 $\pm$    0.7 &   1.84 $\pm$    0.1 \\
 8 &  0 &   7.8 $\pm$   1.9 &   5.32 $\pm$    0.7 &   2.34 $\pm$    0.1 \\
 9 &  0 &  25.4 $\pm$   2.8 &   4.98 $\pm$    0.4 &   1.94 $\pm$    0.0 \\
10 &  0 &  14.6 $\pm$   2.4 &   3.51 $\pm$    0.4 &   2.43 $\pm$    0.1 \\
11 &  0 &  15.6 $\pm$   2.0 &   2.61 $\pm$    0.3 &   2.28 $\pm$    0.1 \\
12 &  0 &  44.9 $\pm$  13.3 &   4.06 $\pm$    0.6 &   1.76 $\pm$    0.1 \\
12 &  1 &  12.8 $\pm$   8.2 &   6.19 $\pm$    4.4 &   2.12 $\pm$    0.3 \\
13 &  0 &  13.7 $\pm$   5.9 &   1.77 $\pm$    0.7 &   1.95 $\pm$    0.2 \\
13 &  1 &   8.8 $\pm$   1.7 &   2.05 $\pm$    0.3 &   2.08 $\pm$    0.1 \\
14 &  0 &  33.1 $\pm$  14.6 &   3.64 $\pm$    1.3 &   2.08 $\pm$    0.2 \\
14 &  1 &  36.2 $\pm$   9.2 &   3.53 $\pm$    0.5 &   2.05 $\pm$    0.1 \\
15 &  0 &  43.7 $\pm$   5.2 &   3.70 $\pm$    0.3 &   1.89 $\pm$    0.0 \\
16 &  0 &  13.6 $\pm$   1.2 &   6.29 $\pm$    0.5 &   1.77 $\pm$    0.0 \\
17 &  0 &  30.7 $\pm$  12.2 &   3.06 $\pm$    0.8 &   1.91 $\pm$    0.2 \\
17 &  1 &  33.3 $\pm$   6.5 &   2.98 $\pm$    0.3 &   1.89 $\pm$    0.1 \\
18 &  0 &  21.8 $\pm$   5.0 &   3.13 $\pm$    0.3 &   2.41 $\pm$    0.1 \\
18 &  1 &  18.2 $\pm$  10.1 &   3.32 $\pm$    1.9 &   2.47 $\pm$    0.2 \\
20 &  0 &  18.8 $\pm$   6.7 &   3.56 $\pm$    0.8 &   2.38 $\pm$    0.1 \\
20 &  1 &  46.6 $\pm$  21.3 &   2.63 $\pm$    1.2 &   2.11 $\pm$    0.2 \\
20 &  2 &  18.4 $\pm$   5.2 &   3.58 $\pm$    0.7 &   2.38 $\pm$    0.1 \\
21 &  0 &  13.6 $\pm$   3.6 &   4.07 $\pm$    0.5 &   2.39 $\pm$    0.1 \\
21 &  1 &  27.6 $\pm$  10.3 &   3.21 $\pm$    0.8 &   2.19 $\pm$    0.1 \\
26 &  0 &  13.2 $\pm$   1.4 &   3.10 $\pm$    0.4 &   1.92 $\pm$    0.1 \\

\hline
\multicolumn{5}{l}{$^a$ Uncertainties estimated via Monte Carlo} \\
\end{tabular}
\end{table}

\subsection{Conduction dominated loss} \label{s:condloss}

The relative importance of conductive and radiative loss can be established by comparing the corresponding decay timescales for the estimated plasma density.  The process with the shorter timescale proceeds faster and thus controls the dynamics.  To estimate it, we introduce and compute a Conductive Loss Importance Factor (CLIF),

\begin{equation}
CLIF = \log\left.\frac{\tau_R}{\tau_C}\right|_{n_{e}} \,.
\end{equation}

The decay timescales that correspond to the computed densities are significantly larger for radiative losses than for conductive losses ($\tau_R \gg \tau_C$; see Table~\ref{t:density}). Thus, we conclude that the dynamics of these brightening events are dominated by conductive loss. To our knowledge, these are the first non-flare events found to be driven by conductive losses in the lower corona.

We note that for typical field strengths of $50$~G, the magnetic energy available over the estimated volumes of these brightenings is $\approx10^{26}$~ergs, substantially greater than the actual deposited energies $\gtrsim10^{24-25}$~ergs (see Table~\ref{t:properties}), This suggests that a large fraction of the deposited energy is not visible in the radiated photons, but is instead conducted away elsewhere, contributing eventually to the thermal content of the corona. The magnitude of these events is similar to that from nano-flare like events, but they do not show the nominal signature of a high-temperature component in the DEM. However, we base our calculations of the DEM in the crucial assumption that the plasma has achieved local equilibrium (both in the neighbourhood of the event and over durations shorter than the lifetimes of the events); if this assumption becomes invalid because of the effects of conduction processes, then a high-temperature component could be present but undetected.

Conduction as a cooling mechanism has been known to be important during the early stages of flares \citep[e.g.,][]{CarMA_1995, RafGMK_2009} and microflares \citep[][]{GupST_2018}. Our estimates of $CLIF$ are consistent with the $\frac{\tau_R}{\tau_C}$ expected from Cargill et al.\ (1995).

Note that the conductive decay timescale $\tau_C$ {\sl} decreases as the plasma temperature increases.  Thus, if our DEM analysis underestimates the high-temperature plasma component, the conductive loss will be more prominent. We assume in our analysis that the brightening events are not subject to continuous heating.  If such a process is operating, the estimated $\tdecay$ is an upper bound on the unforced decay timescale, which also acts to increase $CLIF$. Thus, we find our conclusion that the brightenings are conduction dominated to be robust. Indeed, even to bring parity between $\tau_R$ and $\tau_C$ requires that the radiative power loss is underestimated by nearly two orders of magnitude, well beyond the expected range of atomic data uncertainties; or that the emission measure $EM$ be increased while the plasma density $n_e$ is decreased, which is infeasible since that requires the brightenings to be $~4\times$ larger in size than observed.

\section{Summary and conclusions} \label{s:conc}

 We have studied the energetics of a larger sample (27) of small scale dot-like brightenings recorded by the Hi-C instrument by employing DEM diagnostics. These brightenings were predominantly located near the footpoints of a fan-loop system. Since DEM diagnostics requires intensities measured in different lines or filters, we have made use of the co-spatial and co-temporal observations recorded by AIA. For this purpose, inter-calibration of the images taken using Hi-C and AIA~193{\AA} were performed and summarised below. 

Since our analysis requires pixel to pixel matching of Hi-C and AIA images, we have binned the Hi-C data by different factors and found that a 6x6 binning mimics the AIA scaling. Next, an automated identification procedure was used to identify all the EUV intensity enhancements in the Hi-C FOV (Figure~\ref{f:brloc}). Out of all the identified events, 27 dot-like brightenings that fall near the fan loop footpoint region were chosen for further detailed analysis. The visual inspection of AIA 193~{\AA} images for these dot-like Hi-C brightenings reveal that these events were barely distinguishable. Traces can be observed for most of these events if we specifically look for each of them individually in AIA observations in comparison with Hi-C. However, all the 27 events are clearly identified when the automated identification procedure is applied over the corresponding AIA 193~{\AA} images.

We compared the AIA and Hi-C light curves of all the events and found that the Hi-C light curves show narrow peaks with a higher peak to background intensity ratio. Hi-C background intensity was always approximately half the AIA background intensity. The multiple peaks in Hi-C light curves revealed the bursty nature of energy deposition in the solar corona, which was not captured by AIA observations, though the AIA observations do capture the overall evolution of the event. A combination of PSF and the observed high background intensities possibly reduces the efficiency of AIA in distinguishing these small scale intensity enhancement events.

The achieved spatial/temporal resolution and the PSF of Hi-C and AIA 193~{\AA} images differ substantially.  We accounted for the effect of these differences when comparing the light curves.  First, the Hi-C images were convolved with the AIA PSF and re-binned to match their scale, and the intensities were then corrected to account for an observed drift in the ratio of spatially average intensities over the common field of view. This drift was attributable to the Hi-C, and the convolved, binned, and intensity-corrected Hi-C light curves were found to match well with AIA light curves for all the events.

We then computed Differential Emission Measures (DEMs) for AIA filter intensities using an MCMC-based algorithm for each of the brightenings.  We found that the the average $\log{T} (K) \approx 6.2 - 6.6$, based on averaging the DEM over a temperature range of $0.3-10$~MK.  The estimated plasma densities in these events were found to be a few times $10^{9}$~cm$^{-3}$, similar to the coronal density range.

The radiative fluxes emitted by these events were estimated to be $\approx10^6$ ergs s$^{-1}$ cm$^{-2}$ and the total radiative energy associated with these events were $\approx10^{24} - 10^{25} $~ergs. \cite{regnierhic14}  estimated the amount of magnetic energy associated with these events to be $10^{26}$ ergs and suggested that magnetic energy budget can be higher than the radiated energy, which is consistent with our estimate. 

We found that the dominant cooling mechanism is conductive loss, suggesting that only a small fraction of the energy deposited into the plasma becomes visible as radiation.  This is consistent with the total magnetic energy available in this environment.  Brightenings such as these could form an energetically important channel for the conversion and deposition of magnetic energy into the coronal plasma.

In summary, we have identified apparent low lying brightenings, which are dominated by conduction processes. Such events are important as the large amount of energy involved in these events could be contributing to coronal heating, which has not been taken into account so far. Hence, these events need further investigation both theoretically and observationally: their dynamical evolution must be modelled in detail through hydrodynamic simulations; and possible signatures of high-energy non-thermal processes should be verified and substantiated with high-resolution hard X-ray imagers.

\begin{acknowledgements}
We thank the referee for the constructive comments. S. S. acknowledges that part of this work was carried out at IUCAA during her post doc tenure. V.L.K. was supported through NASA Contract NAS8-03060 to the Chandra X-ray Centre. Part of this work was done during collaborative visits that were part of the ClassACT - an Indo-US Centre for Astronomical Object and Feature Characterisation and Classification, sponsored by the Indo-US Science and Technology Forum (IUSSTF).  V.L.K. also thanks the personnel at the Inter-University Centre for Astronomy \& Astrophysics (IUCAA) for their hospitality during the visits. DT thanks the Centre for Astrophysics (CFA), Harvard for the hospitality during the visit. DT acknowledges the Max-Planck Partner group on Coupling and Dynamics of the Solar Atmosphere of MPS at IUCAA.   We acknowledge the High resolution Coronal Imager instrument team for making the flight data publicly available. MSFC/NASA led the mission and partners include the Smithsonian Astrophysical Observatory in Cambridge, Mass.; Lockheed Martin's Solar Astrophysical Laboratory in Palo Alto, Calif.; the University of Central Lancashire in Lancashire, England; and the Lebedev Physical Institute of the Russian Academy of Sciences in Moscow. The AIA data used are provided courtesy of NASA/SDO and the AIA science teams. The AIA data have been retrieved using the Stanford University's Joint Science Operations Centre/Science Data Processing Facility. CHIANTI is a collaborative project involving George Mason University, the University of Michigan (USA) and the University of Cambridge (UK). Armagh Observatory and Planetarium is funded by the N. Ireland Department for Communities.
\end{acknowledgements}
%
%
\bibliographystyle{aa}
\bibliography{hic}
\begin{appendix}

\section{Pixel size of the Hi-C}
\label{ap0}

\begin{figure}[htp!]
\centering
\hspace{-1.2cm}
\includegraphics[trim=0cm 0cm 0cm 0cm, clip=true,scale=0.5]{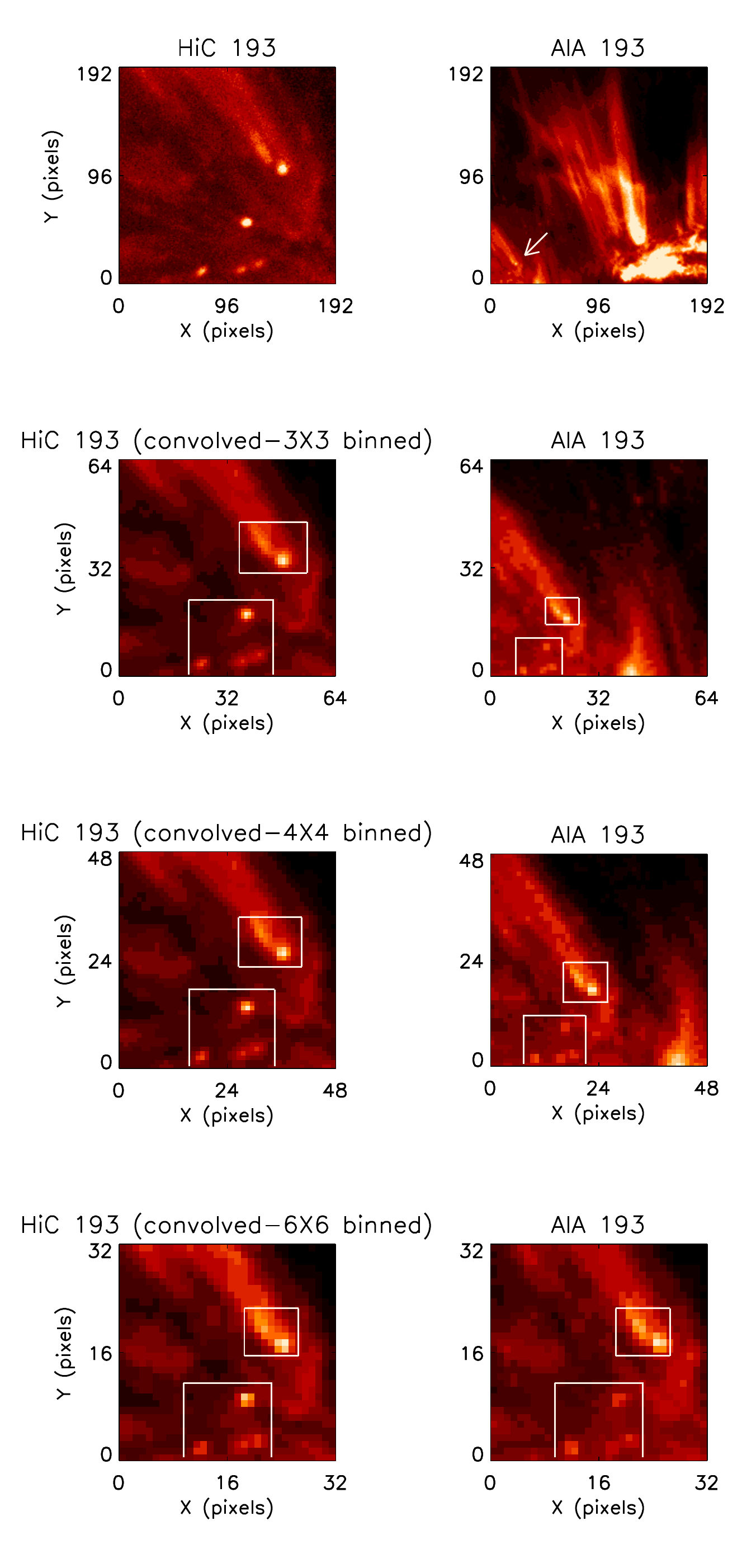}
\caption{To decide the appropriate Hi-C pixel size, we compare zoomed in Hi-C FOI images (left column) at different binning factors ({\sl top row:} no binning; {\sl first row:} $2\times2$; {\sl second row:} $4{\times}4$; {\sl last row:} $6{\times}6$) against correspondingly sized AIA images (right column). Recognisable features are marked with rectangular boxes or arrows. The $6{\times}6$ binning provides the best match between the Hi-C and AIA.} \label{f:HiC_pixsiz}
\end{figure}

The small-scale Hi-C EUV brightenings have a size comparable to the resolution limit of the AIA~193~{\AA}.  Since our work relies on identifying and matching the Hi-C events to the AIA data, an accurate pixel-wise cross-correlation is crucial to our study. For this purpose, postage-stamp FOIs (inner box in Figure~\ref{f:AIA_HiC_FOV_FOI}) that cover the fan loop system of interest were cut-out from both Hi-C images (convolved with AIA PSF) and AIA data (Figure~\ref{f:HiC_pixsiz}). The Hi-C cut-out images were binned over different pixel groups as part of trial-and-error test, to match with the AIA~193~{\AA} images, which have a spatial resolution of 1.2\arcsec. The images were binned over 3$\times$3, 4$\times$4 and 6$\times$6 pixel$^2$
(the second, third, and fourth rows of images in Figure~\ref{f:HiC_pixsiz}). The binned images were individually compared with the corresponding AIA images obtained at the same times. The location of the Hi-C FOI is marked with an arrow in the AIA image in the no-binning case (top right panel of Figure~\ref{f:HiC_pixsiz}), and recognisable features at each binning level are marked with boxes. We limit our comparisons to integer pixel-binning schemes to avoid partitioning individual pixels. The Hi-C images that were 6$\times$6 pixel binned show the best spatial match with AIA images, suggesting that one AIA~193~{\AA} resolution element corresponds to 6$\times$6 Hi-C pixels.


\section{Density and decay timescale}
\label{ap1}

As discussed in Section~\ref{s:density}, the plasma density $n_{e}$ can be computed as a root of the cubic equation (see Equation~\ref{eq:necubic}),
\begin{equation}
c_0 ~+~ c_1\,n_{e} ~+~ c_2\,n_{e}^2 ~+~ c_3\,n_{e}^3 = 0 \,,
\end{equation}
with
\begin{eqnarray}
c_0 &=& -\frac{1}{\tdecay} \nonumber \\
c_1 &=& \frac{\varLambda(T)}{3\,k_B\,T} \nonumber \\
c_2 &=& 0 \nonumber \\
c_3 &=& \frac{T^{5/2}}{4{\times}10^{-10}\,EM^2} \,,
\end{eqnarray}
where $\tdecay$ is the observed exponential decay timescale, $T$ is the plasma temperature, $\varLambda(T)$ is the radiative power from collisionally excited optically thin plasma, $EM$ is the plasma emission measure, and $k_B$ is Boltzmann's constant.

We used Maxima (v5.38.1) to solve the cubic equation.  Maxima ({\tt http://maxima.sourceforge.net}) is a computer algebra system descended from the Macsyma system.

\begin{equation}
n_{e} = \left({{\sqrt{{{27\,{\it c_0}^2\,{\it c_3}+4\,{\it c_1}^3 }\over{{\it c_3}}}}}\over{2\,3^{{{3}\over{2}}}\,{\it c_3}}}-{{ {\it c_0}}\over{2\,{\it c_3}}}\right)^{{{1}\over{3}}} -{{{\it c_1} }\over{3\,{\it c_3}\,\left({{\sqrt{{{27\,{\it c_0}^2\,{\it c_3}+4\, {\it c_1}^3}\over{{\it c_3}}}}}\over{2\,3^{{{3}\over{2}}}\,{\it c_3} }}-{{{\it c_0}}\over{2\,{\it c_3}}}\right)^{{{1}\over{3}}}}} \,.
\end{equation}

A cubic equation has three roots, but only one that is certain to be real, and we report this number. In all cases, 
we find that neither of the other solutions are near the real line.
Note that this solution is strictly applicable only to isothermal plasma.

However, the DEM-weighted average $T$ (see Table~\ref{t:properties}) has error bars of size $\approx{10}\%$, and we thus expect that the systematic error made in integrating over the temperature range is ignorable.
We also provide a measure of the stability of $n_{e}$ by propagating the errors in $\log{T}$, $EM$, and $\tdecay$.
We use Monte Carlo simulations to generate 1000 Gaussian random deviates for each quantity and recompute $n_{e}$; the resulting standard deviations are given as error bars in Table~\ref{t:density}.
We find {\sl post hoc} that their effect on the density estimate is small.

\section{Light curves and DEMs}
\label{ap2}

Here we show the AIA and Hi-C light curves (unaltered, convolved-binned, and convolved-binned-corrected) as in Figure~\ref{f:AIAvHiC_7_26_CBC}, and corresponding emission measure loci, DEM solutions, and the expected temperature dependence of 193~\AA\ fluxes as in Figure~\ref{f:AIADEM_7_26} for all of the detected events (except for Br7 and Br26, which are shown in Figures~\ref{f:brlc_7_26},\ref{f:AIAvHiC_7_26_CBC}, and \ref{f:AIADEM_7_26}).

\begin{figure*}[htp!]
\begin{center}
\hspace{-0.57cm}
\includegraphics[trim=0cm 0cm 0cm 0cm, clip=true,width=0.25\textwidth]{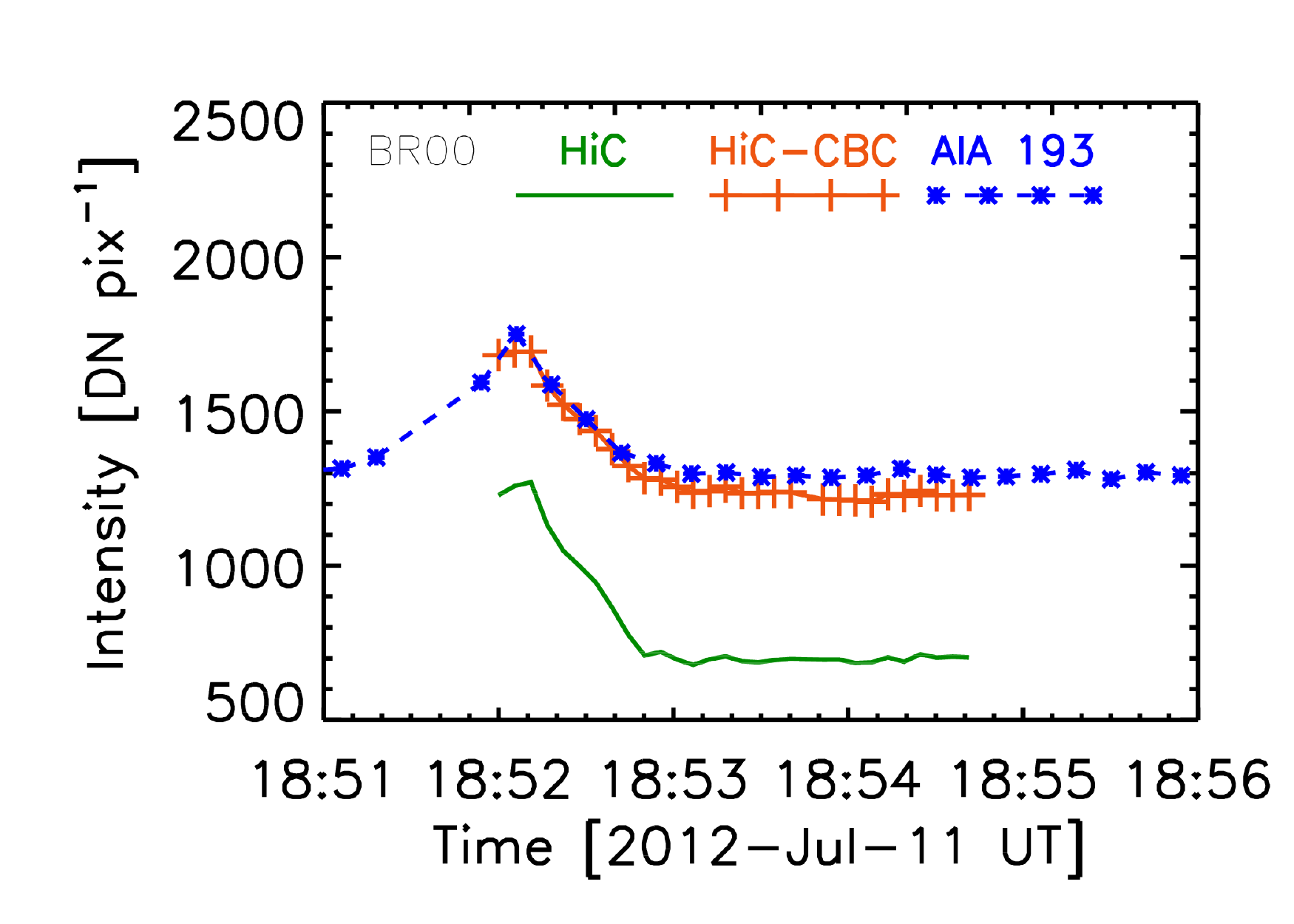}
\includegraphics[trim=0cm 0cm 0cm 0cm, clip=true,width=0.25\textwidth]{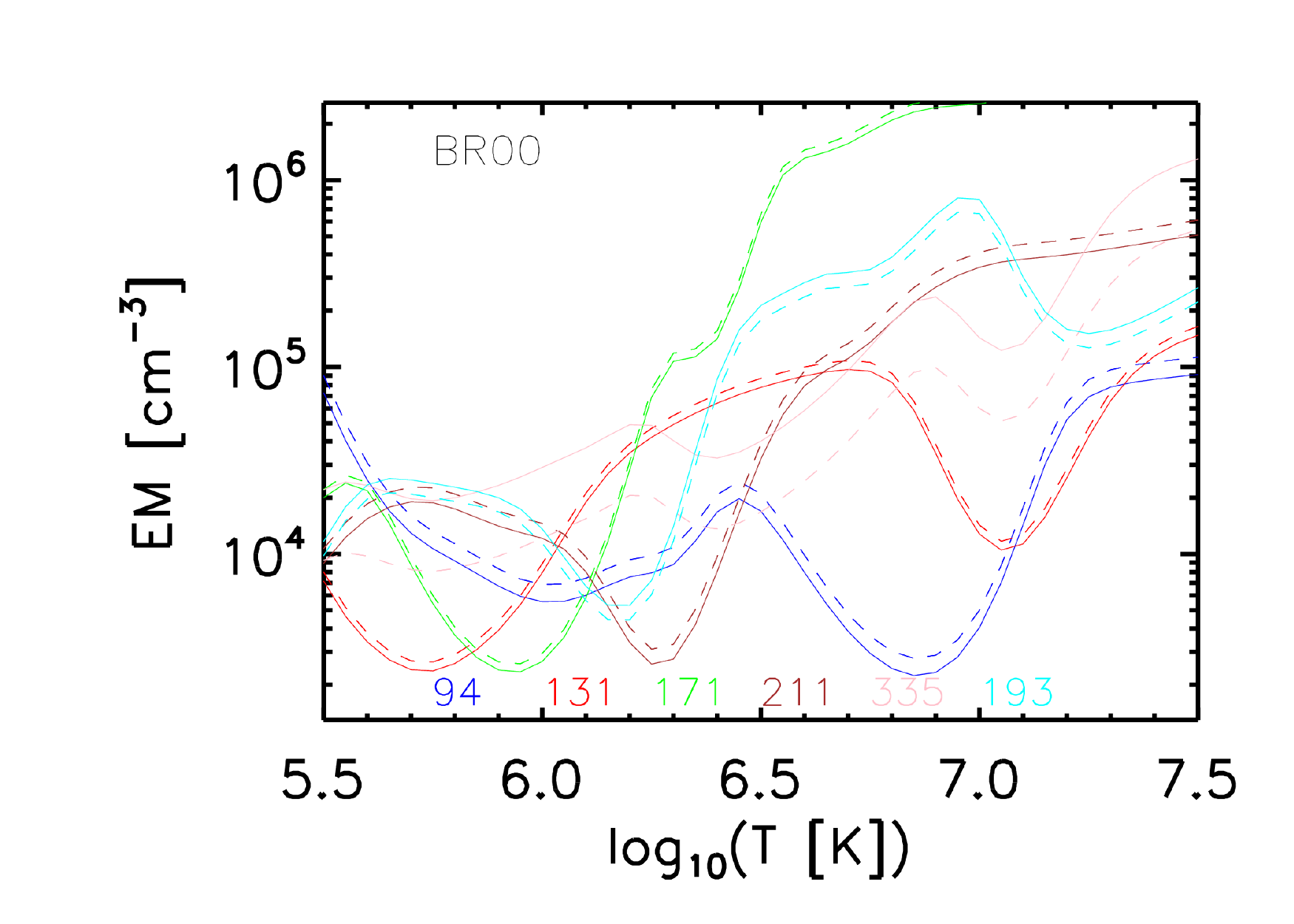}
\includegraphics[trim=0cm 0cm 0cm 0cm, clip=true,width=0.25\textwidth]{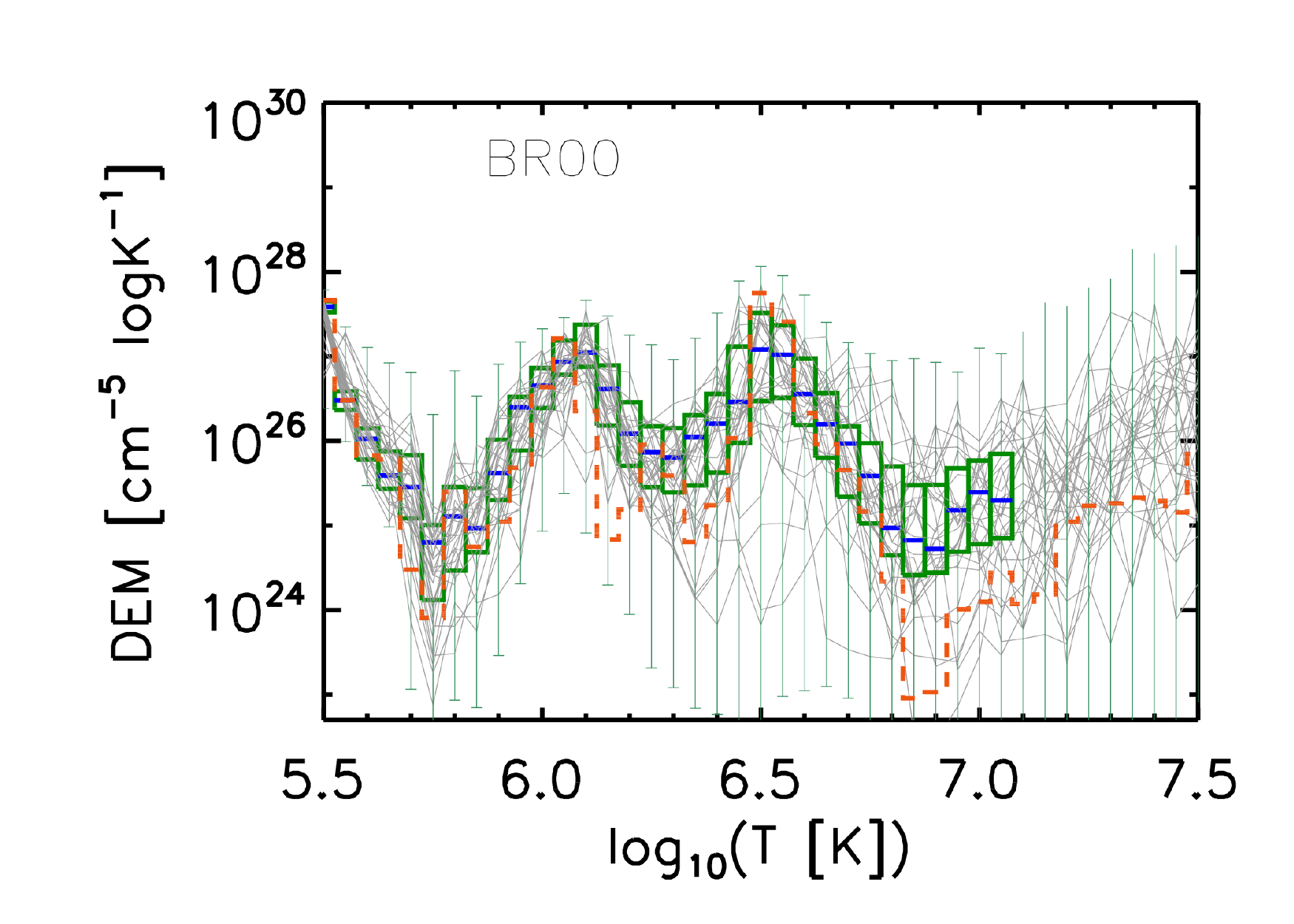}
\includegraphics[trim=0cm 0cm 0cm 0cm, clip=true,width=0.25\textwidth]{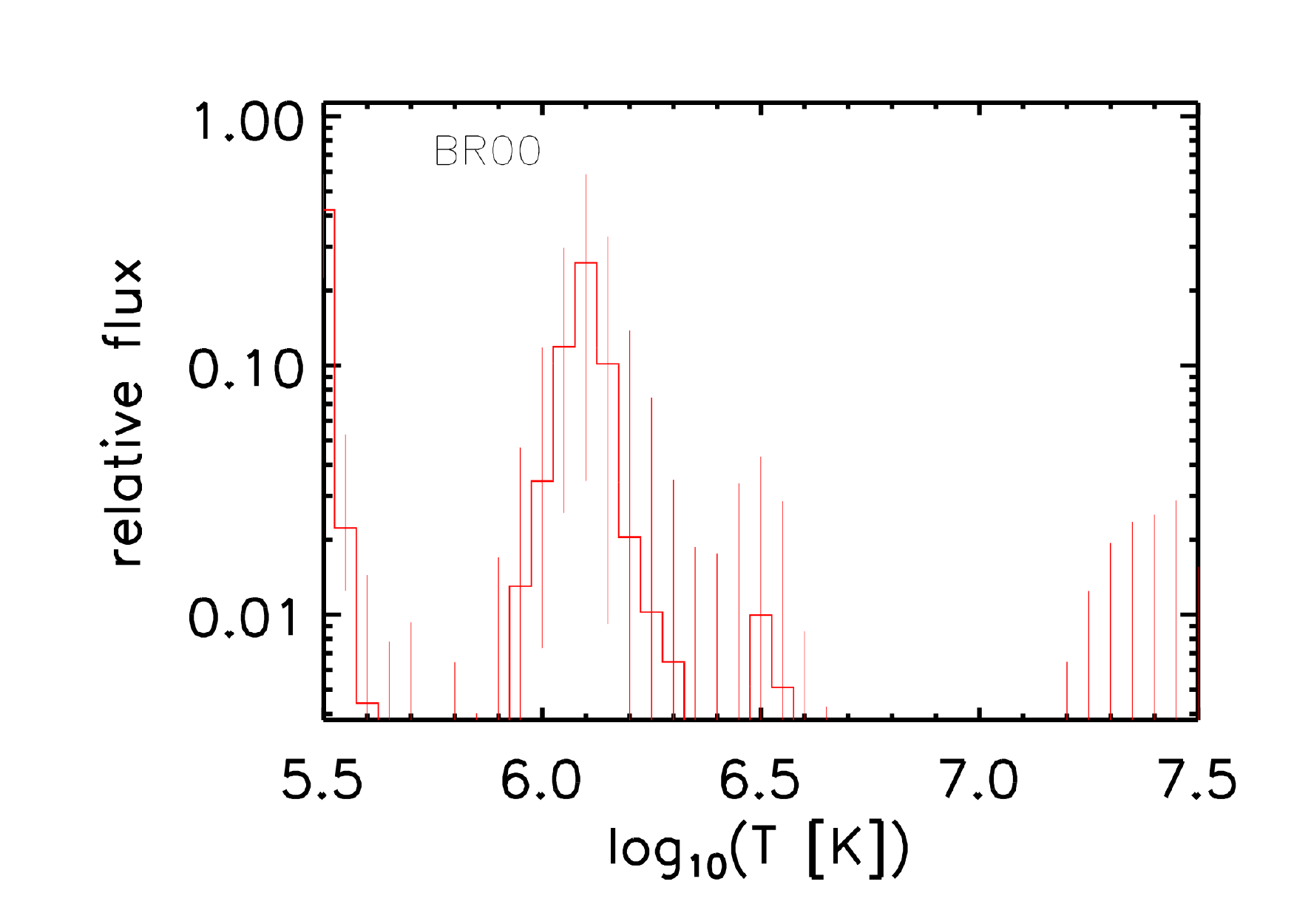}\\
\hspace{-0.57cm}
\includegraphics[trim=0cm 0cm 0cm 0cm, clip=true,width=0.25\textwidth]{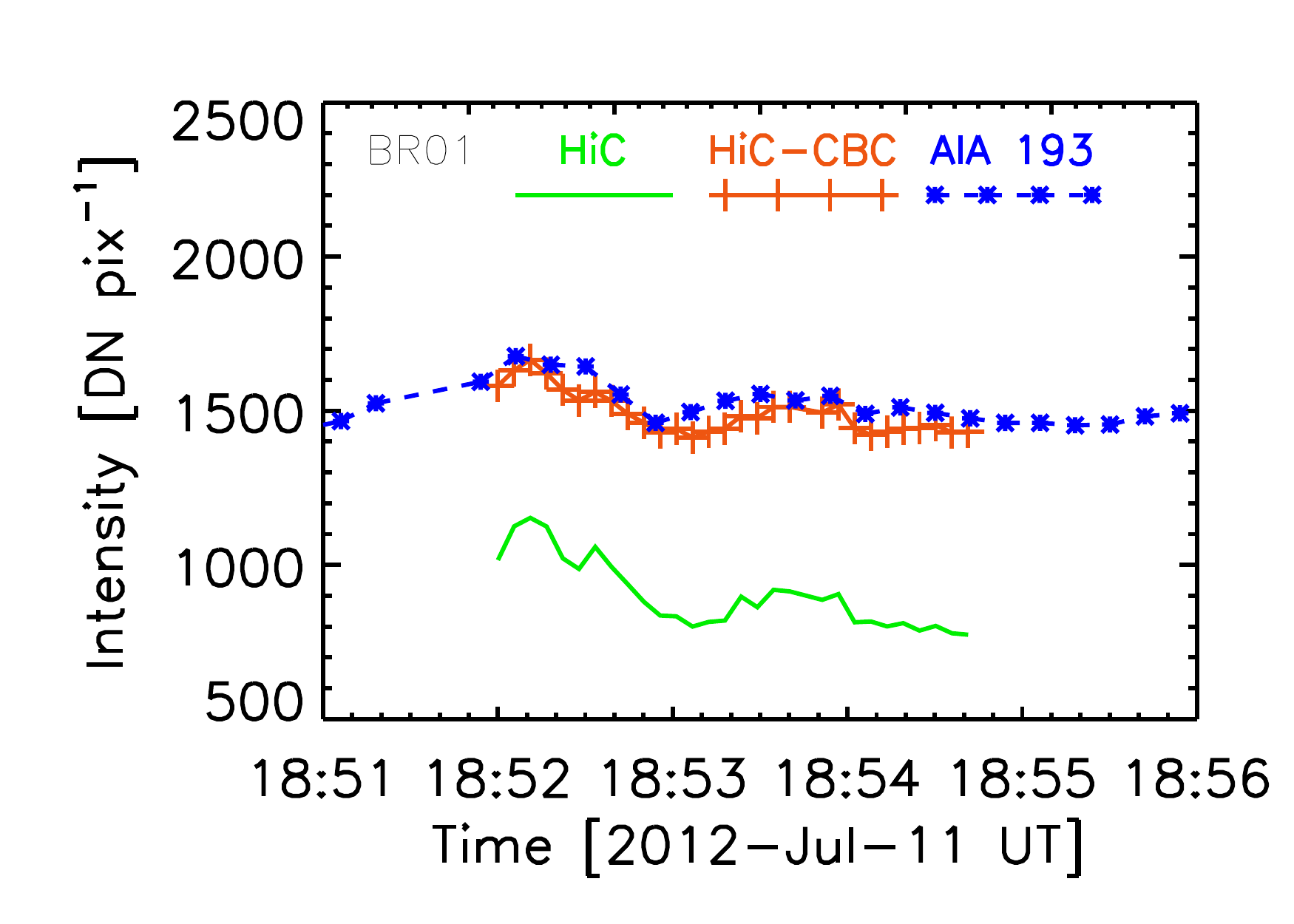}
\includegraphics[trim=0cm 0cm 0cm 0cm, clip=true,width=0.25\textwidth]{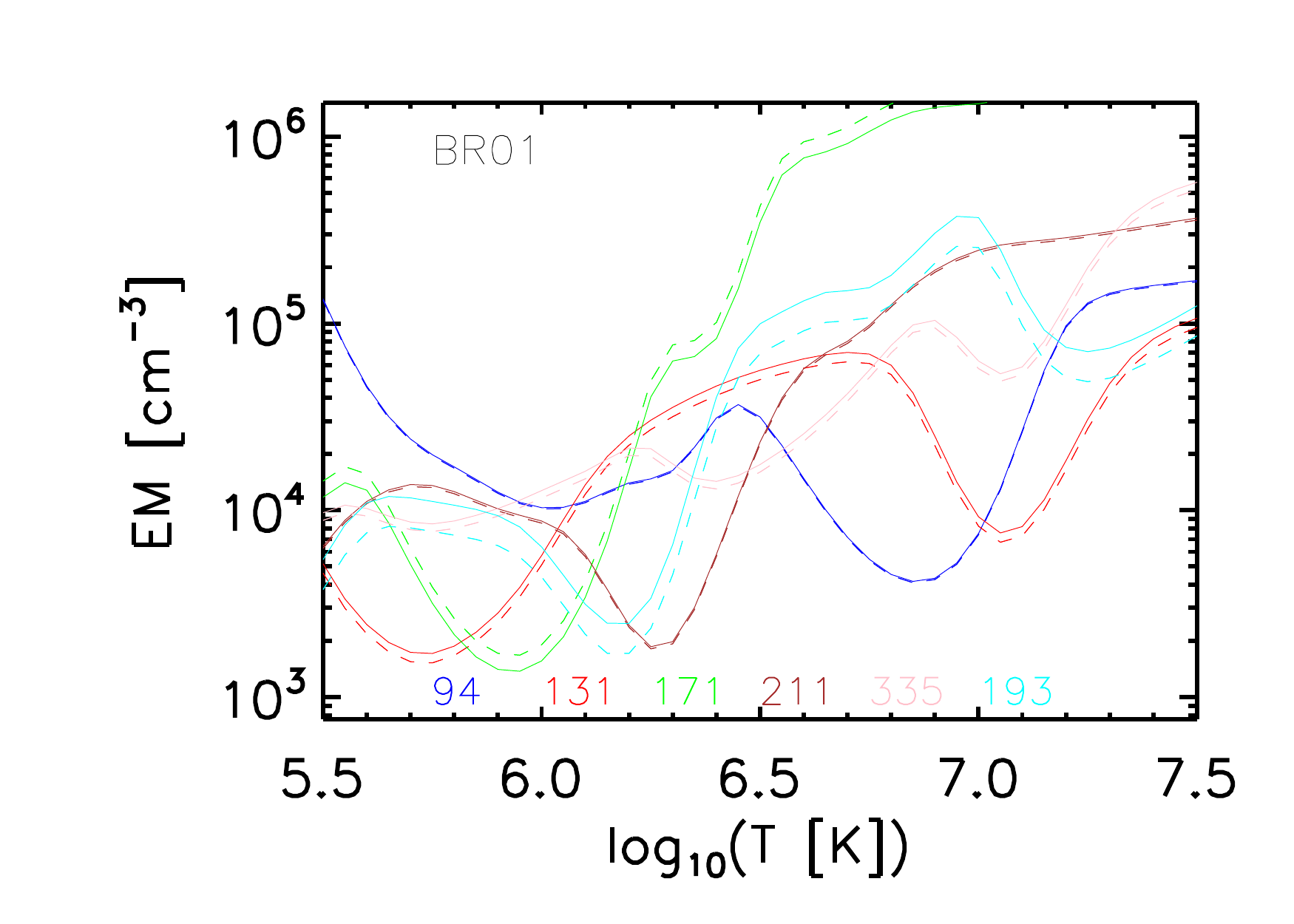}
\includegraphics[trim=0cm 0cm 0cm 0cm, clip=true,width=0.25\textwidth]{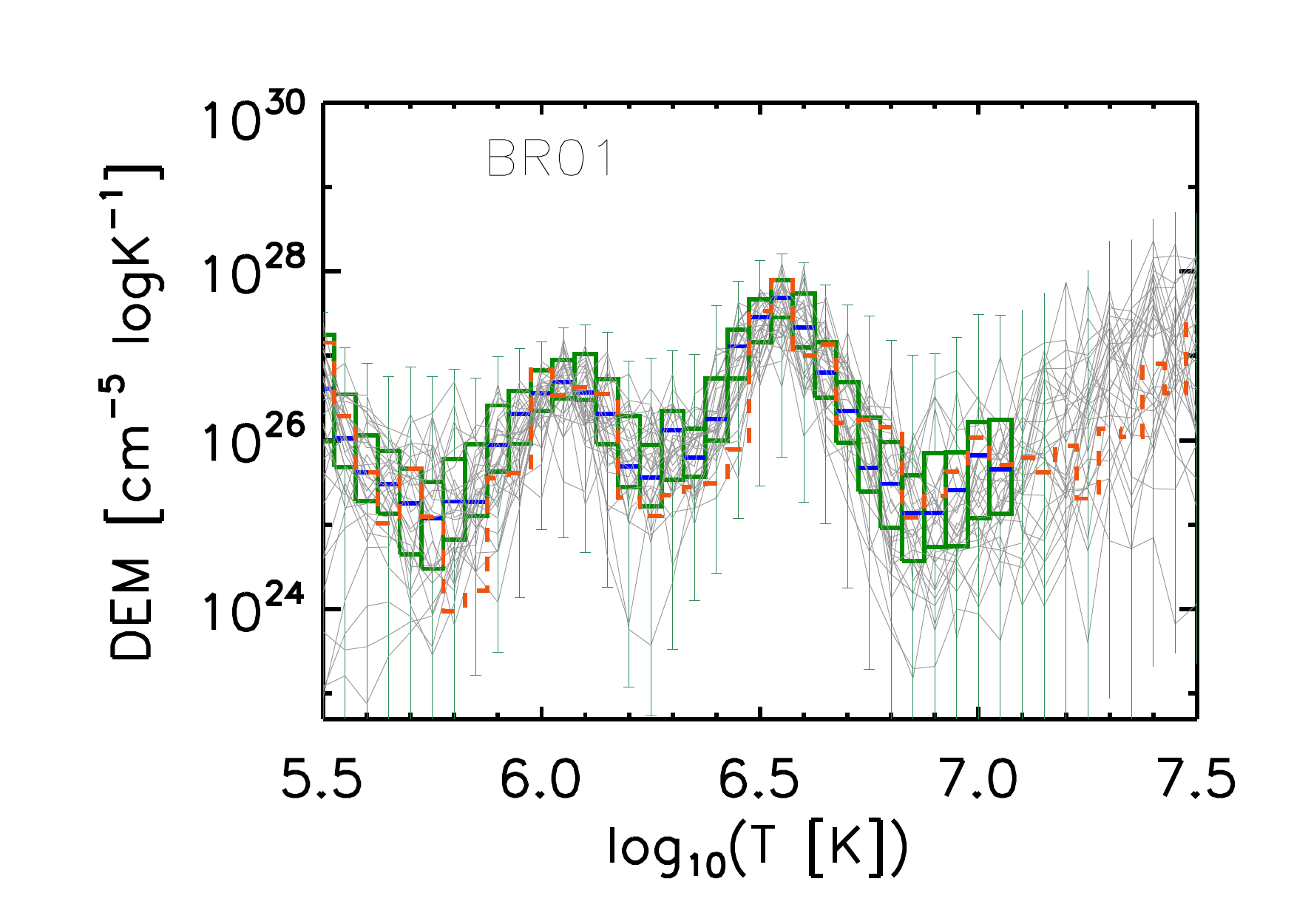}
\includegraphics[trim=0cm 0cm 0cm 0cm, clip=true,width=0.25\textwidth]{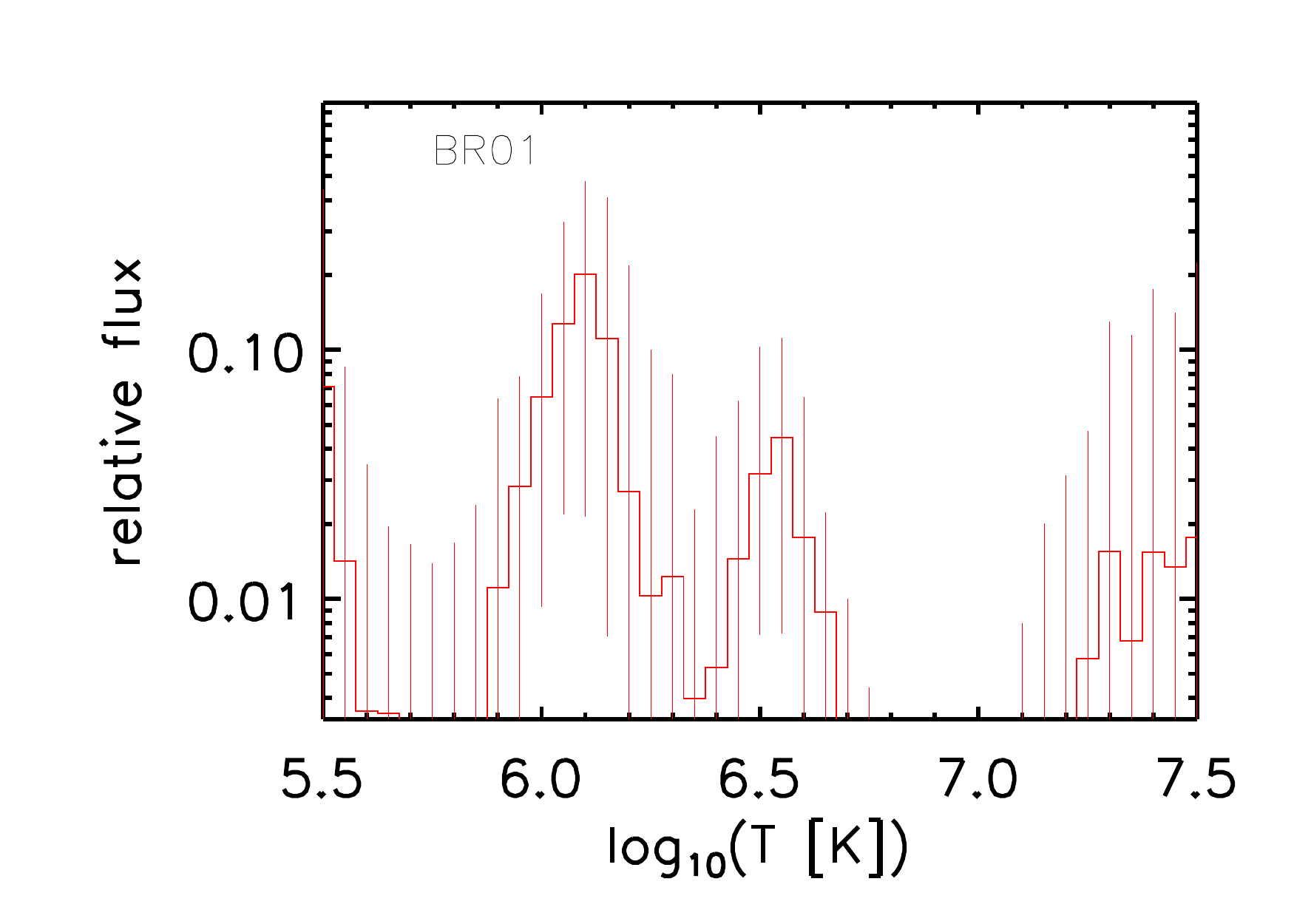}\\
\hspace{-0.57cm}
\includegraphics[trim=0cm 0cm 0cm 0cm, clip=true,width=0.25\textwidth]{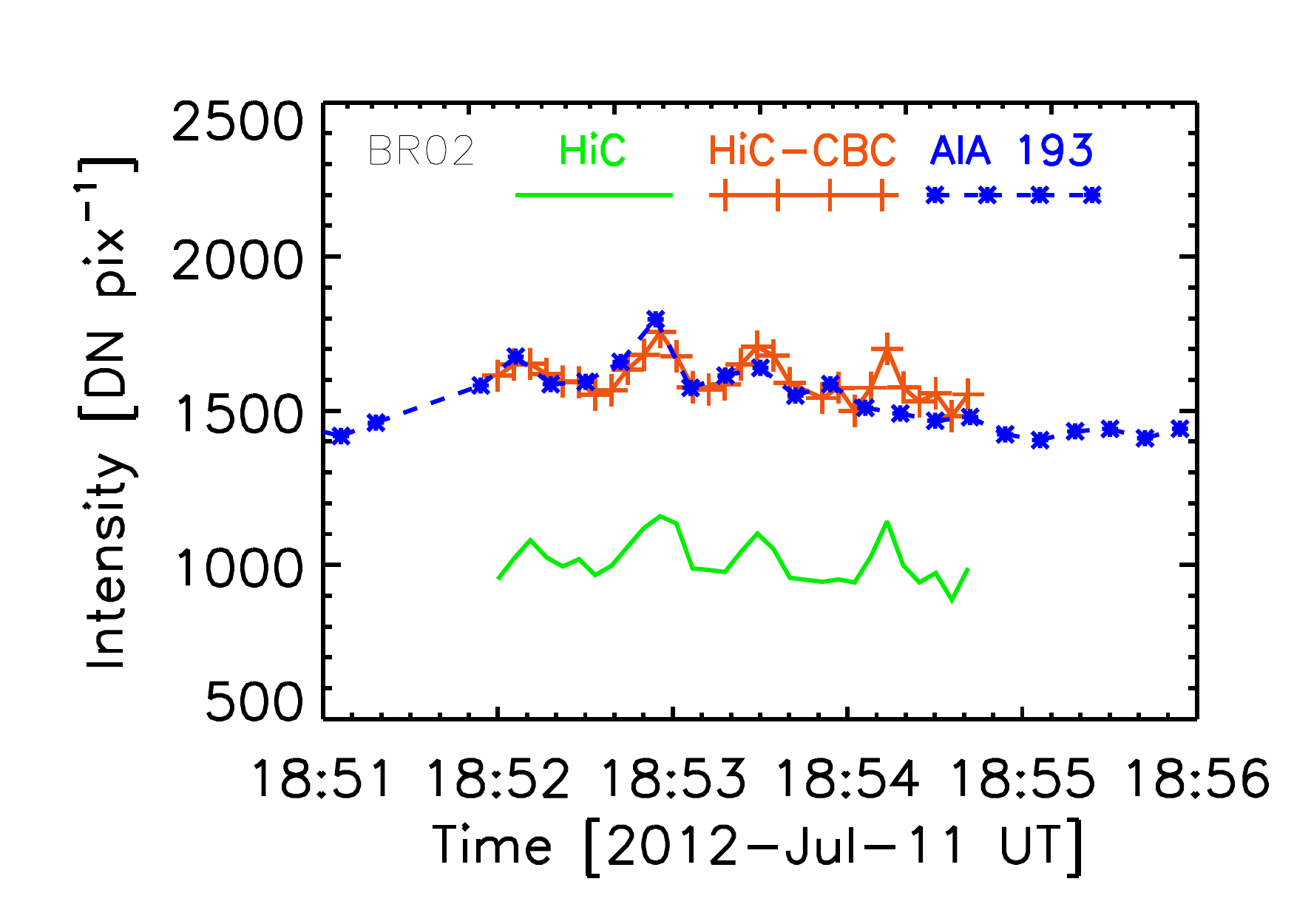}
\includegraphics[trim=0cm 0cm 0cm 0cm, clip=true,width=0.25\textwidth]{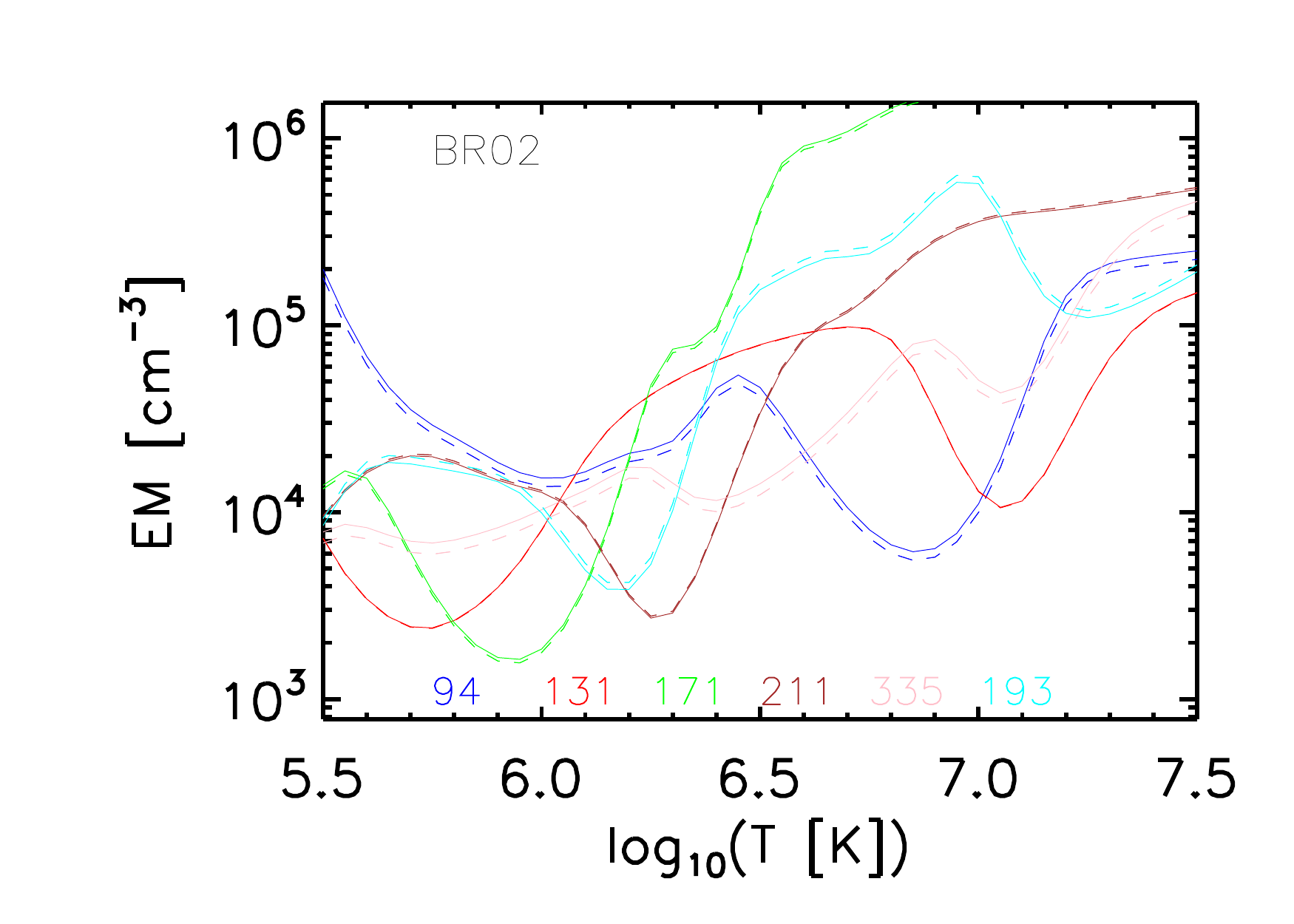}
\includegraphics[trim=0cm 0cm 0cm 0cm, clip=true,width=0.25\textwidth]{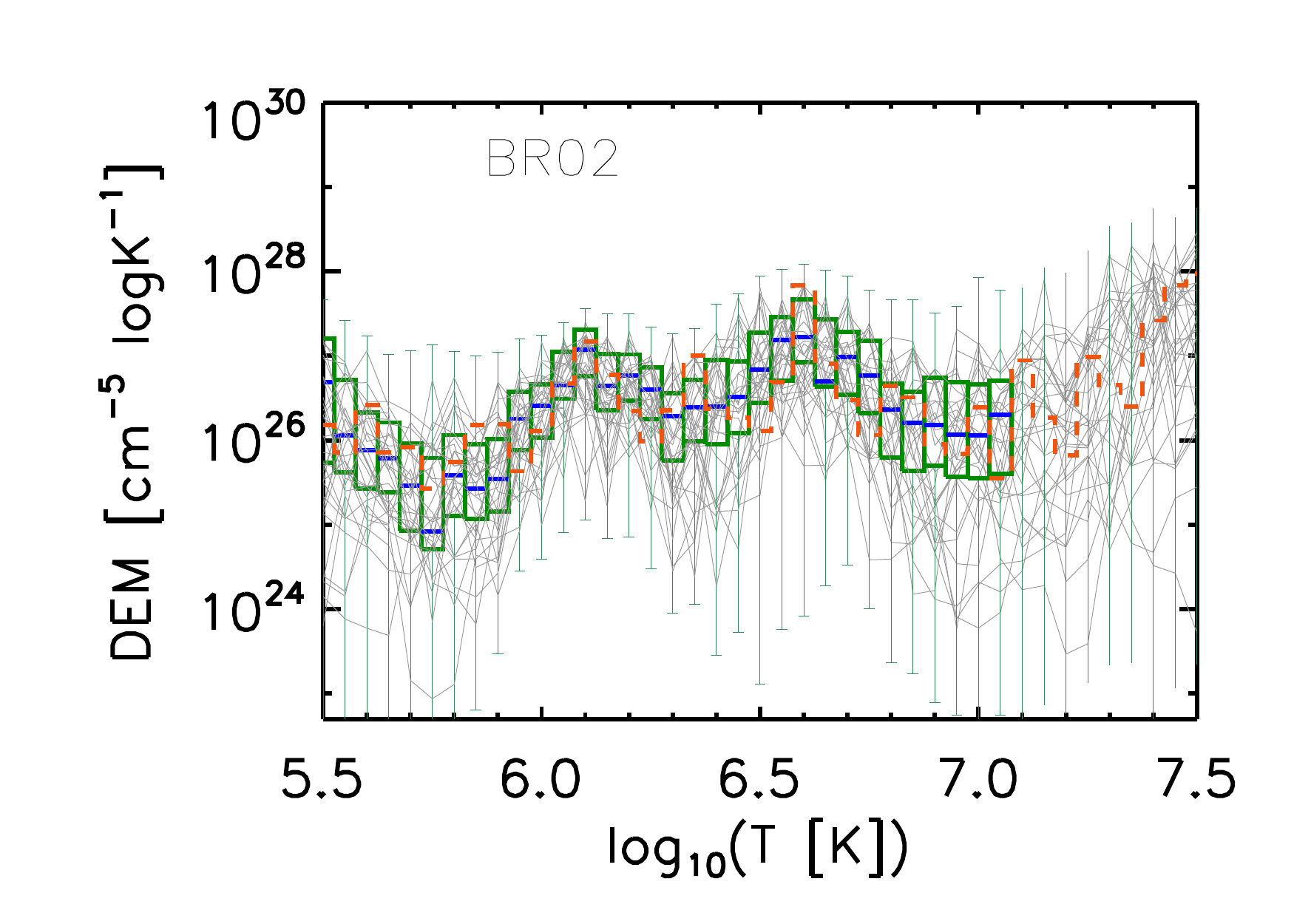}
\includegraphics[trim=0cm 0cm 0cm 0cm, clip=true,width=0.25\textwidth]{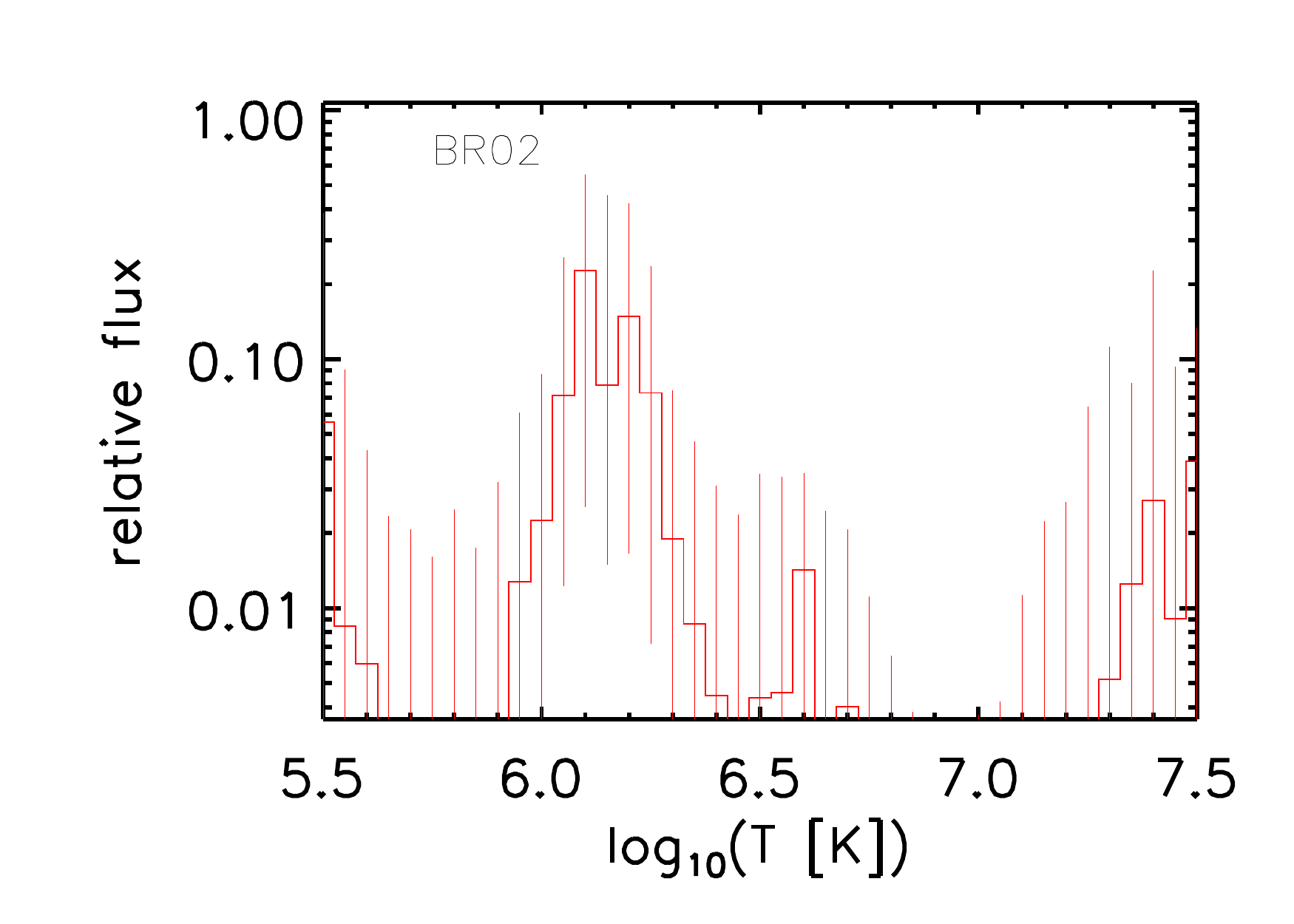}\\
\hspace{-0.57cm}
\includegraphics[trim=0cm 0cm 0cm 0cm, clip=true,width=0.25\textwidth]{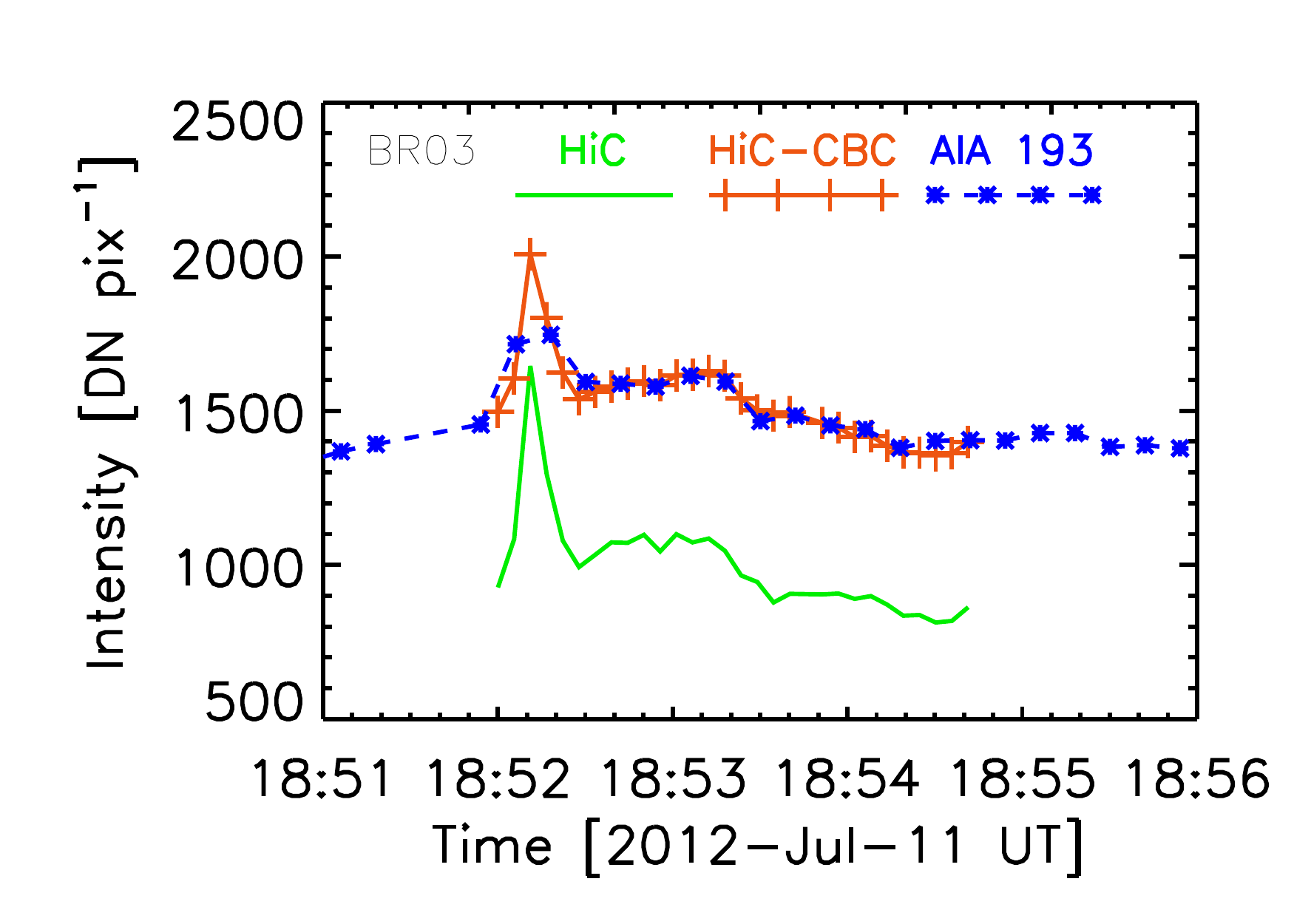}
\includegraphics[trim=0cm 0cm 0cm 0cm, clip=true,width=0.25\textwidth]{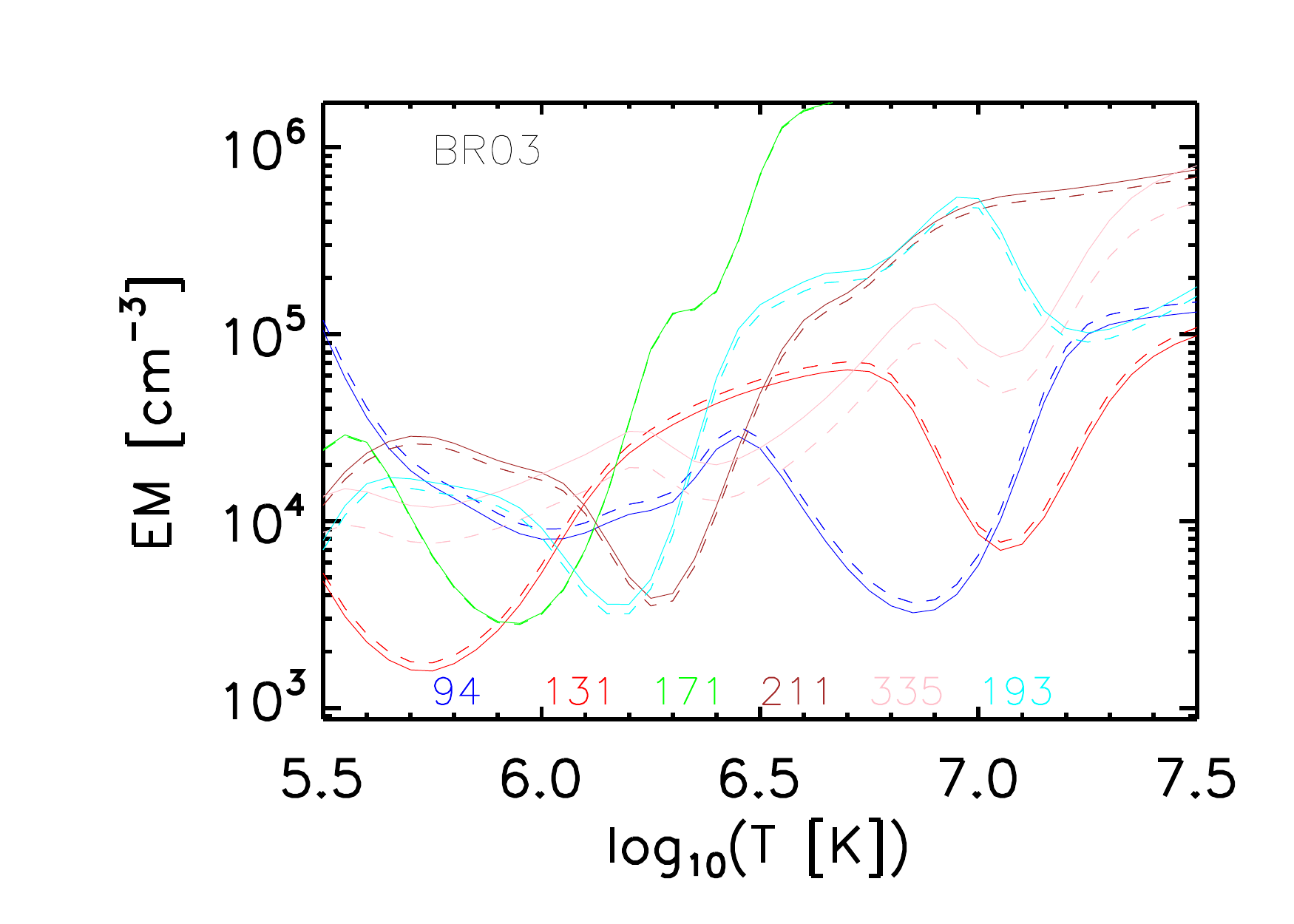}
\includegraphics[trim=0cm 0cm 0cm 0cm, clip=true,width=0.25\textwidth]{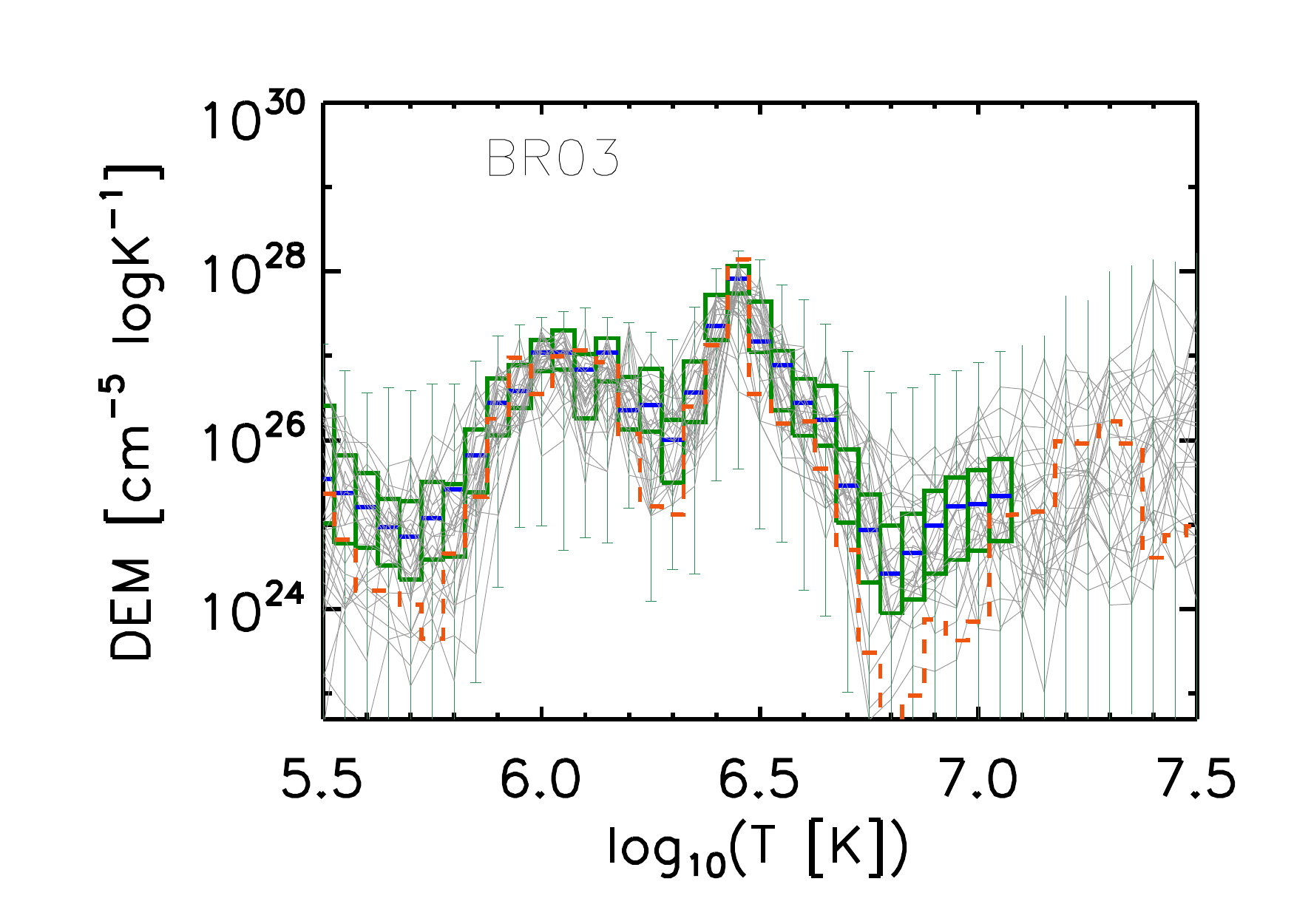}
\includegraphics[trim=0cm 0cm 0cm 0cm, clip=true,width=0.25\textwidth]{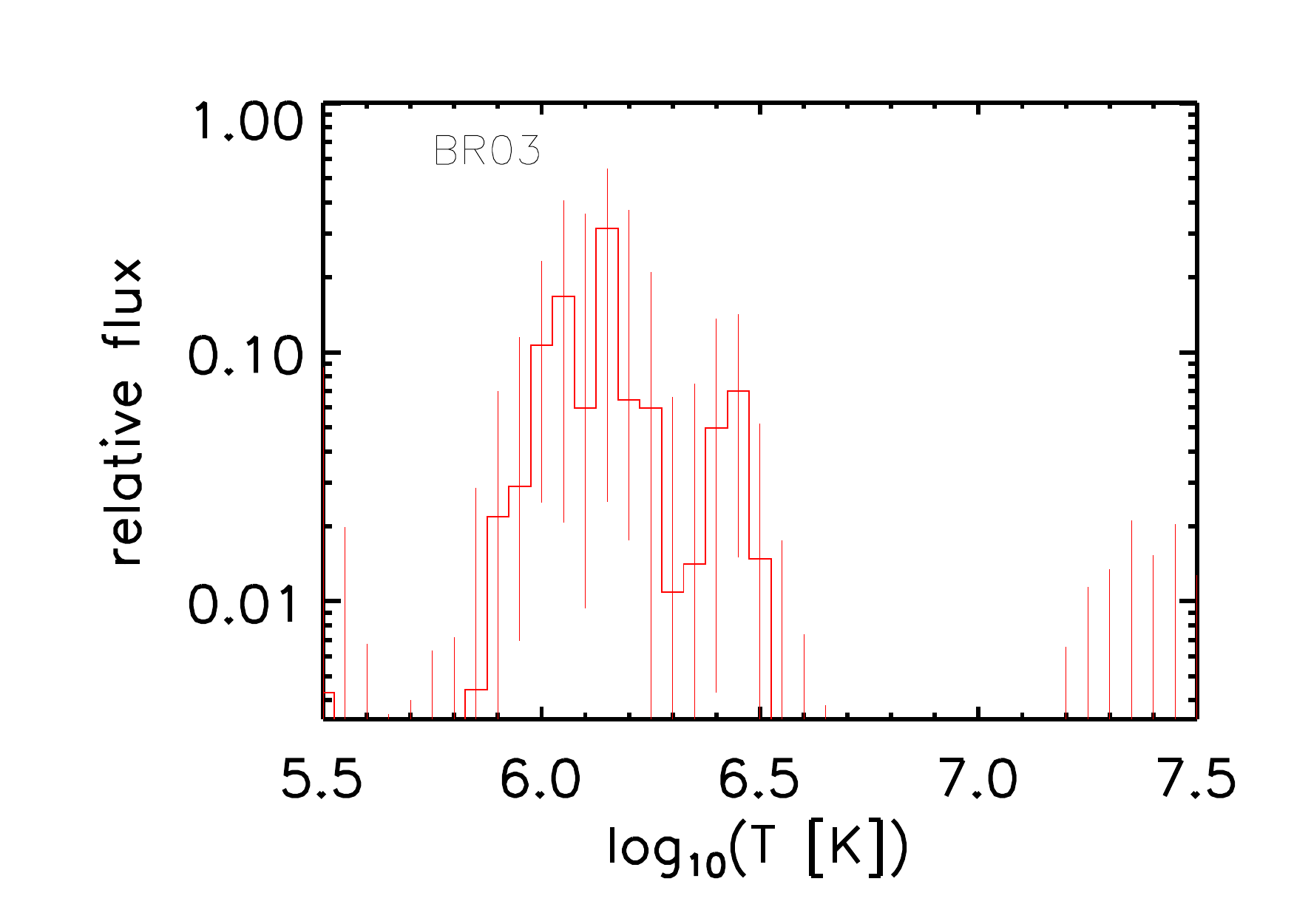}\\
\hspace{-0.57cm}
\includegraphics[trim=0cm 0cm 0cm 0cm, clip=true,width=0.25\textwidth]{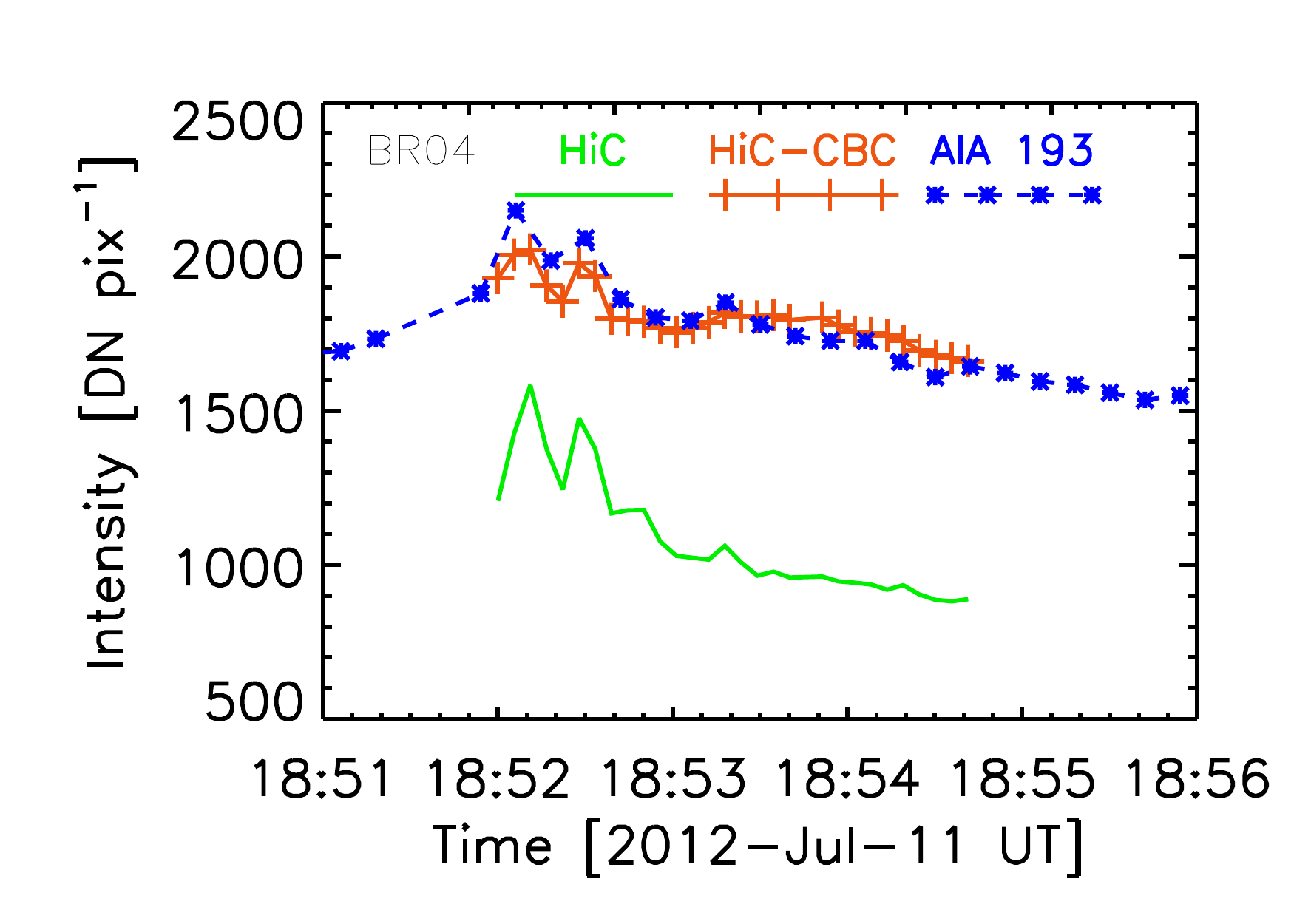}
\includegraphics[trim=0cm 0cm 0cm 0cm, clip=true,width=0.25\textwidth]{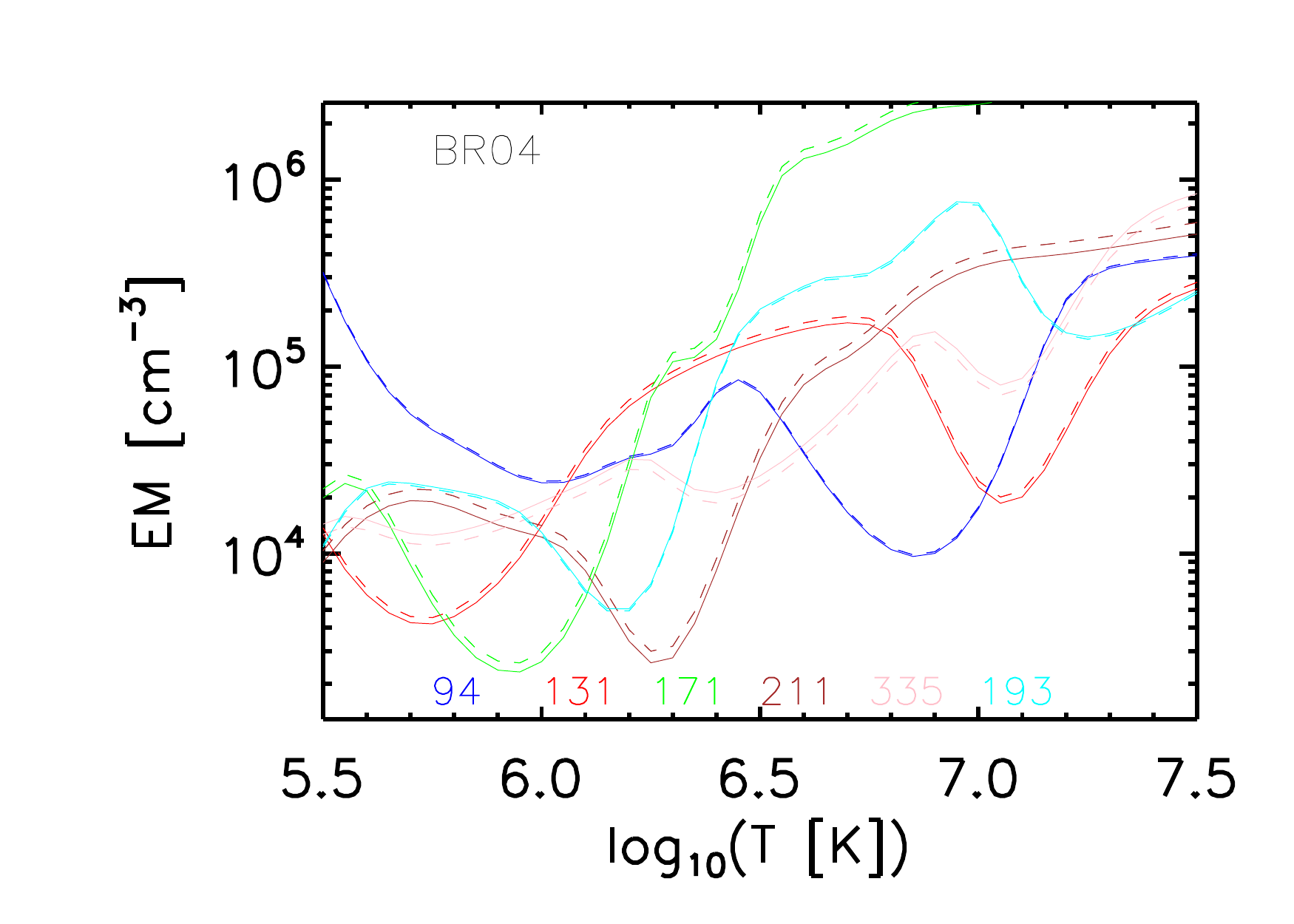}
\includegraphics[trim=0cm 0cm 0cm 0cm, clip=true,width=0.25\textwidth]{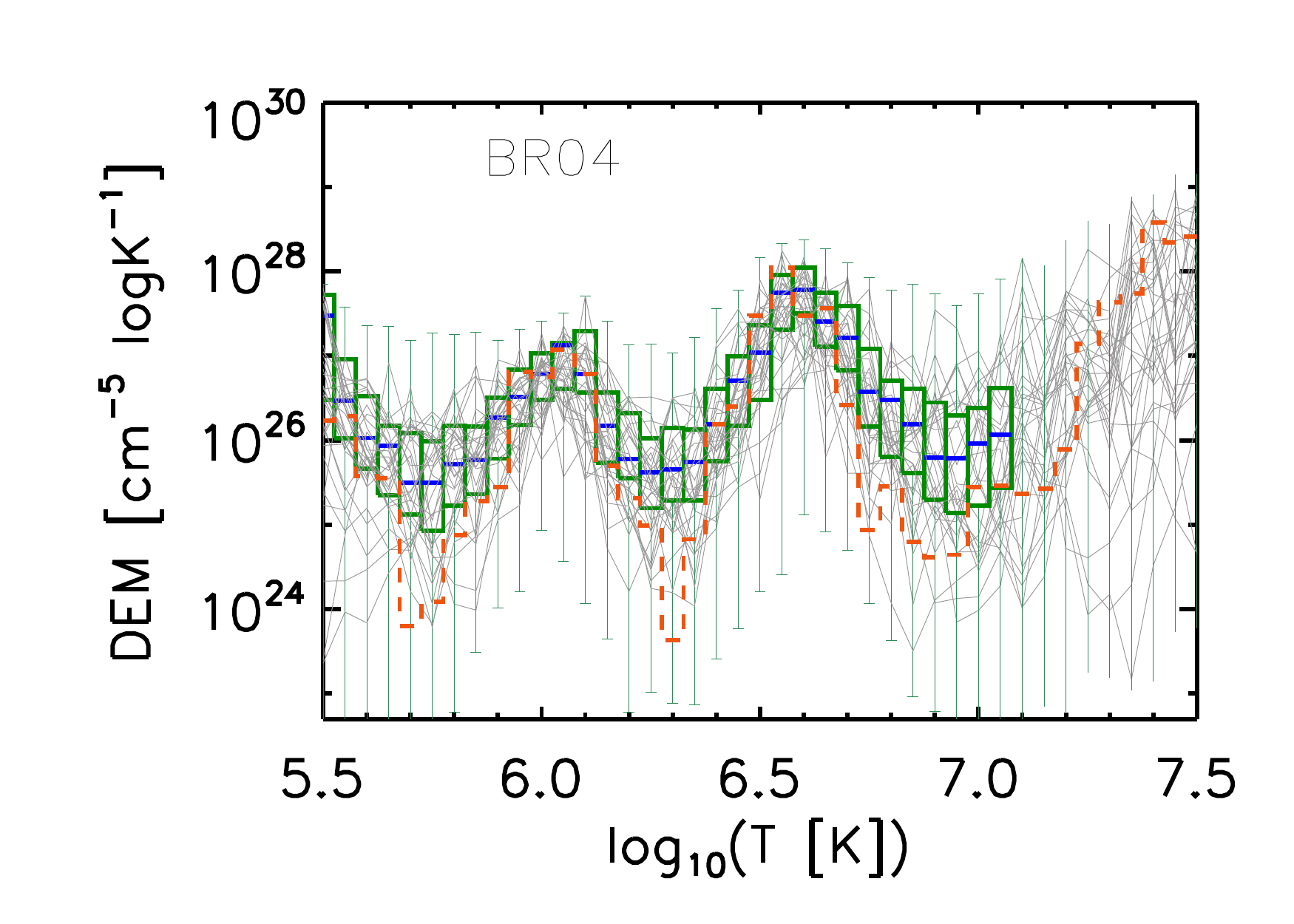}
\includegraphics[trim=0cm 0cm 0cm 0cm, clip=true,width=0.25\textwidth]{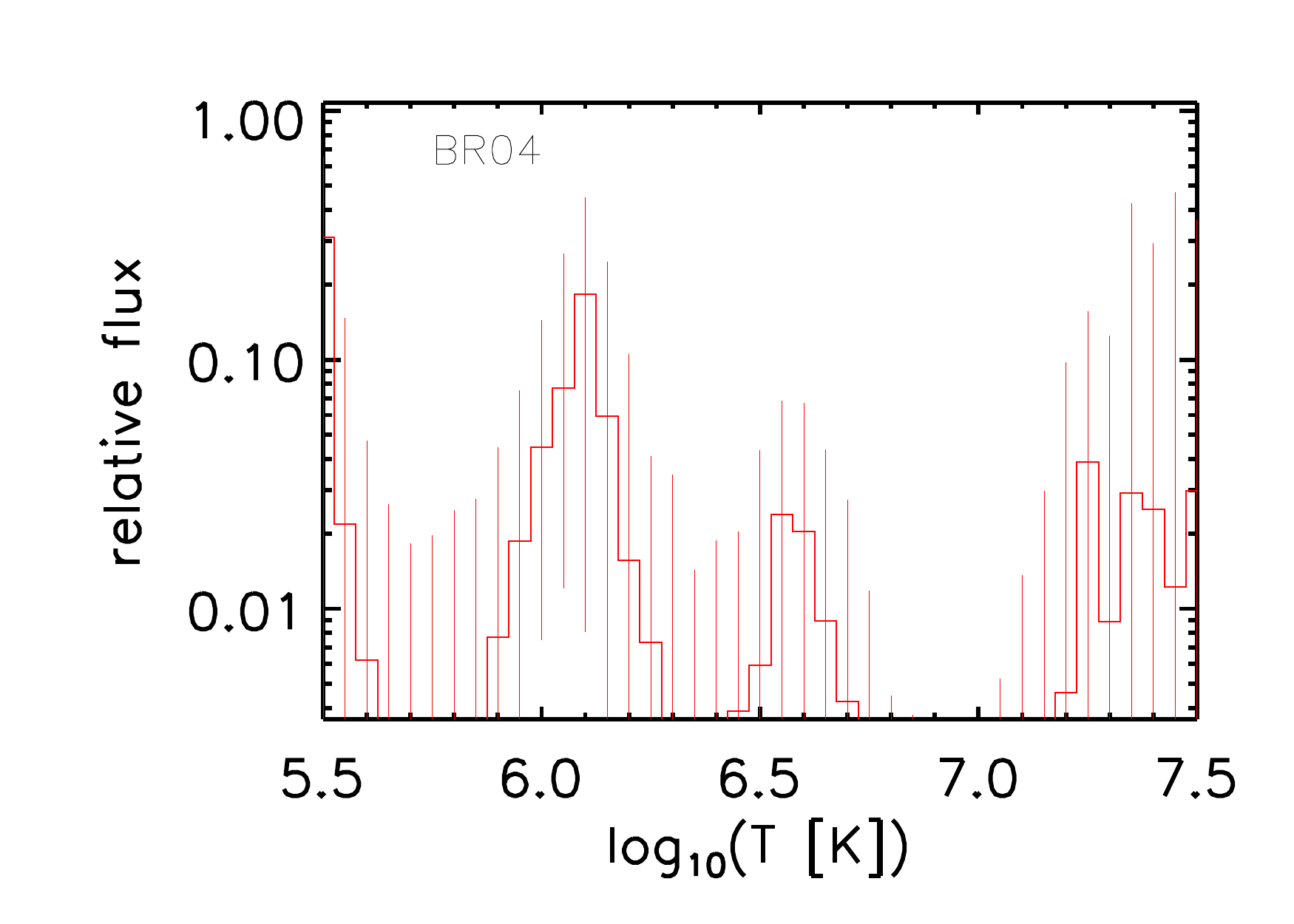}\\
\hspace{-0.57cm}
\includegraphics[trim=0cm 0cm 0cm 0cm, clip=true,width=0.25\textwidth]{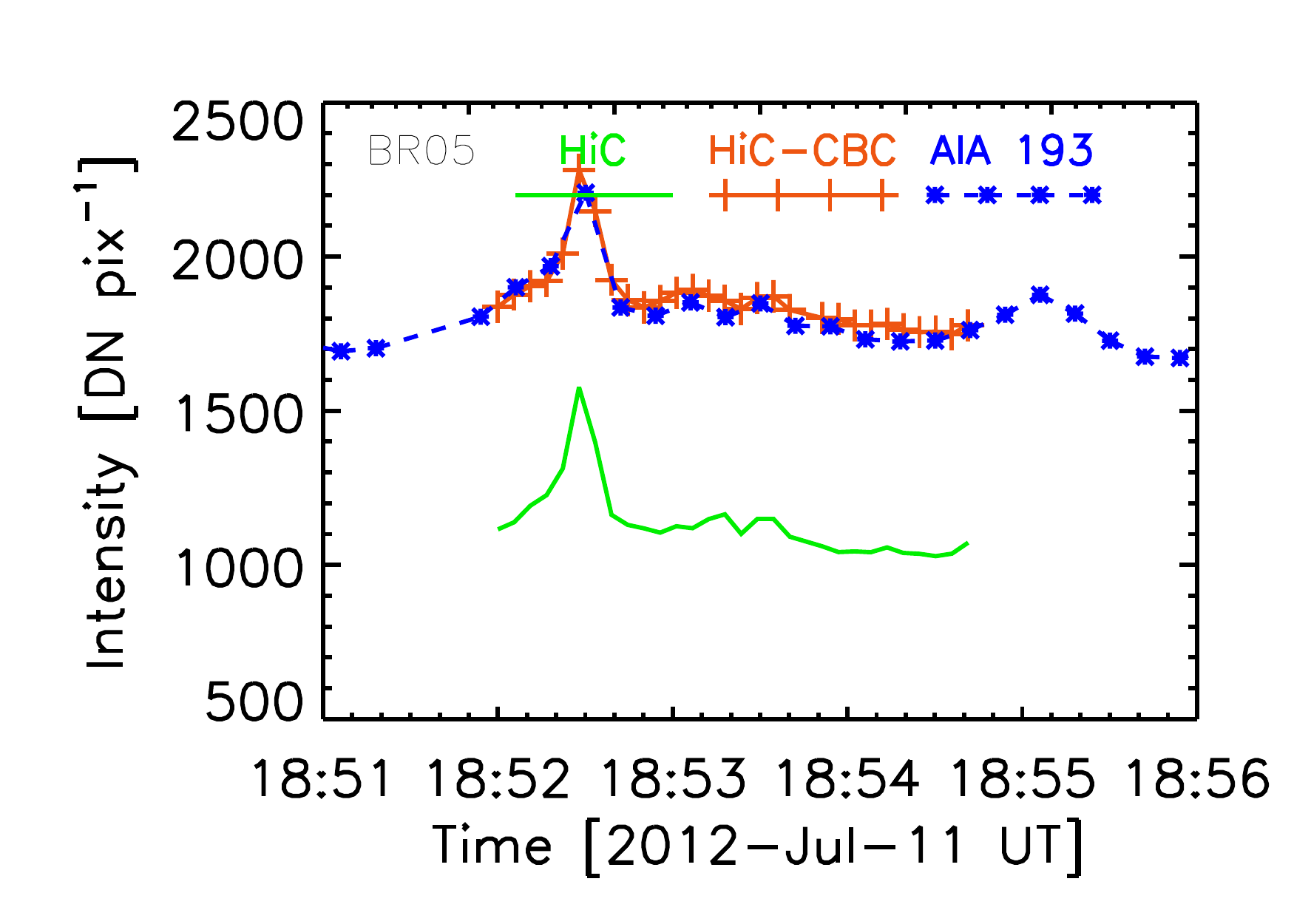}
\includegraphics[trim=0cm 0cm 0cm 0cm, clip=true,width=0.25\textwidth]{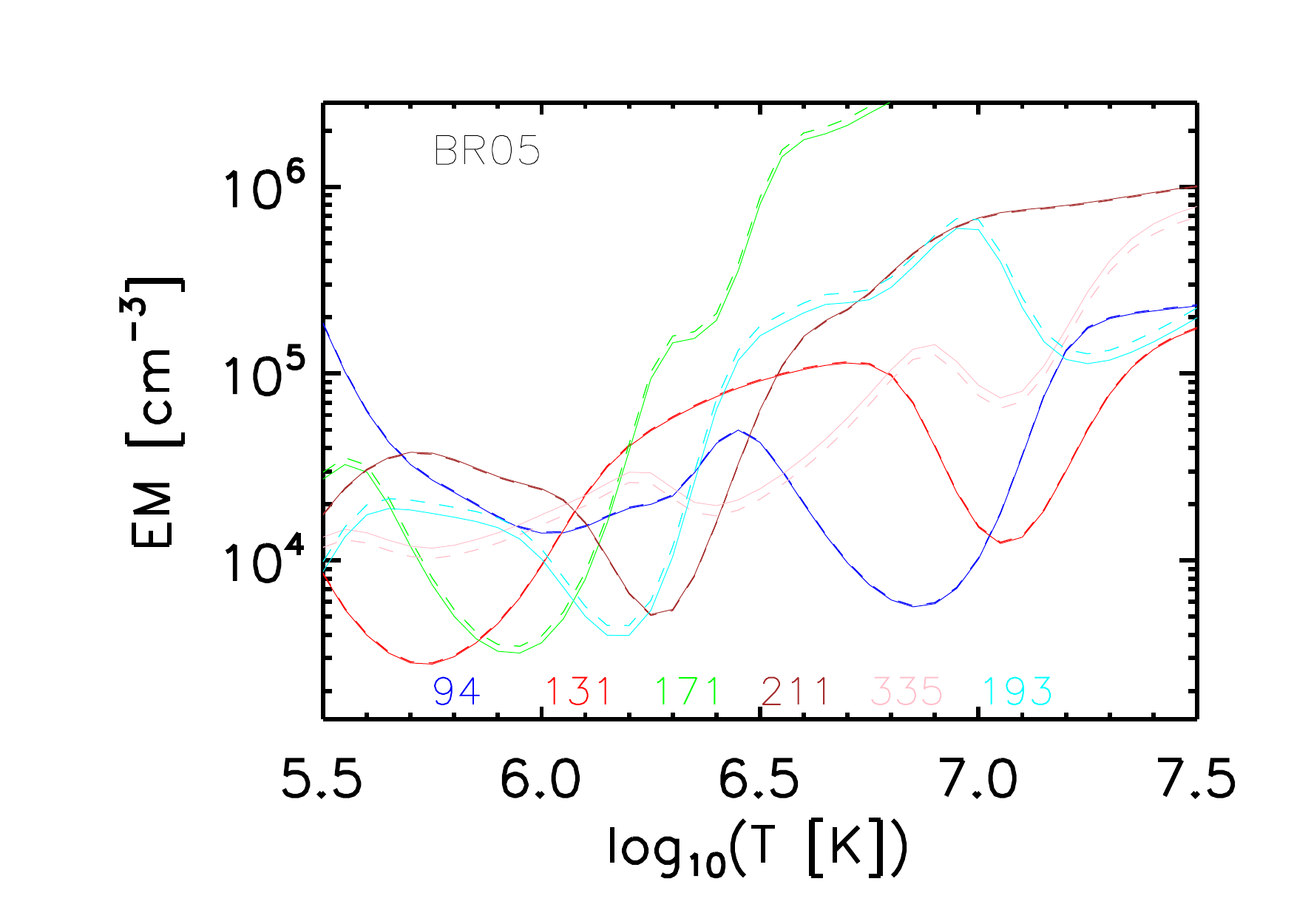}
\includegraphics[trim=0cm 0cm 0cm 0cm, clip=true,width=0.25\textwidth]{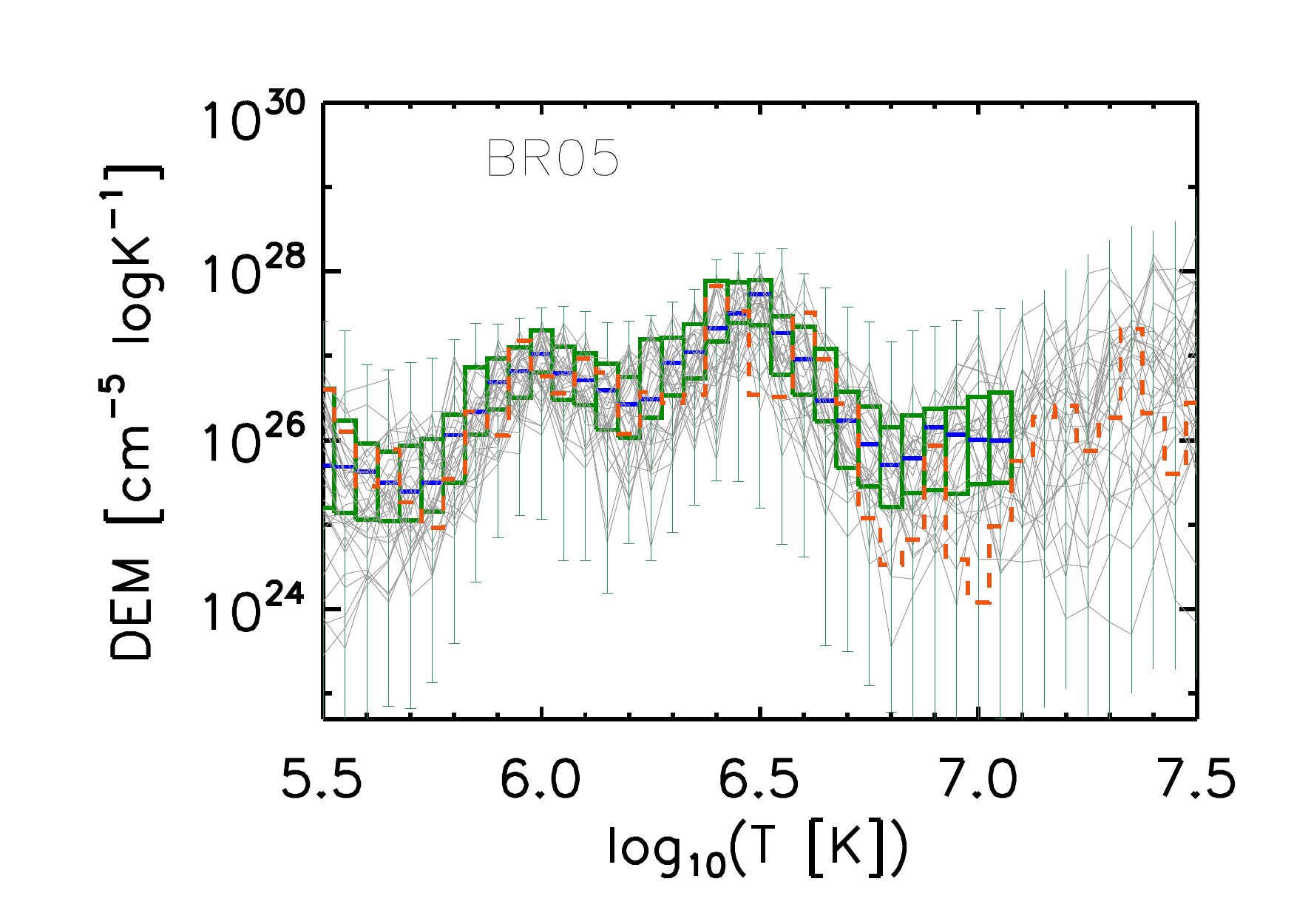}
\includegraphics[trim=0cm 0cm 0cm 0cm, clip=true,width=0.25\textwidth]{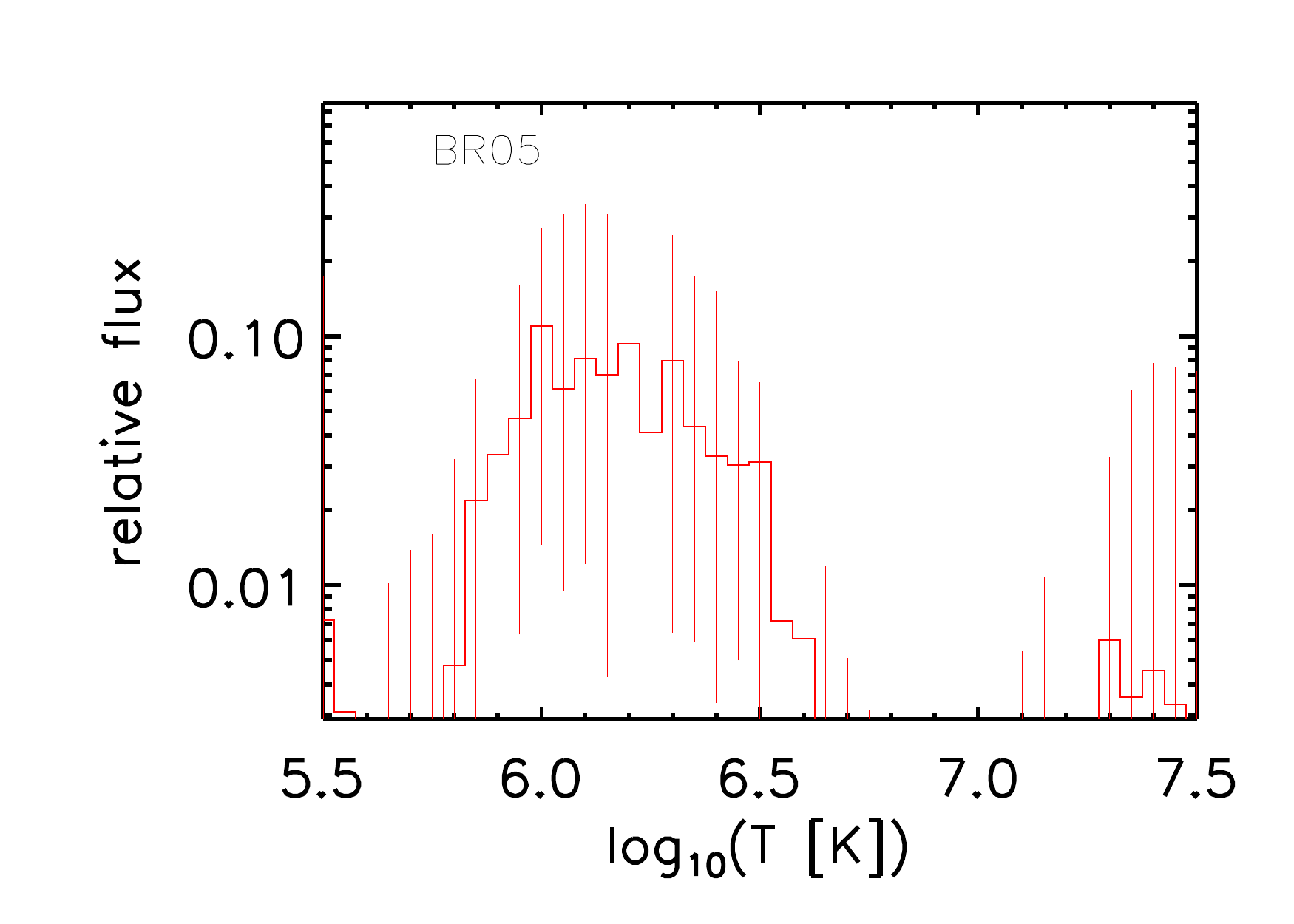}\\
\hspace{-0.57cm}
\includegraphics[trim=0cm 0cm 0cm 0cm, clip=true,width=0.25\textwidth]{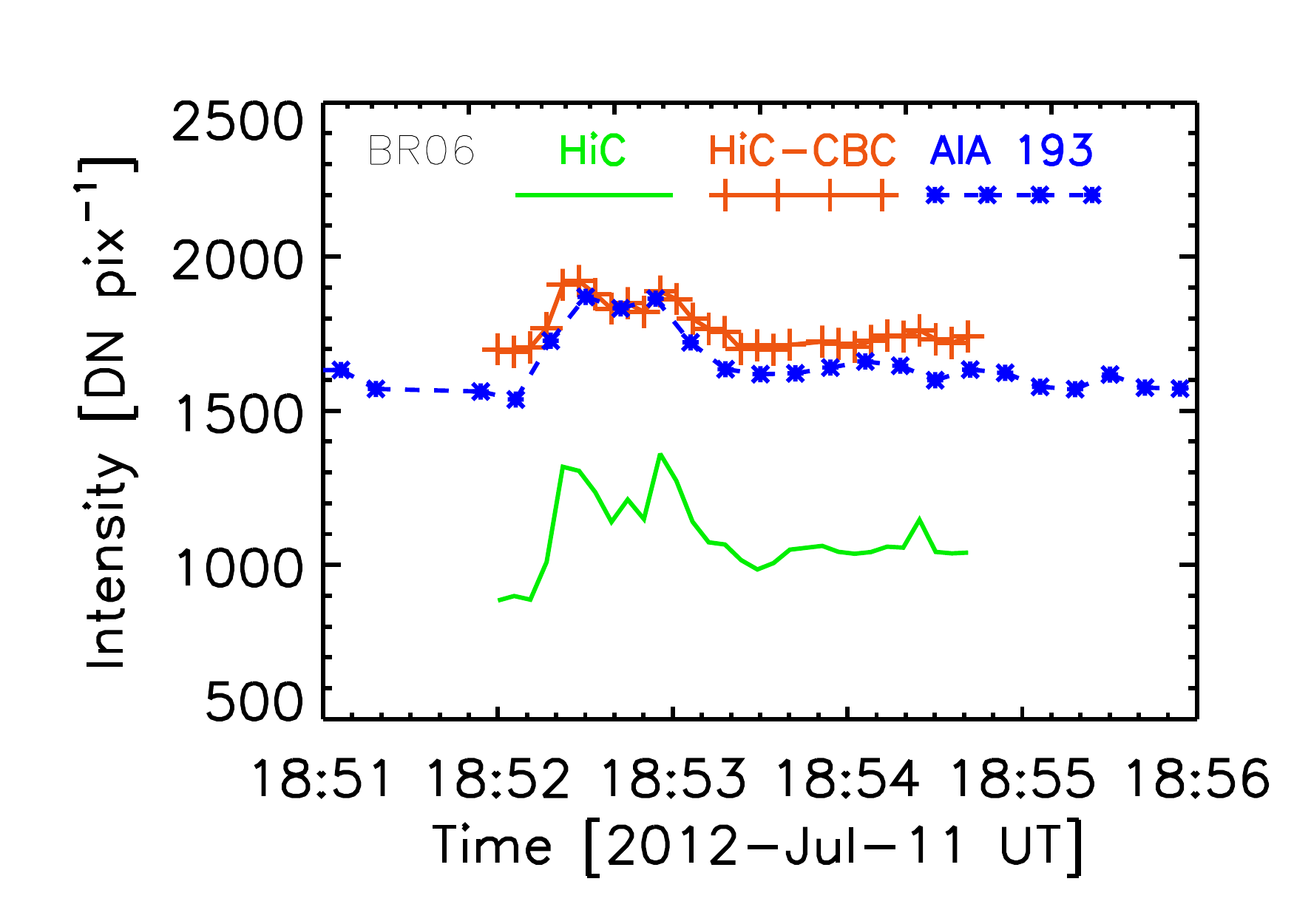}
\includegraphics[trim=0cm 0cm 0cm 0cm, clip=true,width=0.25\textwidth]{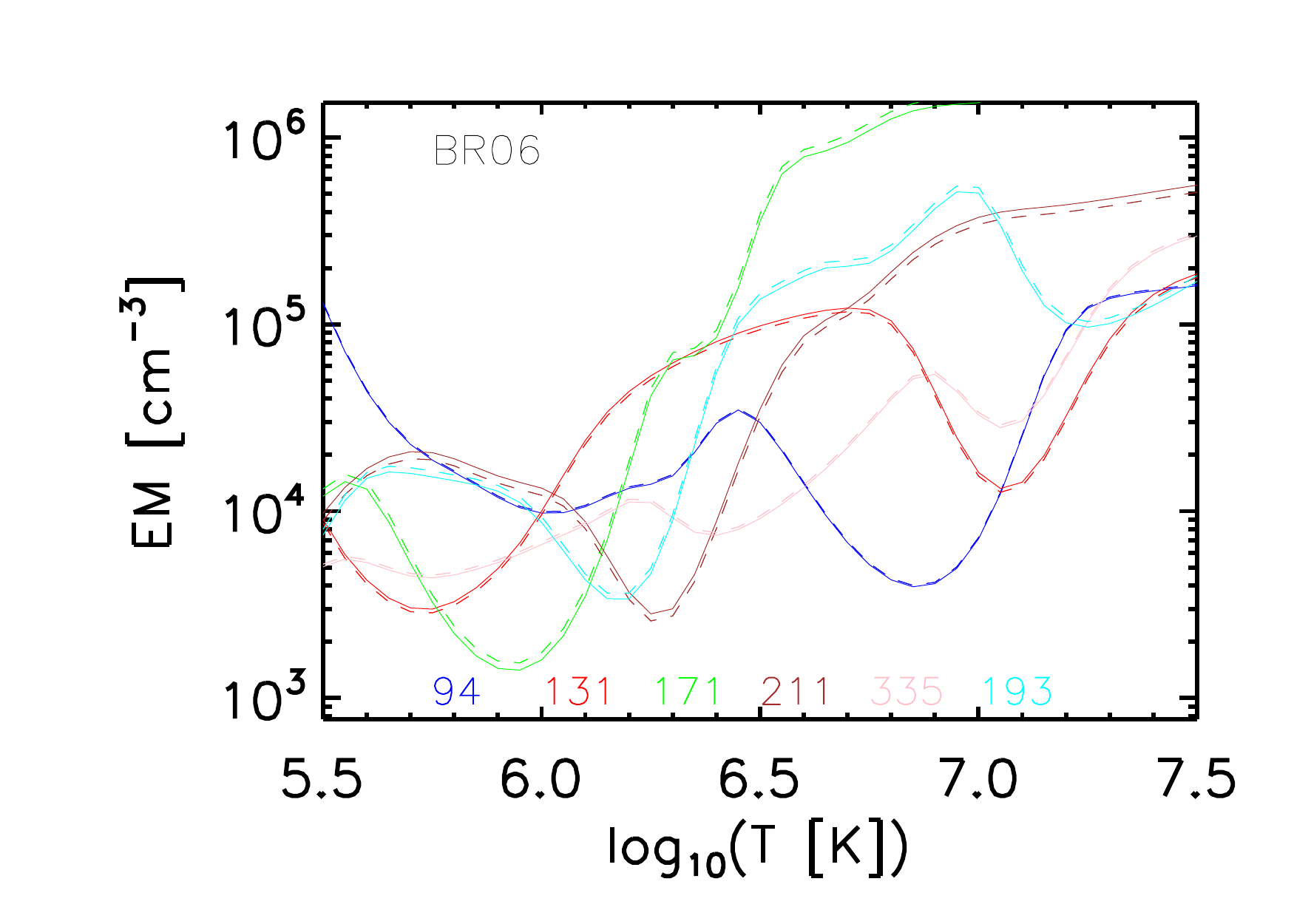}
\includegraphics[trim=0cm 0cm 0cm 0cm, clip=true,width=0.25\textwidth]{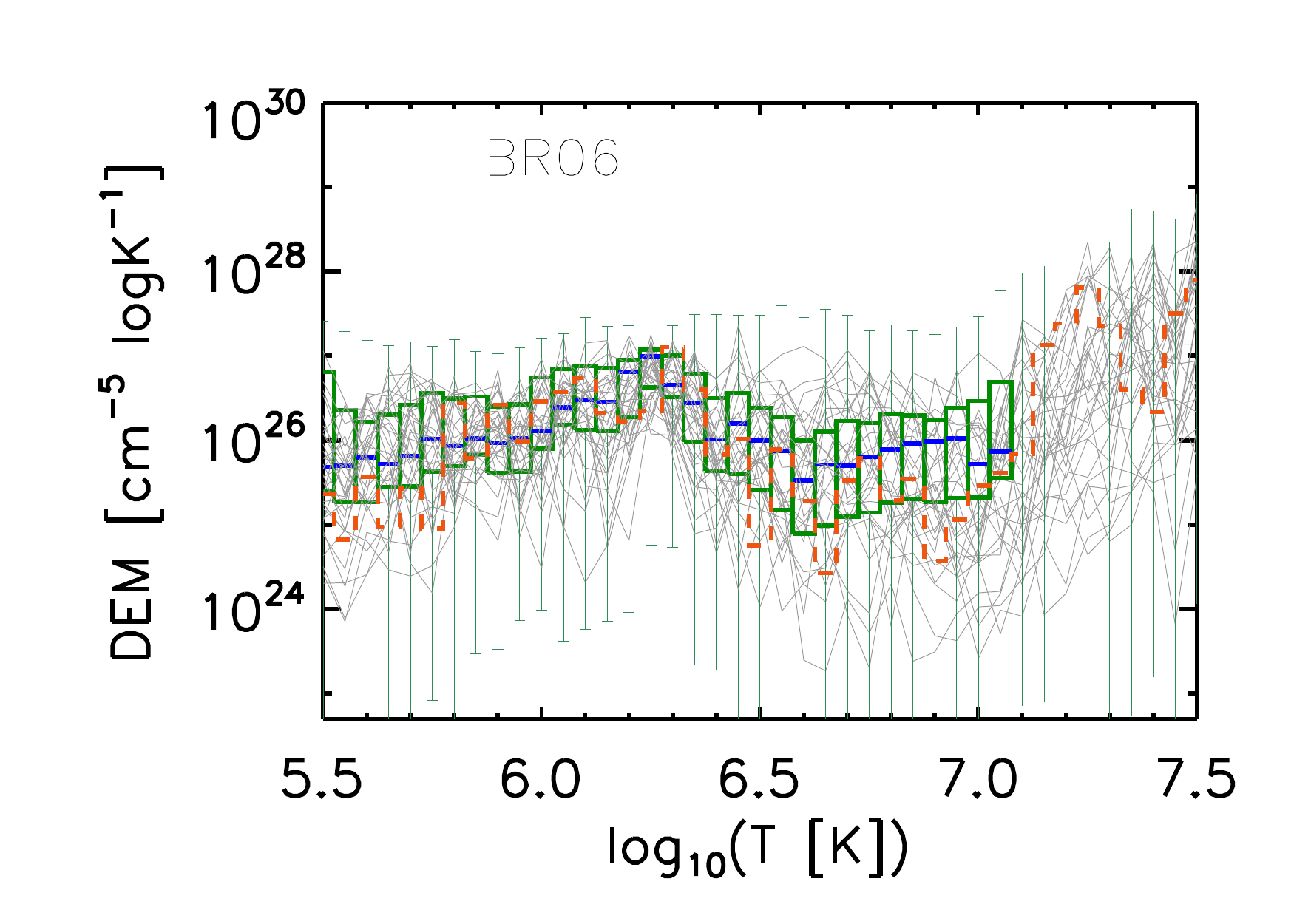}
\includegraphics[trim=0cm 0cm 0cm 0cm, clip=true,width=0.25\textwidth]{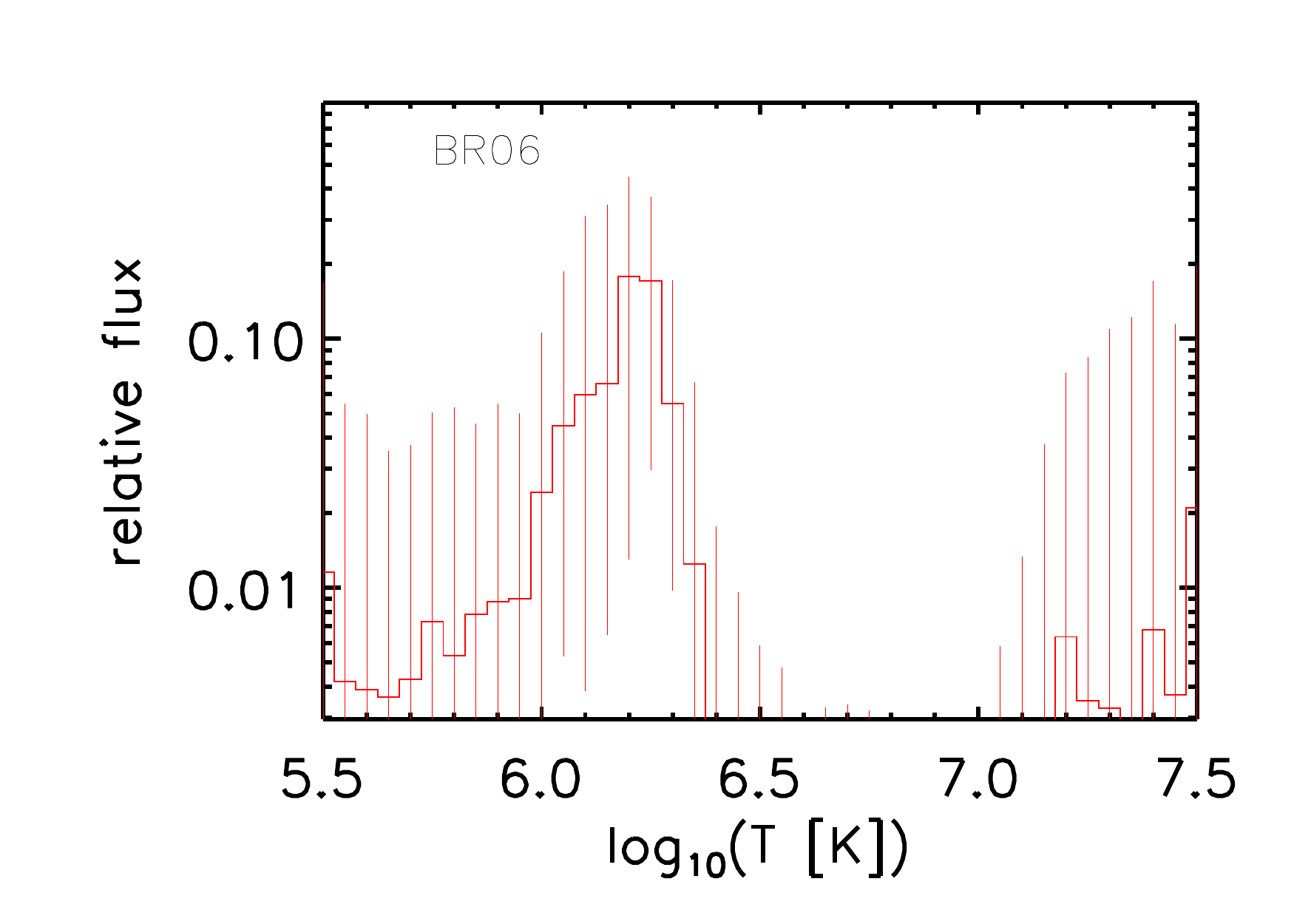}\\
\caption{From left, light curves are in panel~1, EM loci curves in panel~2, DEM reconstructions in panel~3, and contributions of different temperatures to flux in AIA~193~\AA\ in panel~4, for each of the brighhtenings studied here (excluding Br07 and Br26).}
\label{f:app_lc_DEM_1}

\end{center}
\end{figure*}

\begin{figure*}[htp!]
\begin{center}
\hspace{-0.57cm}
\includegraphics[trim=0cm 0cm 0cm 0cm, clip=true,width=0.25\textwidth]{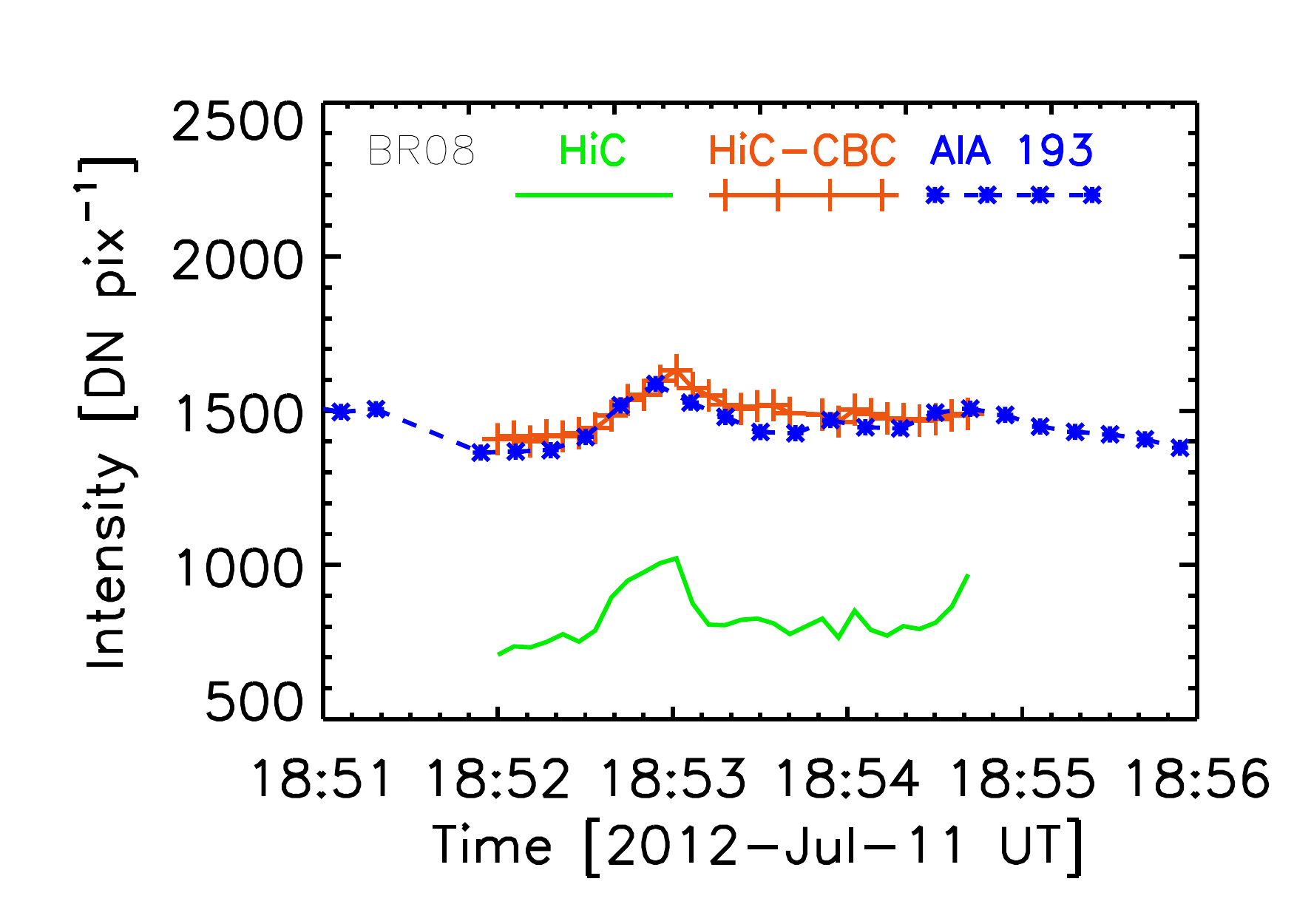}
\includegraphics[trim=0cm 0cm 0cm 0cm, clip=true,width=0.25\textwidth]{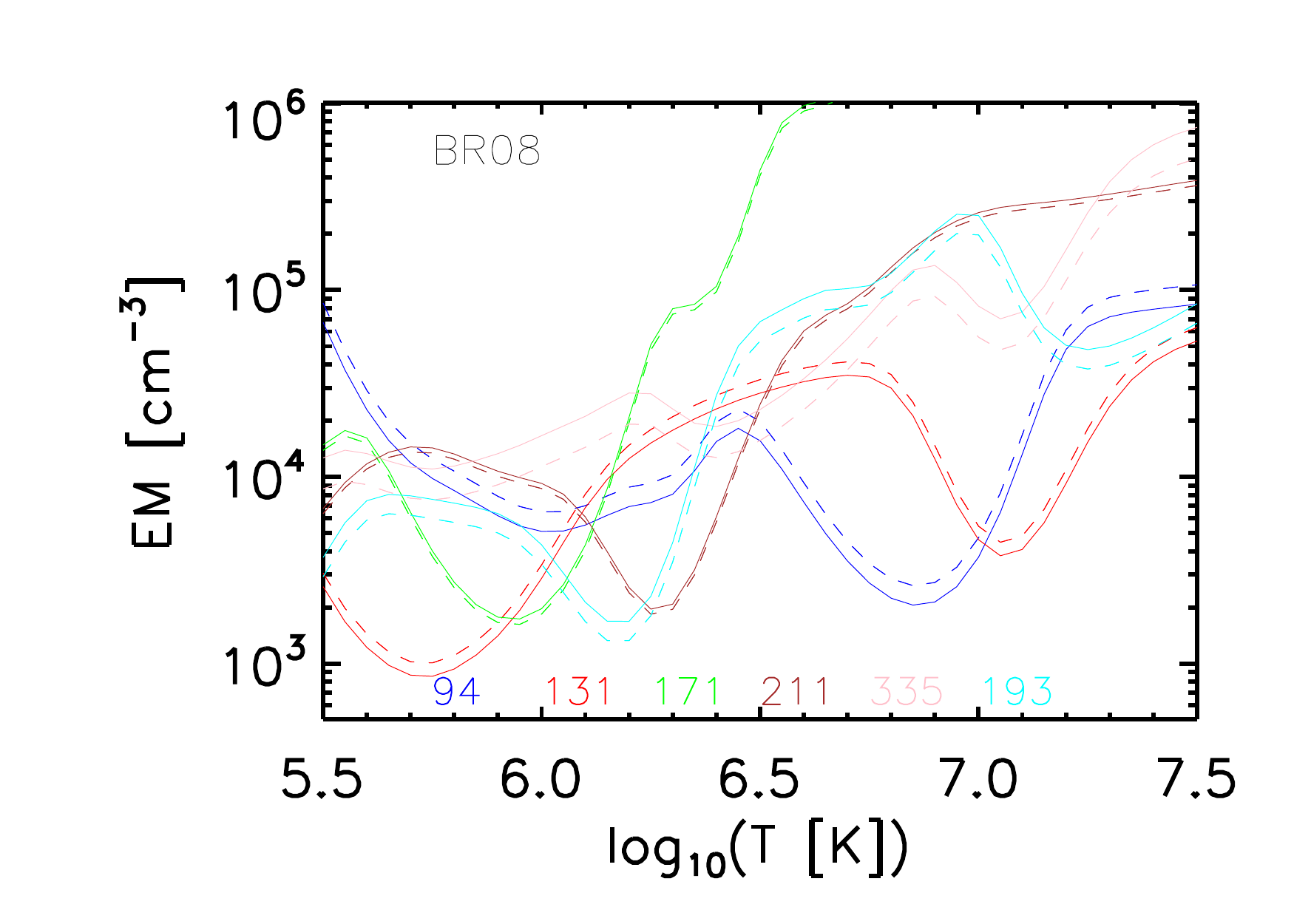}
\includegraphics[trim=0cm 0cm 0cm 0cm, clip=true,width=0.25\textwidth]{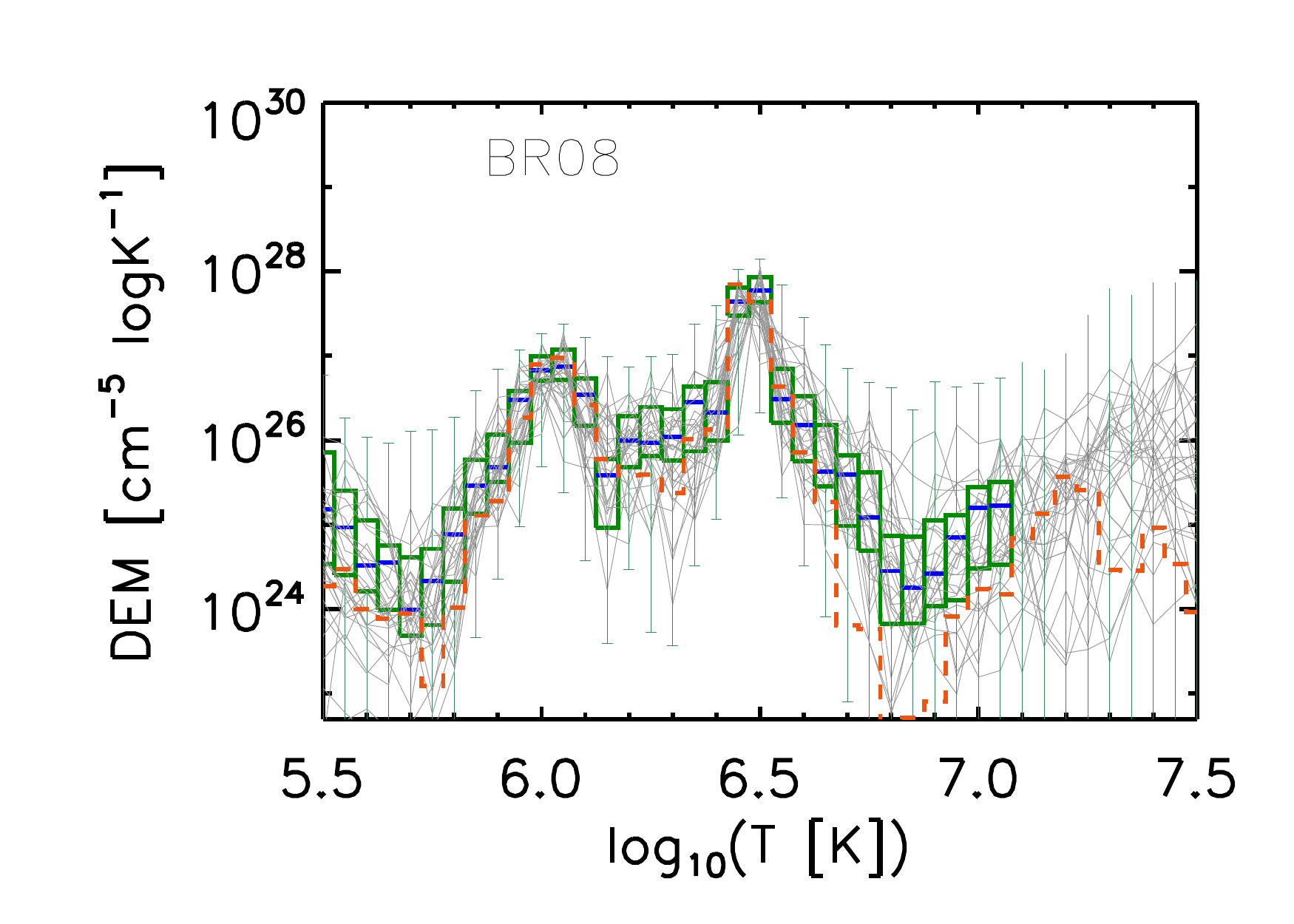}
\includegraphics[trim=0cm 0cm 0cm 0cm, clip=true,width=0.25\textwidth]{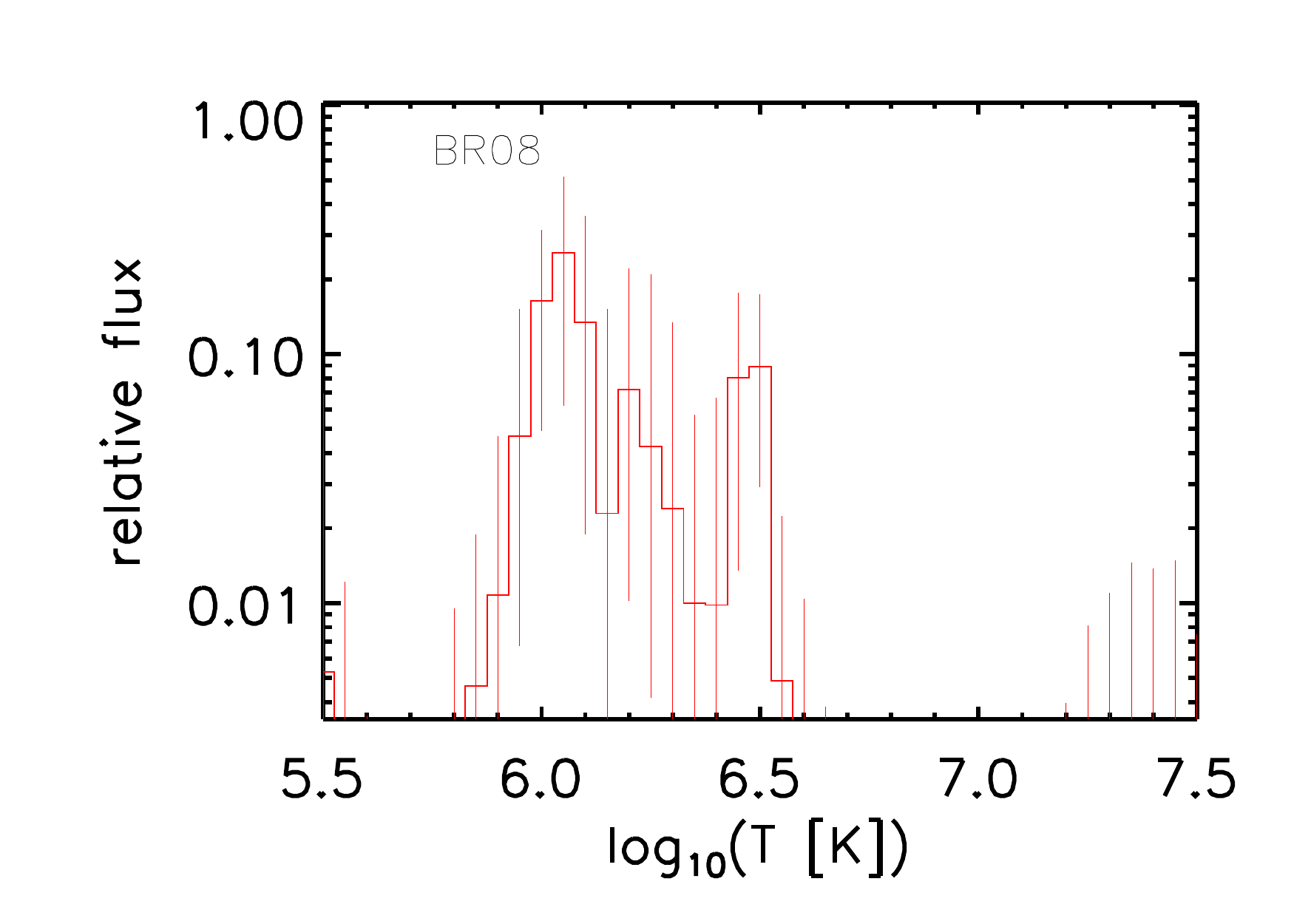}\\
\hspace{-0.57cm}
\includegraphics[trim=0cm 0cm 0cm 0cm, clip=true,width=0.25\textwidth]{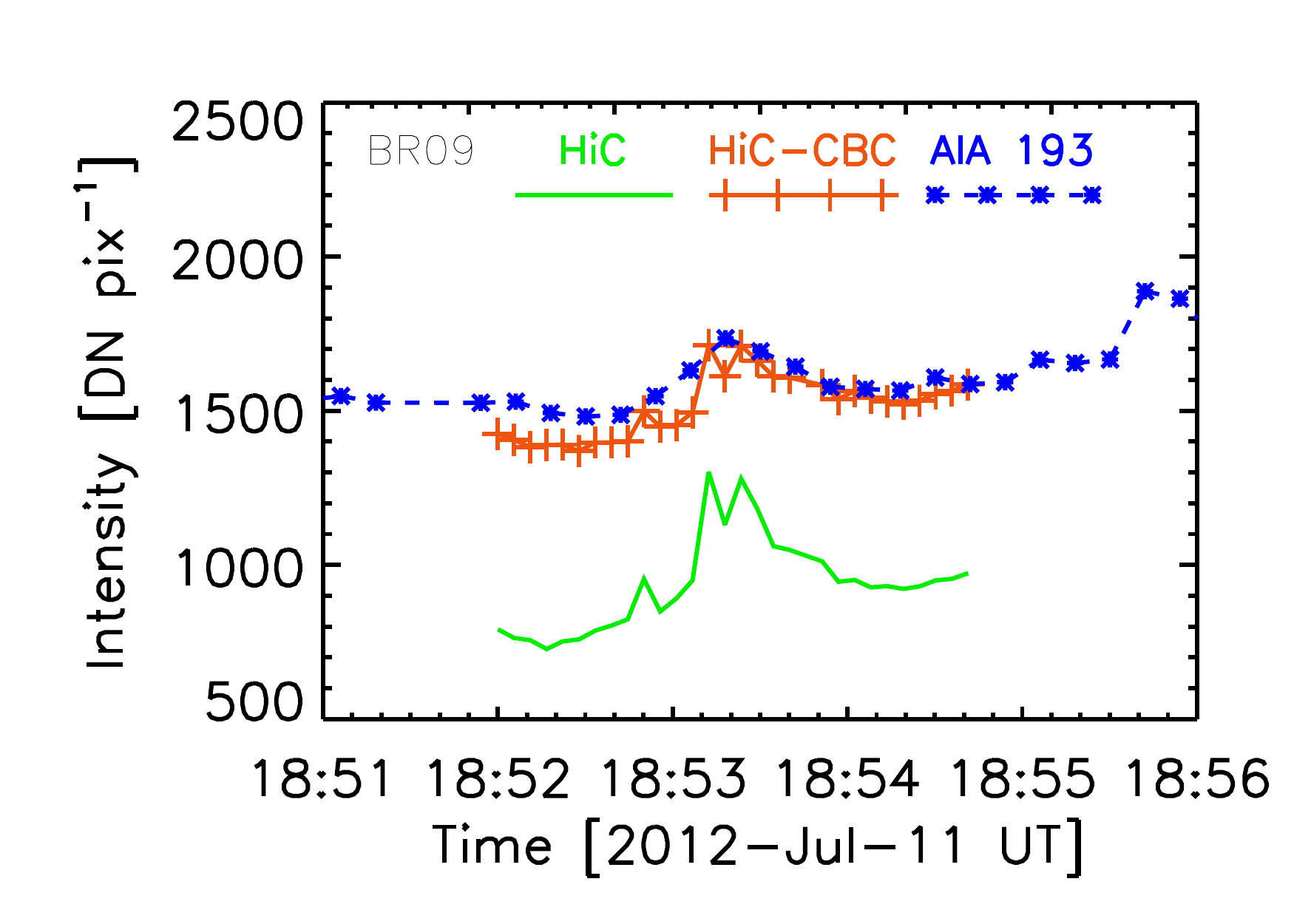}
\includegraphics[trim=0cm 0cm 0cm 0cm, clip=true,width=0.25\textwidth]{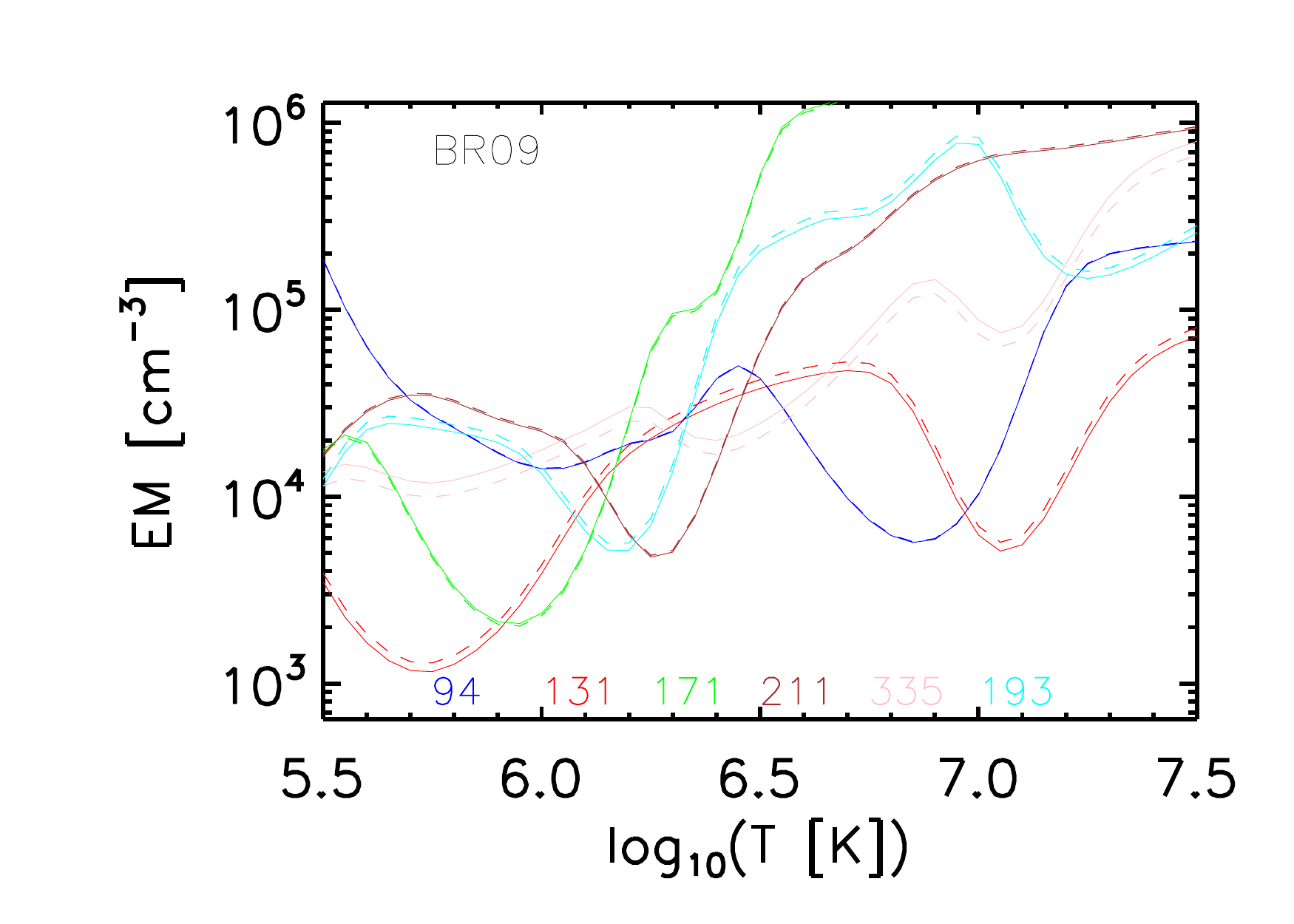}
\includegraphics[trim=0cm 0cm 0cm 0cm, clip=true,width=0.25\textwidth]{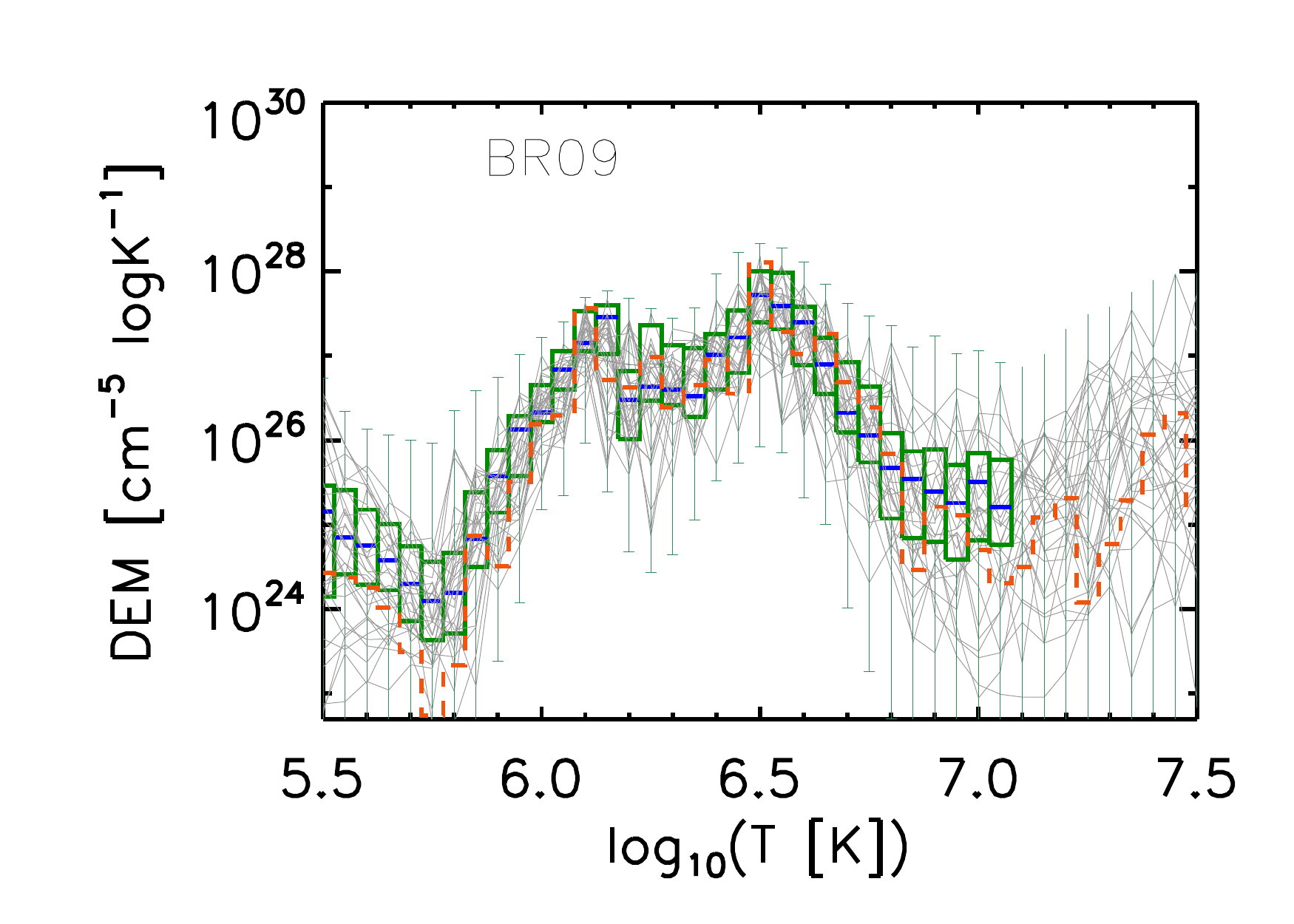}
\includegraphics[trim=0cm 0cm 0cm 0cm, clip=true,width=0.25\textwidth]{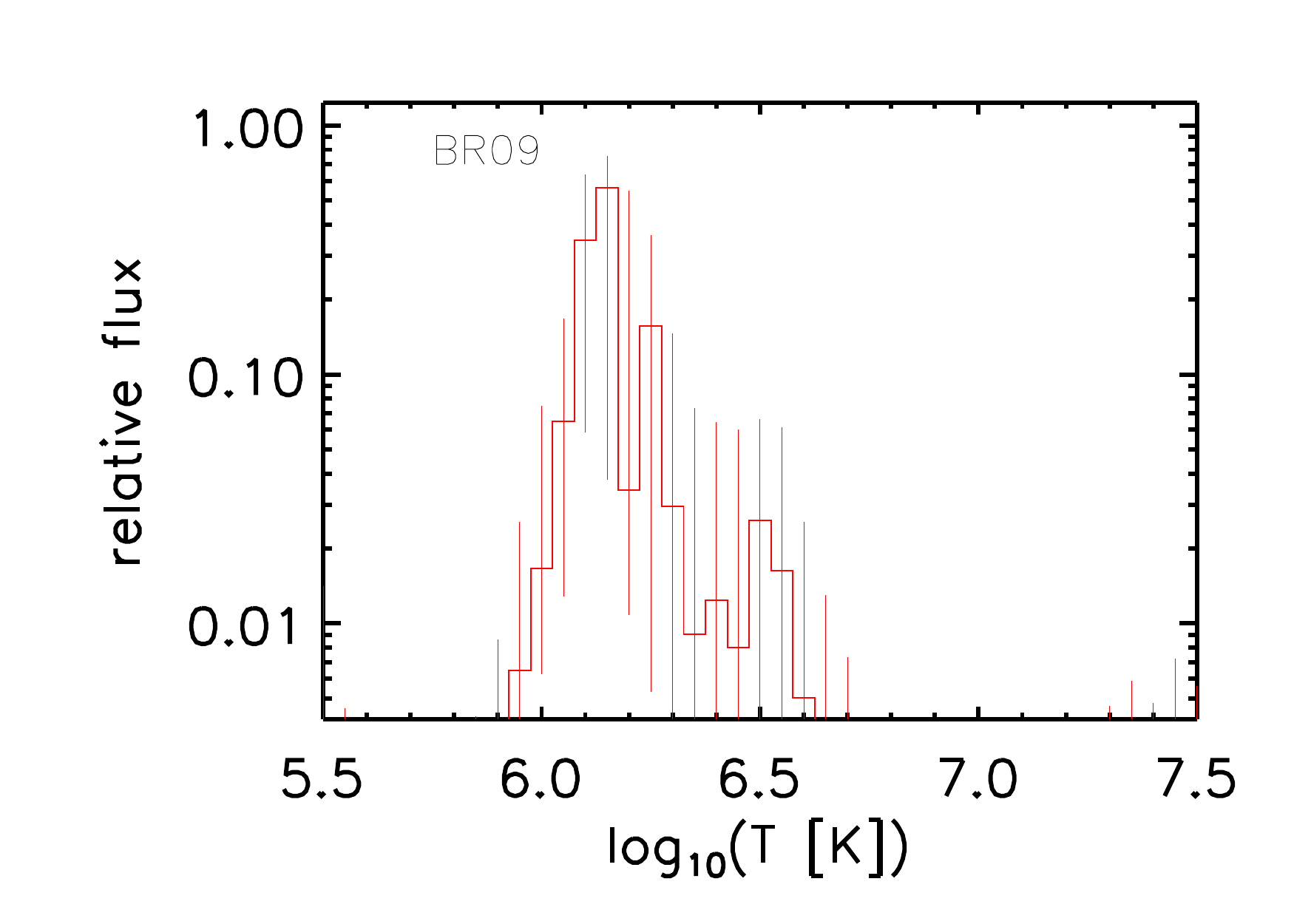}\\
\hspace{-0.57cm}
\includegraphics[trim=0cm 0cm 0cm 0cm, clip=true,width=0.25\textwidth]{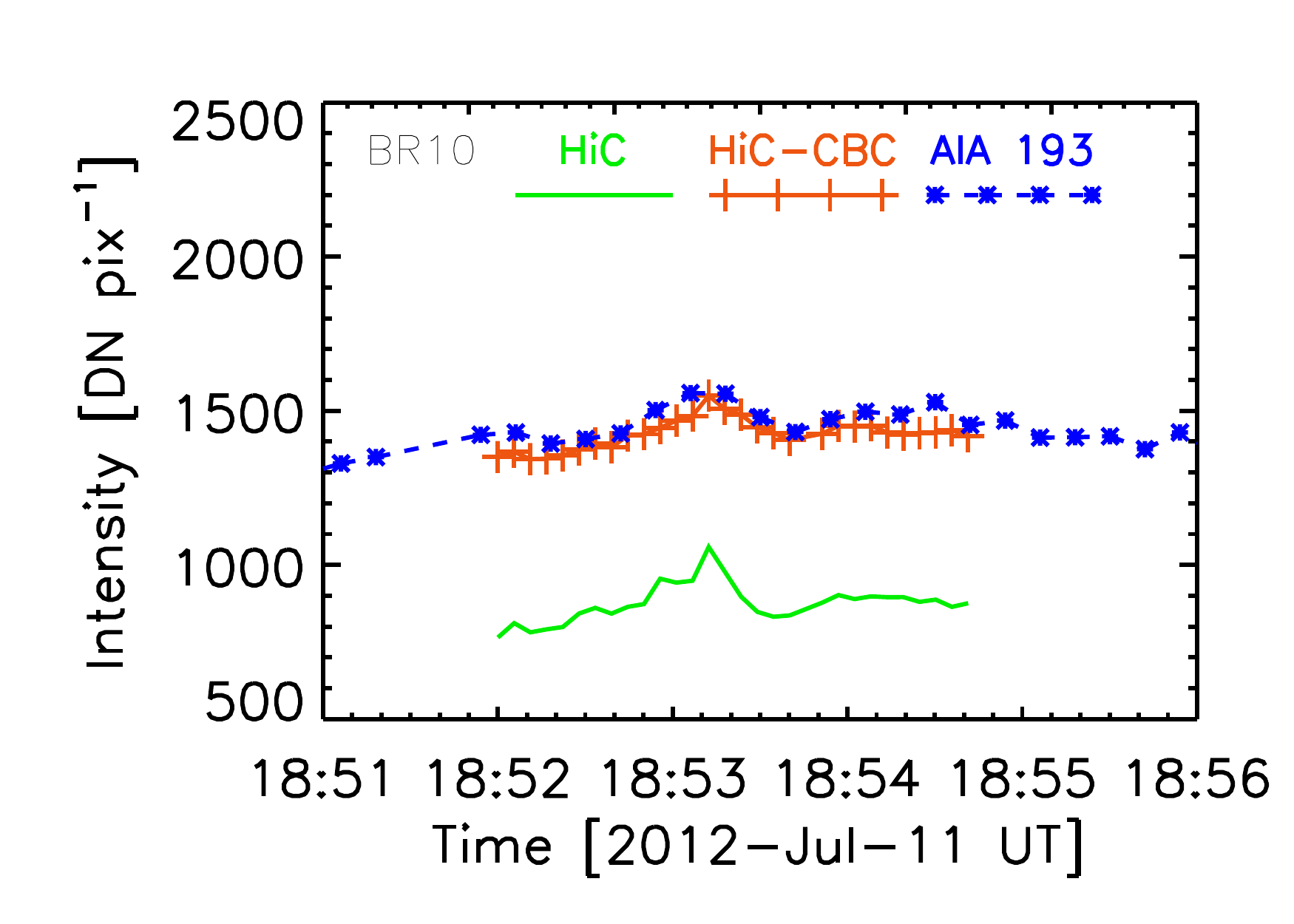}
\includegraphics[trim=0cm 0cm 0cm 0cm, clip=true,width=0.25\textwidth]{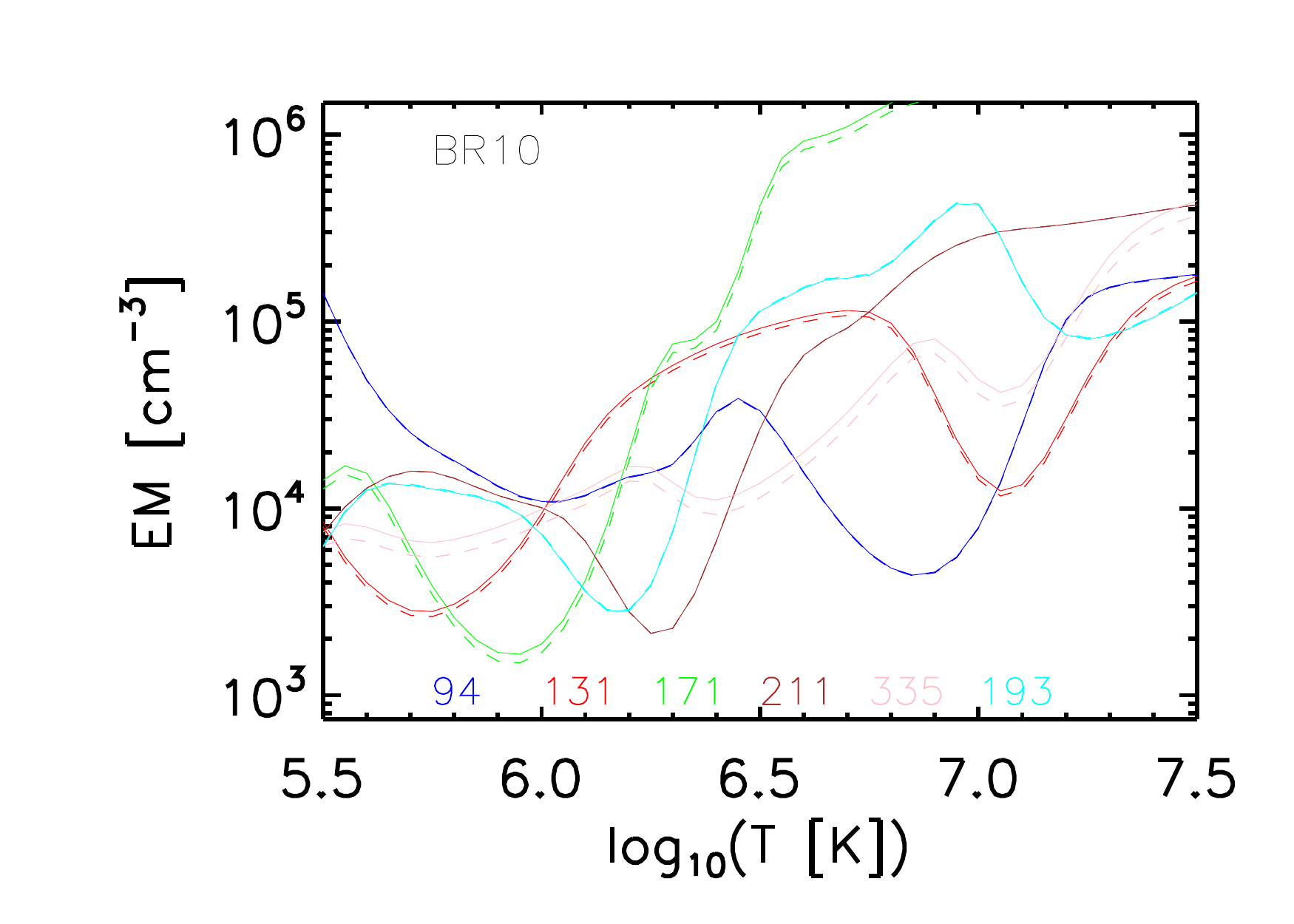}
\includegraphics[trim=0cm 0cm 0cm 0cm, clip=true,width=0.25\textwidth]{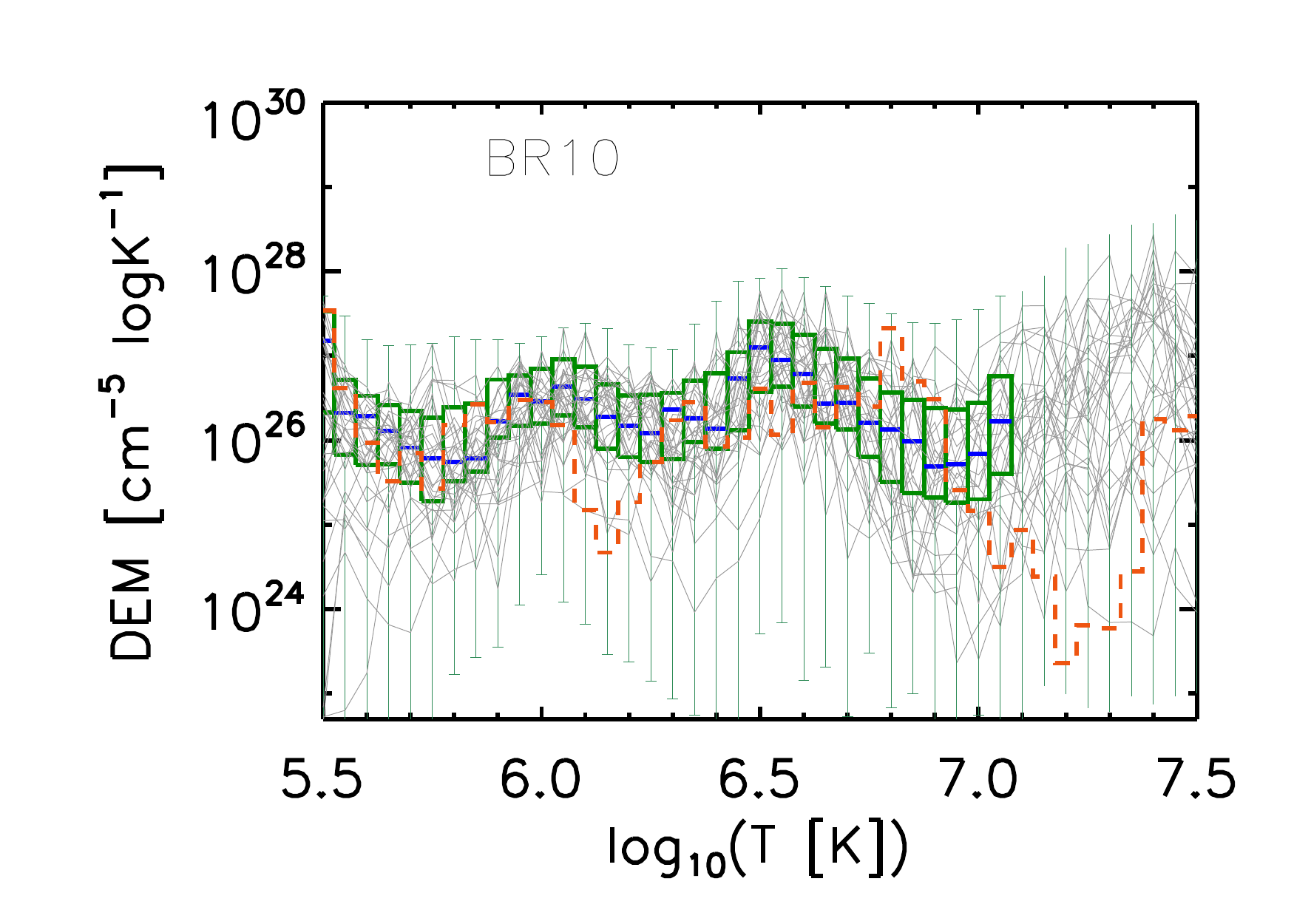}
\includegraphics[trim=0cm 0cm 0cm 0cm, clip=true,width=0.25\textwidth]{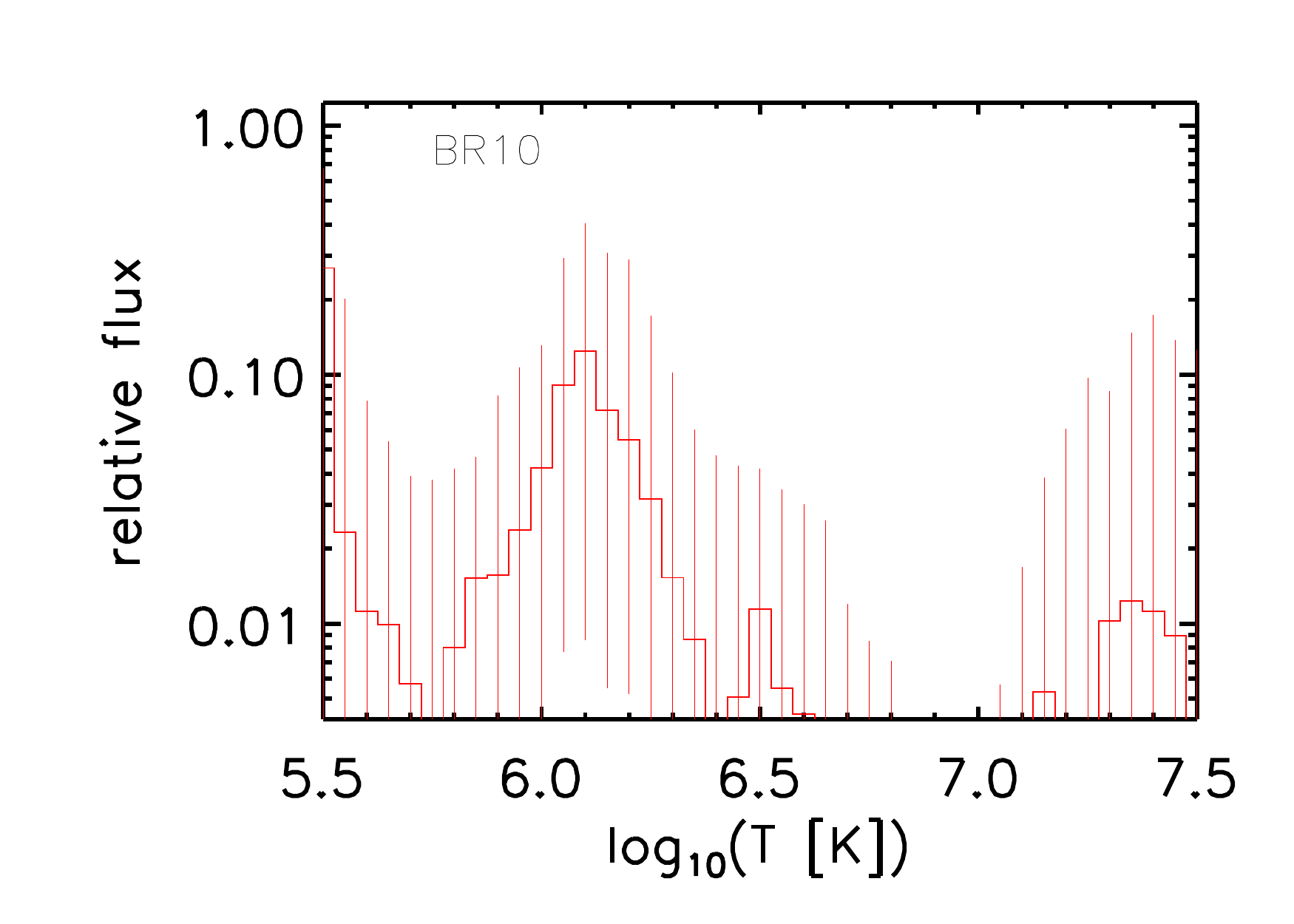}\\
\hspace{-0.57cm}
\includegraphics[trim=0cm 0cm 0cm 0cm, clip=true,width=0.25\textwidth]{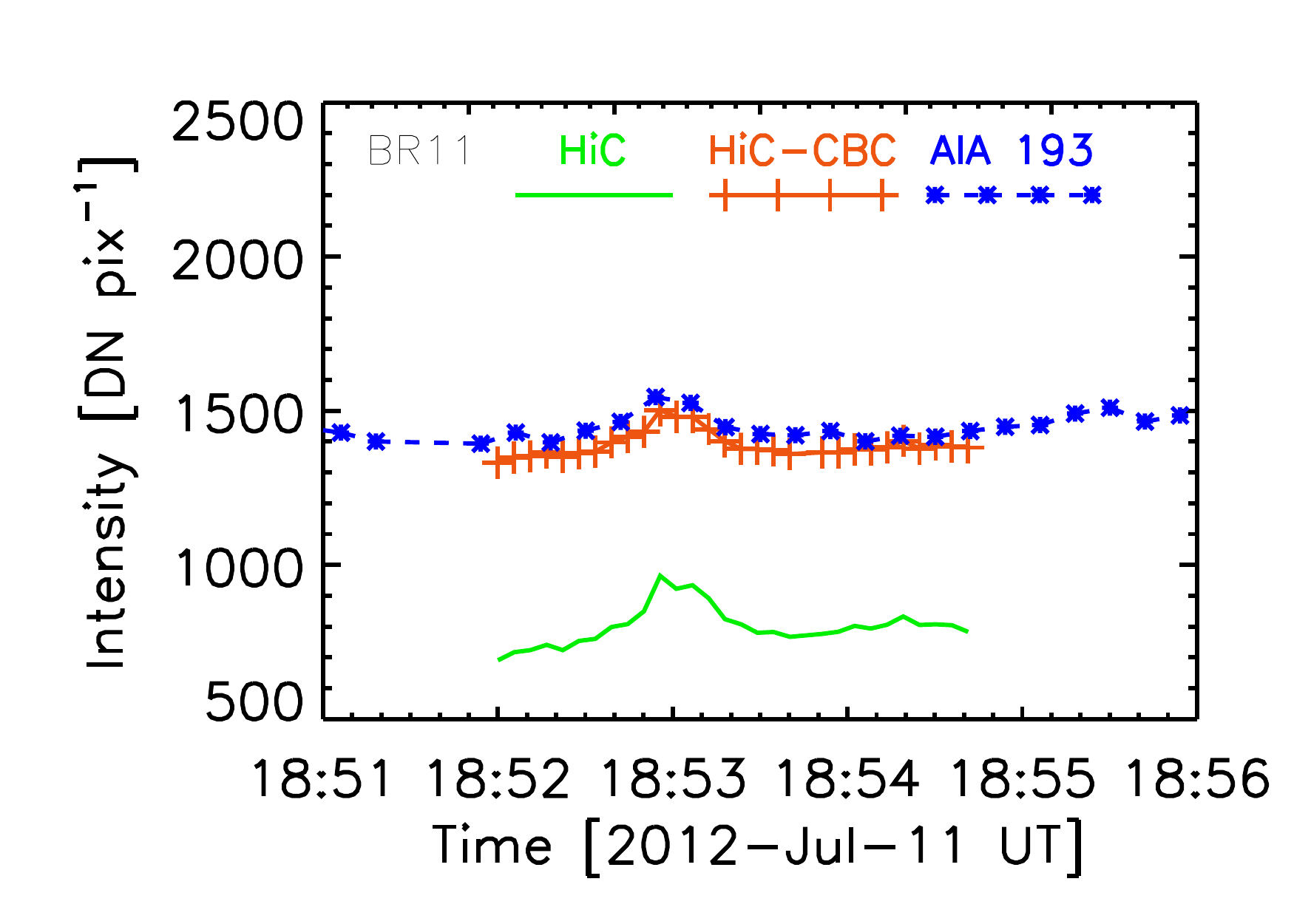}
\includegraphics[trim=0cm 0cm 0cm 0cm, clip=true,width=0.25\textwidth]{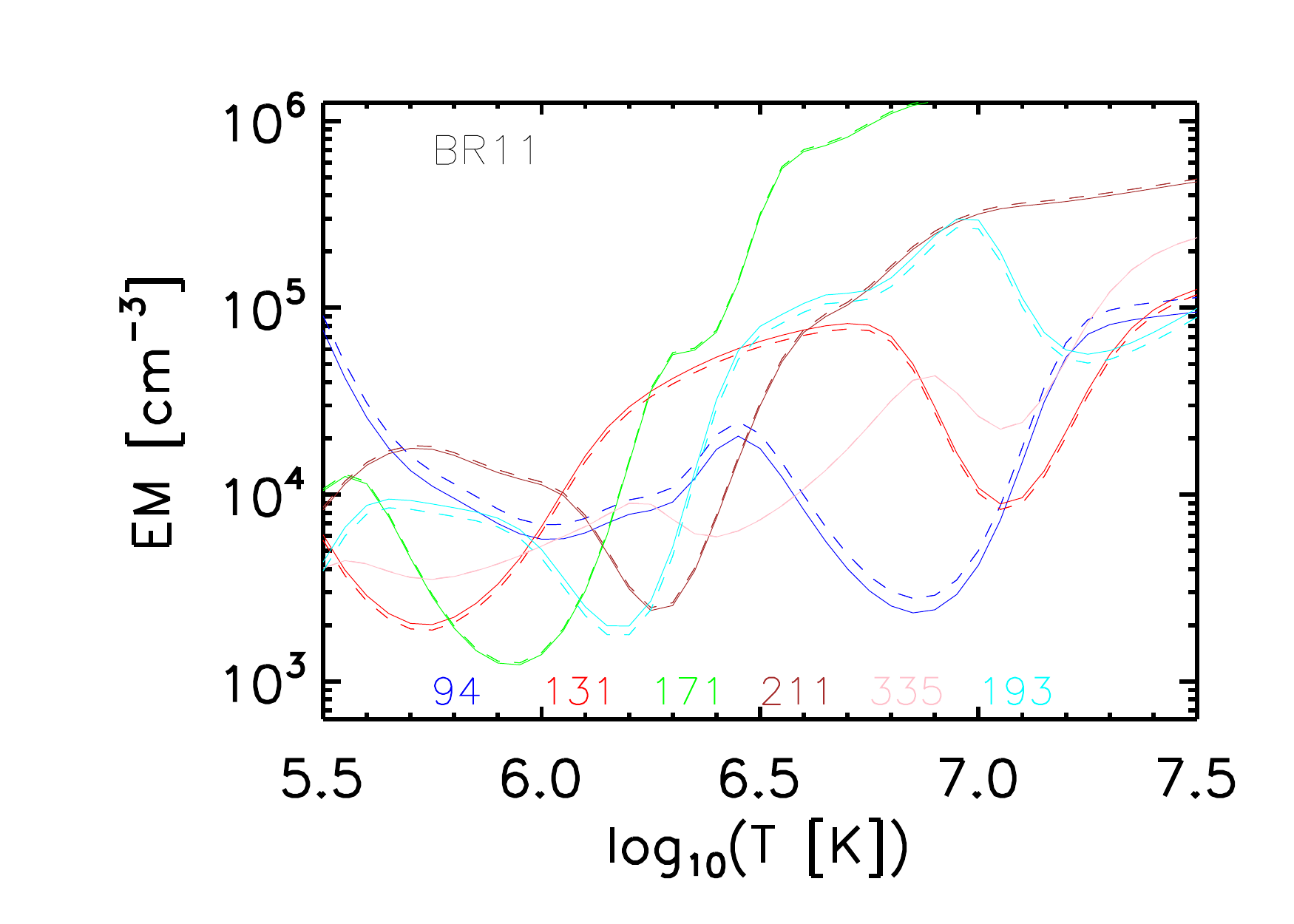}
\includegraphics[trim=0cm 0cm 0cm 0cm, clip=true,width=0.25\textwidth]{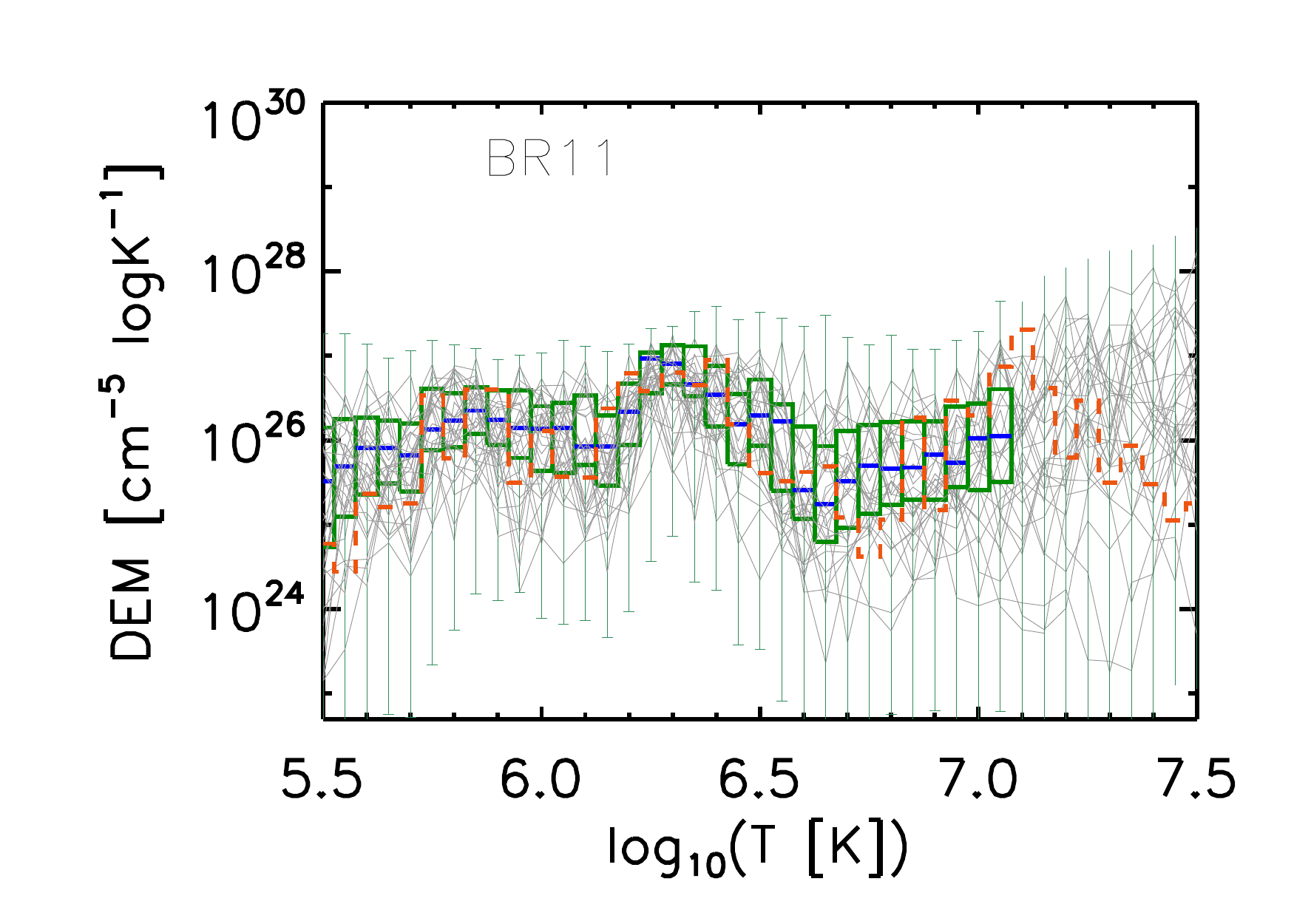}
\includegraphics[trim=0cm 0cm 0cm 0cm, clip=true,width=0.25\textwidth]{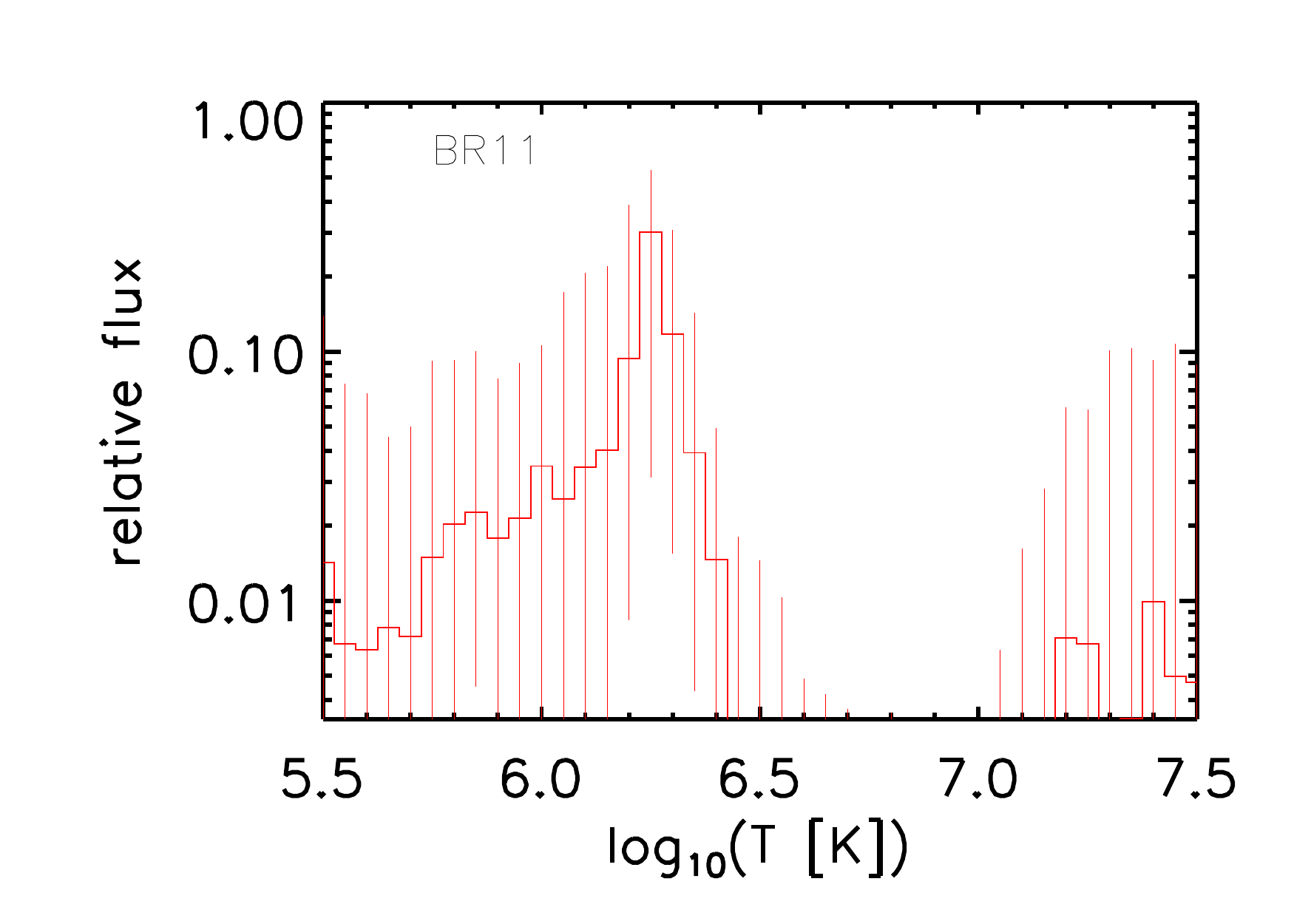}\\
\hspace{-0.57cm}
\includegraphics[trim=0cm 0cm 0cm 0cm, clip=true,width=0.25\textwidth]{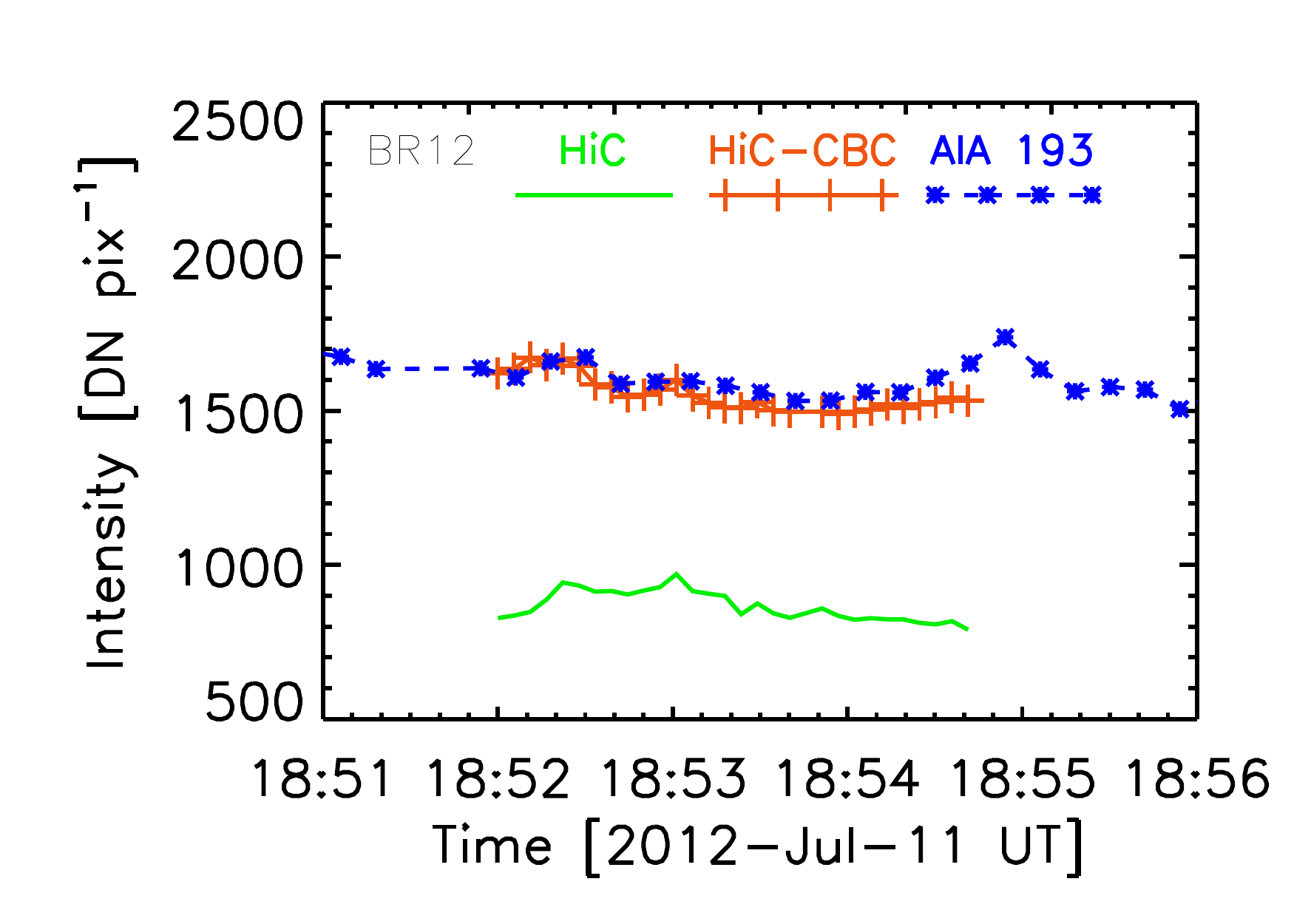}
\includegraphics[trim=0cm 0cm 0cm 0cm, clip=true,width=0.25\textwidth]{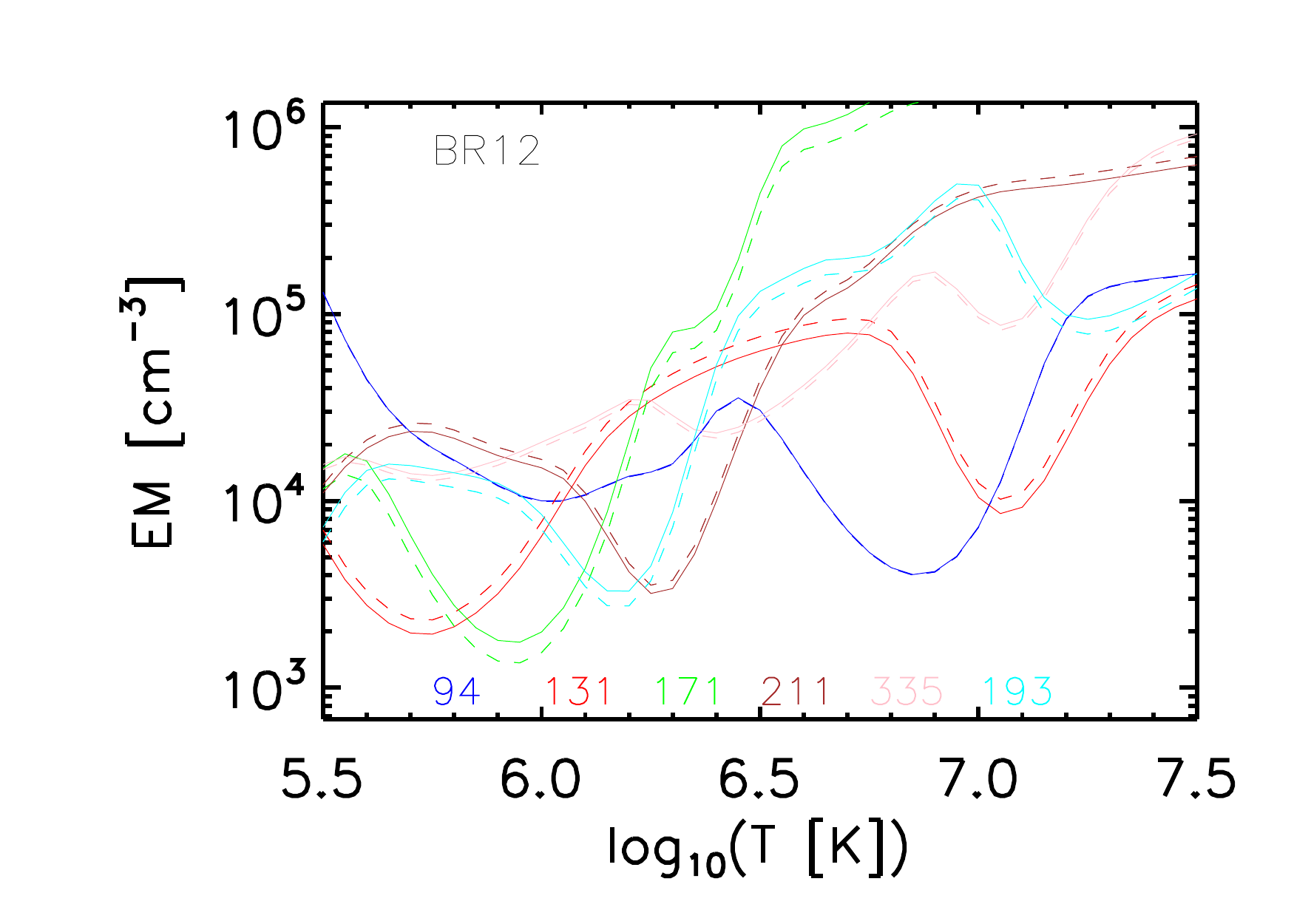}
\includegraphics[trim=0cm 0cm 0cm 0cm, clip=true,width=0.25\textwidth]{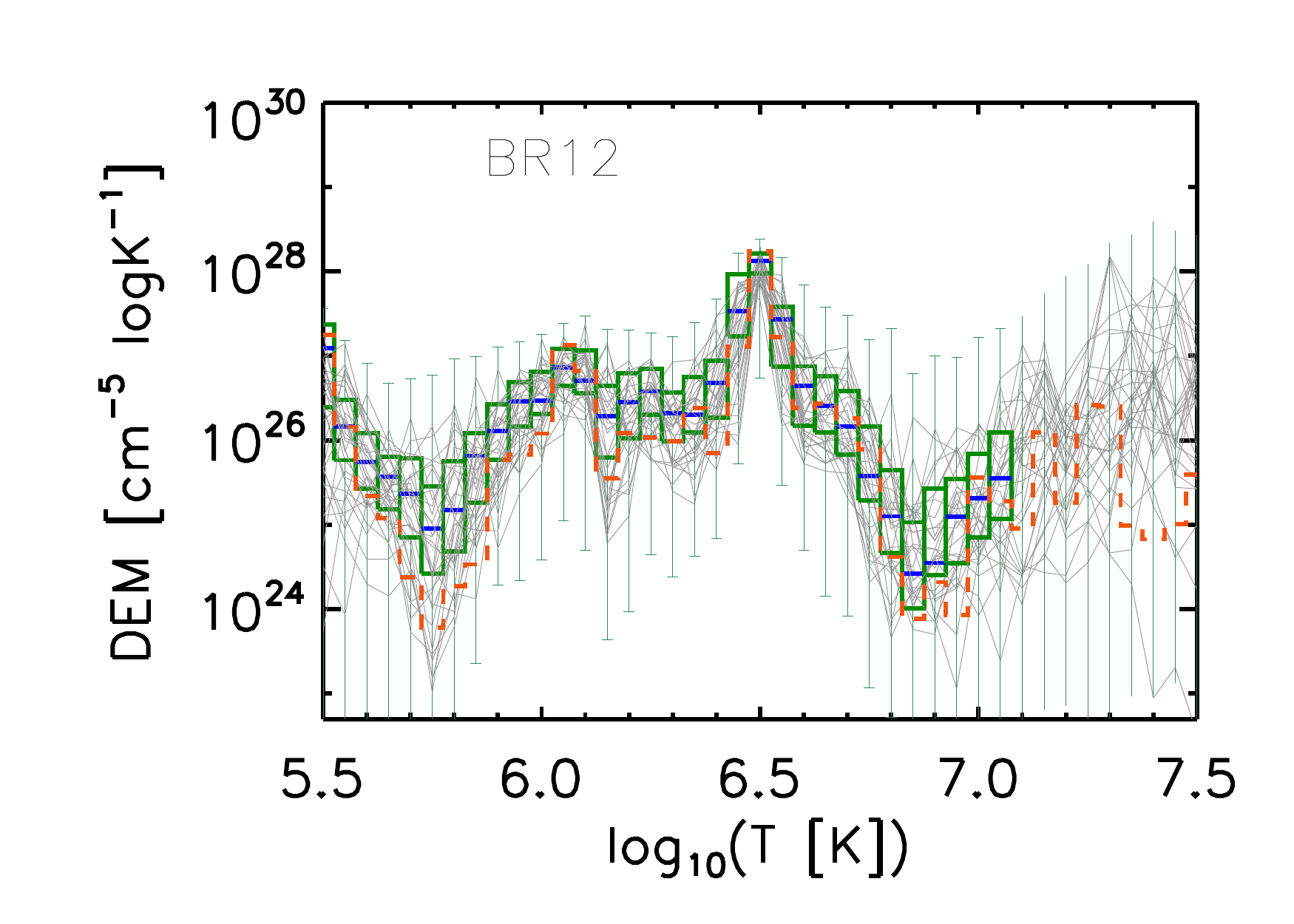}
\includegraphics[trim=0cm 0cm 0cm 0cm, clip=true,width=0.25\textwidth]{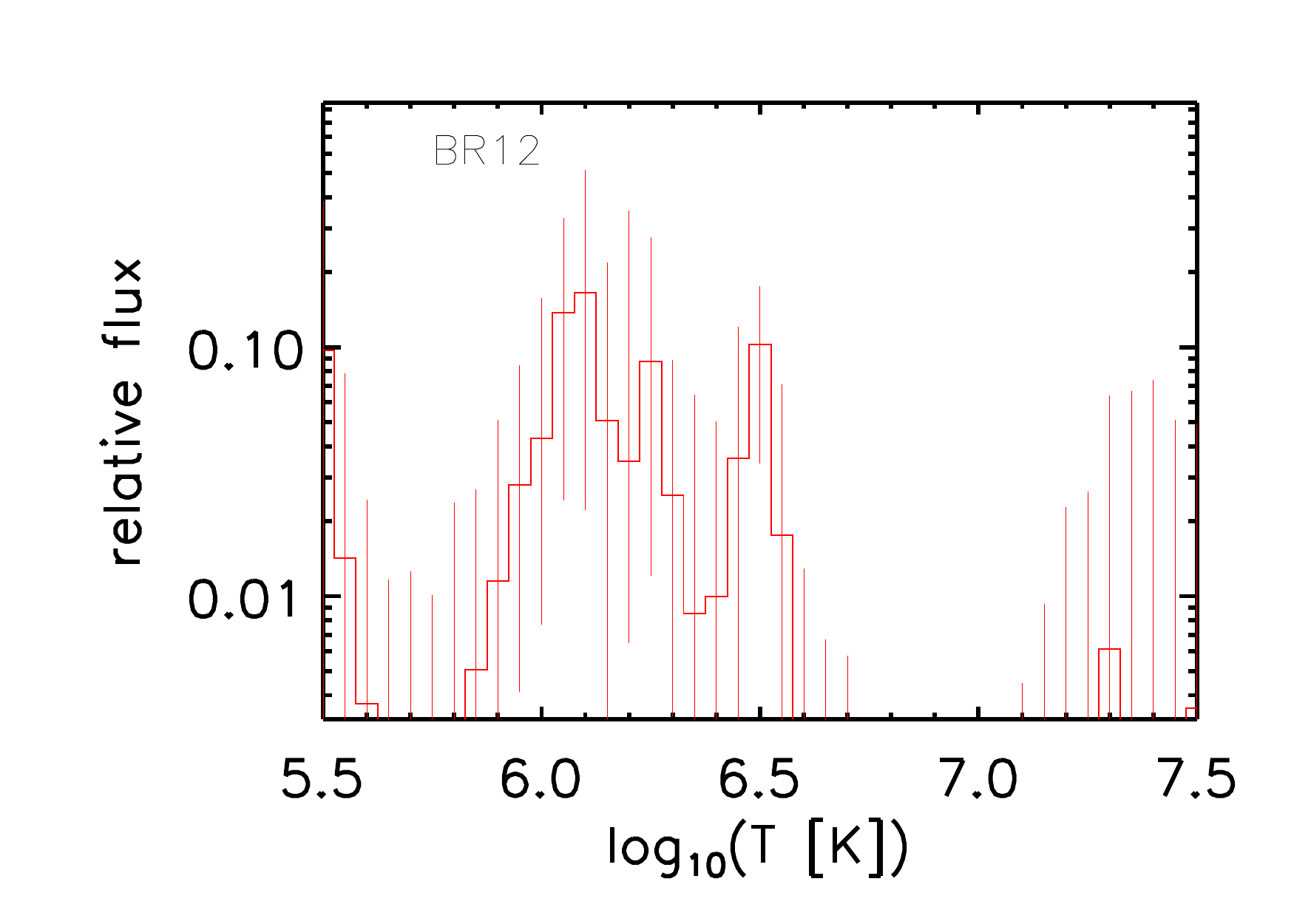}\\
\hspace{-0.57cm}
\includegraphics[trim=0cm 0cm 0cm 0cm, clip=true,width=0.25\textwidth]{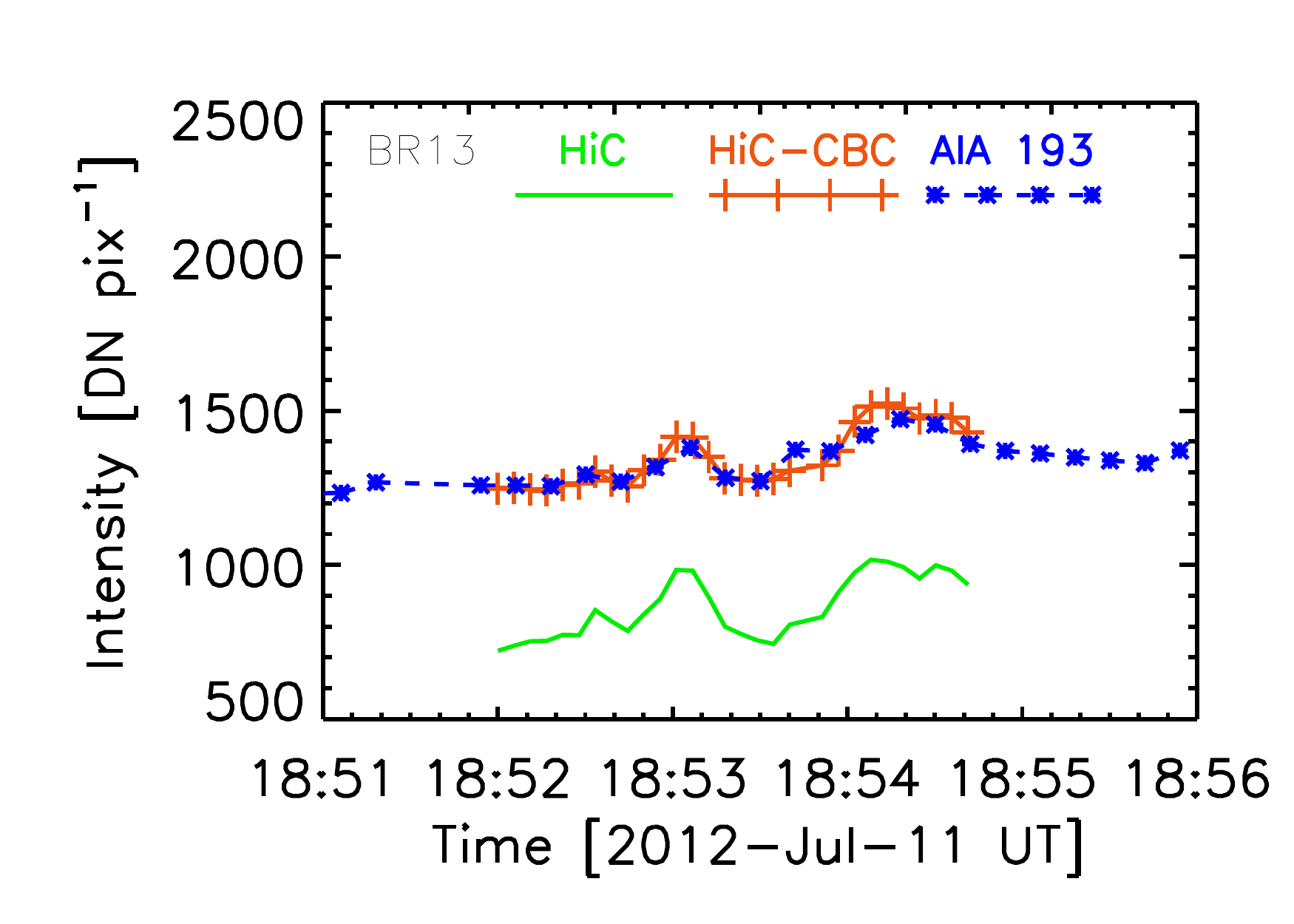}
\includegraphics[trim=0cm 0cm 0cm 0cm, clip=true,width=0.25\textwidth]{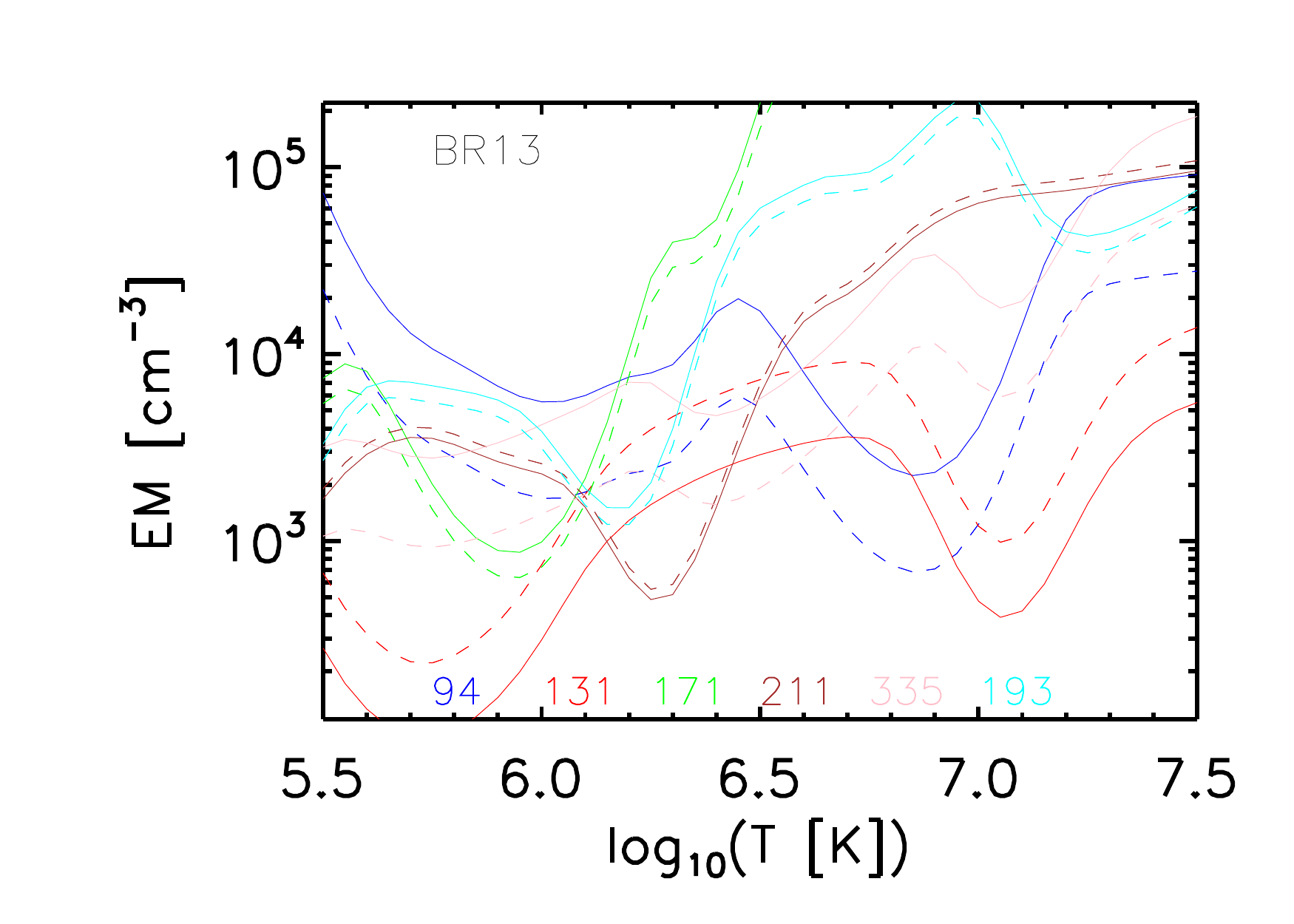}
\includegraphics[trim=0cm 0cm 0cm 0cm, clip=true,width=0.25\textwidth]{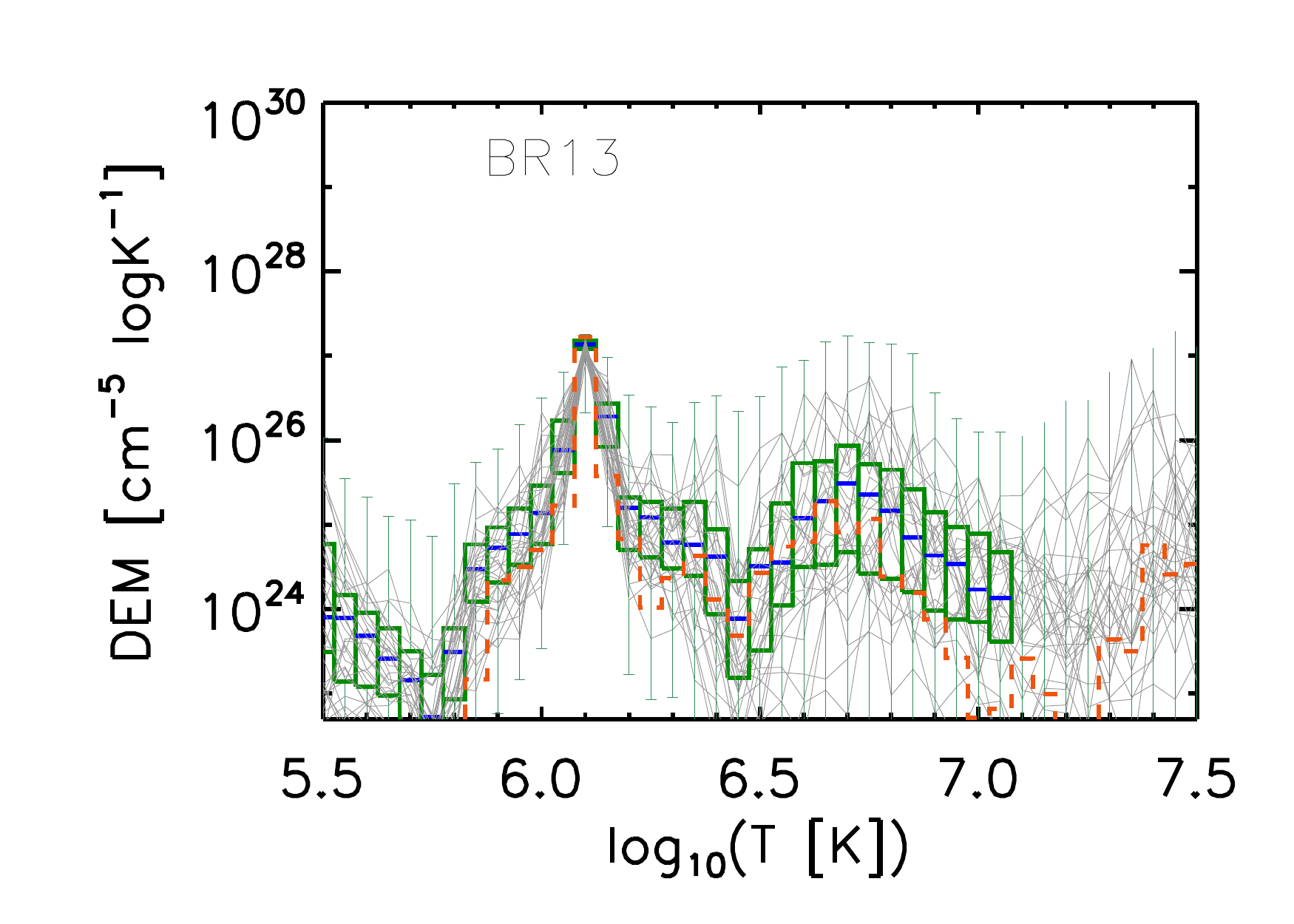}
\includegraphics[trim=0cm 0cm 0cm 0cm, clip=true,width=0.25\textwidth]{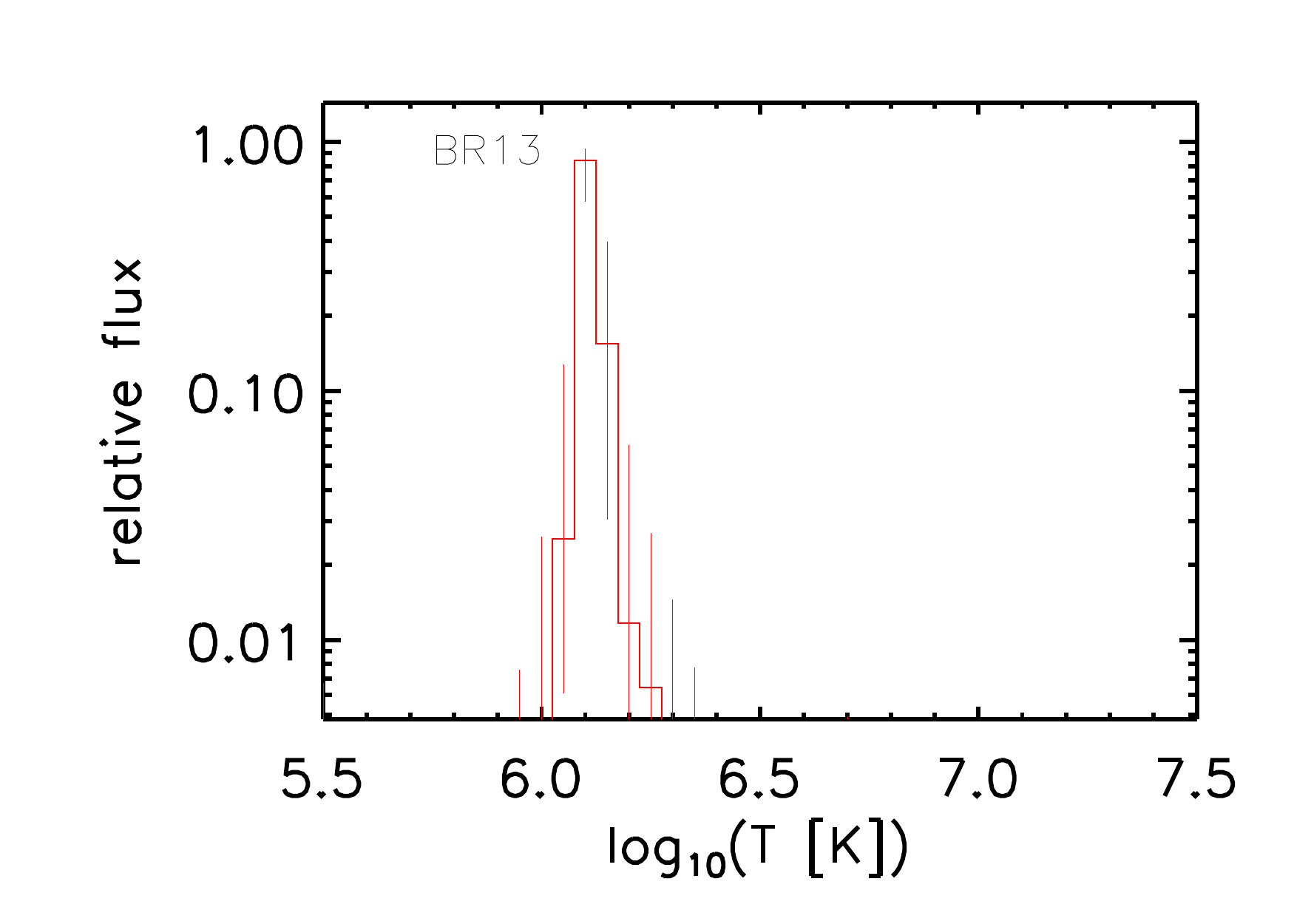}\\
\hspace{-0.57cm}
\includegraphics[trim=0cm 0cm 0cm 0cm, clip=true,width=0.25\textwidth]{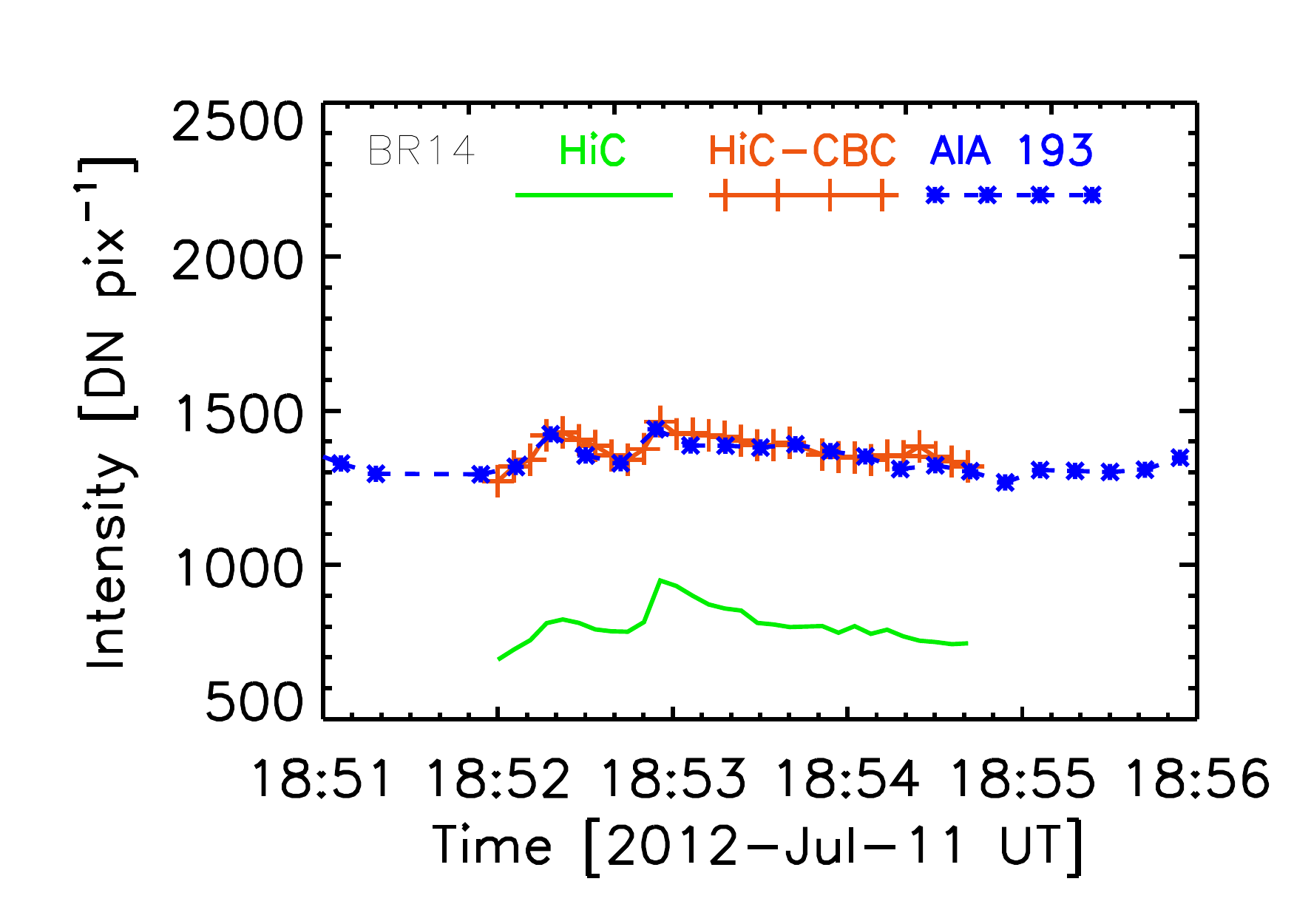}
\includegraphics[trim=0cm 0cm 0cm 0cm, clip=true,width=0.25\textwidth]{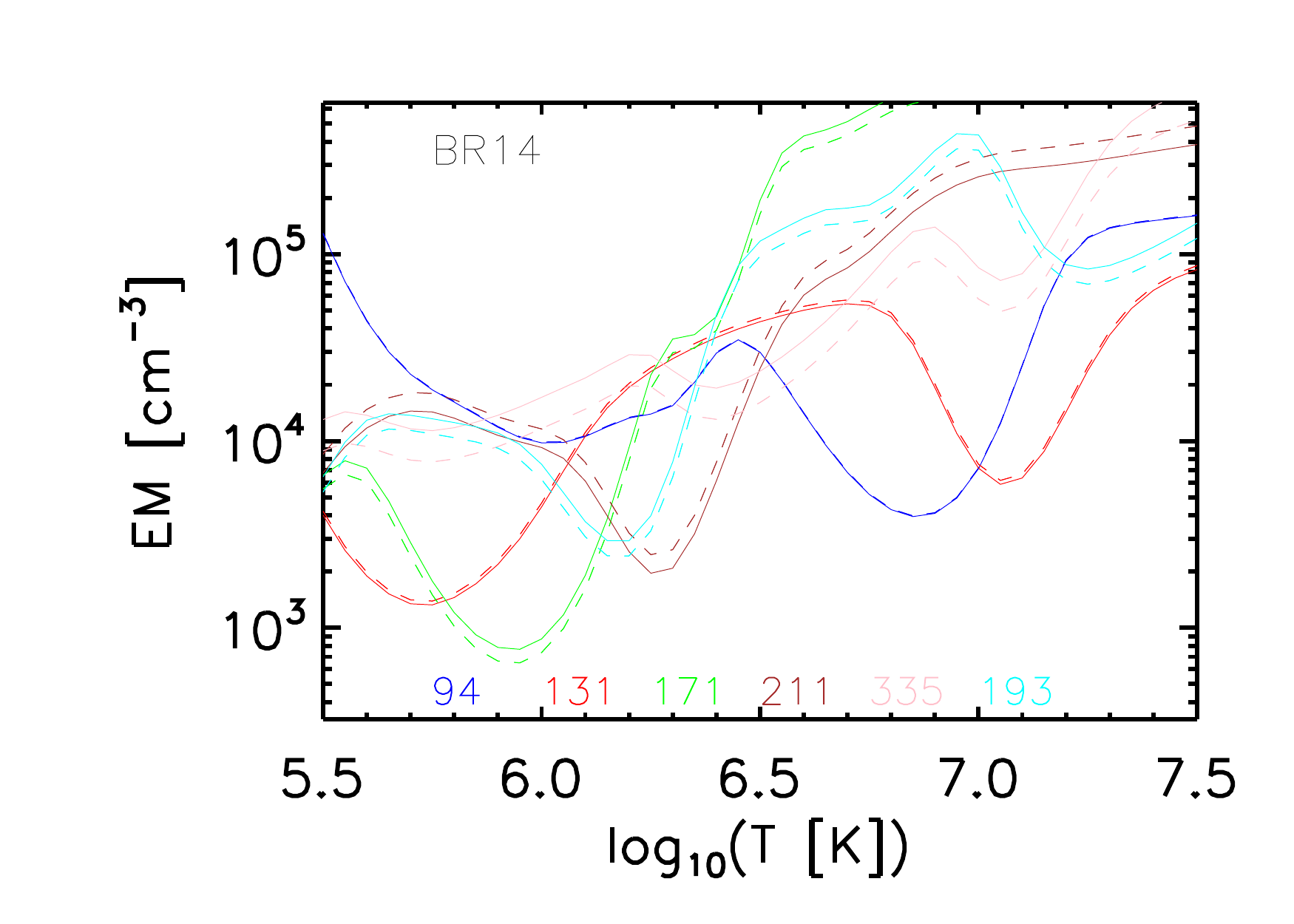}
\includegraphics[trim=0cm 0cm 0cm 0cm, clip=true,width=0.25\textwidth]{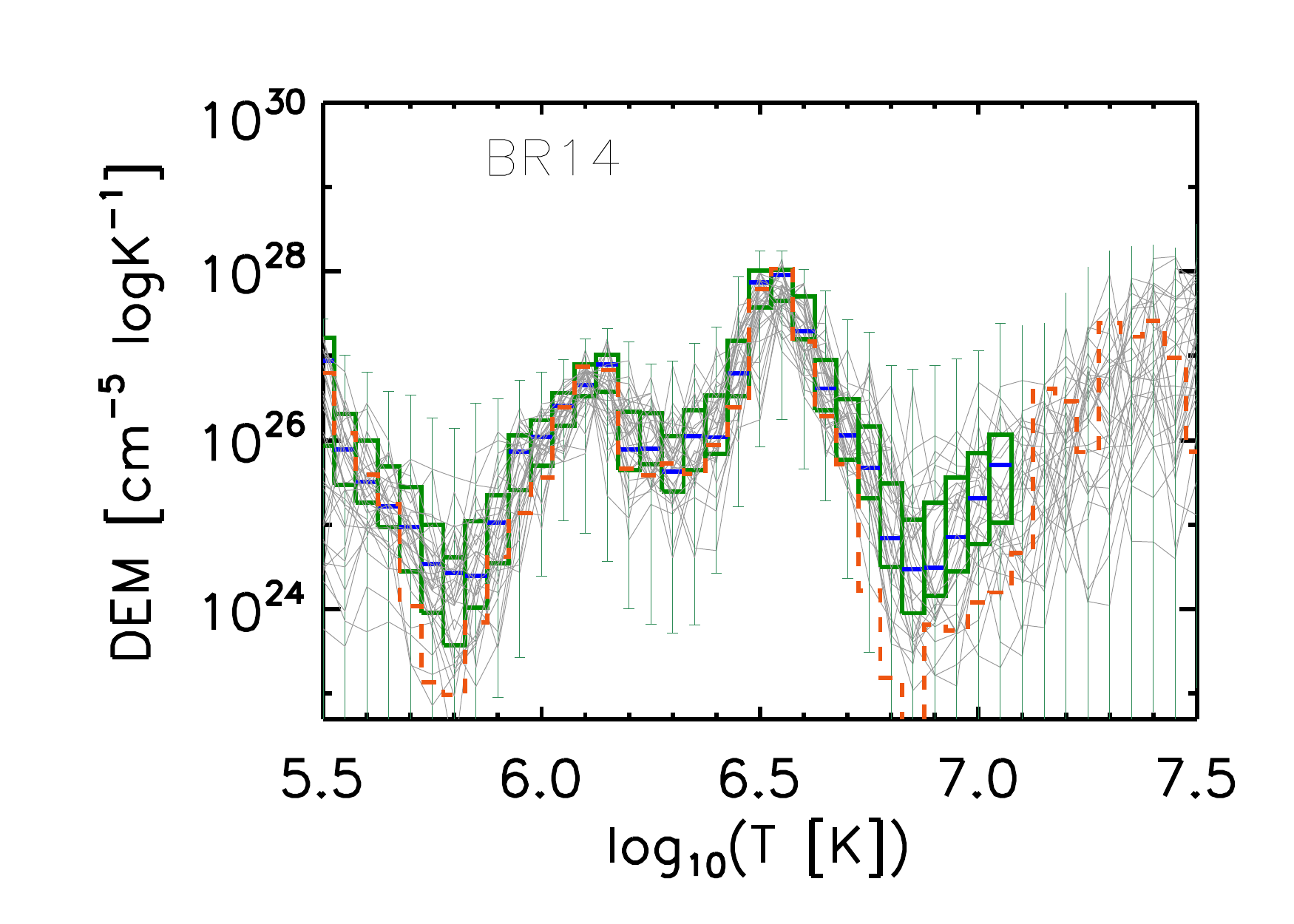}
\includegraphics[trim=0cm 0cm 0cm 0cm, clip=true,width=0.25\textwidth]{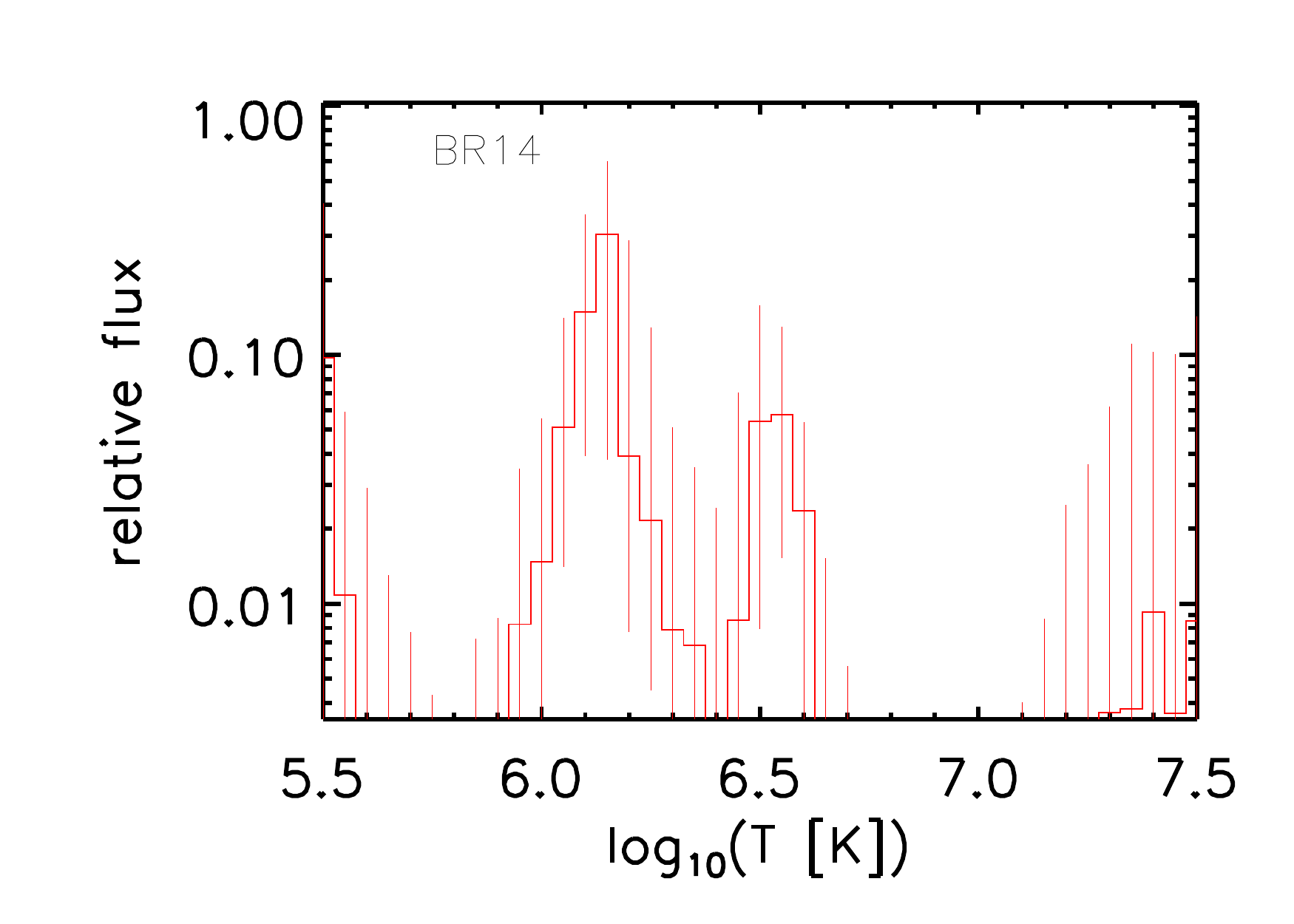}\\
\caption{Light curves, EM loci curves, DEM reconstructions, and contributions of different temperatures to flux in AIA~193~\AA\ for brightenings, as in Figure~\ref{f:app_lc_DEM_1}.}
\label{f:app_lc_DEM_2}
\end{center}
\end{figure*}

\begin{figure*}[htp!]
\begin{center}
\hspace{-0.57cm}
\includegraphics[trim=0cm 0cm 0cm 0cm, clip=true,width=0.25\textwidth]{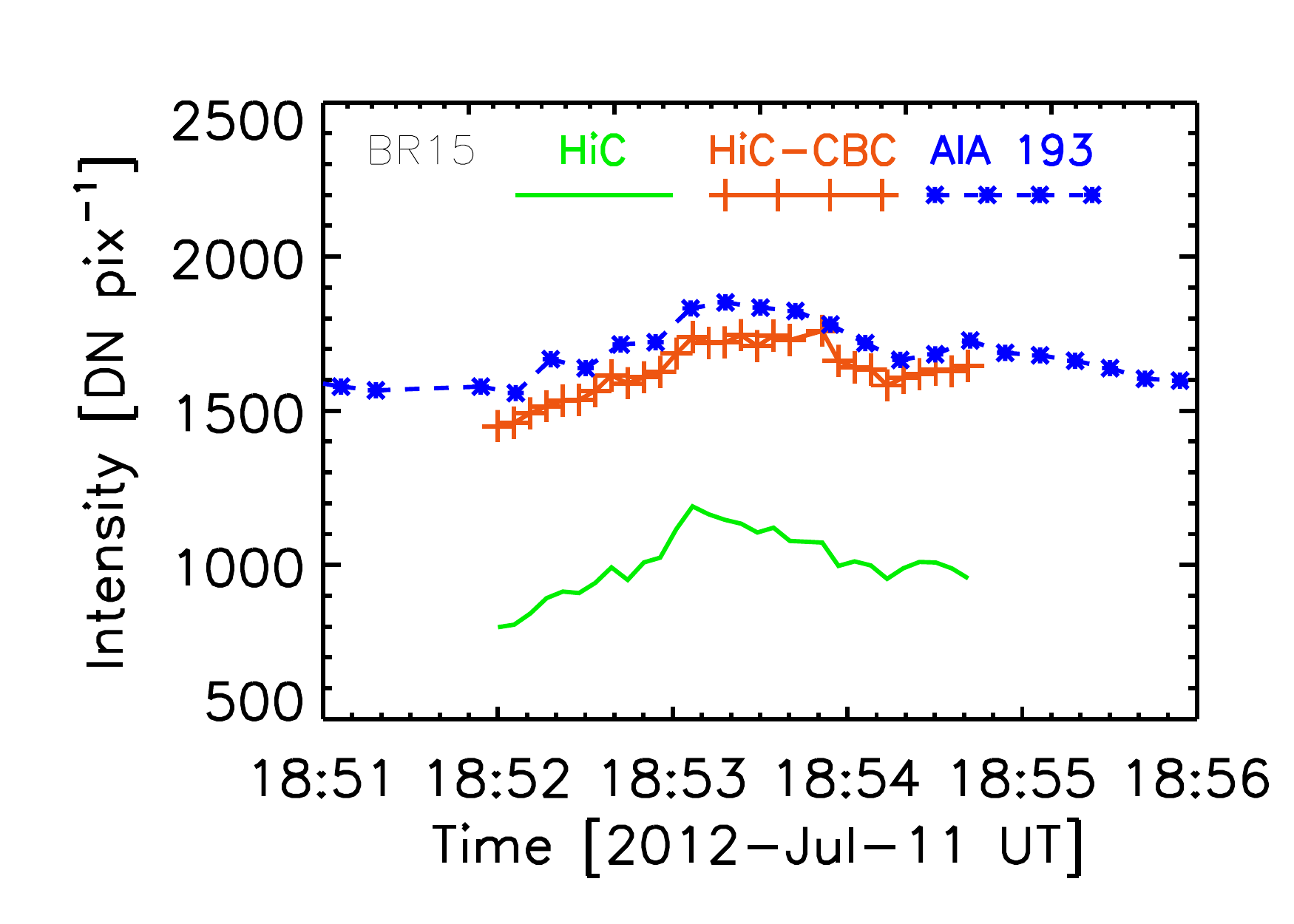}
\includegraphics[trim=0cm 0cm 0cm 0cm, clip=true,width=0.25\textwidth]{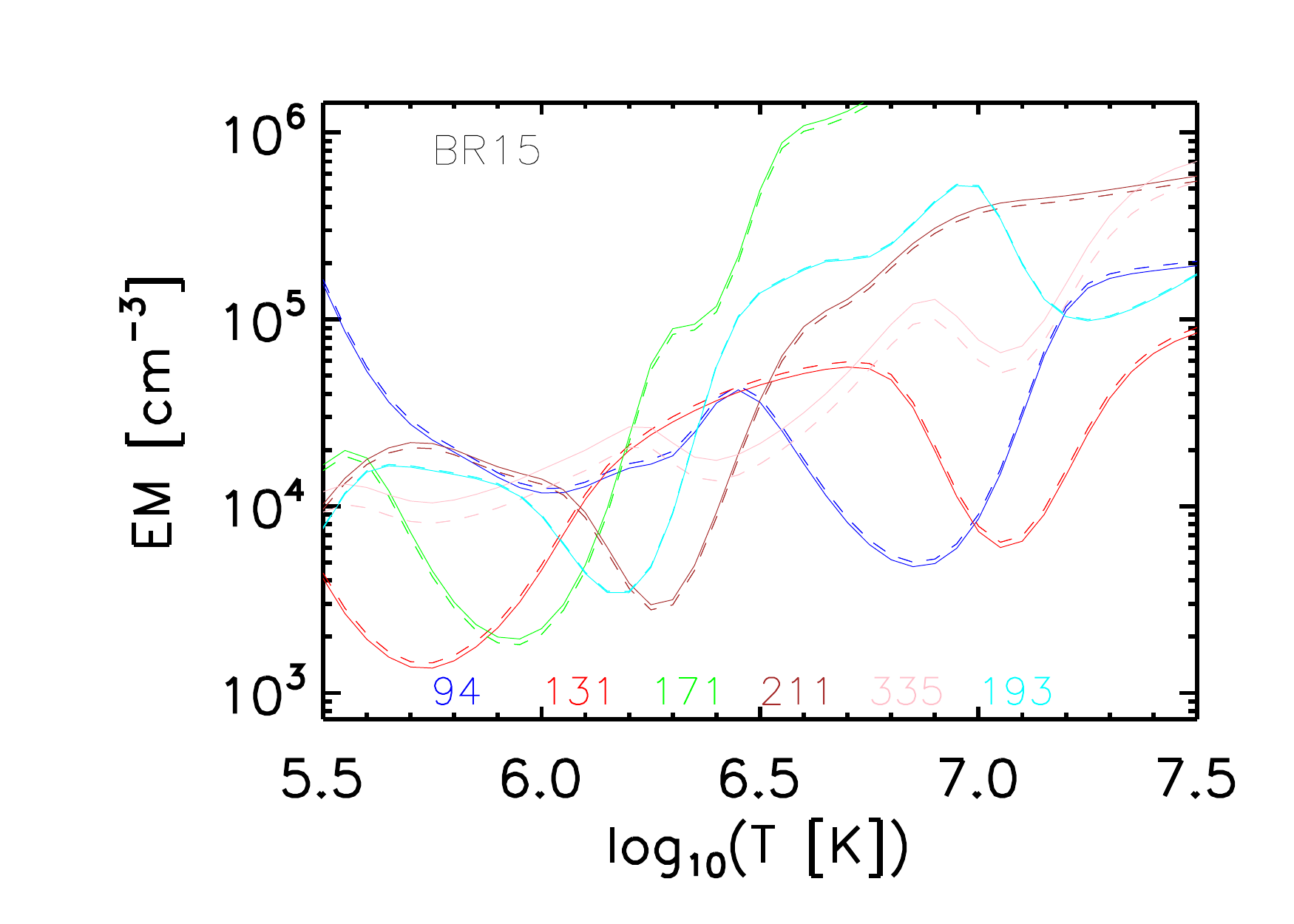}
\includegraphics[trim=0cm 0cm 0cm 0cm, clip=true,width=0.25\textwidth]{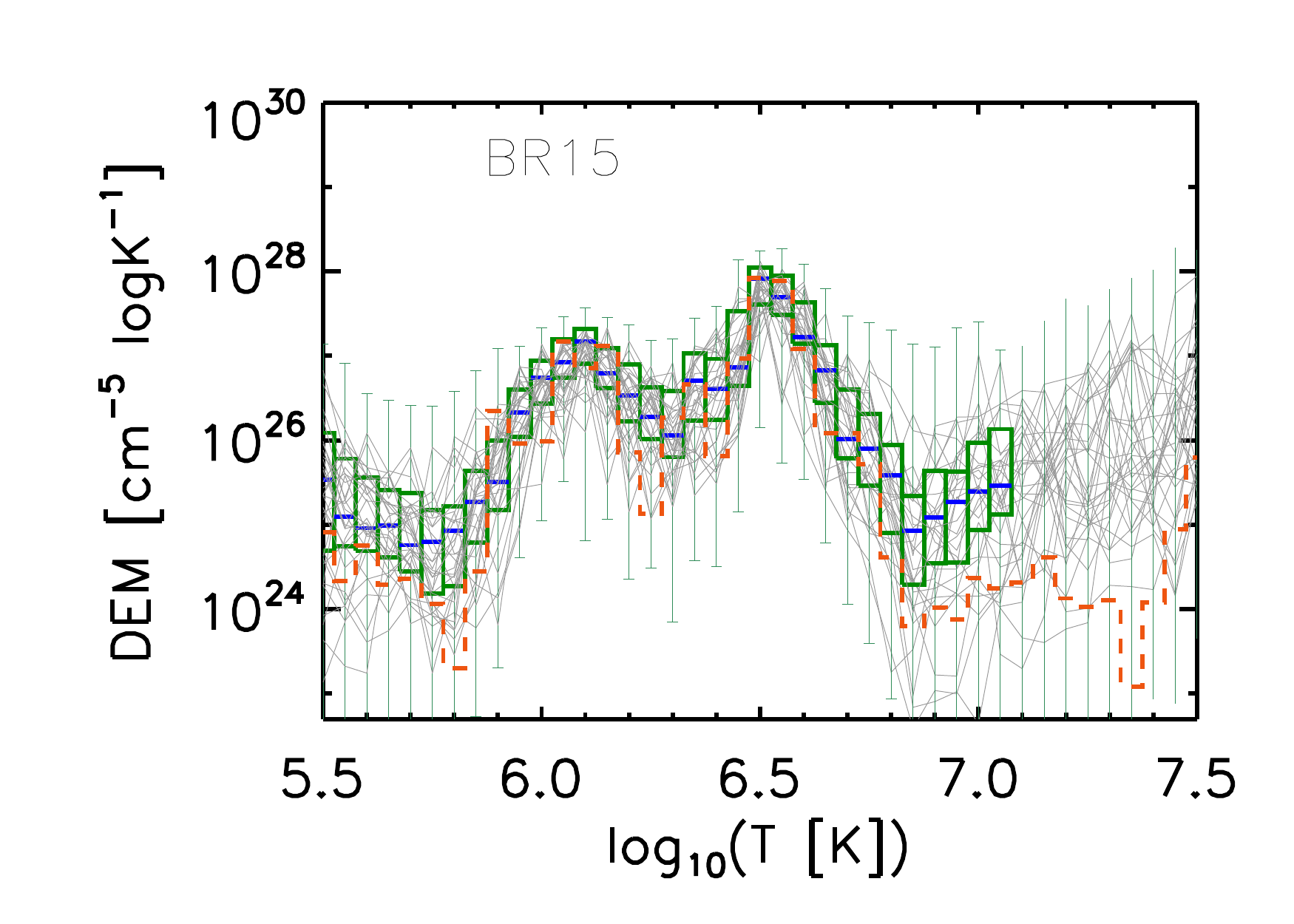}
\includegraphics[trim=0cm 0cm 0cm 0cm, clip=true,width=0.25\textwidth]{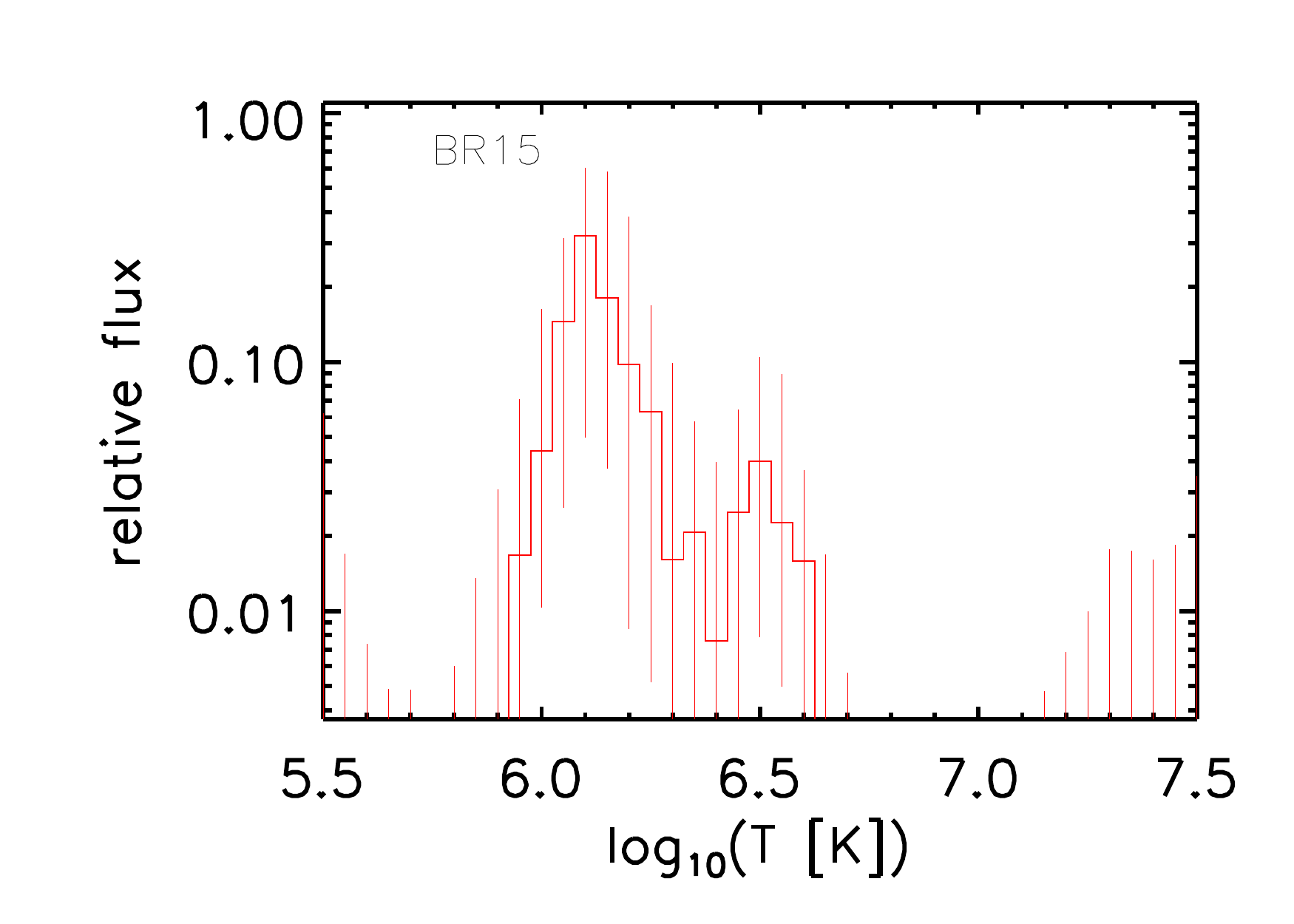}\\
\hspace{-0.57cm}
\includegraphics[trim=0cm 0cm 0cm 0cm, clip=true,width=0.25\textwidth]{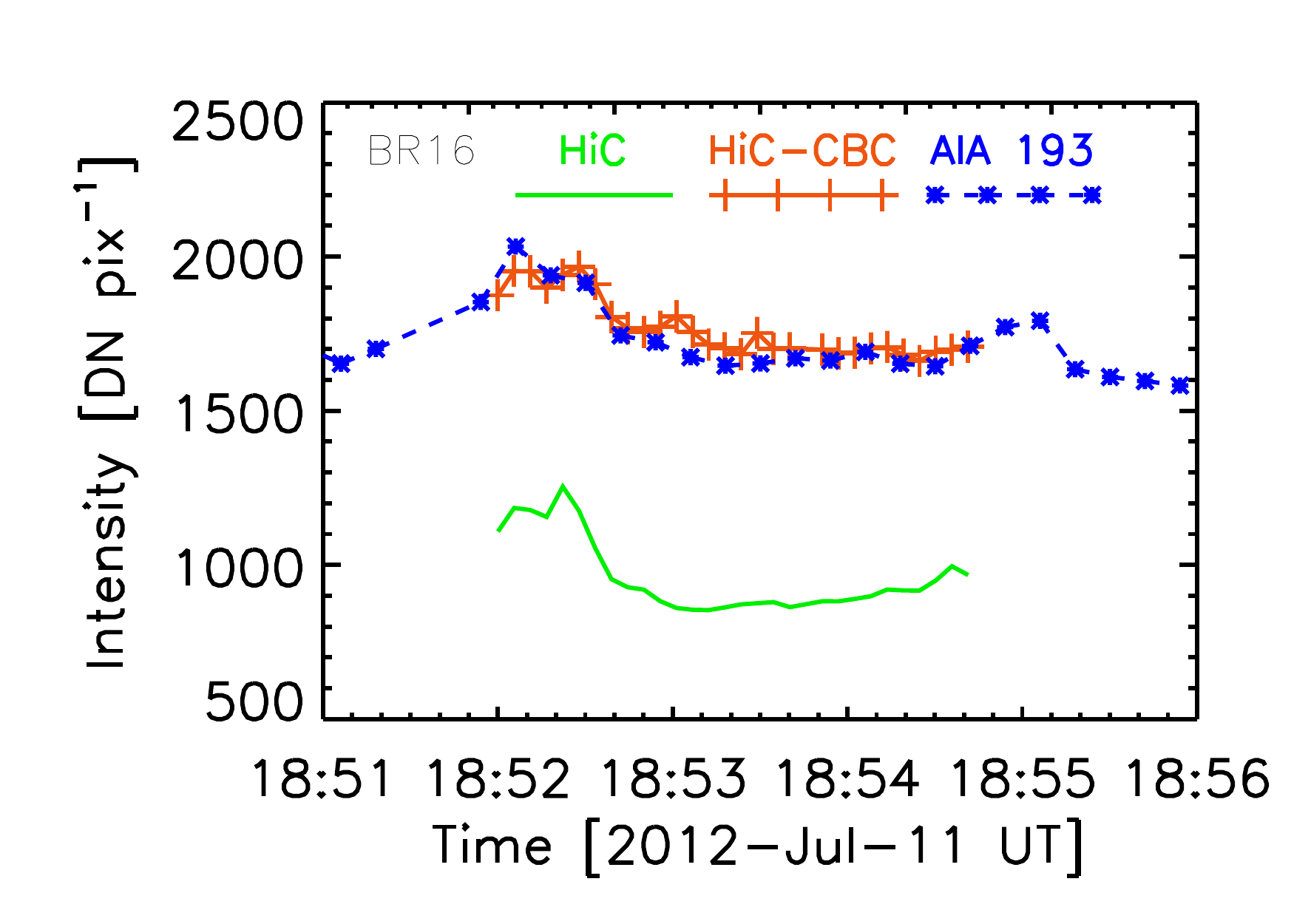}
\includegraphics[trim=0cm 0cm 0cm 0cm, clip=true,width=0.25\textwidth]{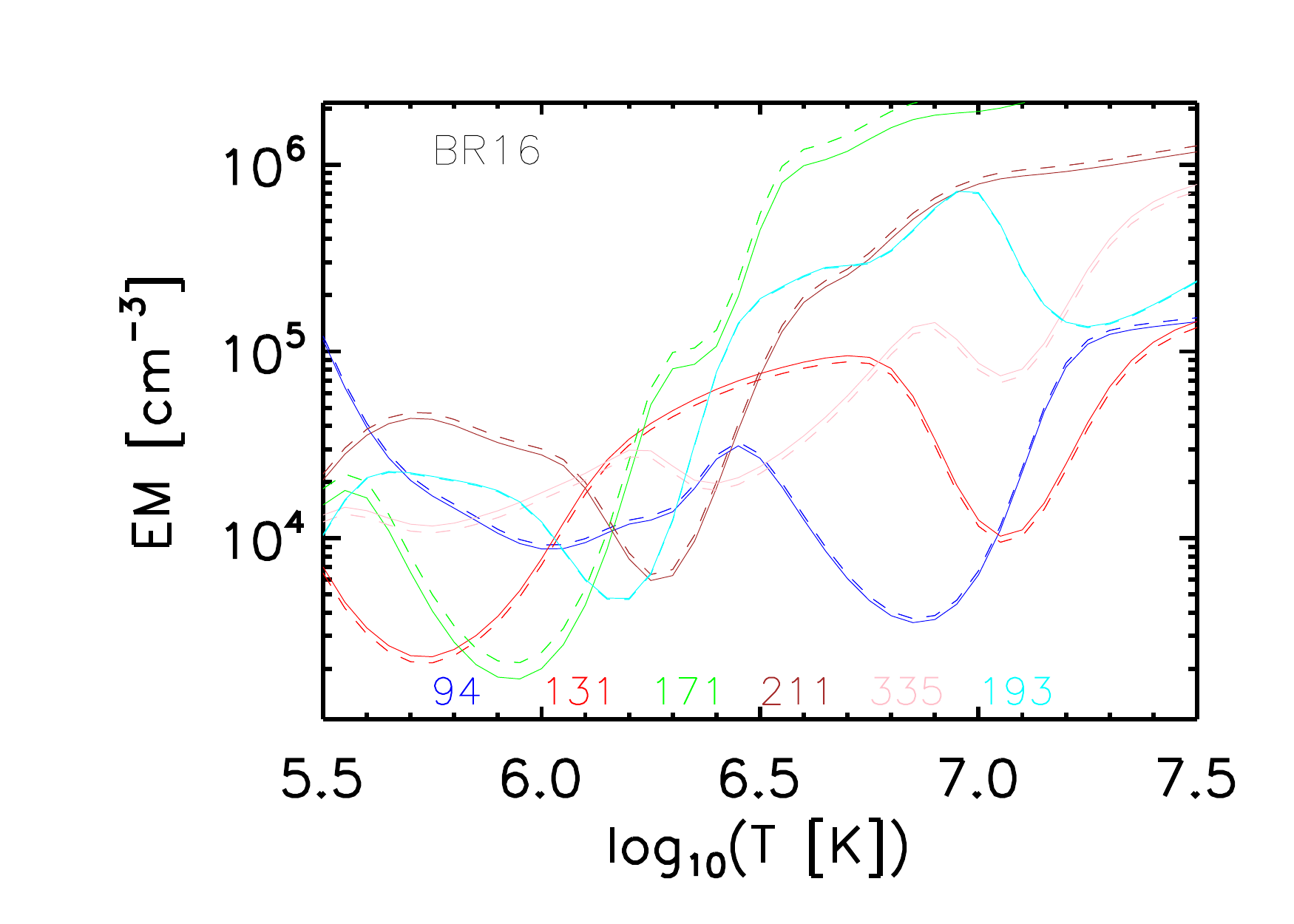}
\includegraphics[trim=0cm 0cm 0cm 0cm, clip=true,width=0.25\textwidth]{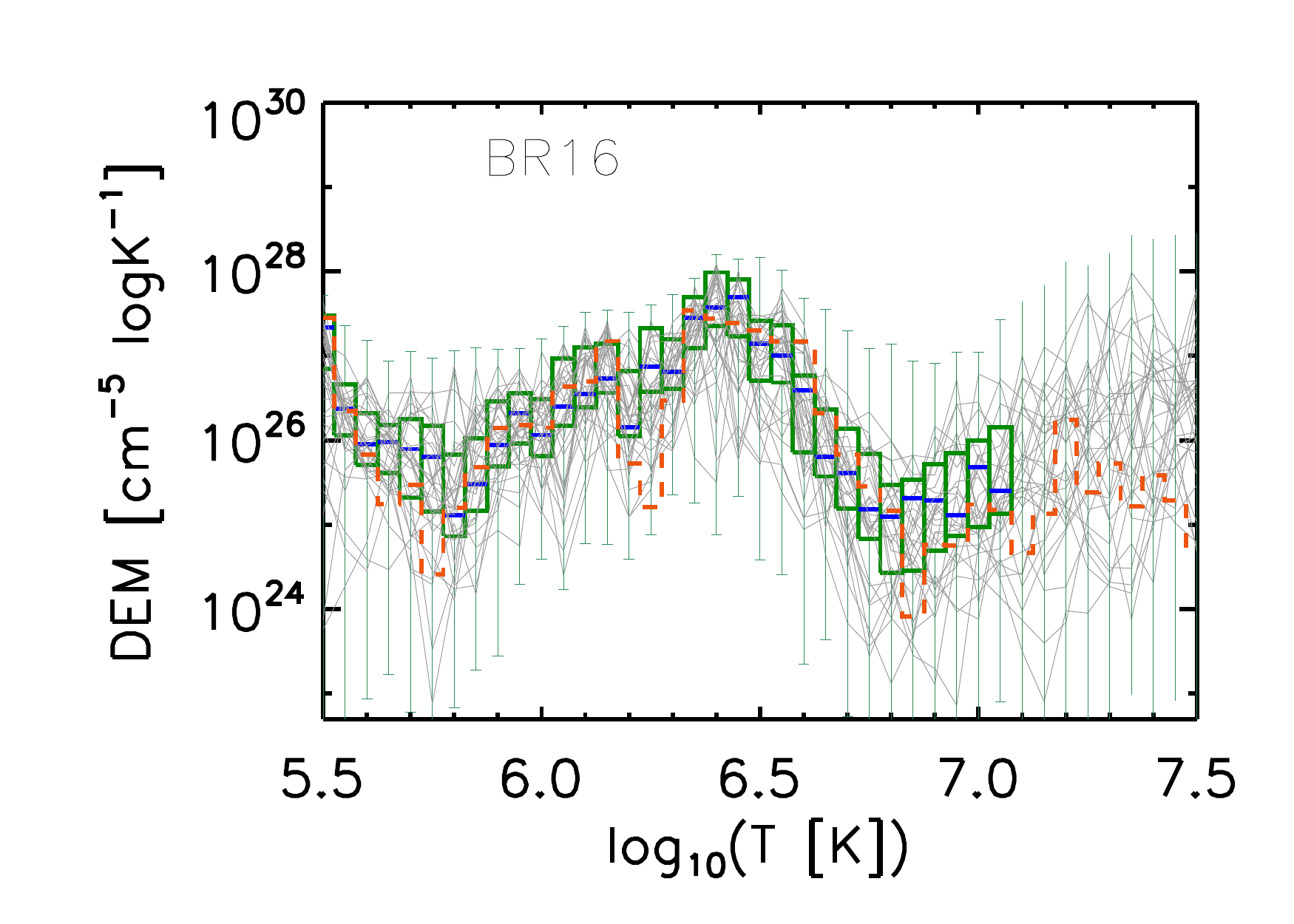}
\includegraphics[trim=0cm 0cm 0cm 0cm, clip=true,width=0.25\textwidth]{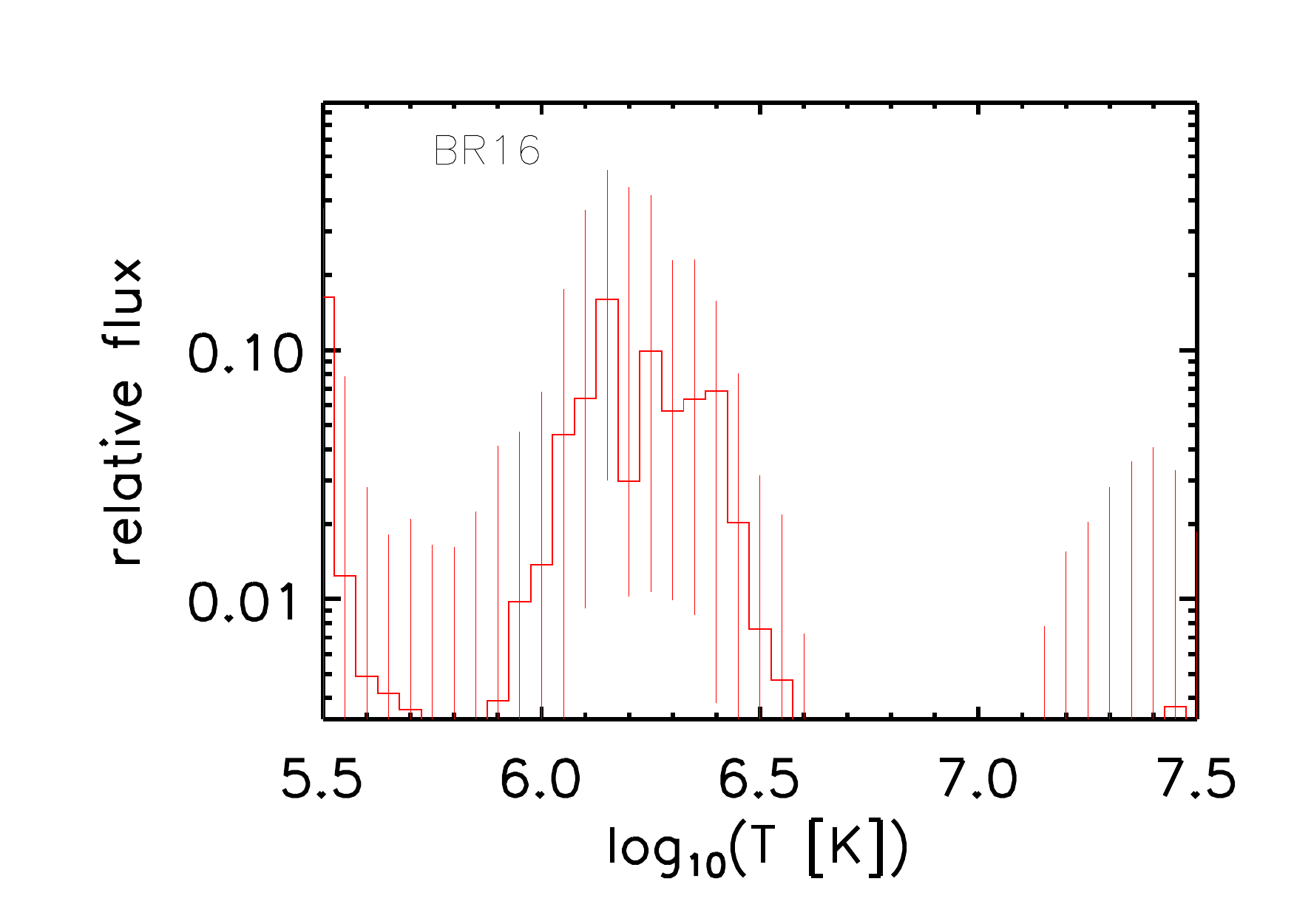}\\
\hspace{-0.57cm}
\includegraphics[trim=0cm 0cm 0cm 0cm, clip=true,width=0.25\textwidth]{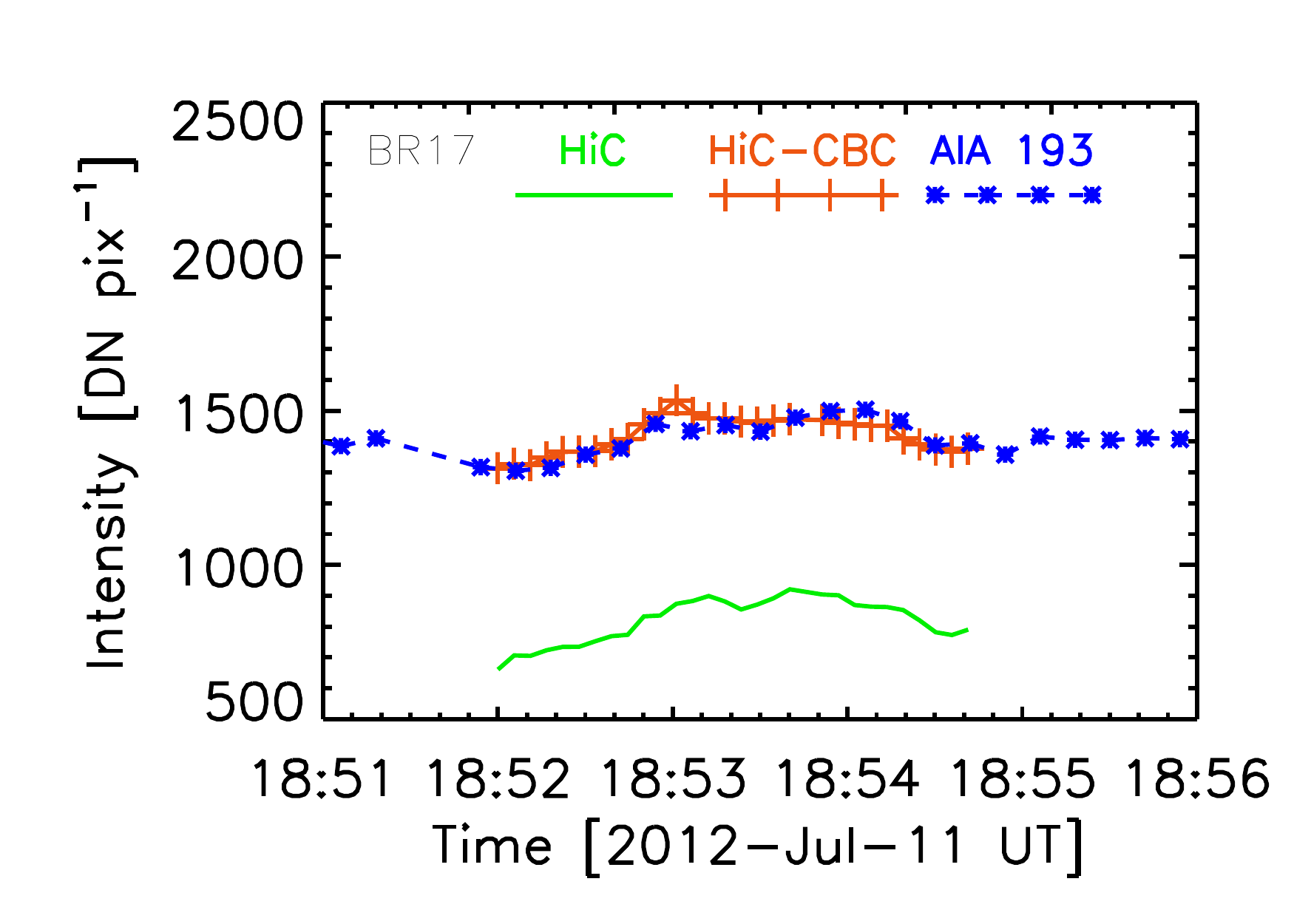}
\includegraphics[trim=0cm 0cm 0cm 0cm, clip=true,width=0.25\textwidth]{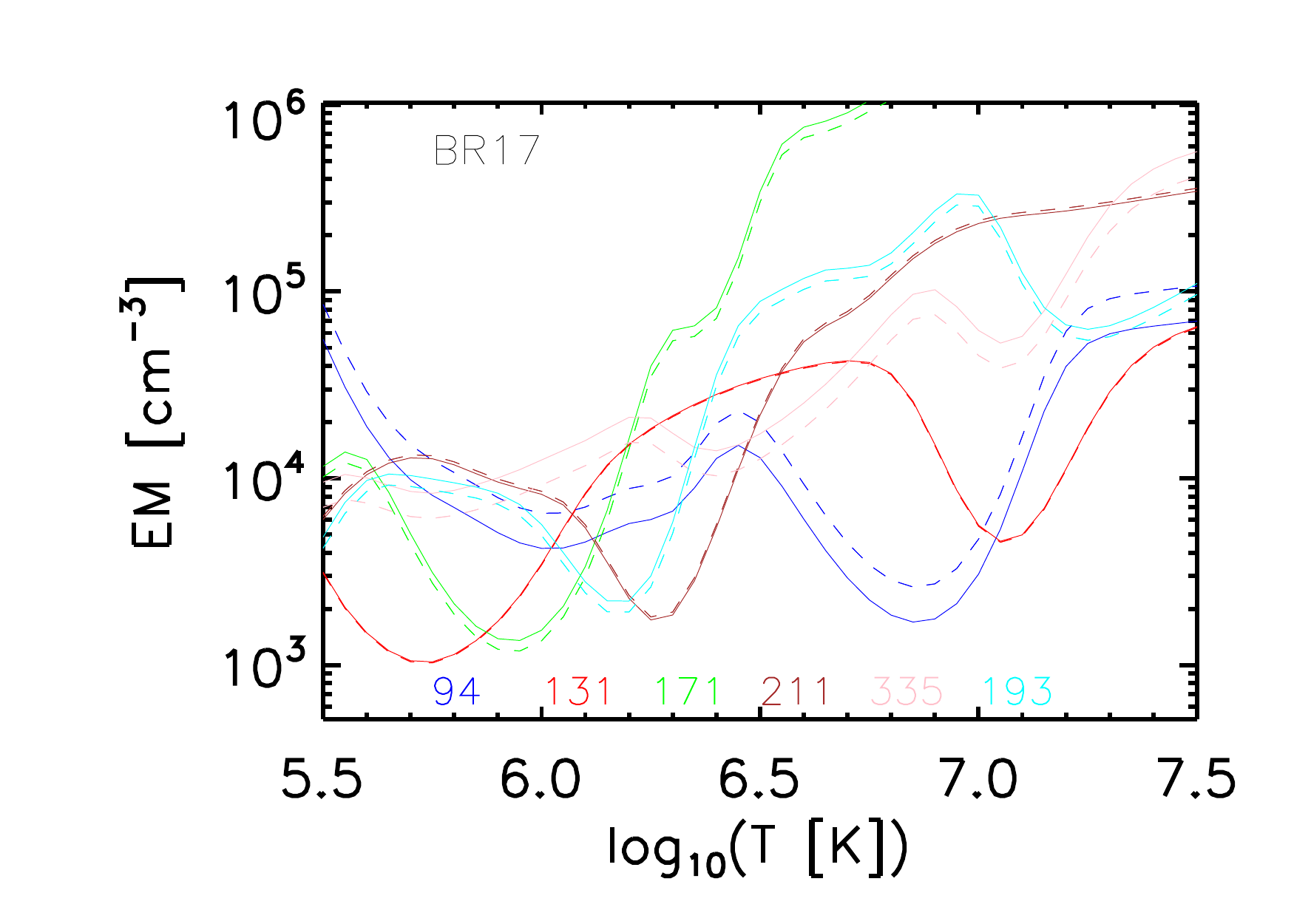}
\includegraphics[trim=0cm 0cm 0cm 0cm, clip=true,width=0.25\textwidth]{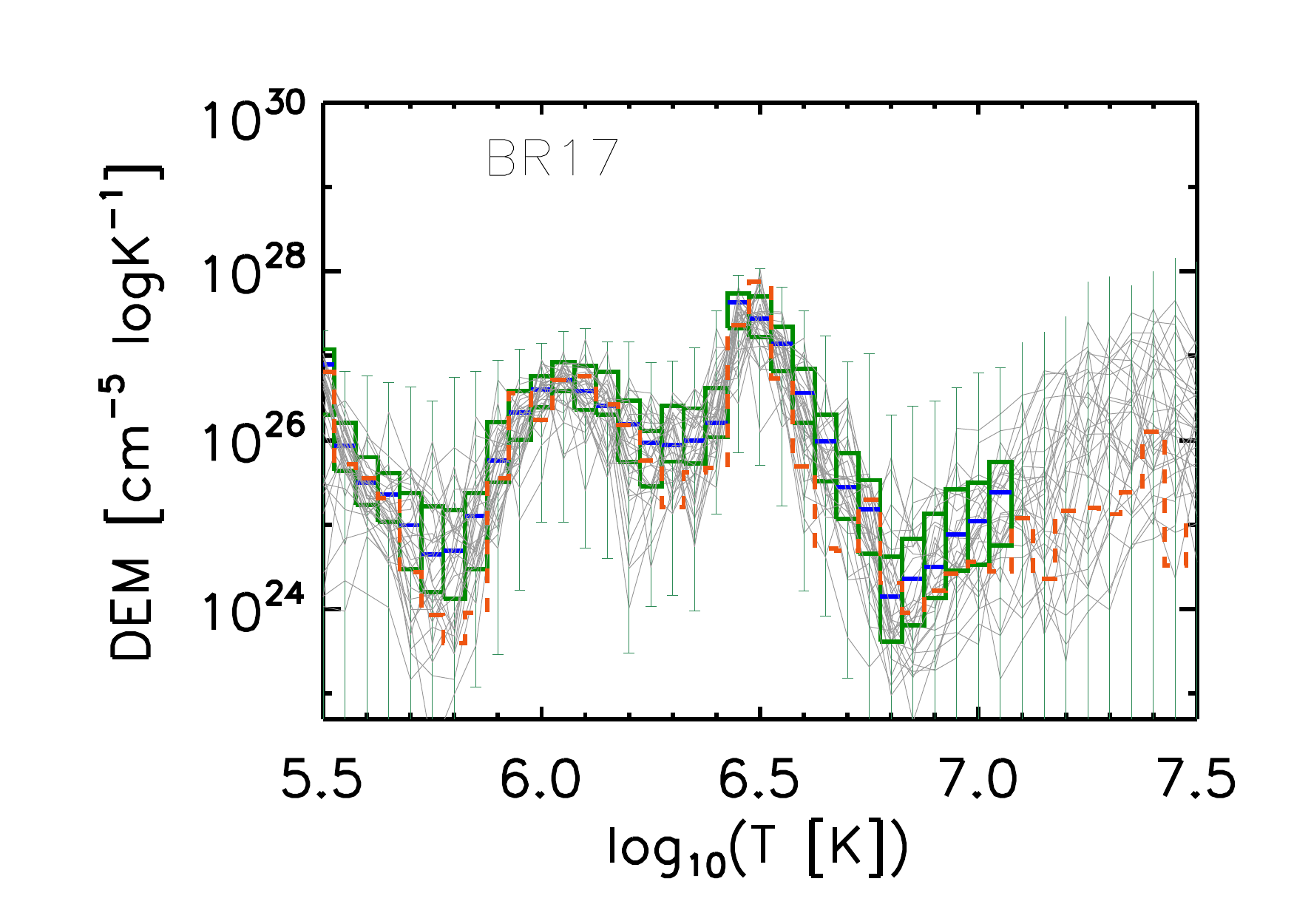}
\includegraphics[trim=0cm 0cm 0cm 0cm, clip=true,width=0.25\textwidth]{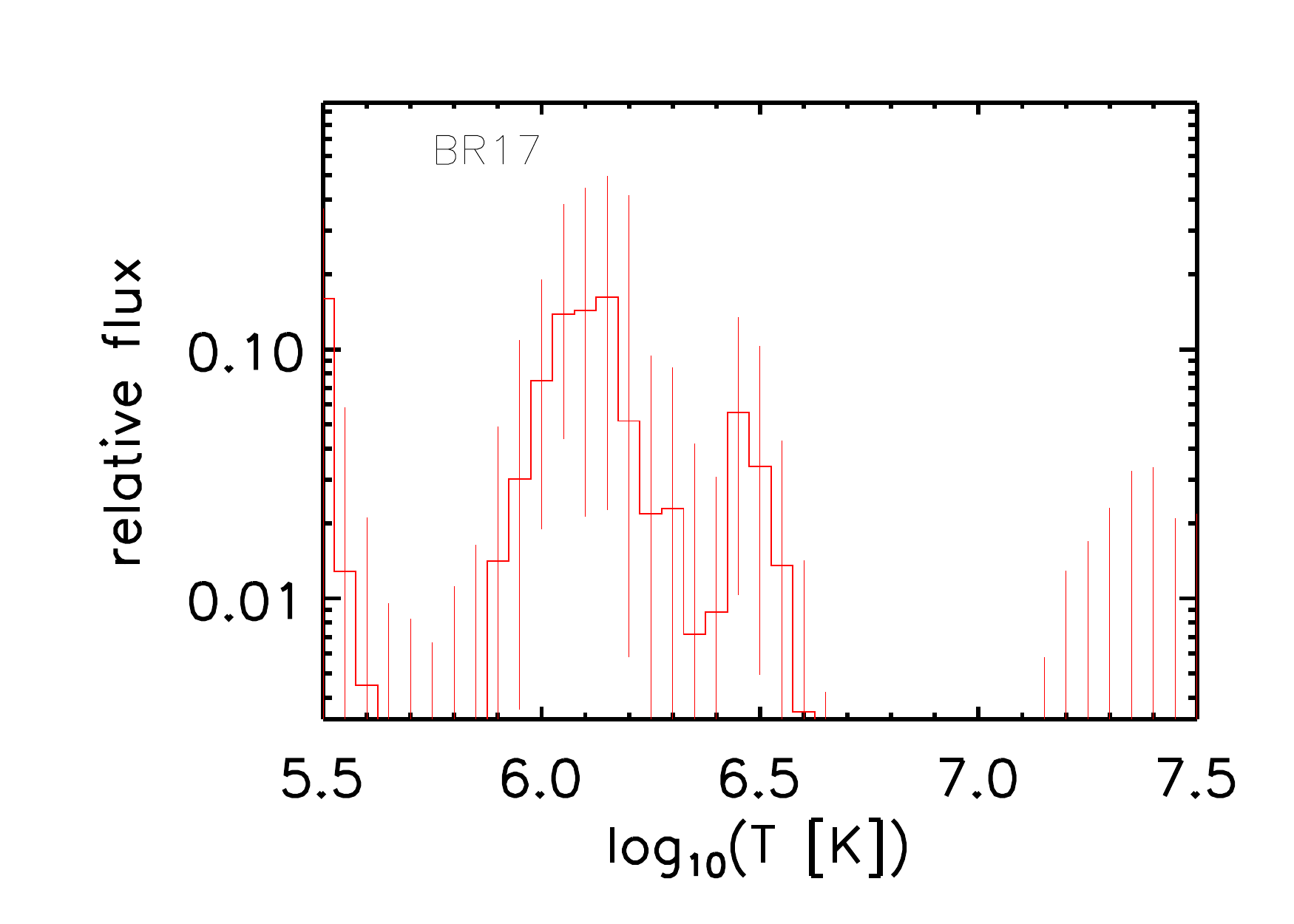}\\
\hspace{-0.57cm}
\includegraphics[trim=0cm 0cm 0cm 0cm, clip=true,width=0.25\textwidth]{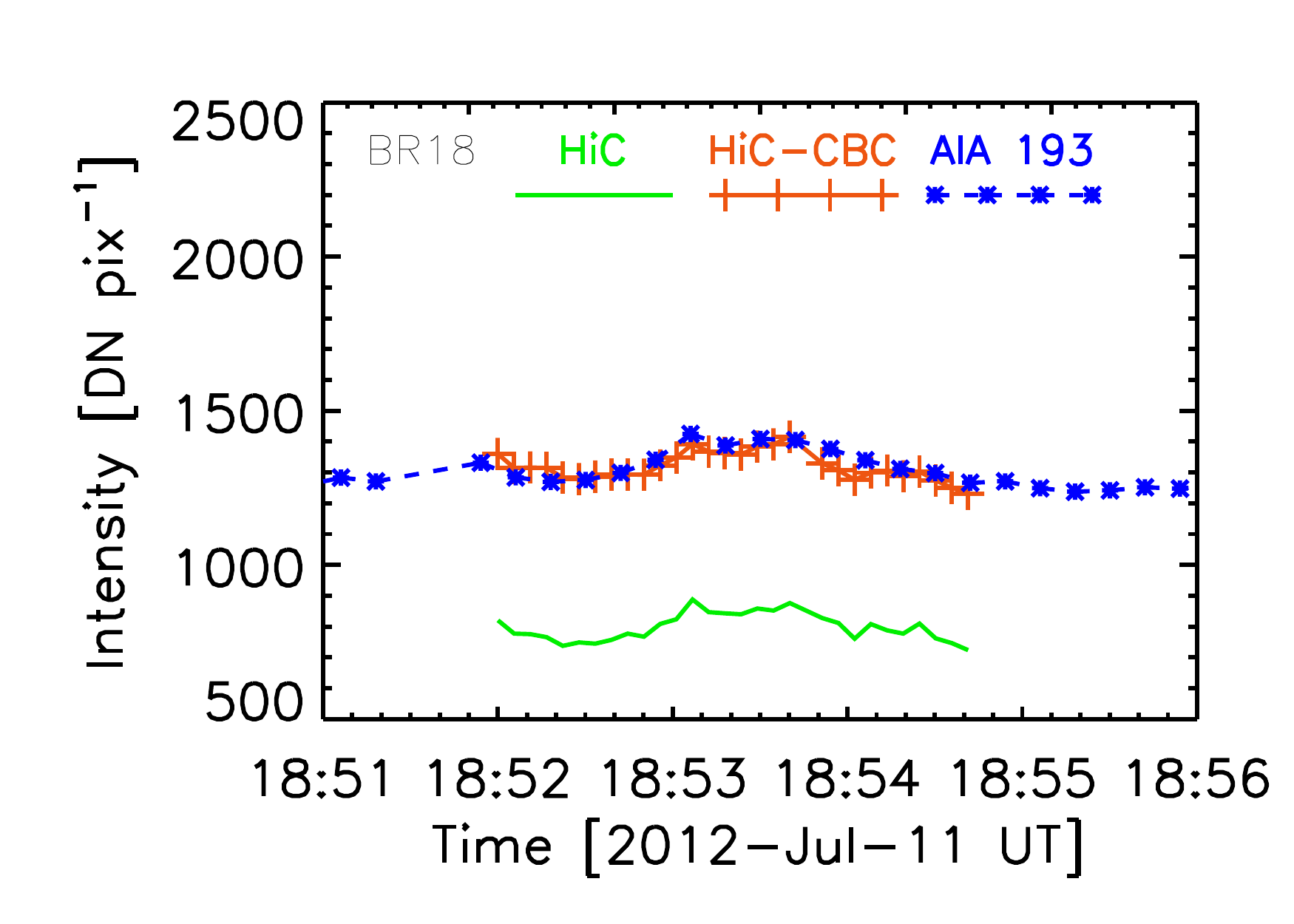}
\includegraphics[trim=0cm 0cm 0cm 0cm, clip=true,width=0.25\textwidth]{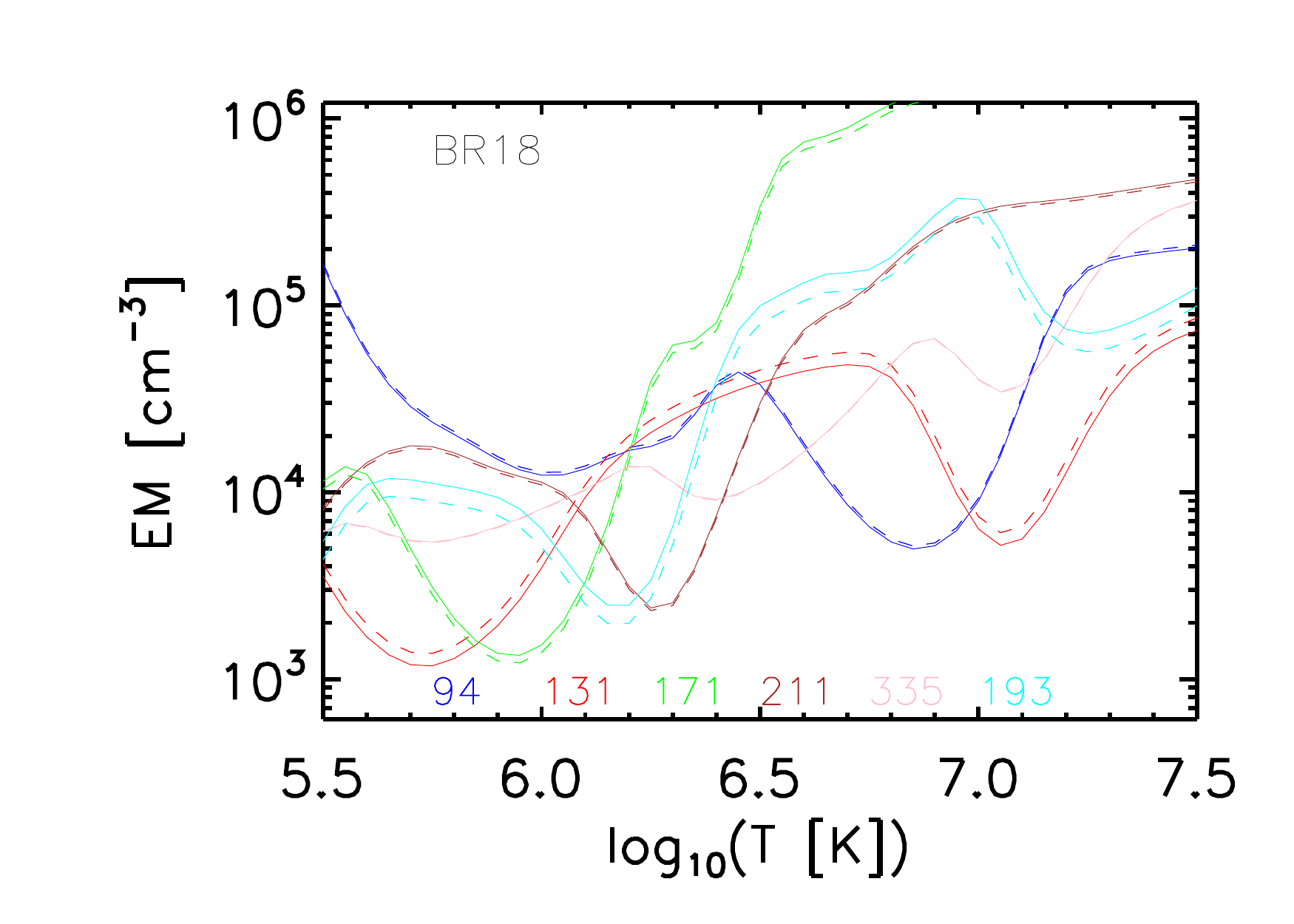}
\includegraphics[trim=0cm 0cm 0cm 0cm, clip=true,width=0.25\textwidth]{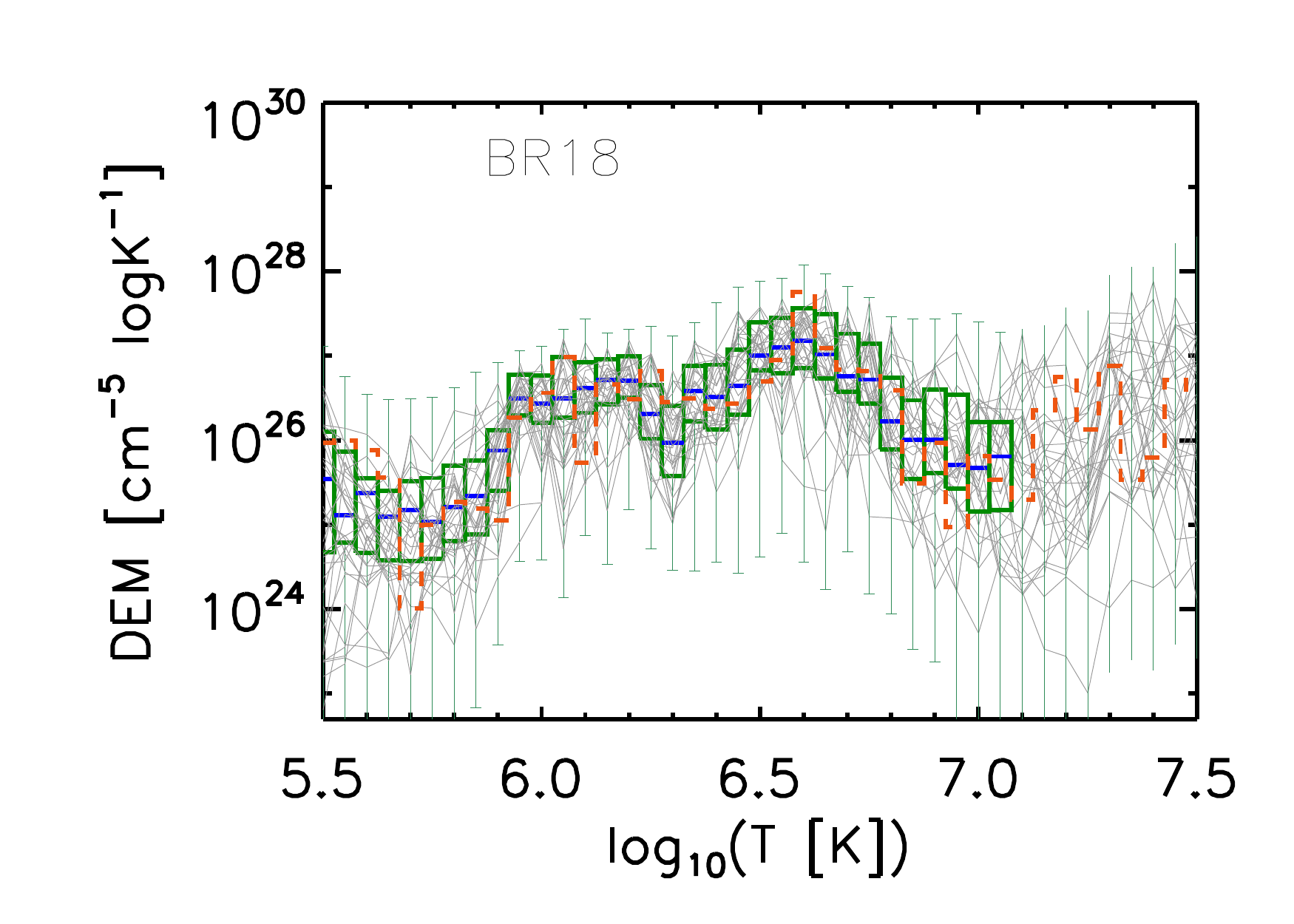}
\includegraphics[trim=0cm 0cm 0cm 0cm, clip=true,width=0.25\textwidth]{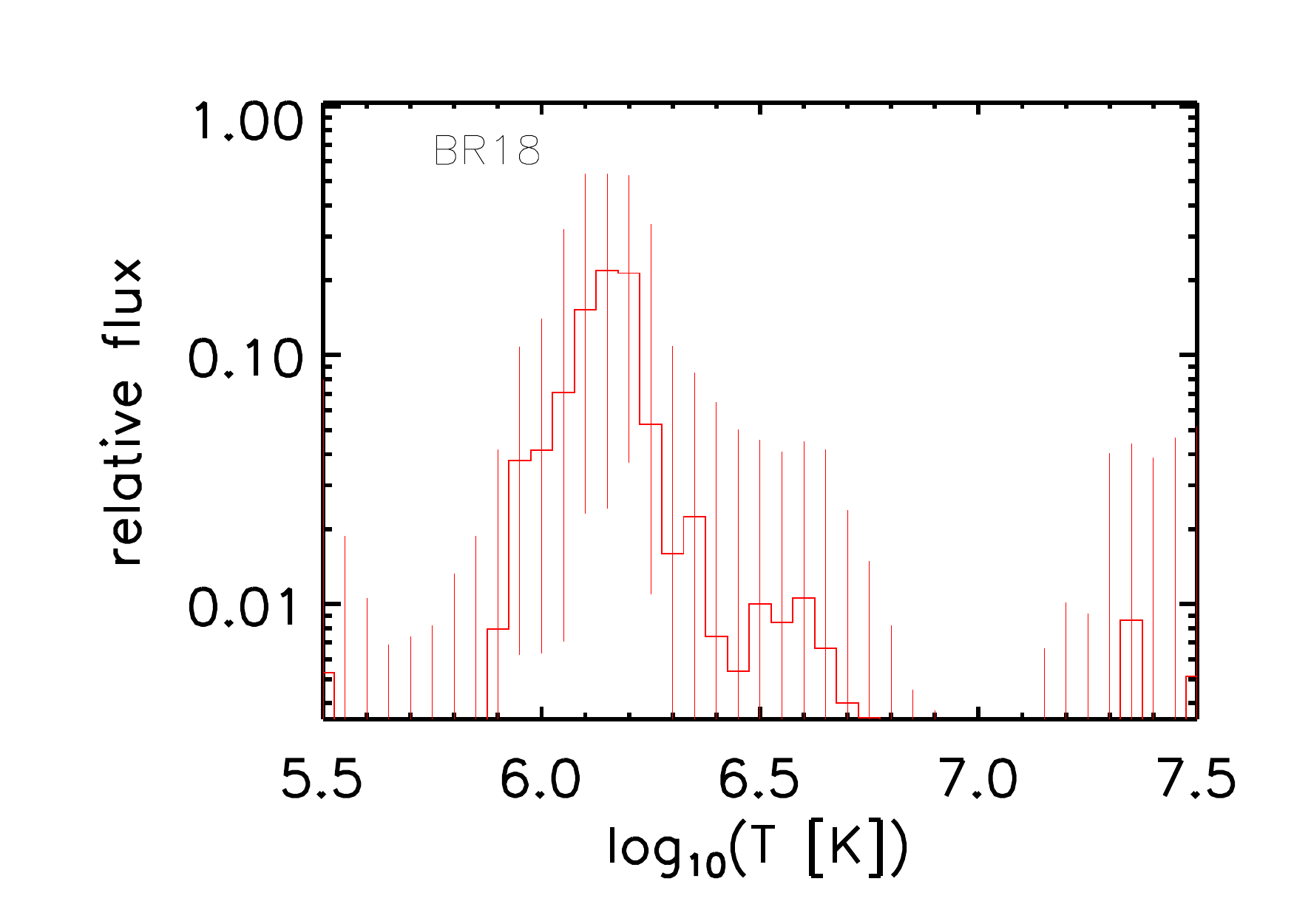}\\
\hspace{-0.57cm}
\includegraphics[trim=0cm 0cm 0cm 0cm, clip=true,width=0.25\textwidth]{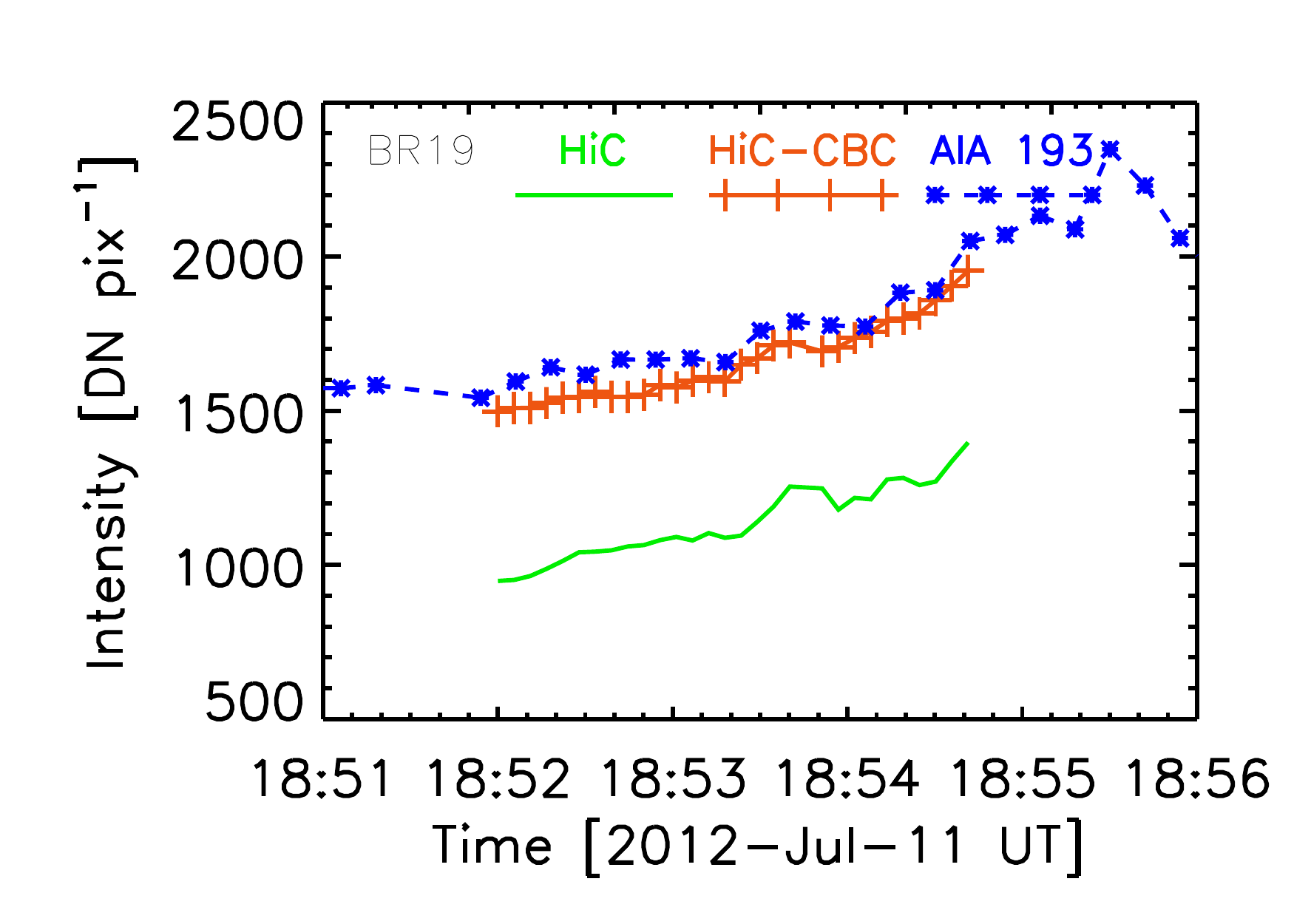}
\includegraphics[trim=0cm 0cm 0cm 0cm, clip=true,width=0.25\textwidth]{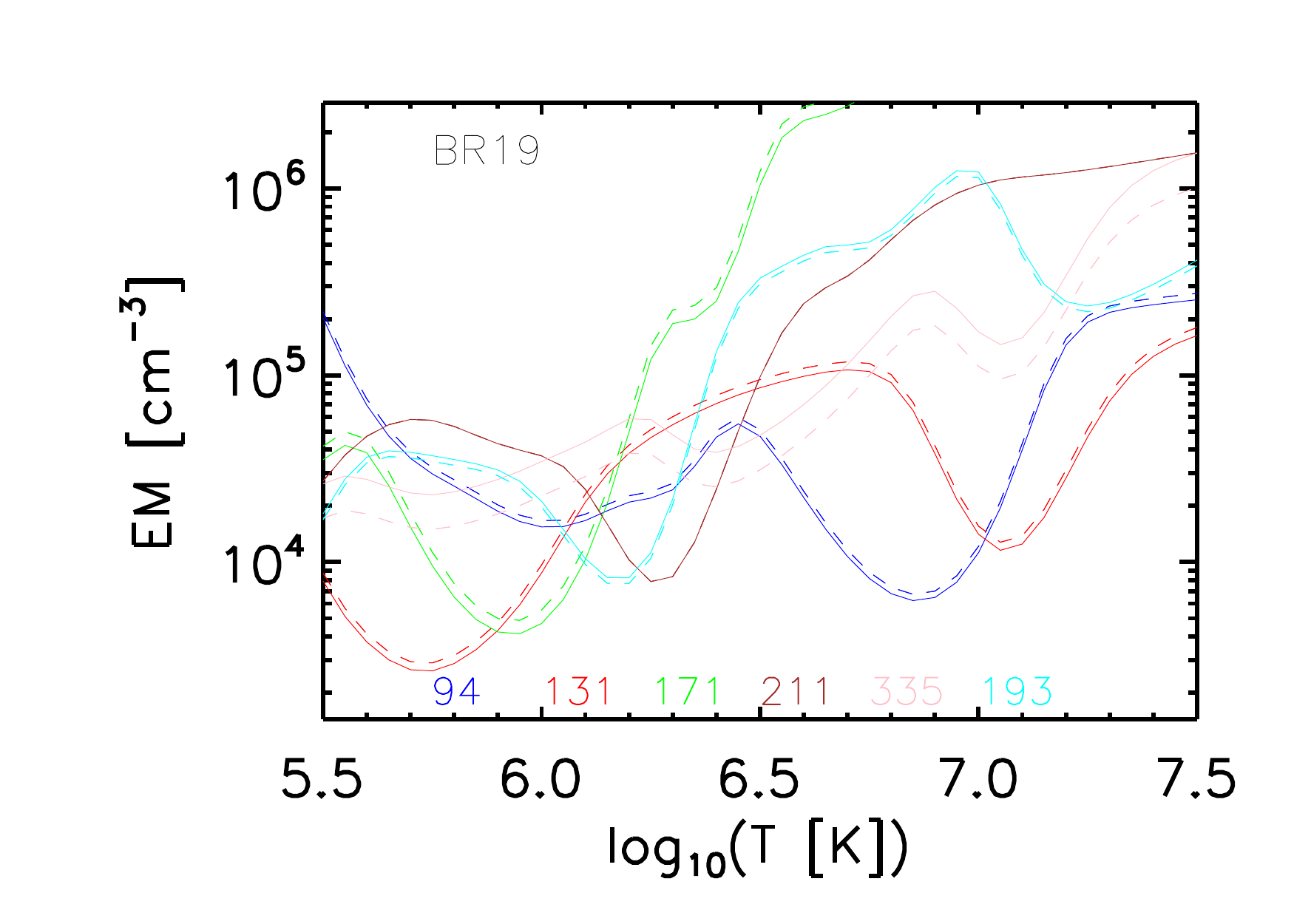}
\includegraphics[trim=0cm 0cm 0cm 0cm, clip=true,width=0.25\textwidth]{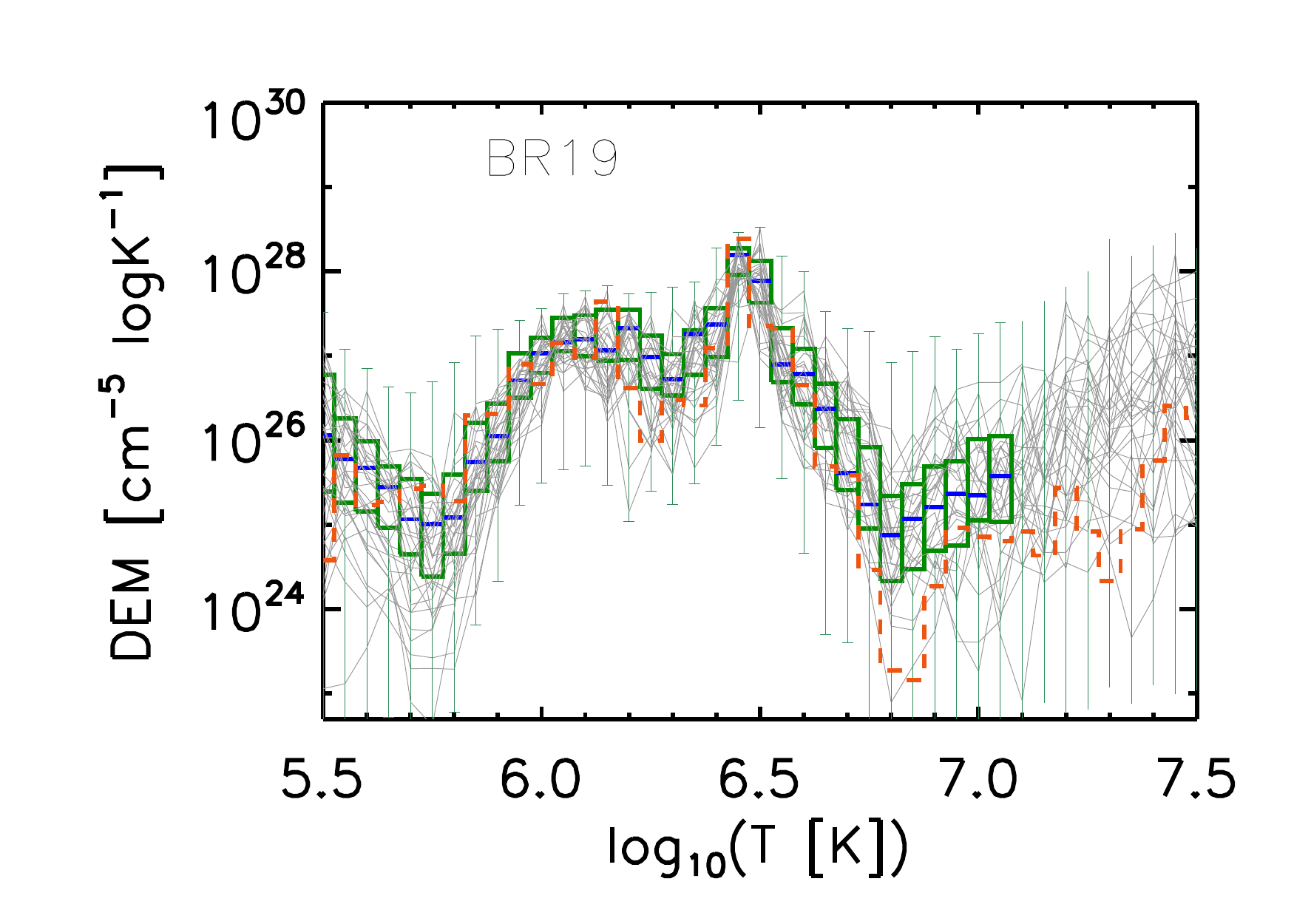}
\includegraphics[trim=0cm 0cm 0cm 0cm, clip=true,width=0.25\textwidth]{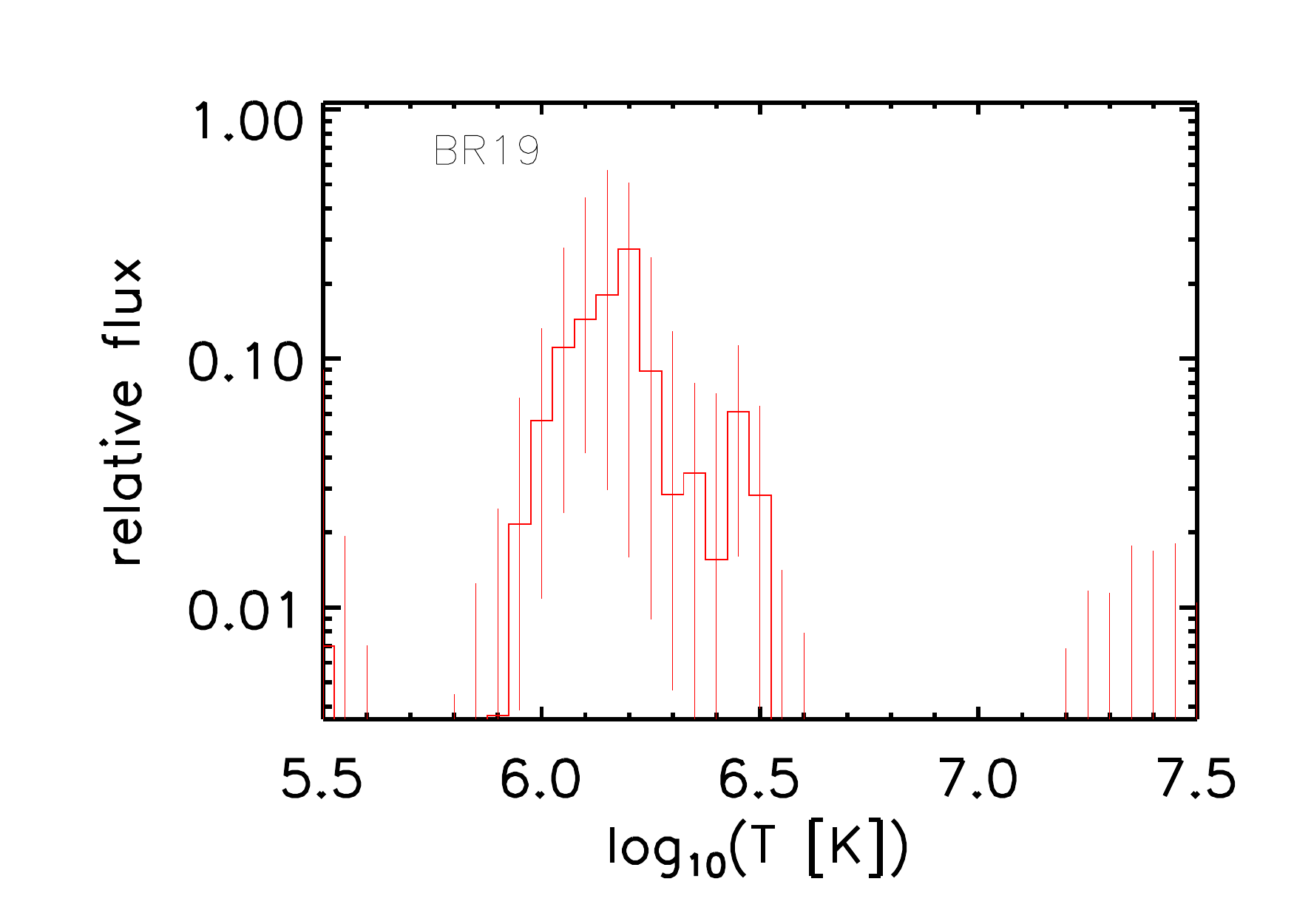}\\
\hspace{-0.57cm}
\includegraphics[trim=0cm 0cm 0cm 0cm, clip=true,width=0.25\textwidth]{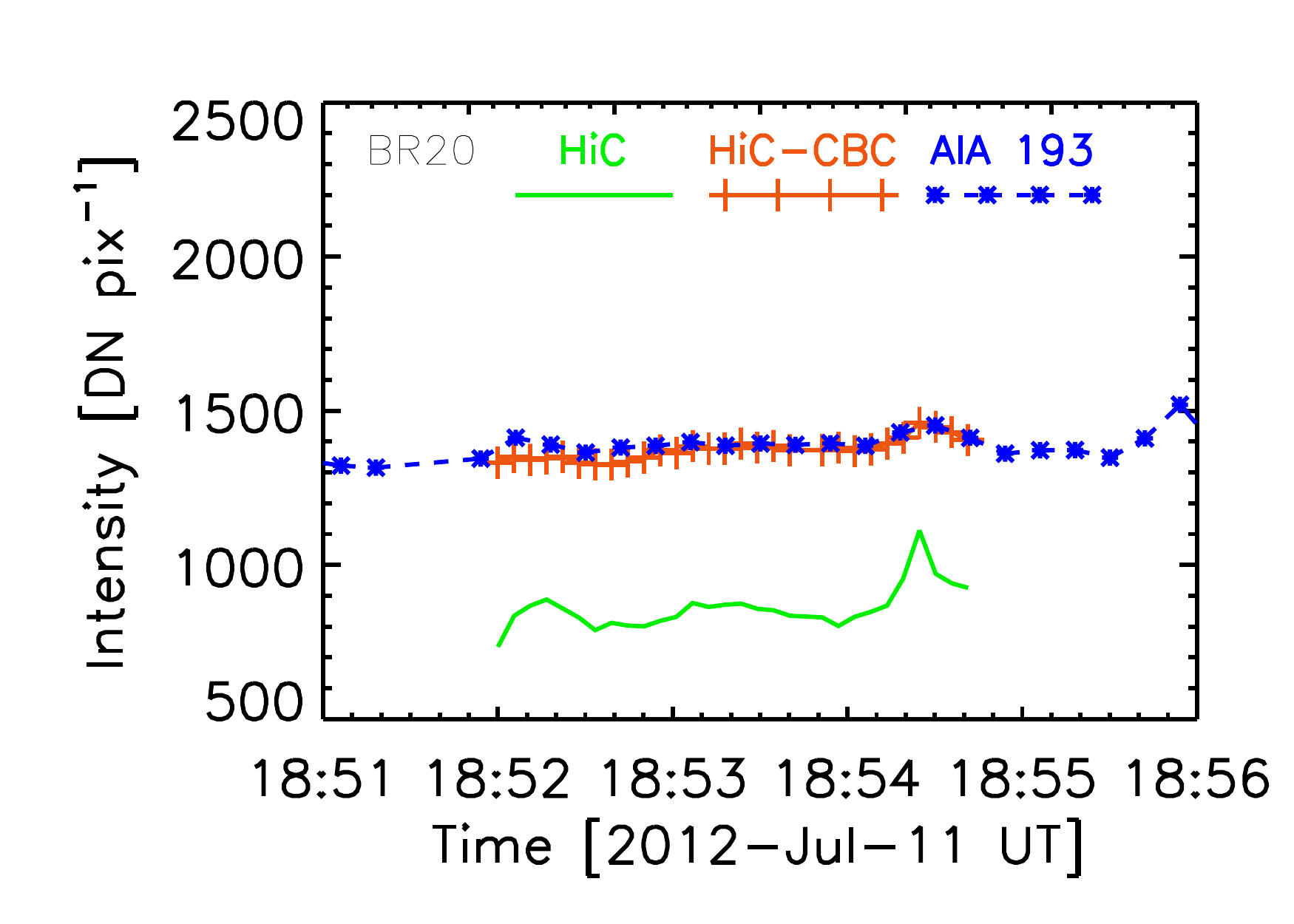}
\includegraphics[trim=0cm 0cm 0cm 0cm, clip=true,width=0.25\textwidth]{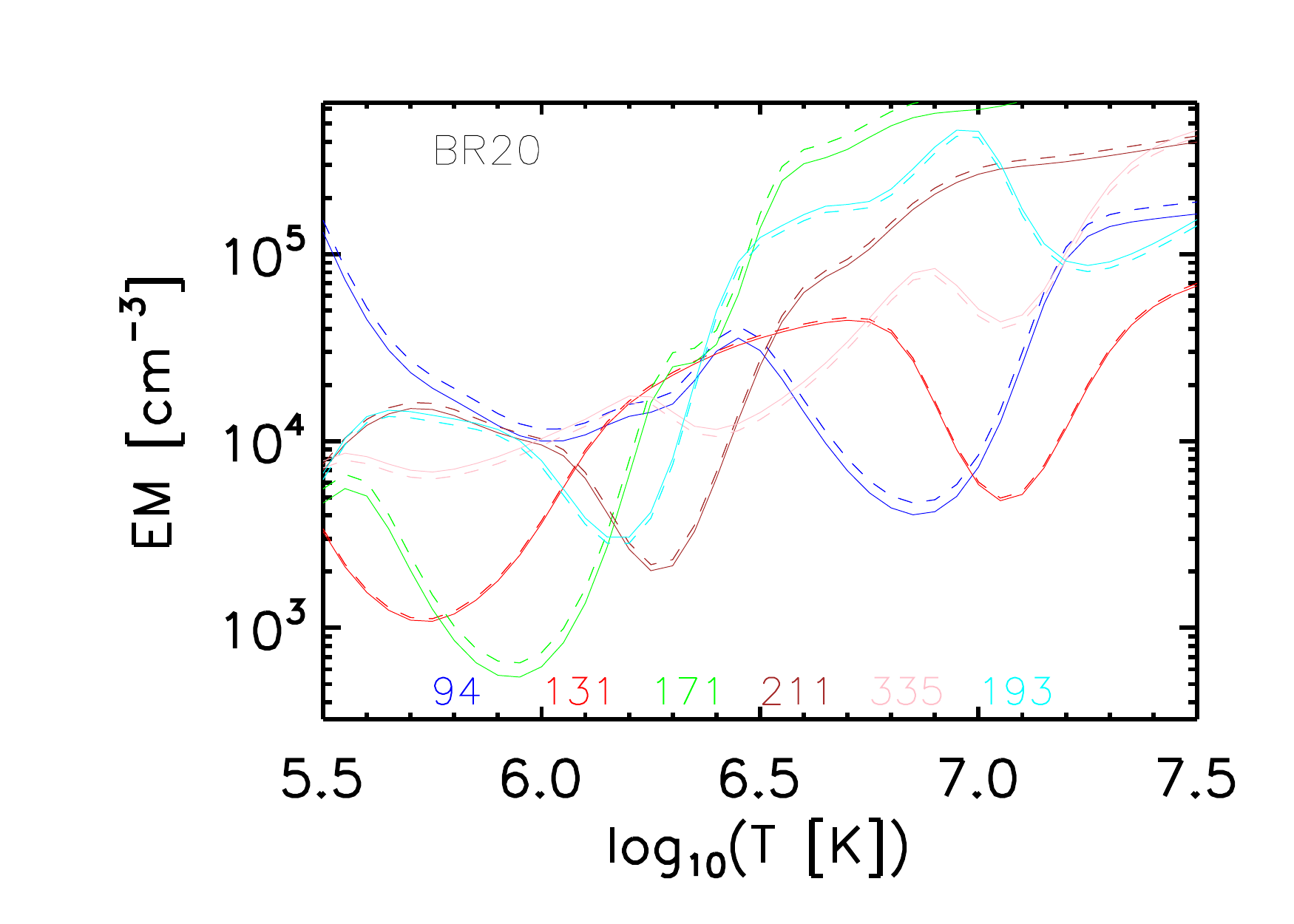}
\includegraphics[trim=0cm 0cm 0cm 0cm, clip=true,width=0.25\textwidth]{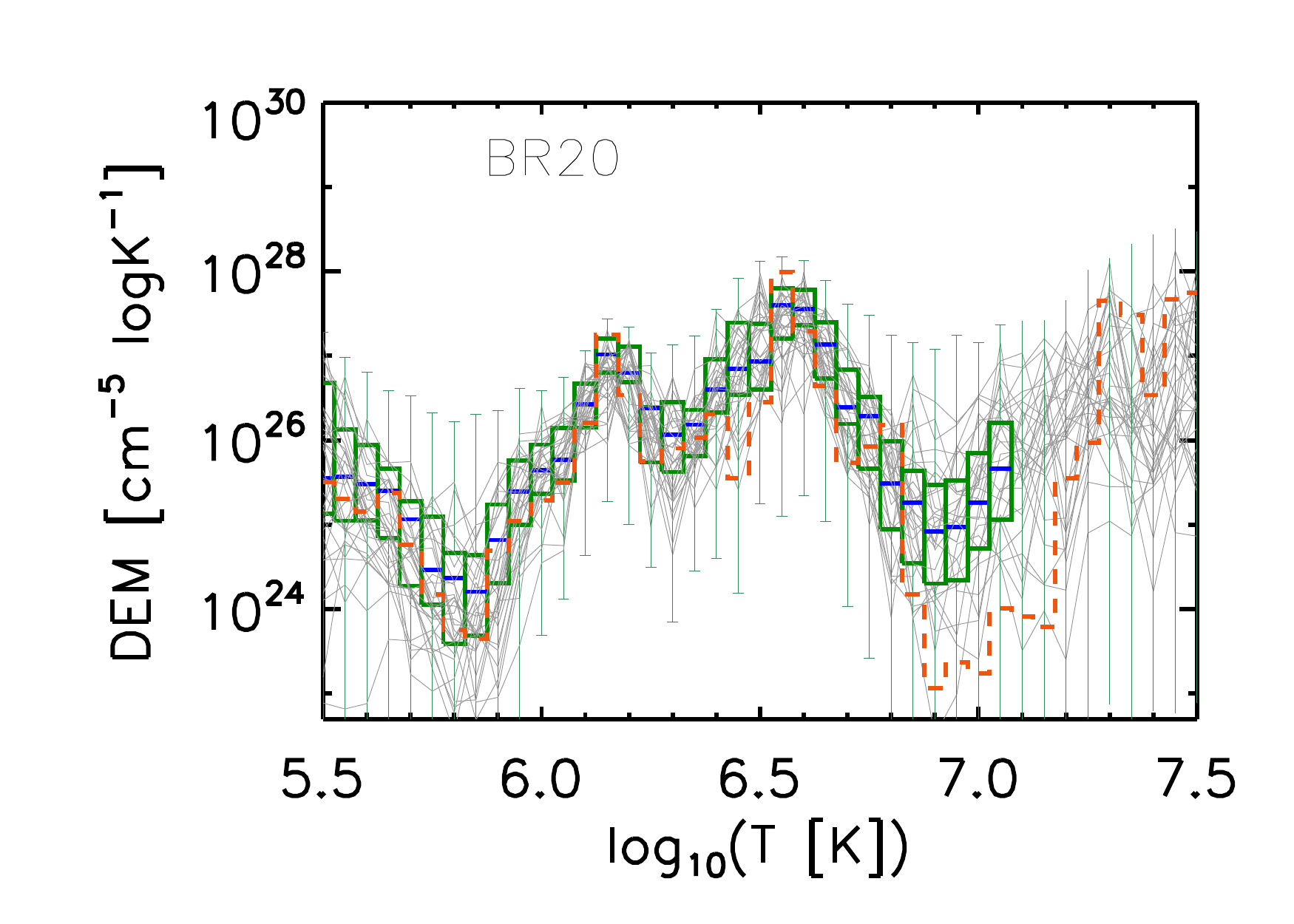}
\includegraphics[trim=0cm 0cm 0cm 0cm, clip=true,width=0.25\textwidth]{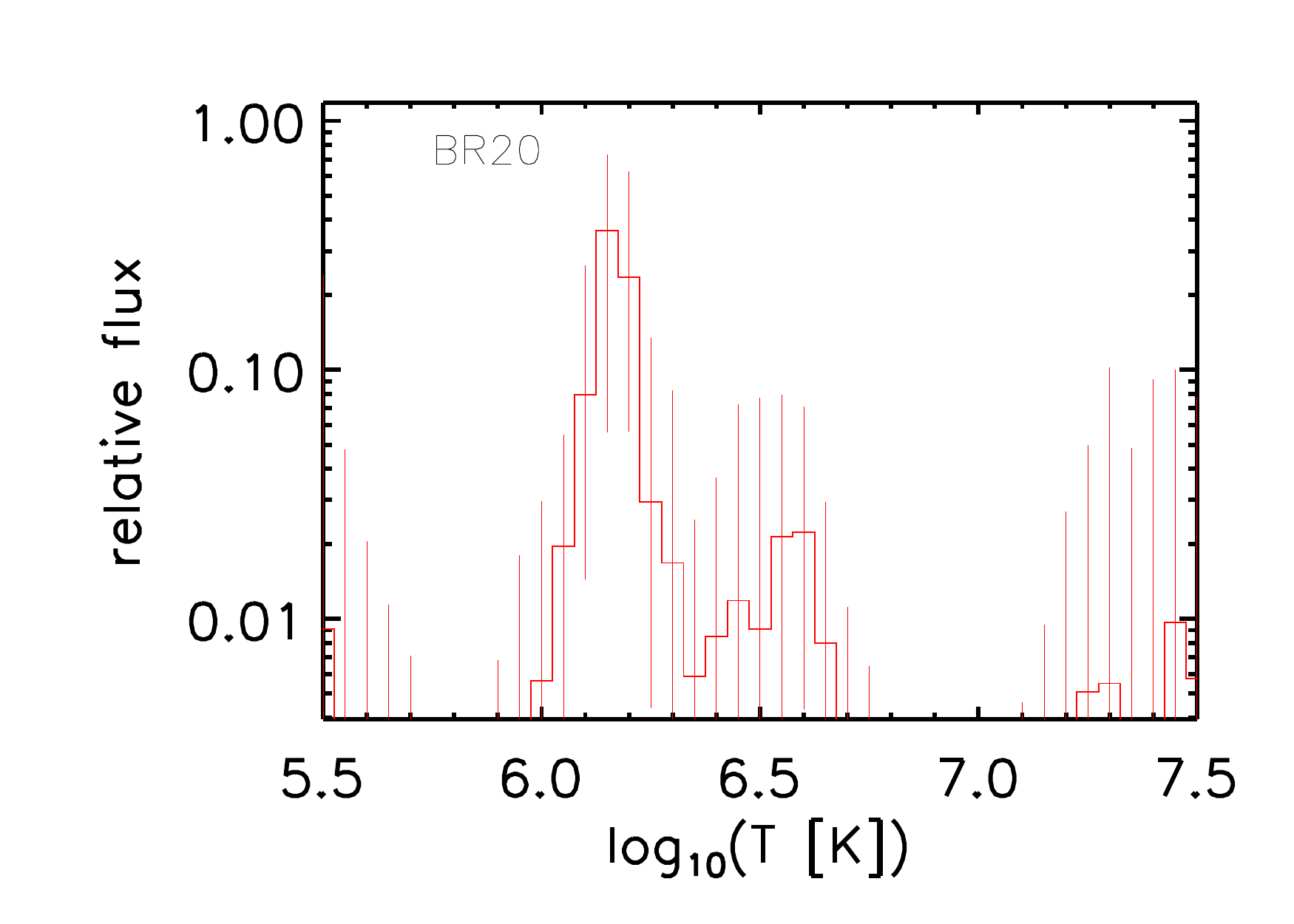}\\
\hspace{-0.57cm}
\includegraphics[trim=0cm 0cm 0cm 0cm, clip=true,width=0.25\textwidth]{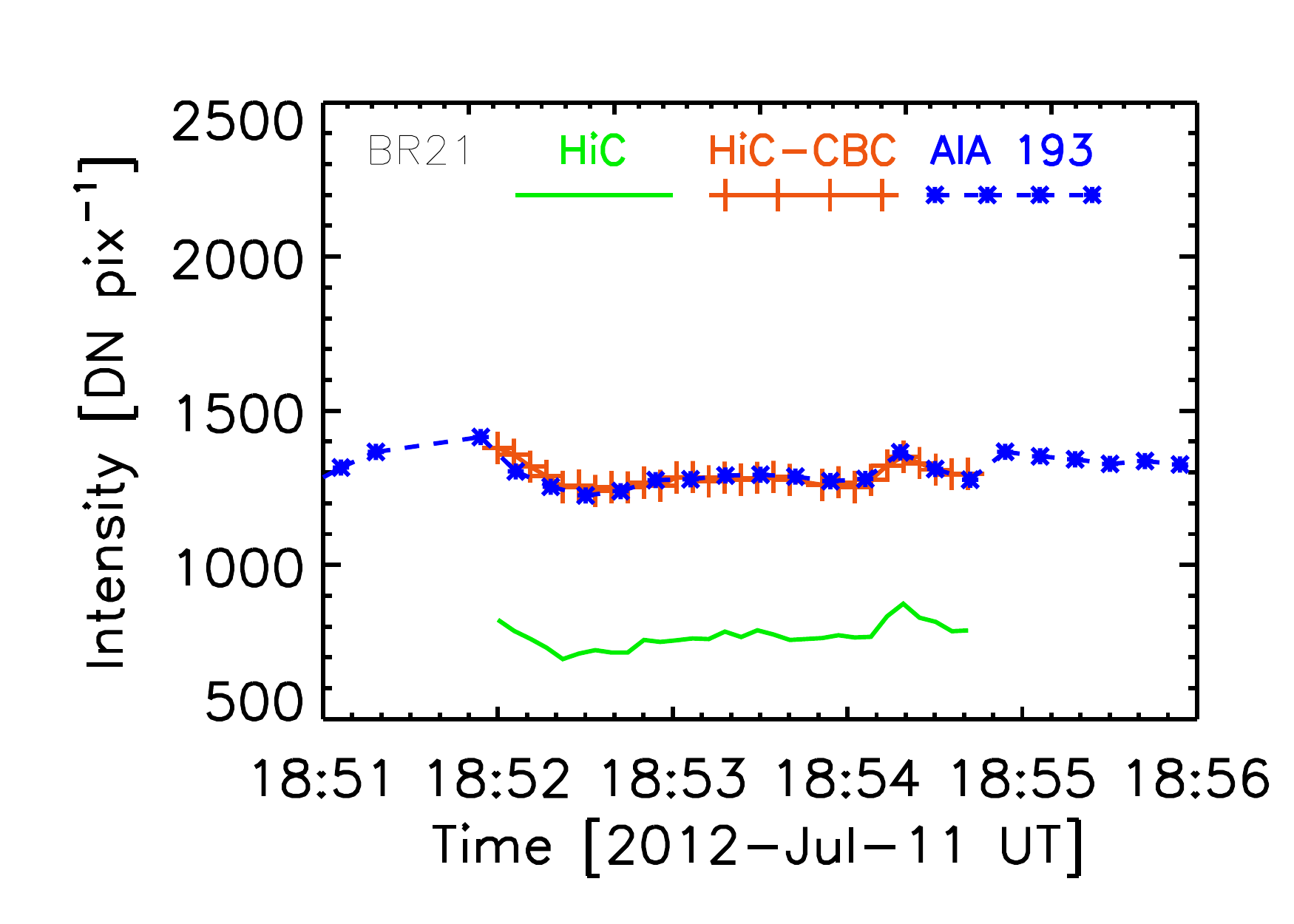}
\includegraphics[trim=0cm 0cm 0cm 0cm, clip=true,width=0.25\textwidth]{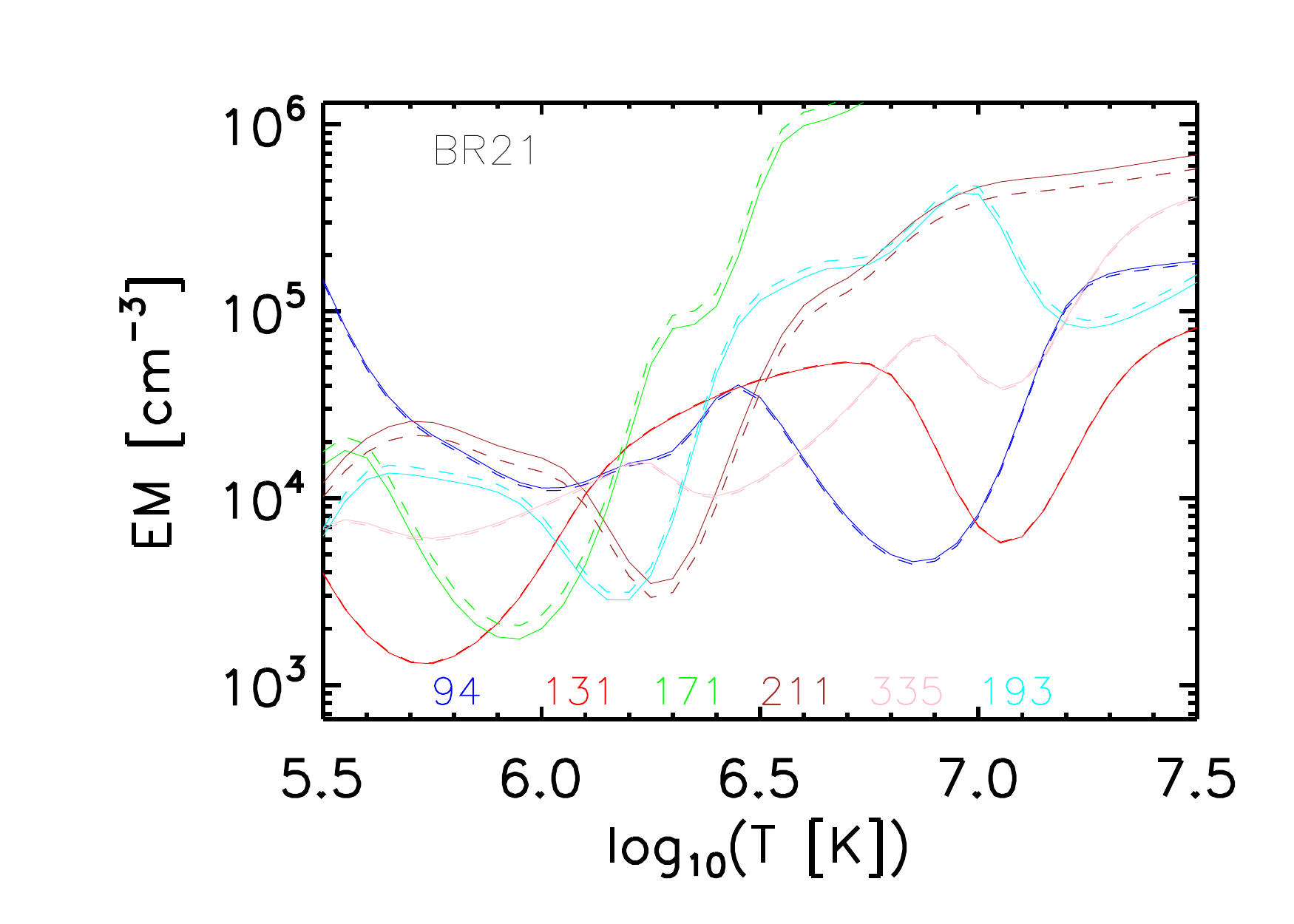}
\includegraphics[trim=0cm 0cm 0cm 0cm, clip=true,width=0.25\textwidth]{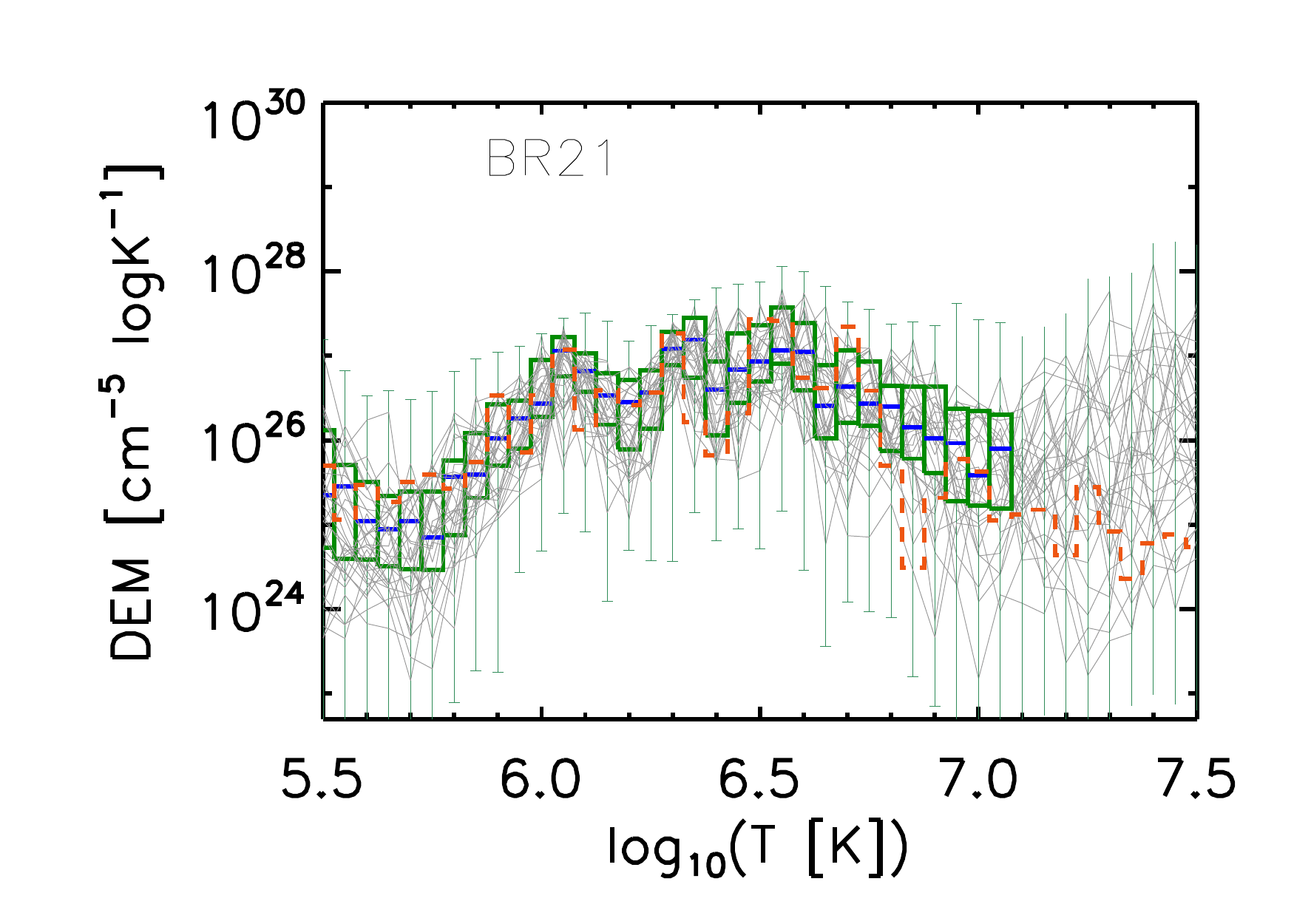}
\includegraphics[trim=0cm 0cm 0cm 0cm, clip=true,width=0.25\textwidth]{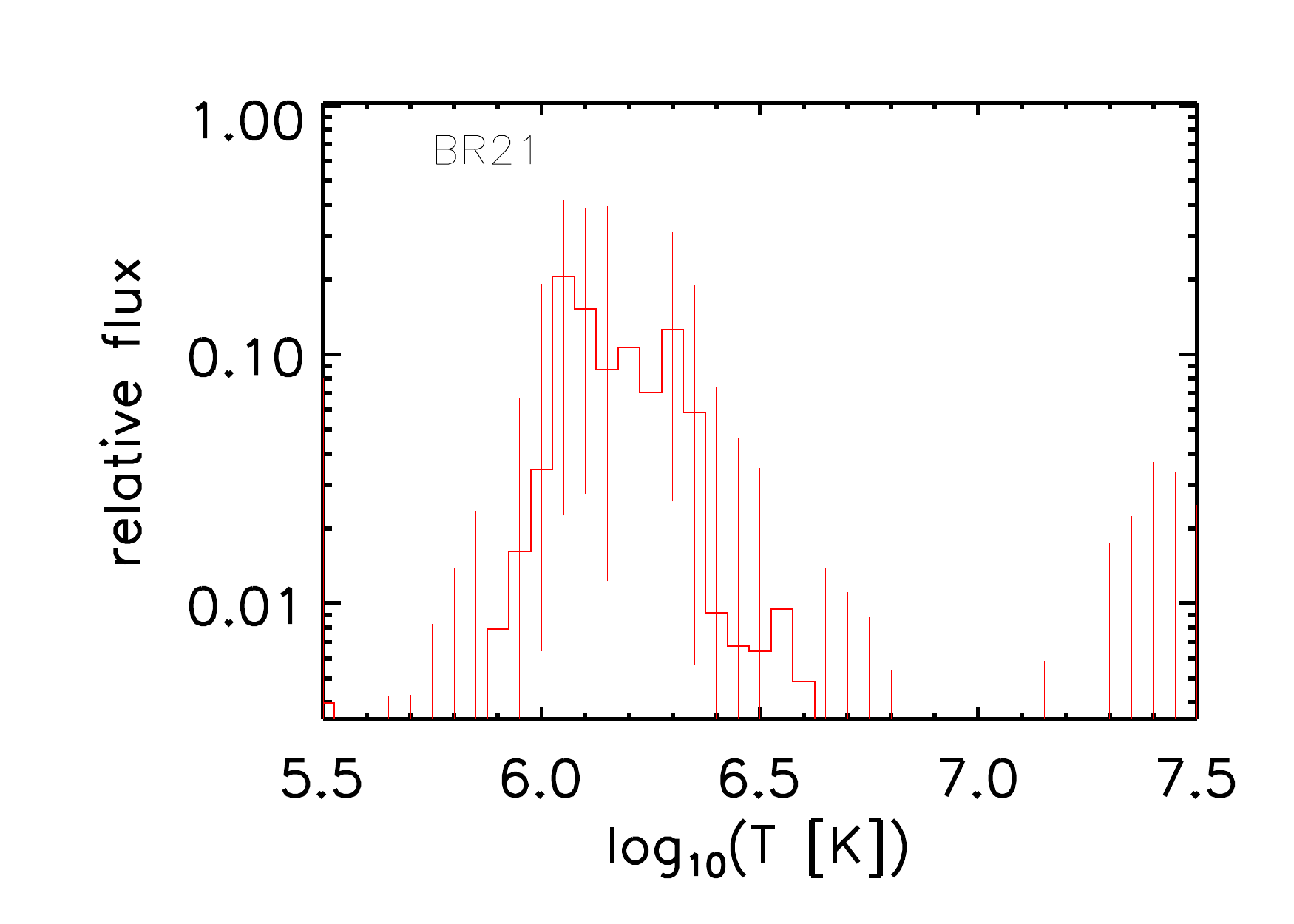}\\
\caption{From left to right: Light curves, EM loci curves, DEM reconstructions, and contributions of different temperatures to flux in AIA~193~\AA\ for brightenings, as in Figure~\ref{f:app_lc_DEM_1}.}
\label{f:app_lc_DEM_3}
\end{center}
\end{figure*}

\begin{figure*}[htp!]
\begin{center}
\hspace{-0.57cm}
\includegraphics[trim=0cm 0cm 0cm 0cm, clip=true,width=0.25\textwidth]{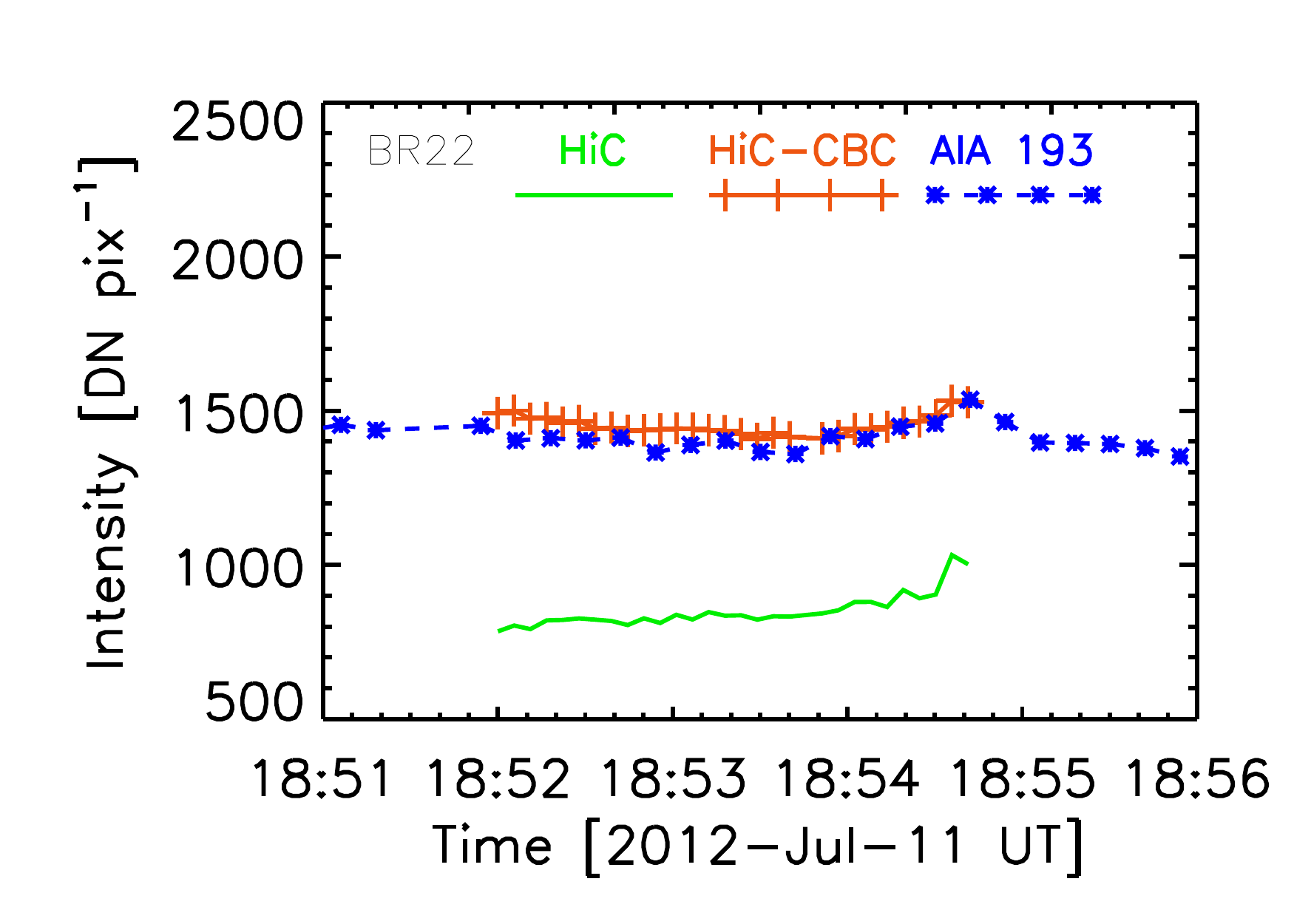}
\includegraphics[trim=0cm 0cm 0cm 0cm, clip=true,width=0.25\textwidth]{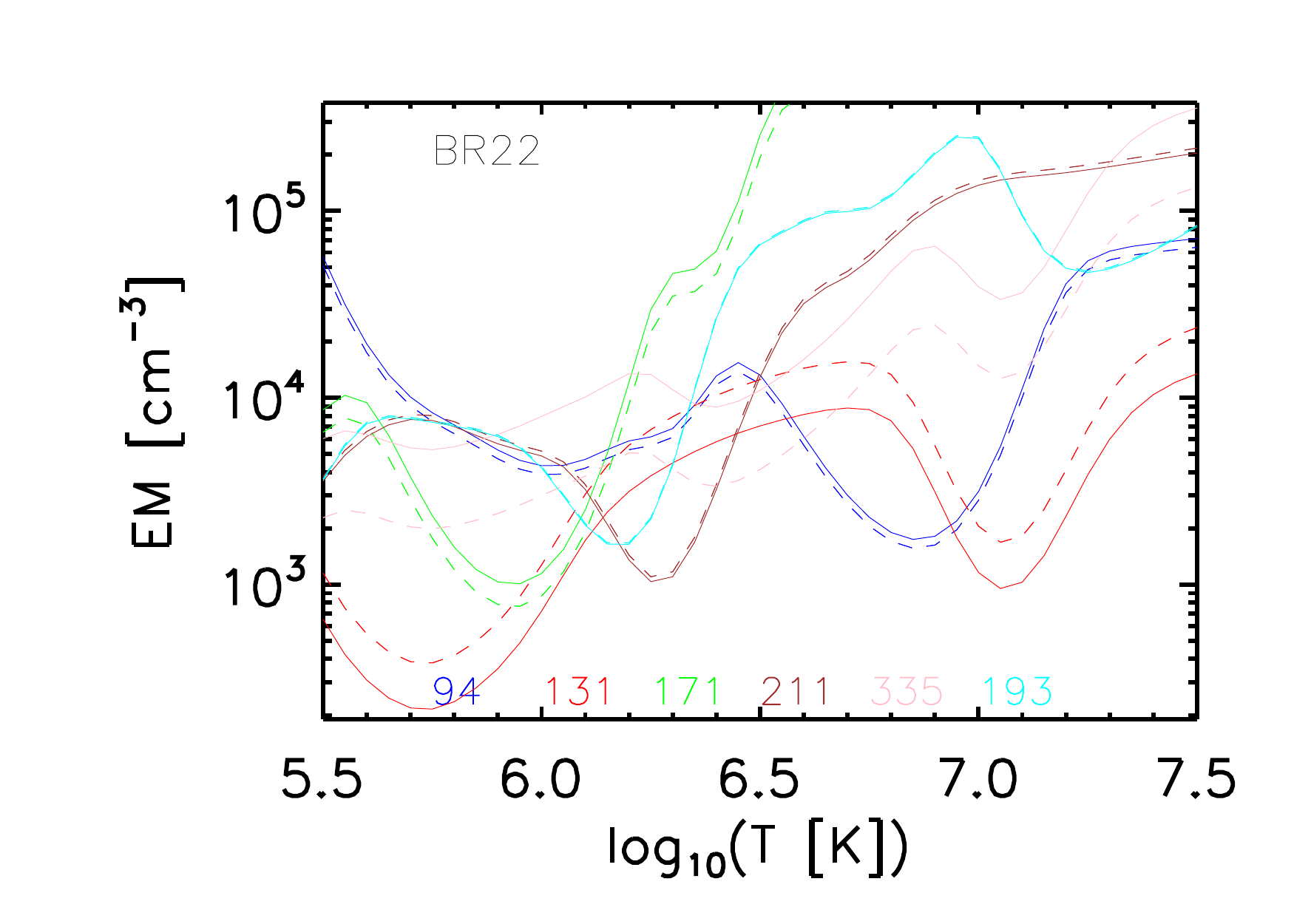}
\includegraphics[trim=0cm 0cm 0cm 0cm, clip=true,width=0.25\textwidth]{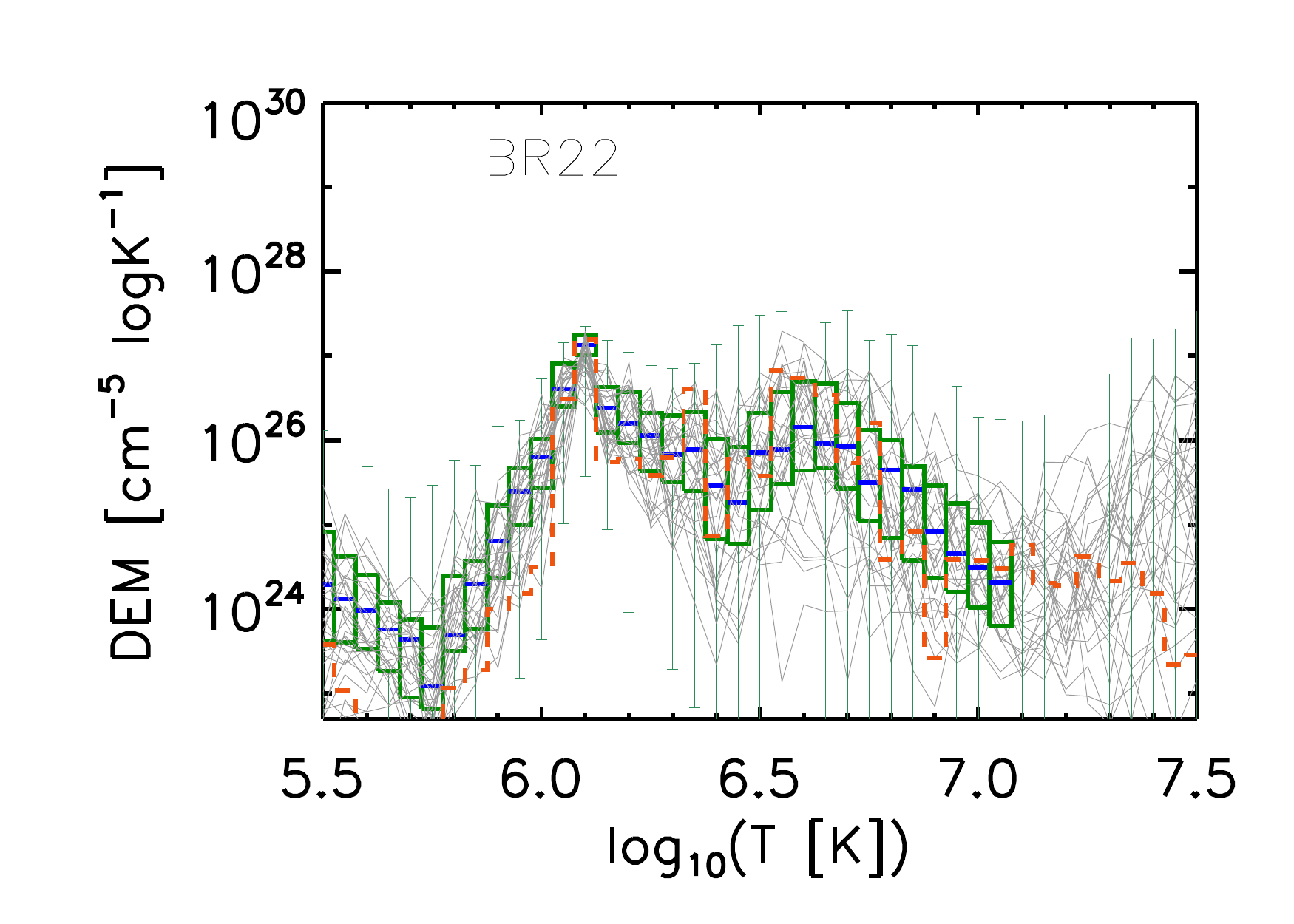}
\includegraphics[trim=0cm 0cm 0cm 0cm, clip=true,width=0.25\textwidth]{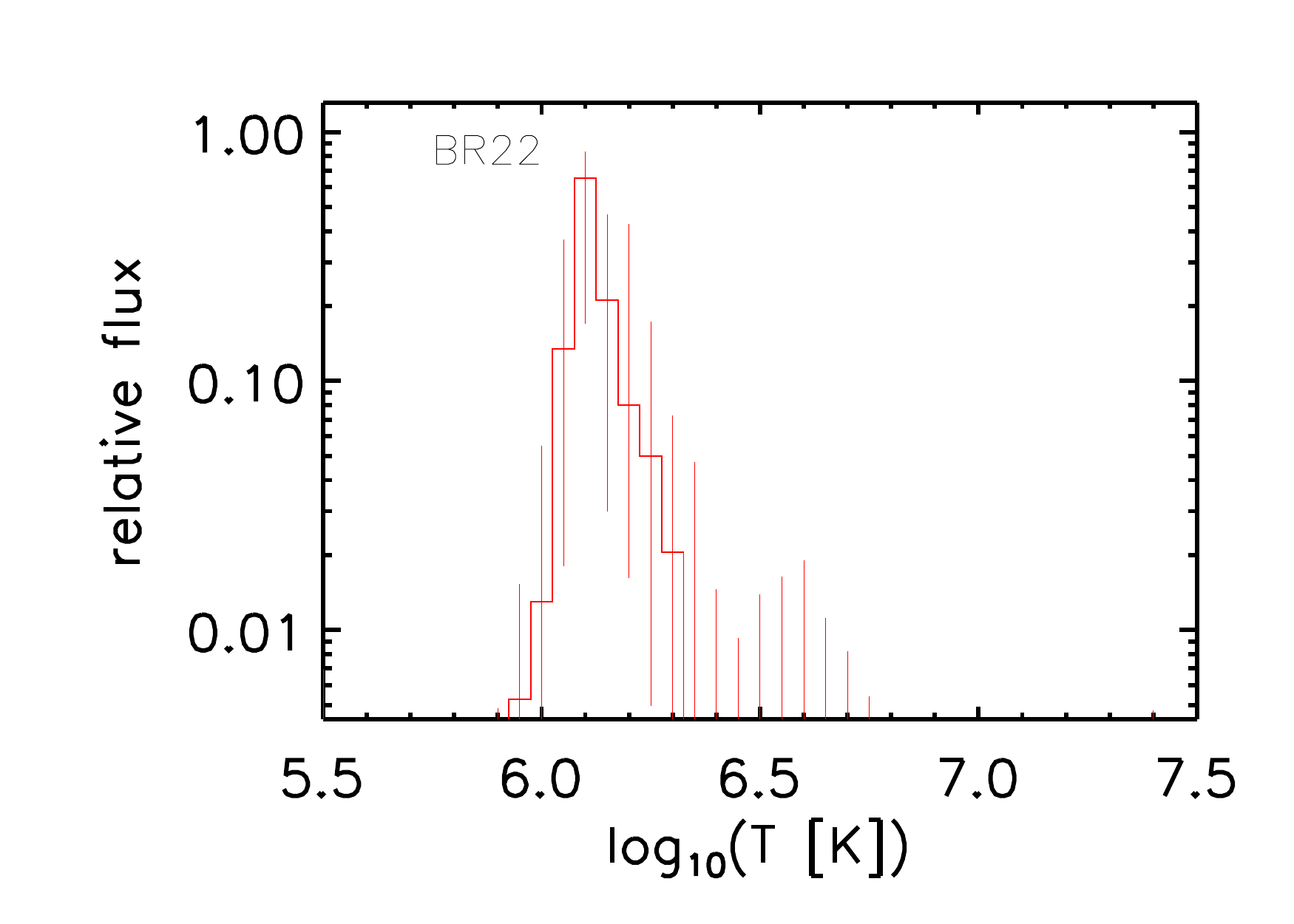}\\
\hspace{-0.57cm}
\includegraphics[trim=0cm 0cm 0cm 0cm, clip=true,width=0.25\textwidth]{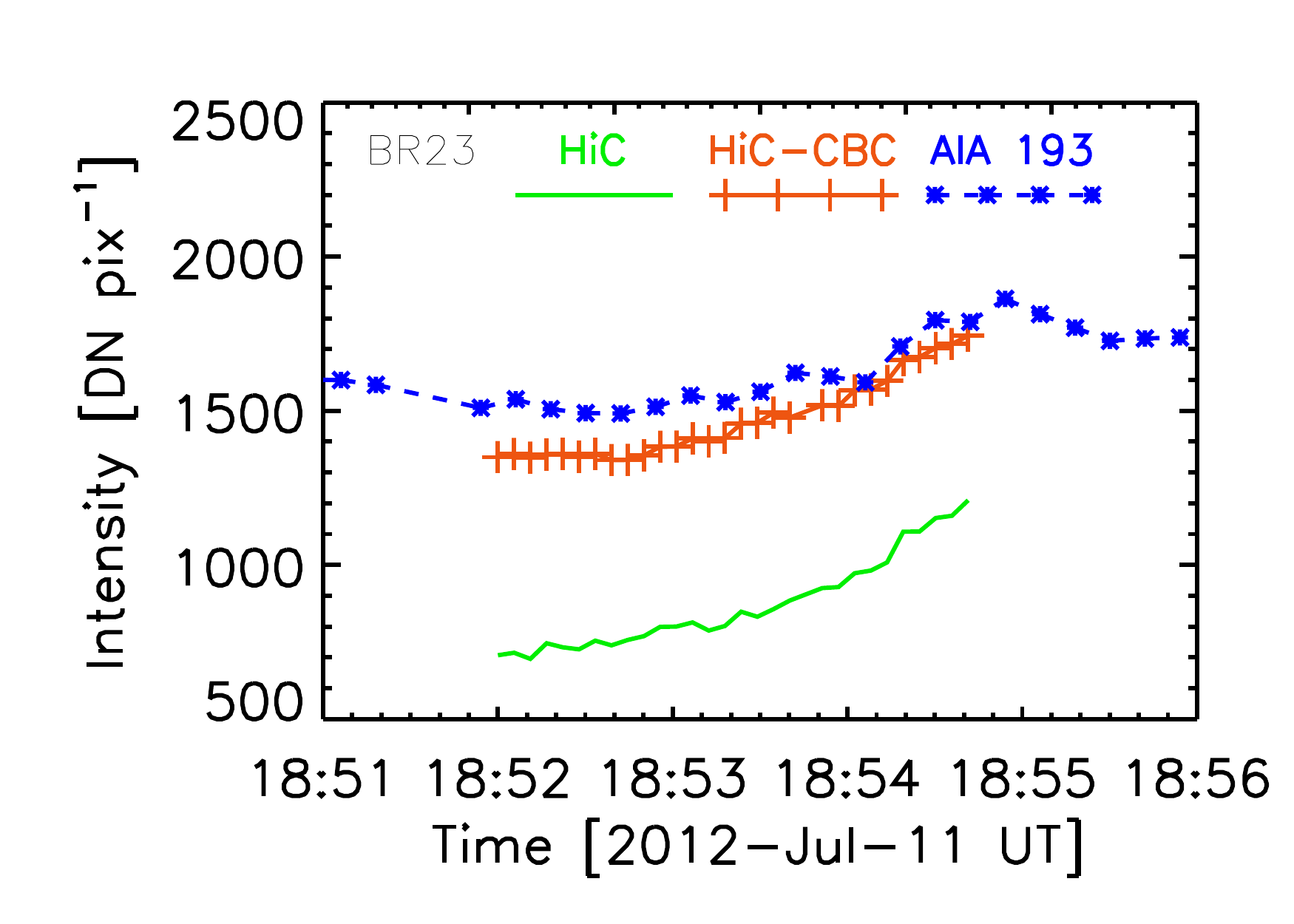}
\includegraphics[trim=0cm 0cm 0cm 0cm, clip=true,width=0.25\textwidth]{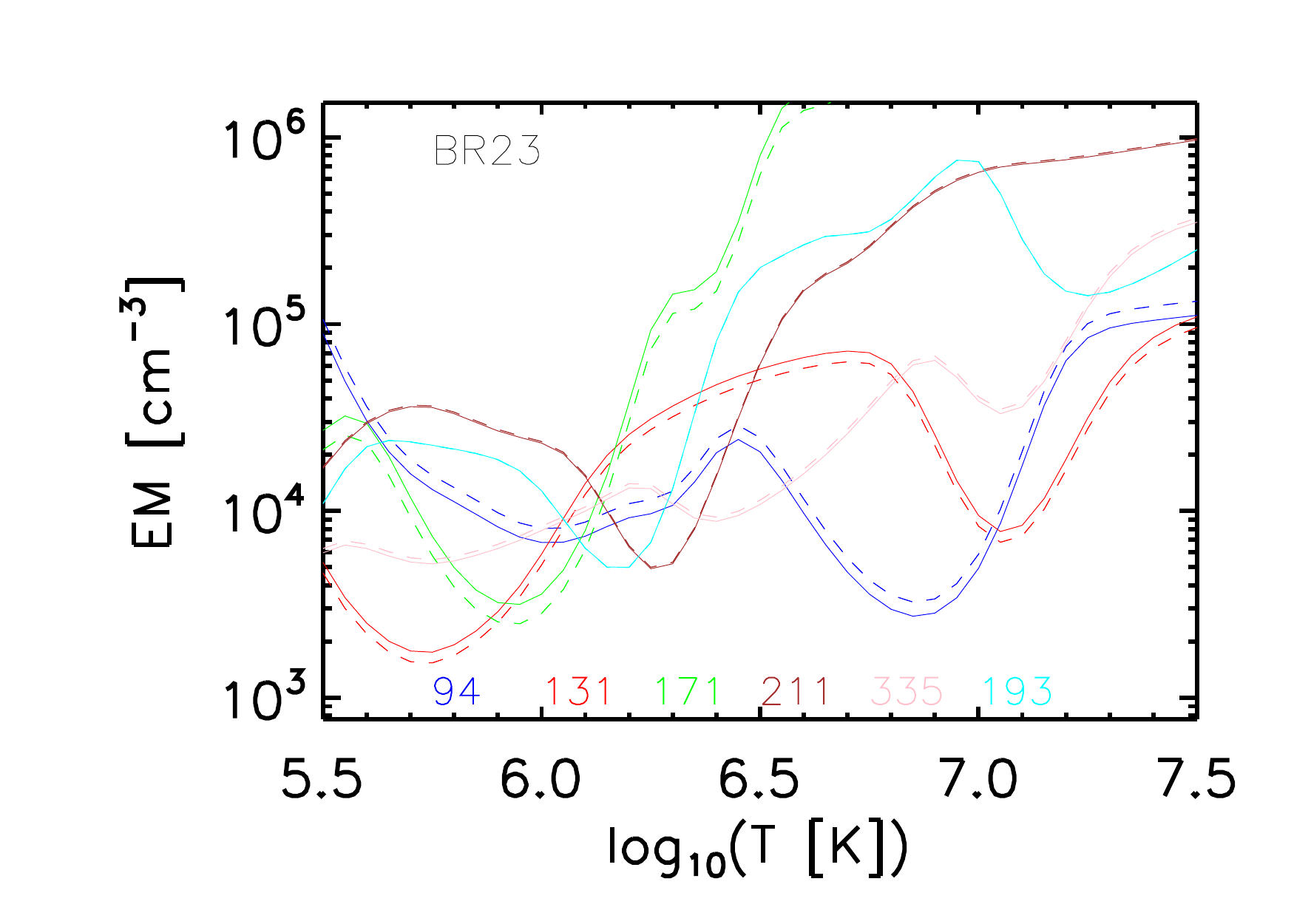}
\includegraphics[trim=0cm 0cm 0cm 0cm, clip=true,width=0.25\textwidth]{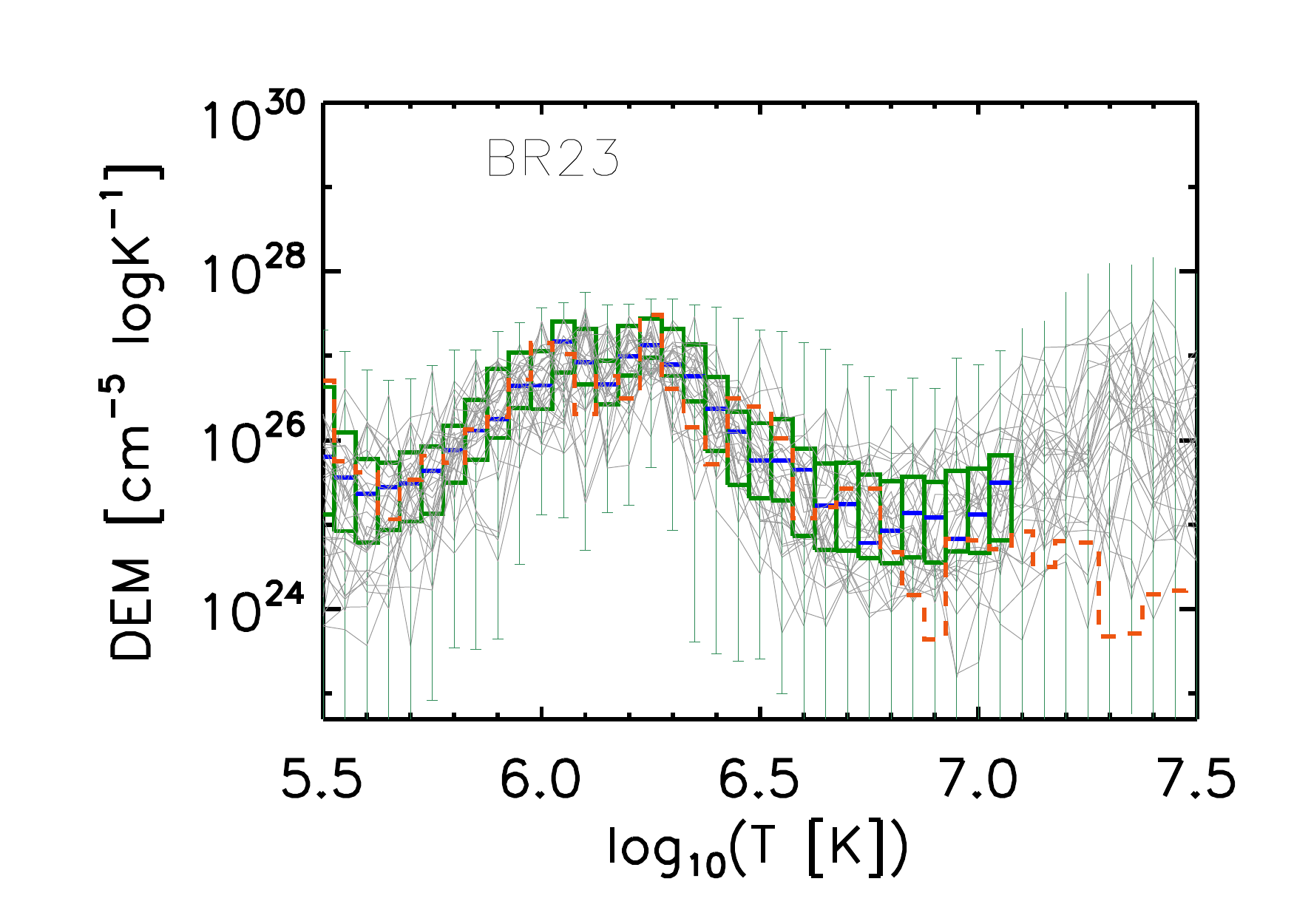}
\includegraphics[trim=0cm 0cm 0cm 0cm, clip=true,width=0.25\textwidth]{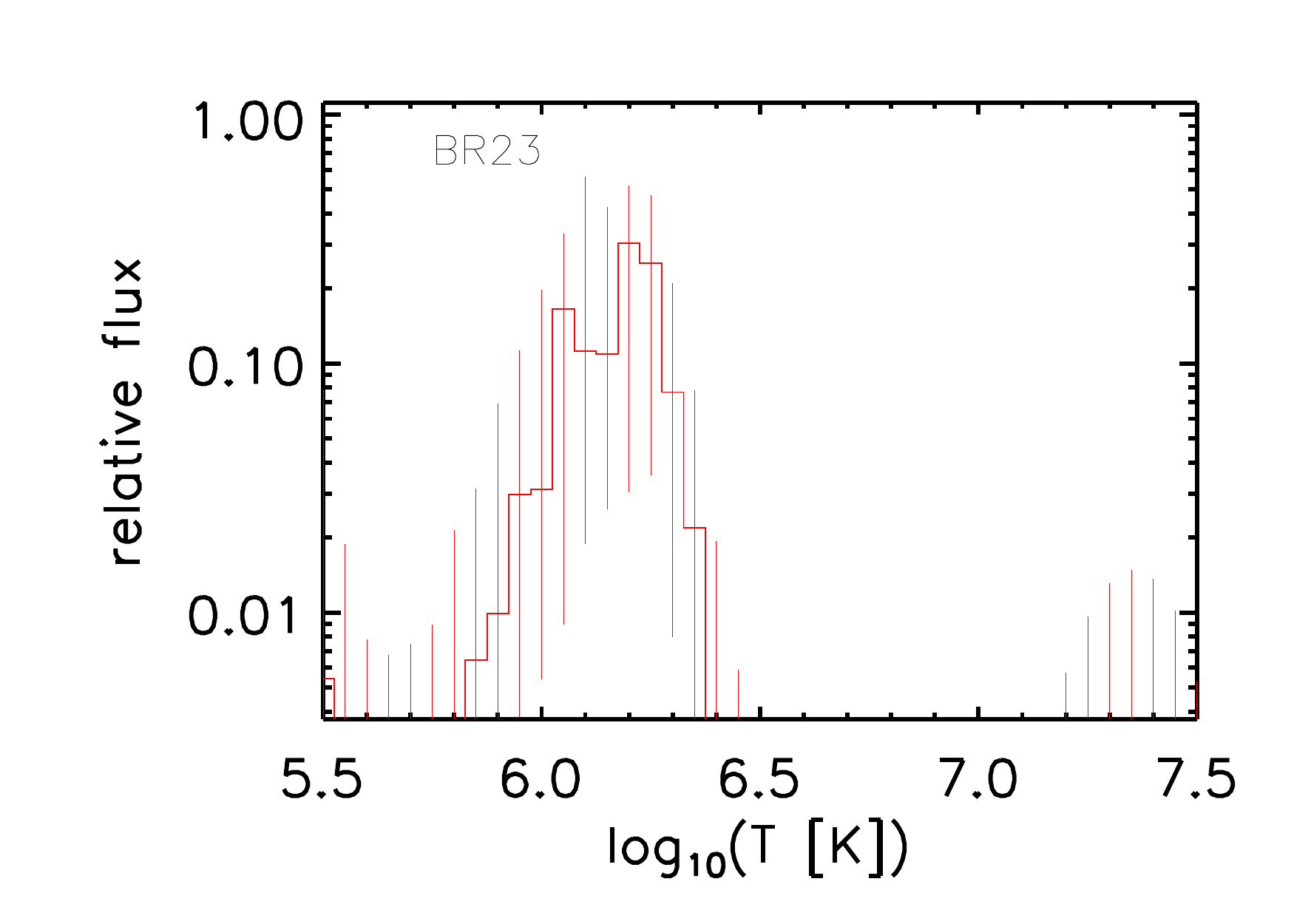}\\
\hspace{-0.57cm}
\includegraphics[trim=0cm 0cm 0cm 0cm, clip=true,width=0.25\textwidth]{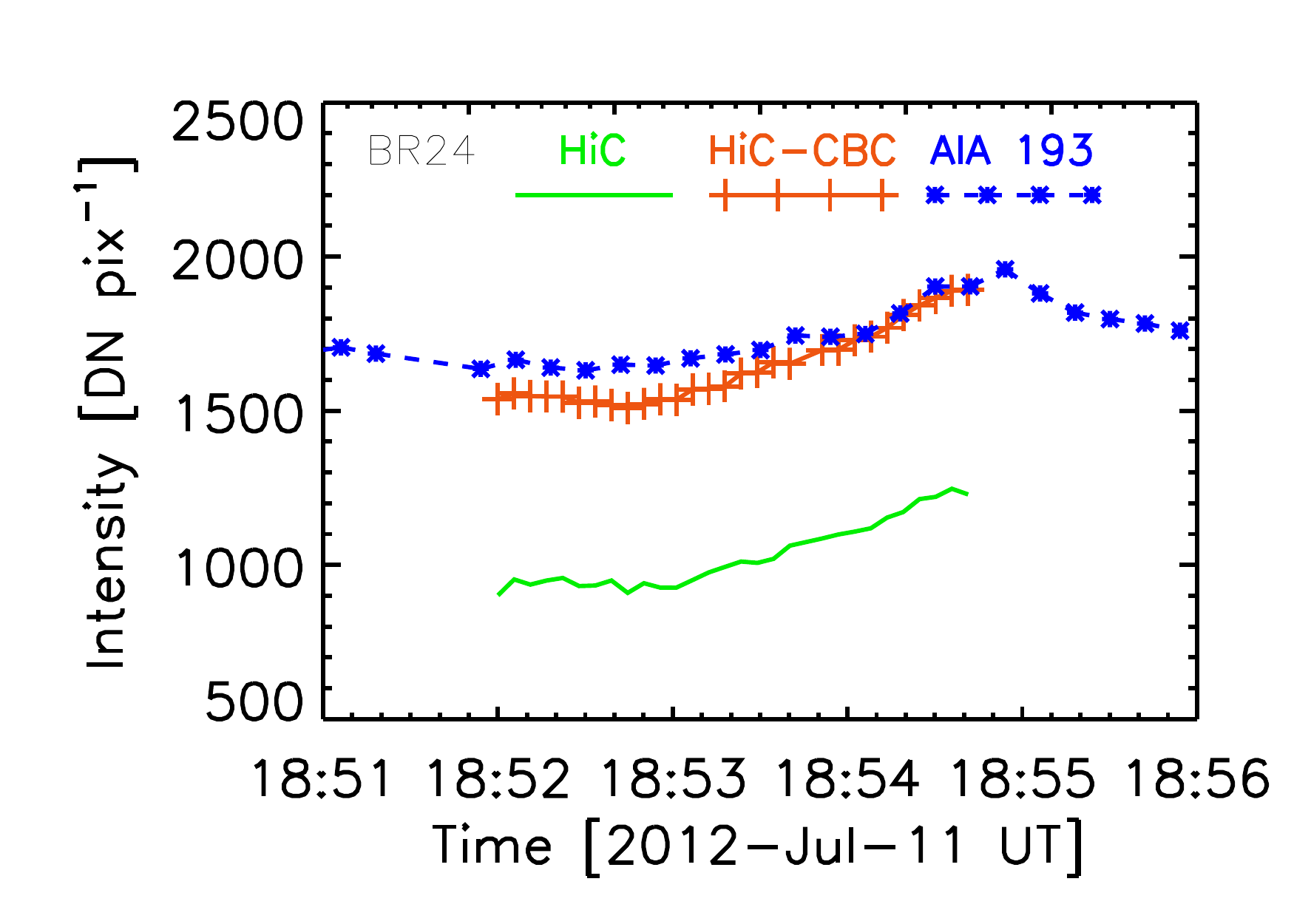}
\includegraphics[trim=0cm 0cm 0cm 0cm, clip=true,width=0.25\textwidth]{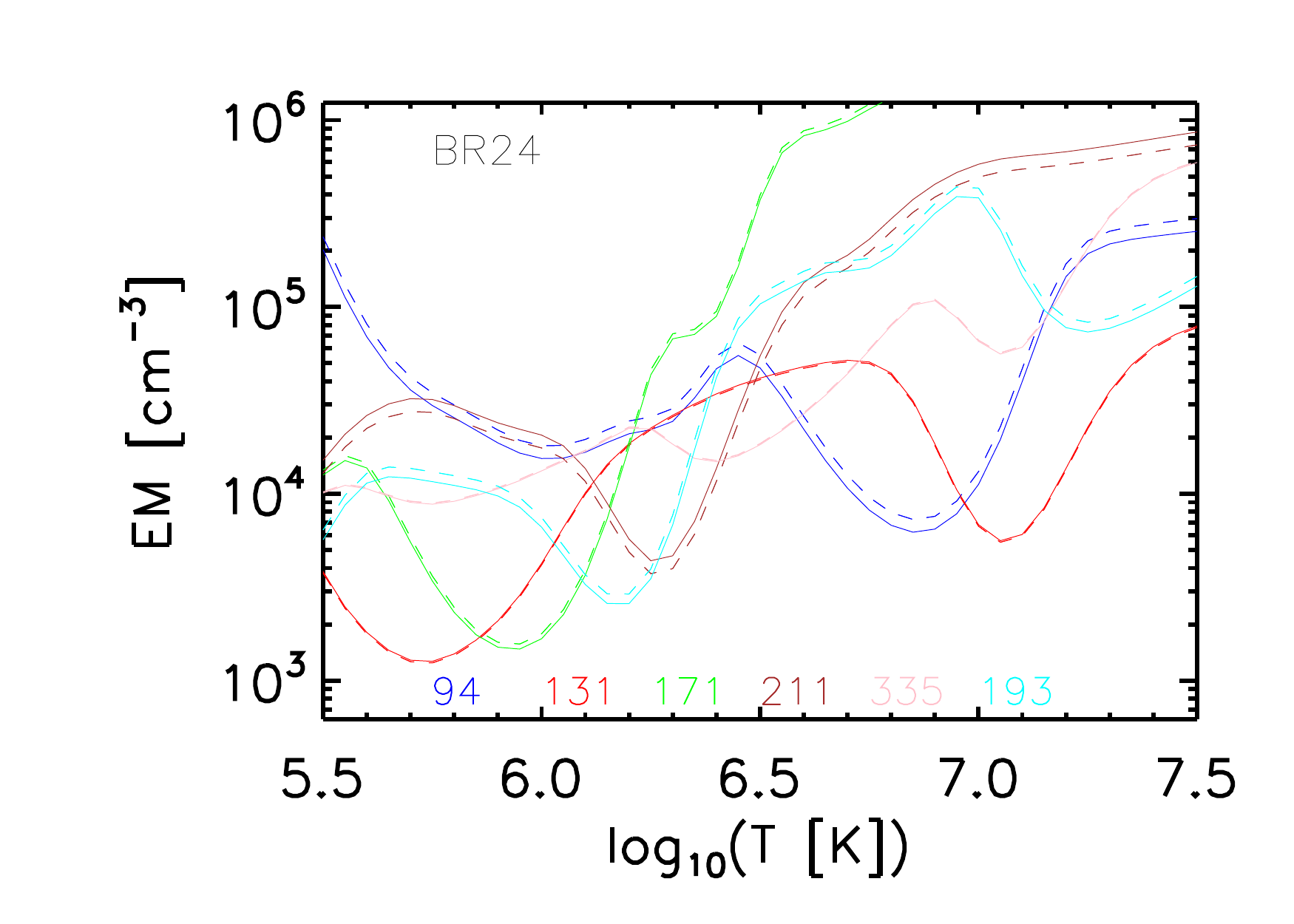}
\includegraphics[trim=0cm 0cm 0cm 0cm, clip=true,width=0.25\textwidth]{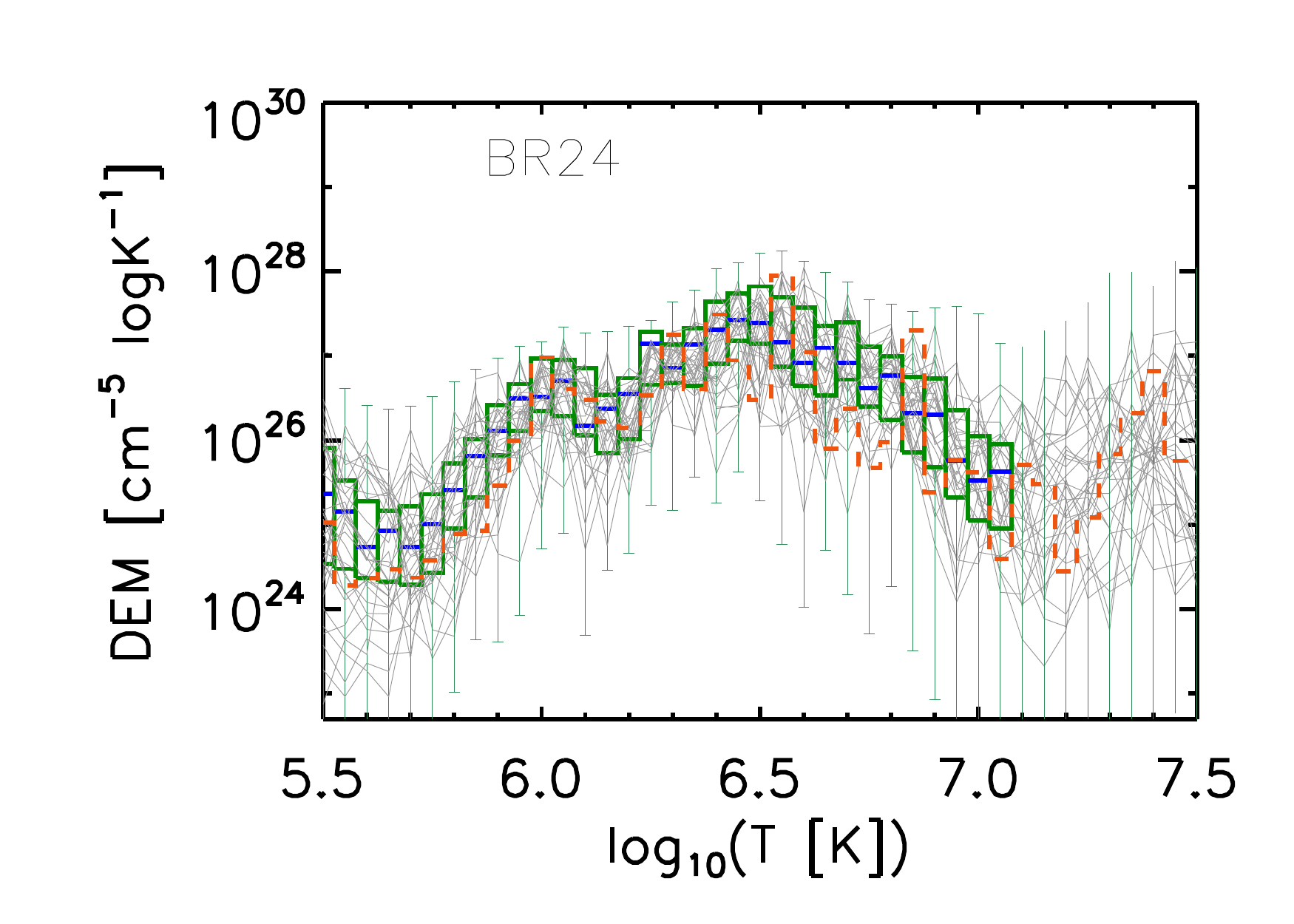}
\includegraphics[trim=0cm 0cm 0cm 0cm, clip=true,width=0.25\textwidth]{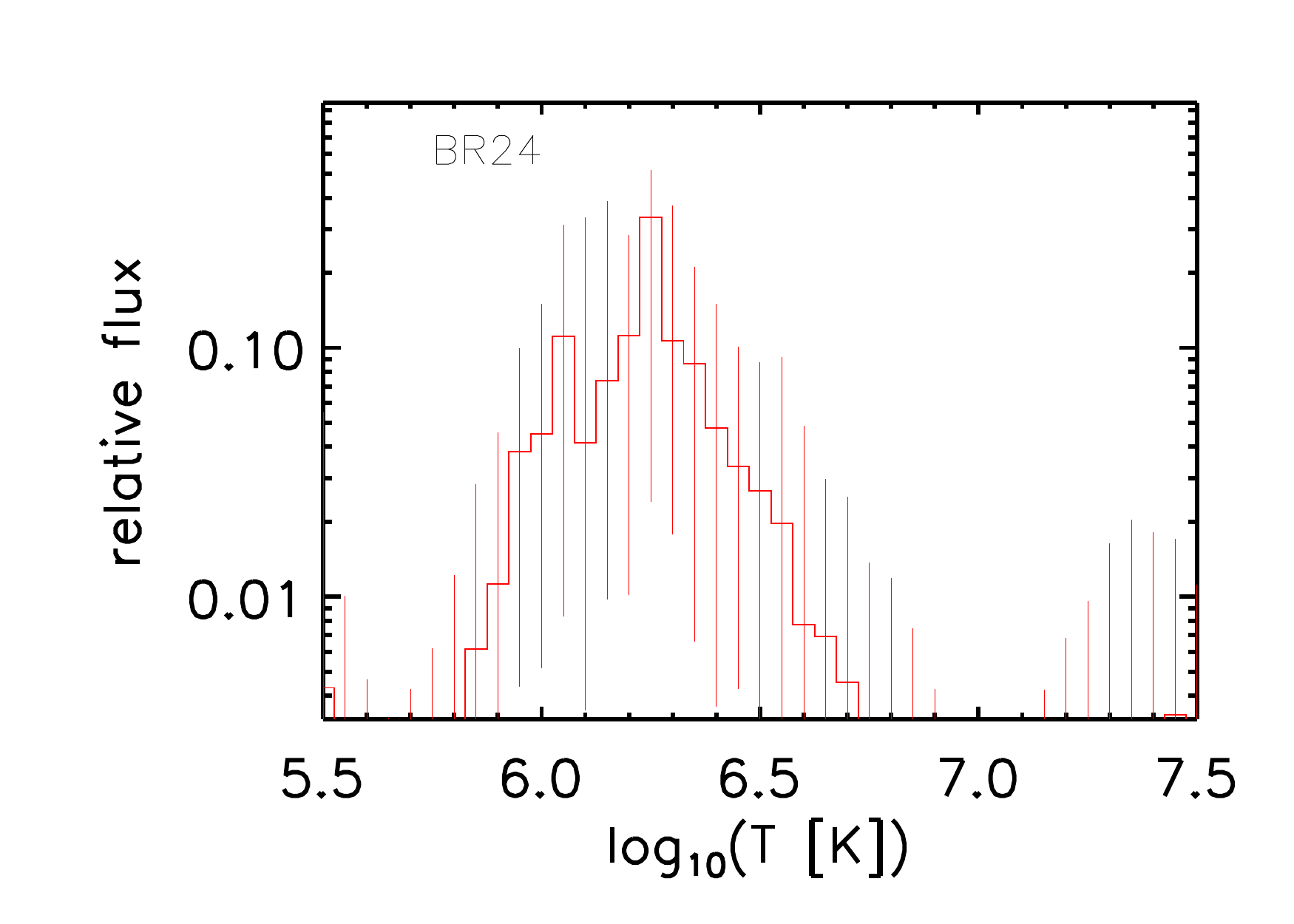}\\
\hspace{-0.57cm}
\includegraphics[trim=0cm 0cm 0cm 0cm, clip=true,width=0.25\textwidth]{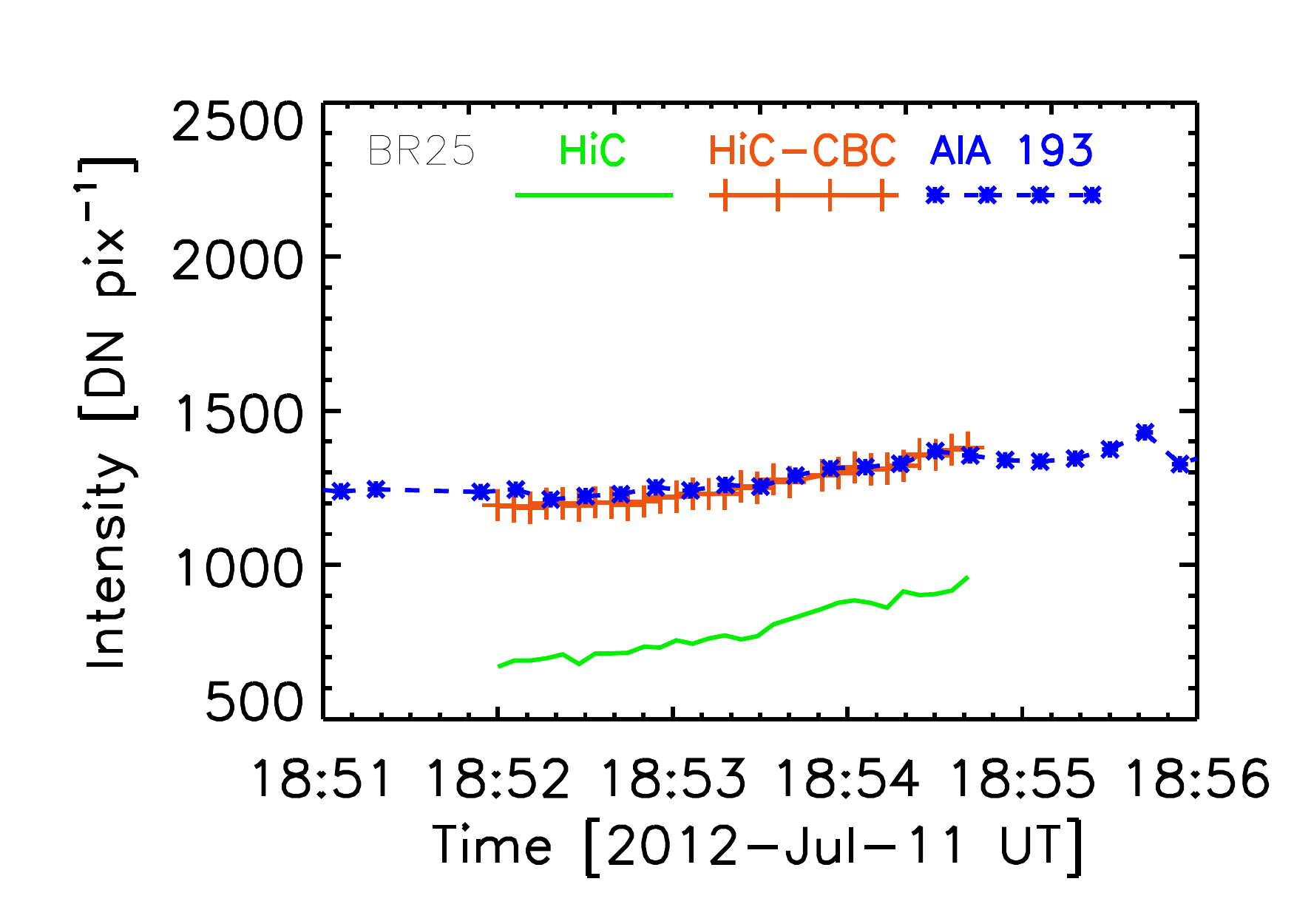}
\includegraphics[trim=0cm 0cm 0cm 0cm, clip=true,width=0.25\textwidth]{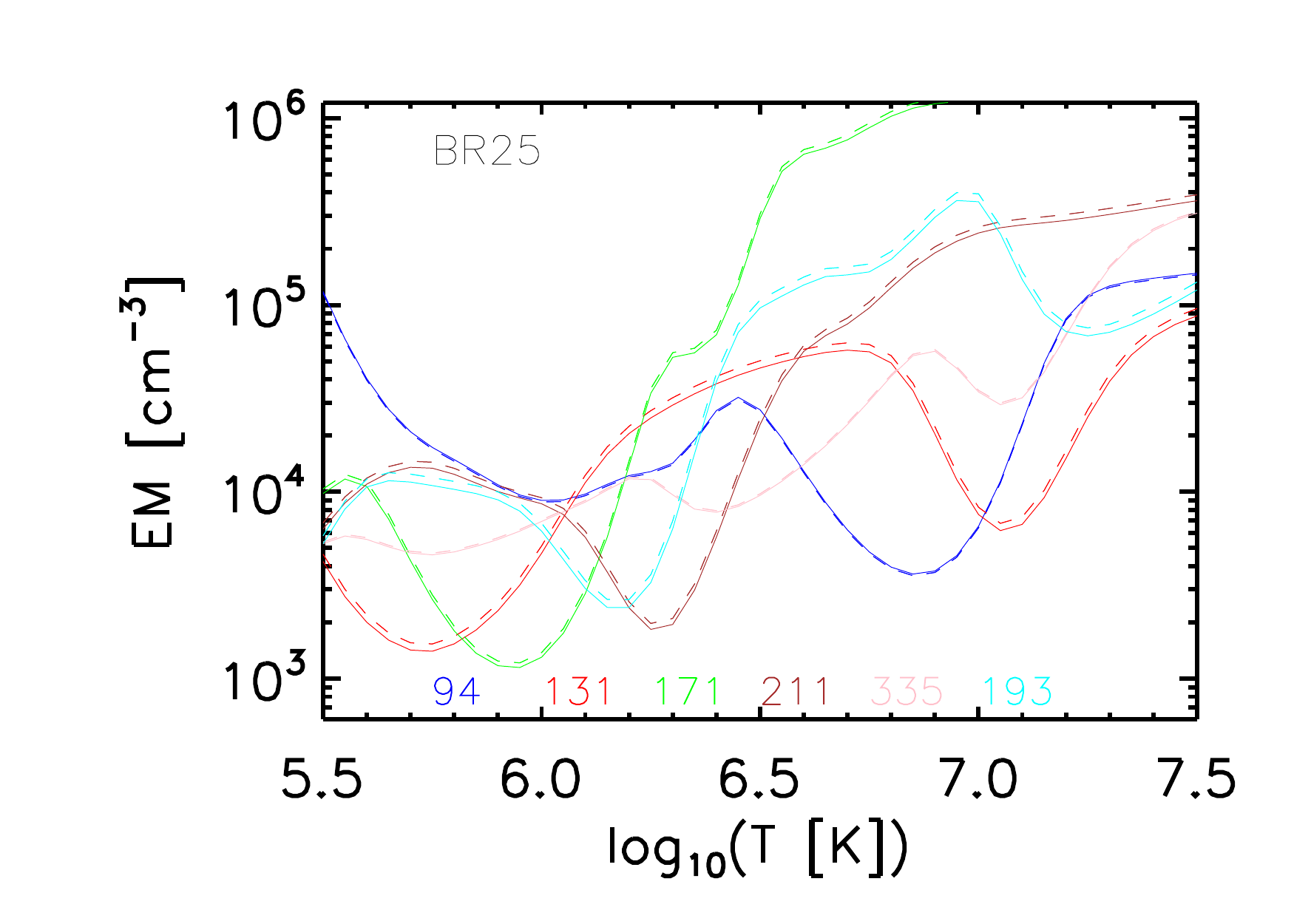}
\includegraphics[trim=0cm 0cm 0cm 0cm, clip=true,width=0.25\textwidth]{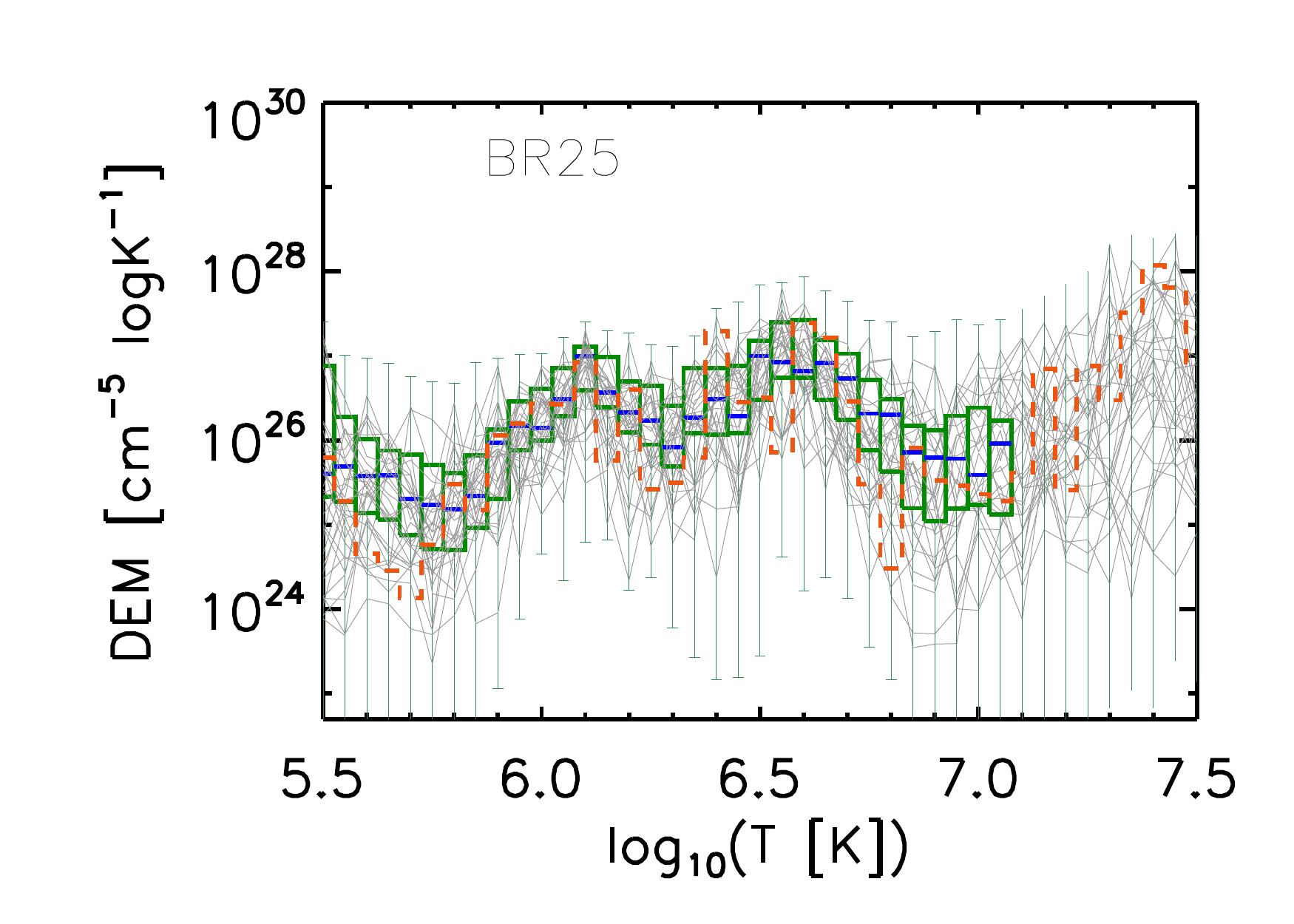}
\includegraphics[trim=0cm 0cm 0cm 0cm, clip=true,width=0.25\textwidth]{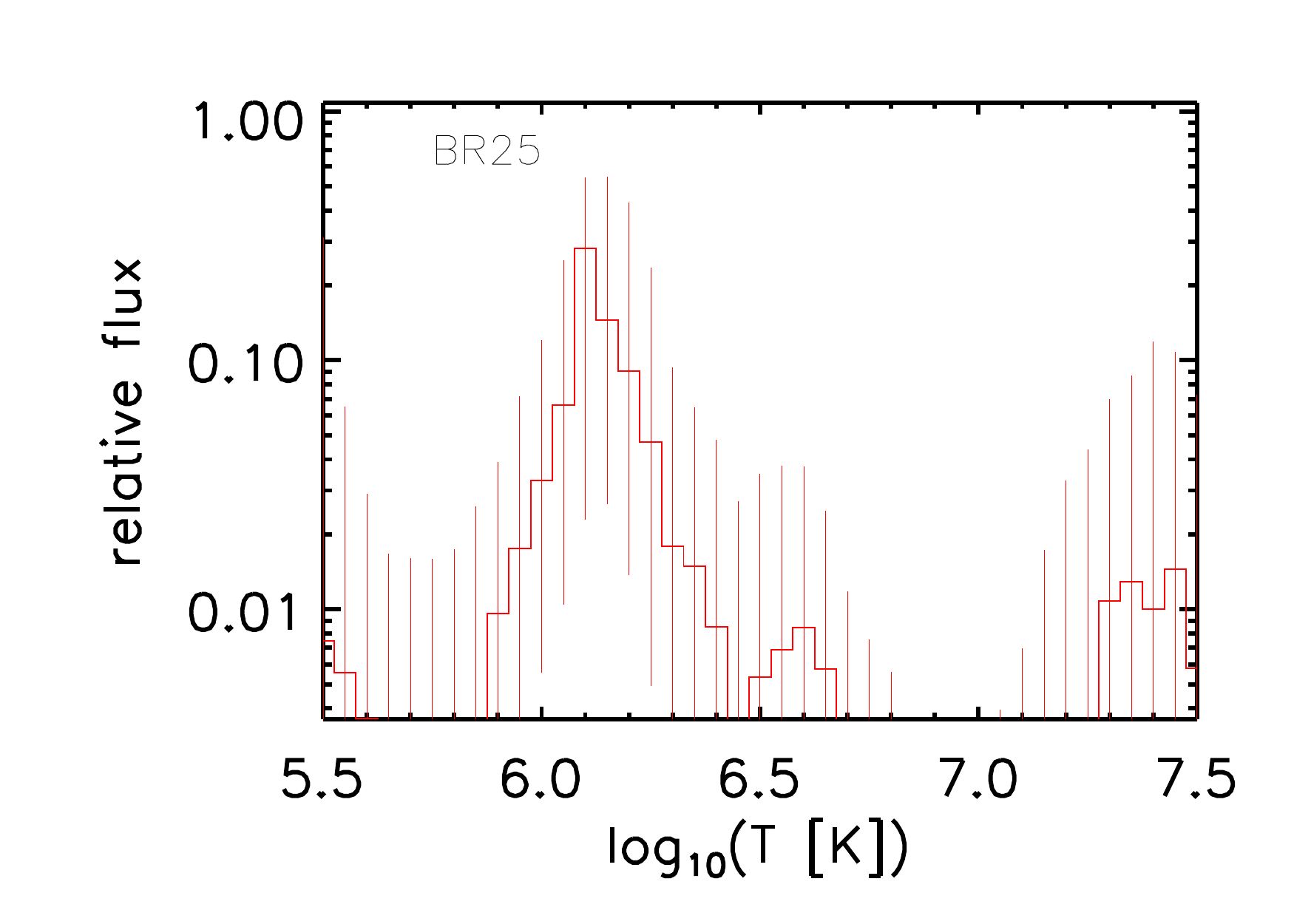}\\
\caption{From left to right: From left to right: Light curves, EM loci curves, DEM reconstructions, and contributions of different temperatures to flux in AIA~193~\AA\ for brightenings, as in Figure~\ref{f:app_lc_DEM_1}.}
\label{f:app_lc_DEM_4}
\end{center}
\end{figure*}


\end{appendix}

\end{document}